\documentclass{article}

\usepackage[usenames,dvipsnames]{xcolor}
\usepackage[T1]{fontenc}
\usepackage{stmaryrd}
\usepackage[english]{babel}
\usepackage[utf8]{inputenc}
\usepackage{amsmath,amssymb,amsfonts,mathrsfs,amsthm}
\usepackage{mathscinet} 
\usepackage{graphicx}
\usepackage{fullpage}
\usepackage{mathtools}
\usepackage{thm-restate}
\usepackage[linkcolor=black,colorlinks=true,citecolor=black,filecolor=black]{hyperref}
\usepackage[capitalise]{cleveref}
\usepackage{xspace}
\usepackage{float}
\usepackage{multirow}
\usepackage{tablefootnote}
\usepackage{comment}
\usepackage{xifthen}
\usepackage{authblk}
\usepackage{MnSymbol} 
\usepackage{ulem} 
\usepackage{makeidx}
\makeindex
\usepackage[totoc]{idxlayout}
\usepackage{complexity}

\usepackage[nottoc,numbib]{tocbibind}
\usepackage{authblk}

\usepackage{fancyhdr}

\let\temperindex\index
\renewcommand{\index}[1]{\temperindex{#1}}
\newcommand{\mindex}[1]{\temperindex{#1|textbf}}

\renewcommand{\A}{\ensuremath{\mathcal{A}}\xspace}
\renewcommand{\L}{\ensuremath{\mathcal{L}}\xspace}
\newcommand{\N}{\ensuremath{\mathbb{N}}\xspace}
\newcommand{\Q}{\ensuremath{\mathbb{Q}}\xspace}

\renewcommand{\C}{\ensuremath{\mathbb{C}}\xspace}
\renewcommand{\R}{\ensuremath{\mathbb{R}}\xspace}
\newcommand{\Z}{\ensuremath{\mathbb{Z}}\xspace}

\DeclarePairedDelimiter\norm{\lVert}{\rVert}

\renewcommand{\epsilon}{\ensuremath\varepsilon}

\renewcommand{\phi}{\ensuremath{\varphi}}

\renewcommand{\epsilon}{\ensuremath{\varepsilon}}

\renewcommand{\theta}{\ensuremath{\vartheta}}

\DeclareMathOperator{\sgn}{sgn}  
\DeclareMathOperator{\GF}{GF}   
\DeclareMathOperator{\BPZ}{\BP^0}   
\DeclareMathOperator{\conv}{conv}  

\newcommand{\EN}{\ensuremath{\exists\mathbb{N}}\xspace}

\newcommand{\EQ}{\ensuremath{\exists\mathbb{Q}}\xspace}
\newcommand{\ER}{\ensuremath{\exists\mathbb{R}}\xspace}
\newcommand{\succER}{\textrm{succ}\ensuremath{\exists\mathbb{R}}\xspace}
\newcommand{\EC}{\ensuremath{\exists\mathbb{C}}\xspace}
\newcommand{\VR}{\ensuremath{\forall\mathbb{R}}\xspace}
\newcommand{\VER}{\ensuremath{\forall\exists\mathbb{R}}\xspace}
\newcommand{\EVR}{\ensuremath{\exists\forall\mathbb{R}}\xspace}
\newcommand{\EVER}{\ensuremath{\exists\forall\exists\mathbb{R}}\xspace}

\newcommand{\DNP}{\textsf{DNP}\xspace}
\newcommand{\FIXP}{\textsf{FIXP}\xspace}

\newcommand{\HQ}{\textup{\textsf{H}}}

\newcommand{\problemname}[1]{\textnormal{\textsc{#1}}\xspace}
\newcommand{\ETR}{\problemname{ETR}}                         
\newcommand{\ETRINV}{\problemname{ETR-INV}}                  
\newcommand{\PETRINV}{\problemname{PLANAR-ETR-INV}}          
\newcommand{\RGETRINV}{\problemname{RANGE-ETR-INV}}          
\newcommand{\SINEQ}{\problemname{STRICT INEQ}}               
\newcommand{\SGE}{\problemname{SGE}}                         
\newcommand{\SEG}{\problemname{SEG}}                         
     
\newcommand{\HTN}{\problemname{H\ensuremath{_2}N}}           
\newcommand{\ISO}{\problemname{ISO}}                      
\newcommand{\SSQR}{\problemname{SSQR}}
\newcommand{\PosSLP}{\problemname{PosSLP}}  
\newcommand{\ReLU}{\problemname{ReLU}}  
\newcommand{\USSR}{\problemname{USSR}}                        
\newcommand{\ETSP}{\problemname{ETSP}}                      

\newcommand{\wordRAM}{\textnormal{word RAM}\xspace}
\newcommand{\realRAM}{\textnormal{real RAM}\xspace}
\newcommand{\trace}{\operatorname{trace}}

\newtheorem{theorem}{Theorem}[section]
\newtheorem{lemma}[theorem]{Lemma}

\newtheorem*{conjecture*}{Conjecture}

\theoremstyle{remark}
\newtheorem*{remark}{Remark}

\newcounter{examples}[section]

\def\sep{\ensuremath{\blacktriangledown}}
\newcommand{\ourref}[2][]{\ifthenelse{\isempty{#1}}{\cref{#2}}{\cref{#2}\index{#1}}}

\newcounter{problemActr}
\crefformat{problemActr}{#2{(A#1)}#3}
\newenvironment{problemA}[1]
{    \refstepcounter{problemActr}%
    \begin{list}{}%
    {%
    \setlength{\itemsep}{-3pt}}%
    \item[{\bf (A\arabic{problemActr}) #1}]%
    \nopagebreak}{\end{list}}

\newcounter{problemAOctr}
\crefformat{problemAOctr}{#2{(A-Open#1)}#3}
\newenvironment{problemAO}[1]
{    \refstepcounter{problemAOctr}%
    \begin{list}{}%
    {%
    \setlength{\itemsep}{-3pt}}%
    \item[{\bf (A-Open\arabic{problemAOctr}) #1}]%
    \nopagebreak}{\end{list}}

\newcounter{problemLctr}
\crefformat{problemLctr}{#2{(L#1)}#3}
\newenvironment{problemL}[1]
{    \refstepcounter{problemLctr}%
    \begin{list}{}%
    {%
    \setlength{\itemsep}{-3pt}}%
    \item[{\bf (L\arabic{problemLctr}) #1}]%
    \nopagebreak}{\end{list}}

\newcounter{problemLOctr}
\crefformat{problemLOctr}{#2{(L-Open#1)}#3}
\newenvironment{problemLO}[1]
{    \refstepcounter{problemLOctr}%
    \begin{list}{}%
    {%
    \setlength{\itemsep}{-3pt}}%
    \item[{\bf (L-Open\arabic{problemLOctr}) #1}]%
    \nopagebreak}{\end{list}}

\newcounter{problemCGctr}
\crefformat{problemCGctr}{#2{(CG#1)}#3}
\newenvironment{problemCG}[1]
{    \refstepcounter{problemCGctr}%
    \begin{list}{}%
    {%
    \setlength{\itemsep}{-3pt}}%
    \item[{\bf (CG\arabic{problemCGctr}) #1}]%
    \nopagebreak}{\end{list}}

\newcounter{problemCGOctr}
\crefformat{problemCGOctr}{#2{(CG-Open#1)}#3}
\newenvironment{problemCGO}[1]
{    \refstepcounter{problemCGOctr}%
    \begin{list}{}%
    {%
    \setlength{\itemsep}{-3pt}}%
    \item[{\bf (CG-Open\arabic{problemCGOctr}) #1}]%
    \nopagebreak}{\end{list}}
    
\newcounter{problemGTctr}
\crefformat{problemGTctr}{#2{(GT#1)}#3}
\newenvironment{problemGT}[1]
{    \refstepcounter{problemGTctr}%
    \begin{list}{}%
    {%
    \setlength{\itemsep}{-3pt}}%
    \item[{\bf (GT\arabic{problemGTctr}) #1}]%
    \nopagebreak}{\end{list}}

\newcounter{problemGTOctr}
\crefformat{problemGTOctr}{#2{(GT-Open#1)}#3}
\newenvironment{problemGTO}[1]
{    \refstepcounter{problemGTOctr}%
    \begin{list}{}%
    {%
    \setlength{\itemsep}{-3pt}}%
    \item[{\bf (GT-Open\arabic{problemGTOctr}) #1}]%
    \nopagebreak}{\end{list}}

\newcounter{problemMLctr}
\crefformat{problemMLctr}{#2{(ML#1)}#3}
\newenvironment{problemML}[1]
{    \refstepcounter{problemMLctr}%
    \begin{list}{}%
    {%
    \setlength{\itemsep}{-3pt}}%
    \item[{\bf (ML\arabic{problemMLctr}) #1}]%
    \nopagebreak}{\end{list}}

\newcounter{problemMLOctr}
\crefformat{problemMLOctr}{#2{(ML-Open#1)}#3}
\newenvironment{problemMLO}[1]
{    \refstepcounter{problemMLOctr}%
    \begin{list}{}%
    {%
    \setlength{\itemsep}{-3pt}}%
    \item[{\bf (ML-Open\arabic{problemMLOctr}) #1}]%
    \nopagebreak}{\end{list}}

\newcounter{problemMDPctr}
\crefformat{problemMDPctr}{#2{(MDP#1)}#3}
\newenvironment{problemMDP}[1]
{    \refstepcounter{problemMDPctr}%
    \begin{list}{}%
    {%
    \setlength{\itemsep}{-3pt}}%
    \item[{\bf (MDP\arabic{problemMDPctr}) #1}]%
    \nopagebreak}{\end{list}}

\newcounter{problemMDPOctr}
\crefformat{problemMDPOctr}{#2{(MDP-Open#1)}#3}
\newenvironment{problemMDPO}[1]
{    \refstepcounter{problemMDPOctr}%
    \begin{list}{}%
    {%
    \setlength{\itemsep}{-3pt}}%
    \item[{\bf (MDP-Open\arabic{problemMDPOctr}) #1}]%
    \nopagebreak}{\end{list}}

\newcounter{problemMiscctr}
\crefformat{problemMiscctr}{#2{(Misc#1)}#3}
\newenvironment{problemMisc}[1]
{    \refstepcounter{problemMiscctr}%
    \begin{list}{}%
    {%
    \setlength{\itemsep}{-3pt}}%
    \item[{\bf (Misc\arabic{problemMiscctr}) #1}]%
    \nopagebreak}{\end{list}}

\newcounter{problemMiscOctr}
\crefformat{problemMiscOctr}{#2{(Misc-Open#1)}#3}

\newcommand{\given}{\item[Given:]}

\newcommand{\question}{\item[Question:]}
\newcommand{\complexity}{\item[Complexity:]}

\newcommand{\comments}{\item[Comments:]}

\newcommand{\universality}{\item[Universality:]}

\newcommand{\also}{\item[Also see:]}
\newcommand{\tags}{\item[Tags:]}
\newcommand{\openq}{\item[Open Questions:]}

\title{The Existential Theory of the Reals as a Complexity Class: A~Compendium}
\author{
Marcus Schaefer\\
DePaul University, Chicago, Illinois, USA,\\
\texttt{mschaefe@depaul.edu}
\and
Jean Cardinal\\
Universit\'{e} libre de Bruxelles (ULB), Brussels, Belgium,\\
\texttt{jean.cardinal@ulb.be}
\and
Tillmann Miltzow\\
Utrecht University, Utrecht, The Netherlands, \\
\texttt{t.miltzow@uu.nl}\\
\bigskip
\textit{The first author was the primary contributor, while the second and third authors contributed equally and are listed alphabetically.}
}

\date{}

\begin{document}

\maketitle

\begin{abstract}
We survey the complexity class \ER, which captures the complexity of deciding the existential theory of the reals.
The class \ER has roots in two different traditions, one based on the Blum-Shub-Smale model of real computation, and the other following work by Mn\"{e}v and Shor on the universality of realization spaces of oriented matroids. 
Over the years the number of problems for which \ER\ rather than \NP\ has turned out to be the proper way of measuring their complexity has grown, particularly in the fields of computational geometry, graph drawing, game theory, and some areas in logic and algebra. \ER\ has also started appearing in the context of machine learning, Markov decision processes, and probabilistic reasoning.

We have aimed at collecting a comprehensive compendium of problems complete and hard for \ER, as well as a long list of open problems.
The compendium is presented in the third part of our survey; a tour through the compendium and the areas it touches on makes up the second part. The first part introduces the reader to the existential theory of the reals as a complexity class, discussing its history, motivation and prospects as well as some technical aspects. 

\end{abstract}

\vfill


\newpage
 
\tableofcontents

\newpage

\newpage
\section*{Introduction}

How many guards are necessary to guard an art gallery---a simple polygon on $n$ vertices---if every location inside the gallery must be seen by at least one of the guards? Chv\'{a}tal~\cite{C75b} famously proved that at most $n/3$ guards are needed and that this bound is tight in general.

\begin{figure}[htbp]
    \centering
    \includegraphics{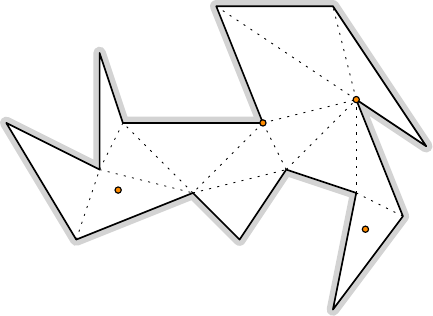}
    \caption{A gallery with $16$ vertices that can be guarded with $4 < \lfloor 16/3 \rfloor$ guards.
    }
    \label{fig:artgalleryn3}
\end{figure}

Figure~\ref{fig:artgalleryn3} shows that for specific layouts fewer than $n/3$ guards may be sufficient to guard a gallery, and we may ask how hard it is to find the smallest number of guards. 
It was shown early on that the problem is \NP-hard~\cite{OR87}, and variants in which guards are stationed at corners 
could be shown to lie in \NP, but membership of the general problem remained elusive. The main difficulty lies in determining the coordinates of the guards, which could (potentially) be 
arbitrary real numbers. How can we handle real coordinates within \NP?

A closer look shows that the art gallery problem is not alone, many computational problems rely on real numbers.
For some of these problems, researchers asked whether they lie in \NP, including the rectilinear crossing number~\cite[Section 9.4]{BMP05} and the collinearity problem~\cite{A11}, but no positive answers were forthcoming. We now know why these questions were so hard to answer: all of these problems are \ER-complete, they are equivalent, up to a polynomial-time reduction, to deciding the existential theory of the reals~\ourref[Existential Theory of the Reals]{p:ETR}. For some problems, this had been known for a while, for the rectilinear crossing number~\ourref[Graph(s)!Rectilinear Crossing Number]{p:rectcross} from a paper by Bienstock~\cite{B91} and for collinearity logic~\ourref[Collinearity Logic]{p:CollLogic} from a paper by Davis, Gotts, and Cohn~\cite{DGC99}; in comparison, the complexity of the art gallery problem~\ourref[Art Gallery Problem]{p:ArtGallery} was not resolved until more than twenty years later in the papers by Abrahamsen, Adamaszek, and Miltzow~\cite{AAM18,AAM22}. 

Since these questions were first asked, the existential theory of the reals has become more established and better known. It has its own Wikipedia page~\cite{W24} and has even found shelter in the complexity zoo~\cite{A24}. The time now seems right to attempt a survey of what we know about this complexity class. There is some material in that direction already, including the first author's computational geometry column~\cite{C15}, Bil\`o and Mavronicolas catalogues~\cite{BM16,BM21} of results on Nash equilibria, and Matou\v{s}ek's introduction to the technical framework underlying \ER~\cite{M14}, but no comprehensive coverage has been published so far. (The last author's unpublished ``The real logic of drawing graphs'', first announced in~\cite{S10}, was never finished, but has served as a quarry for this compendium.) 

Our survey consists of two parts: The first part introduces the existential theory of the reals covering basic definitions, motivation, history, prospects as well as some tools and techniques. The second part is the compendium which lists, to the best of our knowledge, (nearly) all problems that are known to be \ER-complete or \ER-hard. Both parts are indexed and cross-referenced which we hope will help the reader discover interesting old and new problems.

Of course, the compendium is a snapshot in time, but we would like to keep it up to date as we encounter new research, and we would love to hear from readers if they prove new results that should be included in the compendium.

Based on the increase in papers on \ER, see~\Cref{sec:Stats}, and the adoption of \ER in new research communities, we envision that \ER can become a standard tool in theoretical computer science, like \NP\ before it. 

The aim of this compendium is to give an easily accessible overview of current knowledge about \ER, to help find relevant literature, and to serve as an entry point for further study.
%
We hope this compendium will lead to many new and exciting discoveries in the computational world of the theory of the reals.

For the convenience of the reader, we supply this link to our bibliography. 
For the convenience of the reader, we supply this link to our bibliography. 
\begin{center}
\url{https://tinyurl.com/ER-bibliography}
\end{center}

\newpage
\part{Background and Overview}

\section{A Brief Introduction to the Existential Theory of the Reals}

\subsection{What is the Existential Theory of the Reals?}

The easiest way to define the existential theory of the reals as a complexity class is through one of its complete problems; there are several possible candidates including feasibility (of polynomials)~\ourref[Polynomial(s)!feasibility]{p:feasibility} and stretchability (of pseudoline arrangements)~\ourref[Stretchability]{p:Stretch}, two of the original complete problems, but before both of these was ``the existential theory of the reals'' as a formal language.

A (quantifier-free) formula is built from arithmetical terms, consisting of constants $0$, $1$, as well as variables, combined with operations $+, -, \cdot$ (and parentheses); comparing two terms using $<, \leq, =, \geq, >$ leads to atomic formulas, which can be combined using Boolean operators $\wedge, \vee, \neg$; for example, $\varphi(x,y,z) = \neg (x < z) \vee x<y \wedge y < z \wedge x \cdot z = y \cdot y$. Finally, we can quantify free variables in the formula (we can always assume that formulas are in {\em prenex} form, that is, all quantifiers occur at the beginning of the formula). If all variables in the formula are quantified, we obtain a {\em sentence}, e.g.\ 
$\forall x, z \exists y: \varphi(x,y,z)$, where $\varphi$ is the formula defined earlier. If we interpret this sentence over the reals, so variables range over real numbers, it expresses that for every $x < z$ there is a geometric mean $y$ of $x$ and $z$ (which is false, unless we also assume $x > 0$). The quantifier-free part of a quantified formula, in our case $\varphi$, is known as the {\em matrix} of the formula. 

The language \ETR, the existential theory of the reals~\ourref[Existential Theory of the Reals]{p:ETR}, is defined as the set of all existentially quantified sentences which are true if interpreted over the real numbers. For example, $\exists x, y: x\cdot x = y \in$ \ETR. 

From \ETR\ it is a short step to defining the existential theory of the reals as a {\em complexity class} by defining it as the downward closure of \ETR\ under polynomial-time many-one reductions.  We denote this class by \ER (read ``exists R'' or ``ER''). Intuitively, \ER\ contains all decision problems that can be expressed efficiently in \ETR. This is analogous to how \NP\ can be defined from the Boolean satisfiability problem (the existential theory of $\GF[2]$). 

Hardness and completeness are defined as usual for complexity classes: A problem is {\em \ER-hard}, if every problem in \ER\ polynomial-time many-one reduces to it. It is {\em \ER-complete} if it lies in \ER\ and is \ER-hard.

To show that a problem lies in \ER, it is then sufficient to express it in the language of the existential theory of the reals; for that it is useful to know that the definition of \ETR\ can be relaxed to include integer, rational or even algebraic constants, and integer exponents without changing the computational power of the class. To show \ER-hardness of a problem, we need to prove that \ETR\ (or some other problem already established as \ER-hard) reduces to it. We cover proving \ER-completeness in more detail in Section~\ref{sec:PERC}.

\begin{figure}[htbp]
    \centering
    \begin{tabular}{cp{0.2in}cp{0.2in}c}
    \includegraphics[height=1.5in]{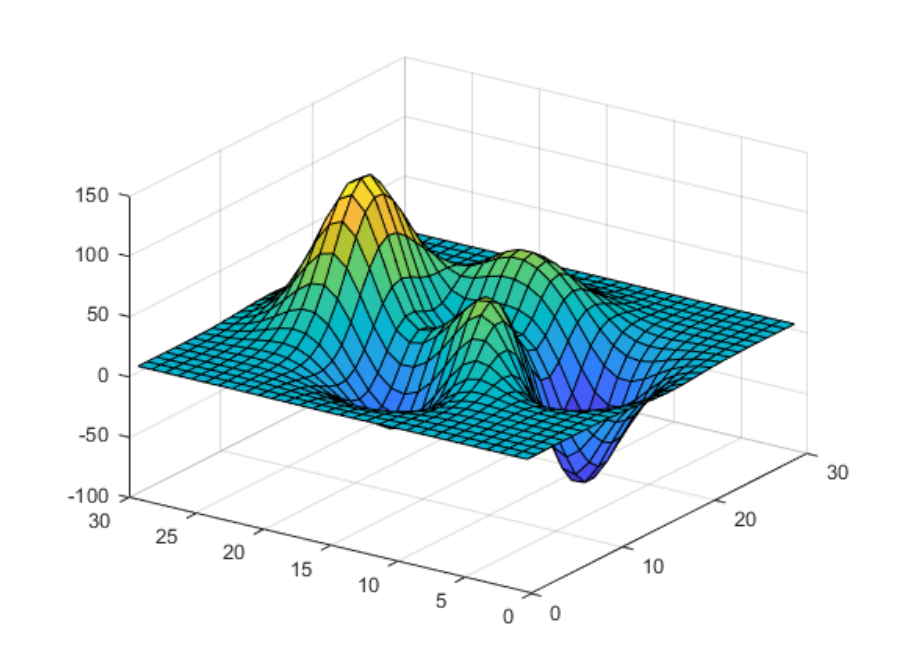}
    &&\includegraphics{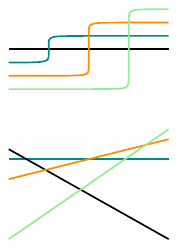}
    &&\includegraphics{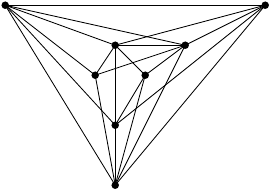}
    \end{tabular}
    \caption{
    {\em Left:} Feasibility problem in 3d; are there $(x,y) \in \R^2$ such that $f(x,y) = 0$? 
    {\em Middle}: A pseudoline arrangement (top) and an isomorphic line arrangement (bottom).
    {\em Right:}  A straight-line drawing of the complete $4$-partite graph $K_{2,2,2,2}$ with $8$ crossings, which is minimal for straight-line drawings. If the drawing does not have the be straight-line, six crossings are sufficient.}
    \label{fig:sampleprob}
\end{figure}

Let us present some sample computational problems which are complete for \ER, see Figure~\ref{fig:sampleprob} for illustrations. 
\begin{description}
    \item[Feasibility~{\ourref[Polynomial(s)!feasibility]{p:feasibility}}.] In the feasibility problem we are given a family of polynomials $(f_i)_{i \in [k]}: \R^n \rightarrow \R^k$ with integer coefficients and we ask whether they have a common zero, that is an $x \in \R^n$ such that $f_i(x) = 0$ for all $i \in [k]$. Given the polynomials, it is easy to express the problem in the existential theory of the reals. Showing hardness is trickier, since \ETR\ allows strict inequalities and Boolean operators, which we need to express using feasibility. This can be done (and was first done in~\cite{BSS89}). For example, the (true) sentence $\exists x,y: x < y \wedge x^2 = 2$ can be expressed as $f_1(x,y,z) := (y-x)z^2 -1$ and $f_2(x,y,z) := 2-x^2$ having a common zero $(x,y,z) \in \R^3$. Note that the term feasibility is often applied to the case $k = 1$ specifically.

    \item[Stretchability~{\ourref[Stretchability]{p:Stretch}}.] The stretchability problem asks whether a \emph{pseudoline arrangement}  -- a family of $x$-monotone curves in the plane so that each pair crosses exactly once -- is {\em stretchable}, that is, if there is a homeomorphism of the plane that turns the pseudoline arrangement into a (straight) line arrangement. Expressing the problem in \ETR\ is again easy, but hardness is difficult; it is a byproduct of Mn{\"{e}}v's universality theorem~\cite{M88,S91,M14}. Via projective duality, the problem amounts to deciding whether an \emph{order type}, a combinatorial description of a set of points giving the orientation of every triple of points, can be realized by an actual point set in the plane~\ourref[Order Type Realizability]{p:ordertype}.
    \item[Rectilinear Crossing Number~{\ourref[Graph(s)!Rectilinear Crossing Number]{p:rectcross}}.] Given a graph and a number $k$, is there a straight-line drawing of the graph with at most $k$ crossings? Membership is not difficult, but cumbersome, worked details can be found in~\cite{D02,S10}. Mirroring what happened with \NP, many papers now skip \ER-membership proofs. Bienstock~\cite{B91} showed how to reduce stretchability, which, as we just saw, is \ER-hard, to the rectilinear crossing number problem, showing that it, in turn, is \ER-hard. 
\end{description}

\medskip

There is another way of defining \ER, which corresponds more closely to how (some) complexity classes have been defined traditionally, and that is through a machine model. For example, \NP\ was first defined as the class of problems solvable by non-deterministic polynomial time Turing machines, and {\em then} satisfiability was shown \NP-complete. The \NP-completeness of satisfiability, now known as the Cook-Levin theorem, laid the foundation of computational complexity. There is a Cook-Levin theorem for \ER\index{Cook Levin theorem for ER@Cook-Levin theorem for \ER} as well, if we define \ER\ in the Blum-Shub-Smale (BSS) model of computation. Roughly speaking, in the BSS model of real computation we work with register machines that can read, store, and write real numbers in a single cell or register (with full precision), and perform exact arithmetic and comparisons on these numbers; registers can be pre-loaded with real constants. As inputs, BSS machines take tuples of real numbers (of arbitrary arity), so a BSS machine accepts a set of tuples of real numbers. 
In the BSS model, we can define polynomial time and non-deterministic polynomial time (real numbers have unit length and operations on real numbers take unit time). This allows a definition of problems that can be decided in polynomial time, $\P_{\R}$, and non-deterministic polynomial time, $\NP_{\R}$. If we restrict the non-deterministic machines so they are {\em constant-free}, that is, only constants $0$ and $1$ are allowed in the program code, and we only consider Boolean input tuples (all entries are $0$ or $1$) we obtain the complexity class $\BP(\NP_{\R}^0)$. The original paper by Blum, Shub, and Smale~\cite{BSS89} already showed that the feasibility problem (with real coefficients) is complete for $\NP_{\R}$, this is the Cook-Levin theorem in the BSS model since it expresses a machine computation as a (real) satisfiability problem; the real Cook-Levin theorem easily implies that the feasibility problem we introduced earlier (with integer coefficients), is complete for $\BP(\NP_{\R}^0)$, and this implies $\ER =\BP(\NP_{\R}^0)$, so the BSS-model is a machine model for \ER.

As we will see in Section~\ref{sec:History} on the history of the existential theory of the reals, these two alternative definitions arose independently.

We have defined \ETR\ as a the existentially quantified subset of true sentences in the theory of the reals. In analogy with the polynomial-time hierarchy building on \NP, we can similarly define a real polynomial hierarchy. The {\em $Q_1\cdots Q_k$-fragment} of the theory of the reals is defined by the true sentences which start with $k$ alternating blocks of quantifiers of the types $(Q_1,\ldots, Q_k)$, where $Q_i \in \{\exists,\forall\}$. The corresponding complexity classes are called $\Sigma_k\R$ and $\Pi_k\R$, depending on whether $Q_1 = \exists$ or $Q_1 = \forall$. The most common classes, apart from \ER, are $\VR = \Pi_1\R$, at the first level, and $\EVR = \Sigma_2\R$ and $\VER = \Pi_2\R$ at the second level~\cite{SS23,JJ23}. These classes can also be defined in the BSS-model by considering alternating machines (machines with existential and universal states) or using an oracle model~\cite{M24}. 

Only a few problems have been shown complete for higher levels of the hierarchy (reflecting the situation for the polynomial-time hierarchy), we collect them in Section~\ref{sec:HigherLevels}.

\subsection{Where is the Existential Theory of the Reals?}

We do not know much about the relationship of \ER\ to traditional complexity classes, but we do know a little. Shor~\cite{S91} proved that $\NP\subseteq \ER$ by reducing satisfiability to stretchability; using \ETR\ as the complete problem leads to an easier proof, since satisfiability can be expressed directly in \ETR; for example, $(p \vee \overline{q} \vee r) \wedge (\overline{p} \vee \overline{r})$ can be expressed as $\exists p, q, r: (p=1 \vee q = 0 \vee r = 1) \wedge (p = 0 \vee r = 0)$ in \ETR. 

The second result we know is due to Canny~\cite{C88,C88b}, who in his thesis showed that $\ER \subseteq \PSPACE$ using his new notion of roadmaps.
\PSPACE\ is the class of problems decidable by (traditional) Turing machines in polynomial space. 
Complete problems for \PSPACE\ include quantified Boolean formulas, and many game-related problems, such as Sokoban. 

And this is all we know about \ER\ with respect to traditional complexity classes. 
The status of \ER\ with respect to \PH, for instance, is wide open.
The possibility that $\NP = \ER$ cannot yet be ruled out, but is usually considered unlikely. Natural polynomial-sized witnesses of positive instances of \ER-complete problems remain elusive. Furthermore, a number of possibly simpler problems which are not known to lie in \NP\ are contained in \ER. 
We review such problems now.\\


{\bfseries\noindent Inside \ER.}
\ER\ contains several problems which are unlikely to be \ER-complete but are often relevant to \ER-complete problems. We discuss three of these problems here: the sum-of-square-roots problem, \PosSLP and convex programming.

The {\em sum-of-square-roots problem} (\SSQR)\mindex{sum of square roots}\index{SSQR@\SSQR|see {sum of square roots}} asks whether
$\sum_{i \in [n]} a_i^{1/2} \geq k$ for positive integers $a_1, \ldots, a_n$ and $k$. 
Clearly, $\SSQR \in \ER$, but this innocent-looking special case of \ER\ has proved difficult to handle. 
There is a highly non-trivial upper bound on \SSQR, via \PosSLP which we discuss next, at the third level of the counting hierarchy (which is expected to be lower than \PSPACE, but way above \NP), see~\cite{ABKPM08}. 
This is a serious first obstacle in proving $\NP=\ER$, since we cannot even prove good upper bounds on \SSQR, which belongs to the quantifier-free part of \ER. 
The $\SSQR$-problem is powerful enough to solve some interesting problems in computational geometry: the complexity class  $\P^{\SSQR}$ contains problems such as $3$-connected planar graph realizability for triangulations~\cite[Theorem 1]{CDR07}, as well as minimum-link path problems~\cite{KLPS17}. 

Also in \ER\ and not known to lie in \NP, is the {\em Euclidean traveling salesperson problem} (\ETSP)\mindex{Euclidean traveling salesperson}\index{ETST@\ETSP|see {Euclidean traveling salesperson}}, in which we are given $n$ cities in the plane and ask for a shortest tour through all cities (in the Euclidean metric). It is often asked whether this problem is \NP-complete; the obstacle to placing \ETSP\ in \NP\ is the computation of the length of the route, which seems to require the solution of an \SSQR-problem. The best upper bound on the Euclidean traveling salesperson is $\NP^{\SSQR}$. There are other problems that fall into this class, most prominently minimum weight triangulation, minimum dilation graphs, and the Steiner Tree problem~\cite[Section 10.4]{PS02}; all of these problems are non-trivially \NP-hard, see~\cite{P77,MR08,GGJ76,GKKKM10}.

If the $n$ cities the Euclidean traveling salesperson must visit lie on an integer grid of size polynomial in $n$, then the problem lies in a slightly smaller class, $\NP^{\USSR}$, where \USSR\mindex{unary sum of square roots}\index{USSR@\USSR|see {unary sum of square roots}} is the variant of \SSQR\index{sum of square roots} in which the $a_i$ are given in unary (and this applies to grid versions of other problems in $\NP^{\SSQR}$ as well). 
Balaji and Datta~\cite{BD23} recently announced that $\USSR \in \Ppoly$, but this still does not imply that this grid version of the Euclidean salesperson lies in \NP.
It is quite possible that \USSR\ and \SSQR\ are polynomial-time solvable, and many researchers believe so, but their exact complexity remains open.

Let us turn to \PosSLP, short for positive straight line program. 
A {\em straight line program (SLP)} is a finite sequence of instructions of the form
$a_0 = 1$, and $a_k = a_i \oplus a_j$, where
$i,j<k$ and $\oplus \in \{\cdot, +, -\}$. An SLP computes a sequence of numbers $(a_k)_{k=0,\ldots,n}$. \PosSLP asks whether $a_n >0$, that is, whether the final  number computed by the SLP is positive. (For a related problem, see~\ourref[Polynomial(s)!Feasibility of Univariate Polynomial (Open)]{p:UniFEAS}).
The value of $a_n$ may require exponentially many bits, since SLPs can perform repeated squaring, so \PosSLP does not even obviously lie in \PSPACE, but it does, since \PosSLP lies in \ER.

The best current upper bound on \PosSLP is that, like \SSQR, it lies in the third level of the counting hierarchy~\cite{ABKPM08}. 
One of the most striking results is that $\P^{\PosSLP}$, 
polynomial time
 on the \wordRAM 
with access to a \PosSLP oracle, 
is as powerful as polynomial time on the \realRAM on integer inputs~\cite[Proposition 1.1]{ABKPM08}: the \realRAM can run the SLP for \PosSLP, since it can perform computations on reals, and therefore integers, in constant time; and we can simulate any \realRAM algorithm on the \wordRAM using the \PosSLP oracle whenever the program branches. This also implies that \SSQR can be solved with oracle access to \PosSLP, as the \realRAM can solve \SSQR.

In other words, \PosSLP explains the gap between the \wordRAM and the \realRAM.
The currently best lower bound on \PosSLP is by 
B\"{u}rgisser and Jindal~\cite{BJ24}
and based on a bold conjecture.
The second author conjectures that \PosSLP also explains the gap between \NP{} and \ER:
\begin{conjecture*}
 \[\NP^{\PosSLP} = \ER.\]
\end{conjecture*}

\PosSLP can be solved using semidefinite programming\index{semidefinite programming}~\cite{ABKPM08}, which is a special case of convex programming\index{convex programming}, and both of these belong to \ER. 
Thus the conjecture would also imply that convex programming is not polynomial time solvable unless $\NP = \ER$.
Furthermore, since $\EC \subseteq \ER$ we get a better upper bound on the complexity of \EC.
The current best bound, assuming the generalized Riemann hypothesis, is $\NP^{\RP}$~\cite{Ko96}.
Semidefinite programming also belongs to \VR~\cite{SS23}, so it lies in $\ER \cap \VR$, making it unlikely to be \ER-complete, unless $\VR = \ER$,
which would be quite surprising. 
This also implies that \ER-complete problems, such as the rectilinear crossing number, cannot be solved using semidefinite programming, unless $\ER = \VR$.
Junginger~\cite{JJ23} argues that polynomial identity testing also belongs to $\ER \cap \VR$ (it also lies in \RP, randomized polynomial time).

\bgroup
\def\arraystretch{1.5}
\setlength{\tabcolsep}{4pt}
\begin{table}[htb]
\centering
\begin{tabular}{|l|l|l|l|}\hline
\multicolumn{2}{|c|}{\multirow{2}{*}{}} & \multicolumn{2}{|c|}{nondeterminism} \\ \cline{3-4} 
\multicolumn{2}{|c|}{}   & discrete & real \\ \cline{1-4} 
\multirow{2}{*}{input} & discrete & \NP & \ER \\
   & real & $\DNP_{\R}$ & $\NP_{\R}$ \\ \hline
\end{tabular}
\caption{Complexity classes corresponding to discrete vs real inputs, and discrete (aka digital) vs real nondeterminism}
\label{tab:discretereal}
\end{table}
\egroup

\bigskip

{\bfseries\noindent Above \ER.} If we look upwards from \ER, towards \PSPACE, we are at the start of a hierarchy, the real polynomial hierarchy. The finite levels of this hierarchy, such as $\EVR$ and $\VER$, see~\ourref[Q1Qk fragment of the theory of the reals@$Q_1\cdots Q_k$-Fragment of the Theory of the Reals]{p:quantifiedR}, still lie in \PSPACE~\cite[Remark 13.10]{BPR06}, but the unbounded quantification of the full theory of the reals takes us outside of \PSPACE, to double exponential time (using Collins' Cylindrical Algebraic Decomposition~\cite{C75}). In this survey we restrict ourselves to the finite levels of the hierarchy, but in the BSS-model there is structural research on the whole hierarchy, for example, an analogue of Toda's theorem~\cite{BZ10}. One can also consider looking at mixing real and discrete quantifiers; in the BSS-model discrete quantifiers were studied under the name ``digital nondeterminism'', leading to the class $\DNP_{\R}$, the restriction of $\NP_{\R}$ to quantifiers over $\{0,1\}$, see the survey by Meer and Michaux~\cite{MM97} and Table~\ref{tab:discretereal}. The Vapnik-\v{C}ervonenkis dimension of a family of semialgebraic sets~\ourref[Semialgebraic Set(s)!Vapnik-\v{C}ervonenkis Dimension (Open)]{p:semiVC} is a natural problem that can be captured by mixed quantification~\cite[Section 4]{SS23}.

Another recent development is the introduction, by van der Zander, Bl{\"{a}}ser, and Liskiewicz~\cite{vdZBL23} of the complexity class \succER, a succinct version of \ER, which is allowed to encode arithmetic formulas using circuits. The class \succER\ is the real analogue of \NEXP, just like \ER\ is the real analogue of \NP~\cite{BDLvdZ24}; it has turned out to be useful for studying problems in probabilistic reasoning, see~\cite{vdZBL23,DvdZBL24, BDLvdZ24, IIM24}.

\subsection{Universalities}
\label{subsec:univ}

\ER-completeness theory is closely related to an algebraic geometric notion usually referred to as \emph{universality}. Informally, universality results state that spaces of solutions of simple geometric problems over the reals can behave ``arbitrarily bad''.
The occurrence of irrational numbers, and more generally numbers of arbitrary algebraic degree, or numbers requiring exponential precision, can sometimes be a hint towards universality, and \ER-hardness. 
In fact, algebraic universality can be seen as the strongest geometric statement about the intrinsic complexity of a realizability problem involving real numbers.

\begin{figure}[htbp]
    \centering
    \includegraphics{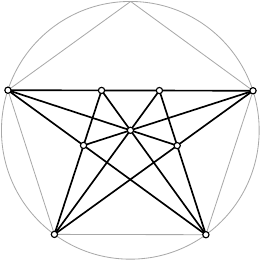}
    \caption{
    The Perles configuration.}
    \label{fig:perles}
\end{figure}

\paragraph{Irrationality.}
The famous Perles configuration, described by Micha Perles in the 1960s, is composed of nine points and nine lines, with prescribed incidences, any realization of which in the Euclidean planes requires at least one irrational coordinate, see Figure~\ref{fig:perles}. 
A much earlier example of such ``nonrational'' configurations with $11$ points was given by MacLane in 1936~\cite{McL36}, using von Staudt's ``algebra of throws''.
Perles also constructed an $8$-dimensional combinatorial polytope that could be realized with vertex coordinates in $\mathbb{Q}[\sqrt{5}]$, but not with rational coordinates only~\cite{G03,Z08}. 
Much later, Brehm constructed three-dimensional polyhedral complexes, every realization of which requires irrational coordinates (see Ziegler~\cite{Z08}).

These three examples can be understood as early indications of the much more general universality property for point configurations, polytopes, and polyhedral complexes. In general, \ER-complete problems often require solutions of arbitrary high algebraic degree. 
Note, however, that this is not always the case; realizations of simple line arrangements, for instance, can always be slightly perturbed, but deciding the existence of one still remains \ER-hard. This leads us to the next topic.
\paragraph{Precision.}

Pursuing the idea that solutions of simple geometric realizability problems may ``behave badly'', let us note that \ETR formulas of size $O(n)$ can encode doubly exponential numbers $x_n=2^{2^{n}}$ as follows 
\[x_0 = 2, \  x_{i} = x_{i-1}^2, \text{ for all } i=1,2,\ldots ,n.\]
For \ER-complete geometric realizability problems, this implies that the number of bits, hence the precision, required to encode the coordinates of a realization, if one exists, can be exponentially large. 

Early papers related to \ER-complete problems emphasized this precision issue. For instance Goodman, Pollack, and Sturmfels~\cite{GPS89,GPS90} showed two things: order types (combinatorial descriptions of point sets giving the orientation of every triple) can always be realized on an integer grid of double exponential size, and there are order types that need what they named ``exponential storage''. The upper bound follows from the work of Grigor{\cprime}ev and Vorobjov~\cite{GV89} (the ``Ball Theorem'', see Section~\ref{subsec:ball}).
Bienstock~\cite{B91} and later Kratochv\'{\i}l and Matou\v{s}ek~\cite{KM94} prove double exponential precision lower bounds for realizations with minimum crossing numbers and segment intersection graphs, respectively.
Similarly, Richter-Gebert and Ziegler~\cite{RGZ95} explicitly state double-exponential lower bounds on coordinates for realizations of polytopes, and McDiarmid and M\"{u}ller~\cite{McDM10,McDM13} actually cast this as the main result for unit disk graphs and disk graphs, while \ER-hardness is only mentioned in passing. 

Like irrationality, the ``exponential storage'' phenomenon is a consequence of the much more general property of algebraic universality.

\paragraph{Algebraic Universality.}

Algebraic universality is the culmination of those partial results on the ``bad behavior'' of polytopes and configurations of points and lines, and it is a property of the so-called \emph{solution}, or \emph{realization} \emph{spaces} of a satisfiability problem over the reals.

In the stretchability problem, for instance, we are given the combinatorial type of an arrangement of $n$ pseudolines, consisting for example in the order, from left to right, of the intersection points of each pseudoline with the other $n-1$ pseudolines. The realization space is the set of all possible coordinates $(a_1,b_1,a_2,b_2,\ldots ,a_n,b_n)\in\R^{2n}$ such that the $n$ lines of respective equations $y=a_ix+b_i$ realize the same arrangement.
The famous universality result by Mn\"{e}v~\cite{M88} states that for
every semialgebraic set $S$ there exists a pseudoline arrangement whose realization space is \emph{stably equivalent} to $S$~\cite{M88,RG99}, where stable equivalence combines rational equivalence with so-called \emph{stable} projections.
Note that stable equivalence preserves the homotopy type and the algebraic number type, see~\cite[Section 4.1]{V23c}. 
In particular, any topology can be preserved in that sense.

Maybe the simplest application of this theorem is to a semialgebraic set $S$ consisting of two points. Mn\"ev's Theorem implies that there exists two arrangements of $n$ lines in the plane, where the orders of intersections are the same, but such that it is impossible to go continuously from one arrangement to the other without changing the combinatorial type in between. So Mn\"ev gives a strong negative answer to Ringel's isotopy conjecture~\cite{R55,RG99}.

This phenomenon of ``disconnected realization spaces'' had been observed before for polytopes~\cite{BG90}. Using \emph{Gale diagrams}, Mn{\"e}v also gave a universality result for polytopes: Any semialgebraic set is stably equivalent to the realization space of a $d$-dimensional polytope with $d+4$ vertices.
This gave rise to a series of results refining and generalizing Mn{\"e}v's proof~\cite{RGZ95,RG96,RG99,PT14,APT15}. Richter-Gebert, in particular, 
dedicated a complete monograph on the universality of realization spaces of $4$-dimensional polytopes~\cite{RG96}, implying that the problem of deciding whether a given lattice is realizable as the face lattice of a $4$-dimensional polytope is \ER-complete. The case $d=4$ is best possible: Steinitz Theorem gives a simple characterization of face lattices of three-dimensional polytopes that is decidable in polynomial time.

The exact definition of stable equivalence used in various contexts has varied over the years; recently it was pointed out that the definition used in Richter-Gebert's text~\cite{RG96} required a fix~\cite{B22d,V23c}.

While algebraic universality and \ER-hardness are closely related, they do not always imply each other: stable equivalences do not have to be effective (e.g.\ Datta~\cite{D03}), and \ER-hardness reductions rarely maintain stable equivalence. In many cases, stable equivalences can be made effective though; on the other hand, \ER-hardness results typically do not automatically yield algebraic universality results, but the proofs often allow the conclusion of weaker universality properties, e.g.\ Bienstock's proof of \ER-hardness result for the rectilinear crossing number shows that every semialgebraic set occurs as the projection ({\em not} stable) of the realization space of a rectilinear crossing number problem. Our compendium indicates if universality results are known. 

\subsection{Proving \ER-Completeness}\label{sec:PERC}

How can we tell whether a problem is a candidate for \ER-hardness? There are no hard-and-fast rules, but there are some tell-tale signs. First of all, there should be solutions to the problem that use real parameters. \NP-hardness, without \NP-membership can be a next clue: why does the problem not lie in \NP, are real numbers the issue? To strengthen a suspicion of \ER-hardness, one can look for evidence of algebraic universality: can one construct instances that require algebraic solutions or rational solutions of exponential bitlength? None of these clues are sufficient, of course, for example, the Euclidean traveling salesperson problem is \NP-hard and does not lie in \NP, but it is unlikely to be \ER-hard, since it belongs to $\NP^{\SSQR}$; and representing a plane graph as a disk contact graph may require algebraic radii, but testing planarity can be decided in linear time. Topological universality can also be misleading, there are problems in \NP{} that exhibit topological universality~\cite{SW23, BEMMSW22, ST22}. 

\subsubsection*{\ER-Membership}

Since \ETR\ is a very expressive language, it is typically not difficult, though sometimes cumbersome, to establish membership in \ER, and we can even use the result of ten Cate, Kolaitis, and Othman~\cite[Theorem 4.11]{tCKO13} that \ETR\ is closed under \NP-many-one reductions, for example, see ~\ourref[Markov Decision Process!Total Variation Distance]{p:totvardistMDP} and~\ourref[Satisfiability of Probabilistic Languages]{p:satprobL}. 

Compare this situation to \NP. We can show membership by encoding the problem as a satisfiability problem, yet it is often easier to show that the problem can be accepted by a Turing machine in non-deterministic polynomial-time. 
Or we can even work in the word RAM model, which allows random access to an unlimited number of registers storing integers on which we can perform arithmetical operations.

Erickson, van der Hoog, and Miltzow~\cite{EvdHM20}) showed that we have the same option for \ER: if we replace the \wordRAM by a \realRAM, which allows us to guess real, not just discrete, numbers and has unlimited storage for real numbers, we obtain \ER.
The \realRAM model can simplify \ER-membership arguments significantly, as describing an algorithm is much easier than writing down an \ETR{} formula, and researchers can make use of the huge number of \realRAM algorithms already developed in computational geometry and other areas.  


Efficient membership proofs may also lead to algorithmic solutions of problems, e.g.\ Dean~\cite{D02} showed how to express the rectilinear crossing number problem as an optimization problem with linear constraints. 
While solvers for these problems may not yet be as fast as solvers for \NP, where a linear programming formulation for the (topological) crossing number problem has been implemented and can compute crossing numbers of reasonably large graphs on the web~\cite{CW16}, 
this may very well change in the future.

\subsubsection*{\ER-Hardness}

In this section, we discuss two primary techniques to show \ER-hardness, one based on algebra, and the other on geometry. The first technique, we refer to as algebraic encoding, involves encoding real variables, addition and multiplication directly. The second technique, known as stretchability encoding, is based on encoding the stretchability problem. 

\paragraph{Algebraic Encoding.}

\begin{figure}[htbp]
    \centering
    \includegraphics
    {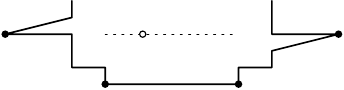}
    \caption{The white guard needs to be placed anywhere on the dotted line segment in order to see all the marked vertices.
    We can interpret the position of the guard as a value in a real-valued interval.
    }
    \label{fig:guard-variable}
\end{figure}

Algebraic encoding work by encoding algebraic constraints on variables, such as $x = y+z$ and $x = yz$; so in a first step, we need to encode real variables. How to do so is problem-specific but often straightforward. For instance, in the art gallery problem 
we can encode variables by using the coordinates of a guard's position, see \Cref{fig:guard-variable} for an illustration.

Similarly, to show stretchability \ER-hard, we can work with the parameters of the lines, or the coordinates of the intersections. 

It is often useful to know that the variables we need to encode can be assumed to be bounded, e.g. \ETRINV~\ourref[ETRINV@\ETRINV]{p:ETRINV} allows us to assume that all variables belong to the interval $[1/2,2]$. 
Bounded variables can be encoded in the coordinates of a geometric object with limited range, such as a guard in an art gallery. One can even reduce the range to much smaller intervals, such as $[0,n^{-200}]$, where $n$ is the size of the input. This may seem counter-intuitive as it leaves very little wiggle room for the variables, but because we can encode variables with very high precision, it is sufficient.

Secondly, we encode addition, of the form $x= y+ z$. 
This is typically simple as it is a linear relation between three variables. For geometric problems, addition can often be encoded by forcing two geometric objects to lie next to each other, e.g.\ two pieces in the packing problem, with the combined object encoding the sum. 

\begin{figure}[htbp]
    \centering
    
    \includegraphics{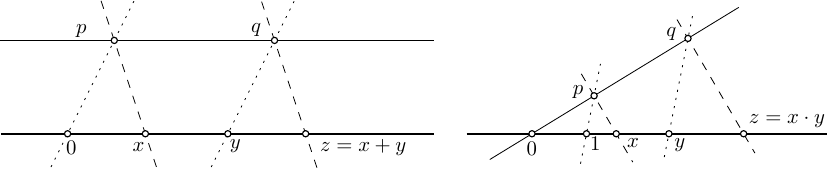}
    \caption{Simplified von Staudt addition gadget, {\em left}, and multiplication gadget, {\em right}. The original gadgets are projective, not Euclidean. 
    }
    \label{fig:vonStaudt}
\end{figure}

\begin{remark}[Addition with von Staudt]
Many proofs work with variants of the von Staudt gadgets originally used by Mn\"{e}v~\cite{M88} to show that order type~\ourref[Order Type Realizability]{p:ordertype}, and thereby stretchability~\ourref[Stretchability]{p:Stretch}, is universal. Figure~\ref{fig:vonStaudt} shows a simplified von Staudt addition gadget on the left. Variables are encoded as length of segments (starting at $0$). Assuming the drawing of the gadget satisfies that $0,y,z$ and $x$ are collinear and $pq \parallel 0x$, $0p \parallel zq$ and $py\parallel qx$, then $x = y+z$. 
\end{remark}

In some rare cases, though, the encoding of addition is trickier, as, for example, in the case of the  boundary-guarding variant of the art gallery problem; addition here is the main obstacle as there is no direct interaction of three guards at all, so it is not obvious how to encode a ternary relation.

Thirdly, we need to encode multiplication, i.e., $x = y \cdot z$. 

\begin{remark}[Multiplication with von Staudt]
Figure~\ref{fig:vonStaudt} shows how multiplication can be encoded using von Staudt gadgets (again simplified, removing the projective setting): as long as $0pq$ and $01zyx$ are each collinear, but distinct, and $p1 \parallel qz$ and $py \parallel qx$, we have $x = y \cdot z$. It follows that geometric problems that can express parallelism and collinearity are \ER-hard. By interpreting the von Staudt gadgets in the projective plane (as von Staudt and Mn\"{e}v's did), both collinearity and parallelism can be expressed using order types. 
\end{remark}

Encoding multiplication directly can be tricky, but it can be simplified by
simulating multiplication through inversion, as is done in \ETRINV~\ourref[ETRINV@\ETRINV]{p:ETRINV}. 
To see how to do this, note that 

\[(y+z)^2 - (y-z)^2 = 4x\]

is equivalent to 

\[x = y\cdot z.\]

Which indicates that multiplication can be simulated by squaring, i.e., $x=y^2$.
Consider

\[\frac{1}{\frac{1}{x-1} - \frac{1}{x}} + x = x^2.\]

This identity shows that inversion, i.e. $x\cdot y = 1$ , is sufficient to encode squaring.
Inversion is often much easier to encode compared to multiplication as it involves only two variables in a symmetric way. 

The inversion constraint can be replaced by almost any non-linear constraint~\cite{MS24} of the form 
\[f(x,y) = 0,\]
see~\ourref[Continuous Constraint Satisfaction Problem]{p:CSSP}.
Although, this is very powerful, we are only aware of one application, namely showing that geometric packing~\ourref[Geometric Packing]{p:geopack}, packing convex polygons into a square container, is \ER-complete~\cite{AMS20}.
The idea of the proof is that any non-linear constraint that is sufficiently smooth can be very well approximated by the first two terms of the Taylor series in a sufficiently small neighborhood.
And the first two terms of the Taylor series are just a quadratic equation.

Finally, we need to connect the variables to the different gadgets. This step, which looks like simple housekeeping, can be surprisingly challenging in many reductions. As Richter-Gebert writes in~\cite{RG96}: 
\begin{quote}
The main difficulty in Mn\"{e}v's proof is to organize the construction in a way such that different basic calculations do not interfere and such that the underlying oriented matroid stays invariant for all instances of a geometric computation.
\end{quote}
To see what the issue is, recall the von Staudt gadget for addition (or multiplication) in Figure~\ref{fig:vonStaudt}. If we want to encode the gadget using order types, we need to know whether $x < y$, $x = y$ or $x > y$. How would we know? And even if we did, how would we know for other, intermediate variables? This problem has been solved using various normal forms; maybe the most elegant solution is due to Richter-Gebert~\cite{RG96}, this is the approach presented by Matou\v{s}ek~\cite{M14}.

It can also be helpful to know that the variable-clause incident graphs are of a restricted type, for example planar, as in \PETRINV~\ourref[Planar ETRINV@\PETRINV]{p:PlanarETRINV}.

\paragraph{Stretchability Encoding.}

The stretchability problem~\ourref[Stretchability]{p:Stretch} asks whether a given pseudoline arrangement can be {\em stretched}, that is, if there is an equivalent line arrangement, that is a line arrangement which has the same combinatorial structure. 

\begin{figure}[htbp]
    \centering
    \includegraphics{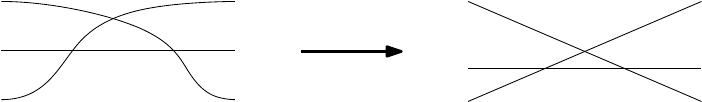}   
    \caption{A (stretchable) pseudoline arrangement on the {\em left}, and an equivalent line arrangement on the {\em right}.
    }
    \label{fig:stretch1}
\end{figure}

\begin{figure}[htbp]
    \centering
    \includegraphics[page = 2]{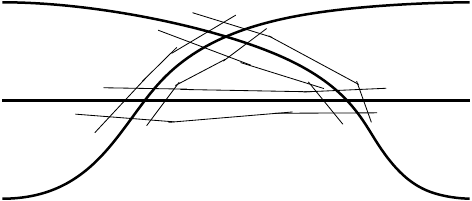}   
    \caption{Forcing the combinatorial structure of a pseudoline arrangement using segment intersections. 
    }
    \label{fig:stretch2}
\end{figure}

Reductions from stretchability are particularly popular in the graph drawing and computational geometry community. The segment intersection graph problem~\ourref[Segment Intersection Graph]{p:SEG} is an early example: we can represent each pseudoline by a segment and surround each pseudoline-segment with additional segments which enforce the combinatorial structure of the pseudoline arrangement, see Figure~\ref{fig:stretch2}.

If there is a representation of the graph using segments then the underlying pseudoline arrangement is stretchable, see~\cite{KM94,S10}. Another, earlier, example due to Bienstock~\cite{B91} shows how to embed stretchable arrangements in the rectilinear crossing number problem.

\subsection{What is next for \ER?}

In this section, we want to point at possible future research
directions that appear of interest to us.
We focus on broad research lines, for specific candidate problems see \Cref{sec:Candidates}.

\paragraph{Breadth.}
    Most of the early \ER-complete problems arose in graph drawing and algebraic complexity.
    As the compendium shows, many more areas like probabilistic reasoning, machine learning, polytope theory etc. have been added since then, and it seems reasonable to believe that this trend continues and that more \ER-complete problems are found in other research areas.

\paragraph{Depth.}

    After a problem has been shown \ER-complete, we can start asking whether more restricted versions of the problem remain \ER-complete. This is particularly important for problems that are common starting points for reductions; for \NP, this is evidenced by the numerous satisfiability variants that have been shown \NP-complete. For \ER, the original \ETR~\ourref[Existential Theory of the Reals]{p:ETR} has been refined in many ways, including the many variants of \ETRINV~\ourref[ETRINV@\ETRINV]{p:ETRINV} which have significantly simplified reductions for many problems.

    Similarly, the partial version of order type realizability~\ourref[Order Type Realizability]{p:ordertype} was introduced in~\cite{S21b} to show that many problems remain hard if some parameters are constant. To give an example, the simultaneous geometric embedding~\ourref[Graph(s)!Simultaneous Geometric Embedding]{p:SGE} is \ER-complete~\cite{CK15}, and it is now known that it remains so, even if the number of input graphs is bounded, indeed at most $31$~\cite{FKMPTV24, S21b}, or all the input graphs are edge-disjoint~\cite{KR23}. There are many examples like this already, and we expect that many more problems with constant natural parameters will be shown \ER-complete.

\paragraph{Parameterized Theory of \ER.}

    Many \NP-complete problems become considerably easier, if 
    some of their parameters are bounded, the point of view taken by the area of parameterized complexity.
    
    For example, we can decide whether a graph on $n$ vertices contains a vertex cover of size at most $k$ in time $O(2^k n^2)$. For fixed $k$, this gives a quadratic-time algorithm for the vertex cover problem. Even if $k$ is not fixed, but small compared to $n$, this running-time is a tremendous speed-up to the general worst-case complexity of $O(c^n)$.
    We are not aware of a single \ER-complete problem with a running time of $f(k)n^{O(1)}$ for some natural parameter $k$.
    It would be interesting to see if such a problem exists and what type of parameterization is required.
    On the other hand, we are aware of some \ER-complete problems for which the parameterized version is $W[1]$-hard~\cite{ArtParaHard}.
    It would be interesting to be able to find the correct parameterized complexity class for those problems.

\paragraph{Structural Complexity.}

    The study of structural properties of complexity classes has matured since \NP\ was first introduced, see for instance the book by Arora and Barak~\cite{arora2009computational}. Structural questions have been studied widely in the BSS-model of computation, starting with the book by Blum, Cucker, Shub and Smale~\cite{BCSS98}; for a more recent survey, see Baartse and Meer~\cite{BM13}.
    Yet, for many classical results, we do not have analogue results for \ER\ or the real hierarchy.
    To list some examples:
    \begin{itemize}
    \item Does $\VR = \ER$ imply that the real hierarchy collapses? This would follow from showing that $\VR\cap \ER$ is {\em low} for \ER, that is $\ER^{\VR \cap \ER} = \ER$; to make these statements meaningful, we need to work with a model for \ER\ that allows for oracles, see~\cite{M24}.
    \item Fagin's theorem characterizes \NP\ as the set of properties captured by monadic second-order logic. Can we characterize \ER\ in a similar fashion? There is a similar theorem for $\NP_{\R}$, see~\cite[Section 7]{MM97}.
    \item Toda's theorem, that the polynomial-time hierarchy reduces to \PP, probabilistic polynomial time, was a breakthrough result in classical complexity theory~\cite[Theorem 17.4]{arora2009computational}. Basu and Zell~\cite{BZ10} showed that the result can be adopted to the BSS-model. Is there a discrete  analogue for the real hierarchy?
    \item Are there oracles relative to which $\NP = \ER \neq \PSPACE$ or $\NP \subsetneq \ER$?
    \item What are the connections between \ER and other complexity classes defined to capture working with continuous solution spaces? Specifically, how does \ER relate to CLS~\cite{CLS-intro} and FIXP~\cite{EY10}? This can be studied by looking at finding a Nash Equilibrium problem which is either \ER-complete or FIXP-complete, depending on the technical details of the definition.
    \end{itemize}
    We expect positive answers to many of these questions, and some of them may not be that difficult.

\paragraph{Counting Complexity.}

    A counting version of \ER\ was introduced by B\"{u}rgisser and Cucker~\cite{BC06} as $\BP(\#\P_{\R}^0)$, which we could call $\#\R$; they show that counting the number of points in a semialgebraic set is complete for this class, and computing the Euler-Yao characteristic of a semialgebraic set is complete for this class under Turing reductions. The paper leaves many interesting open questions, not least of it being the question of where $\#\R$ is located with respect to traditional complexity classes.

\paragraph{Better Techniques to Show Hardness.}

    Reductions for \ER-hardness tend to start with a relatively small set of problems. 
    Maybe the three most prominent starting problems
    are simple stretchability~\ourref[Simple Stretchability]{p:Stretch}, feasibility~\ourref[Polynomial(s)!feasibility]{p:feasibility} and \ETRINV~\ourref[ETRINV@\ETRINV]{p:ETRINV}.
    
    Simple stretchability is useful as a very geometric problem that has turned out to be very adaptable to different contexts; one advantage it has (particularly over stretchability) is that its solution space is open, which is the case for many geometric problems.  
    \ETRINV is quite versatile and can be adapted to many situations: the range of variables can be bounded, multiplication can be replaced by inversion. 
    Feasibility has served as the starting point for nearly all problems from algebraic geometry and analysis; that may not be that surprising, since it is the oldest and best-studied problem.

    These results can be refined even further.
    Instead of Stretchability, we can now use pseudo-segment stretchability~\ourref[Stretchability!Pseudo-segments]{p:SegStretch}. This makes the reduction more tedious, but it allows to bound certain parameters by a constant~\cite{S21}.
    Similarly, the inversion in \ETRINV can be replaced by virtually any curved function~\cite{MS24}.
    In analogy with \NP, we expect that the core problems will be further refined to become more flexible, easier to apply, and yield stronger results.


\paragraph{Real Hierarchy.}

    There is a small number of papers that study the real polynomial time hierarchy starting with a paper by B\"{u}rgisser and Cucker~\cite{BC09} in the BSS-tradition. 
    Some fundamental problems like the Hausdorff distance are complete for the second level of this hierarchy~\cite{JKM22}.
    Studying the second level of the hierarchy is more challenging as many standard techniques that can be used for the first-level are not yet known to work for the second-level.

    Developing techniques for higher levels of the hierarchy and identifying additional natural problems captured by these levels appear of interest to us. There are also structural questions, for example related to the power of exotic quantifiers that have not been answered yet. 

\paragraph{Unifying Algorithmic Techniques.}

    Although, \ER is a relatively new complexity class, 
    the problems that are complete for it have a very long and rich history.
    Yet, the study of different algorithms for \ER-complete problems is almost independent. For instance, neural networks with billions
    of parameters are routinely trained in industry applications; the insights and techniques developed in this application of gradient descent to neural networks
    seem to have little effect on the way that the graph drawing community is drawing graphs. See~\cite{Stress-Minimization, GD-squared} for some first work to transfer knowledge from machine learning to graph drawing.
    For \NP-complete problems algorithmic techniques 
    that work well for one problem also work well for another problem.
    Therefore, those problems are often studied together.
    Yet the graph drawing, packing, and machine learning community are almost disjoint.

    It may be the case that the problem-specific differences are more important for \ER-complete problems and 
    thus a more unified approach might just not be very fruitful.
    Even then it would be nice to capture this in some way.

\section{A Toolbox for \ER}

In this section we review some tools that have shown themselves useful in various aspects of \ER-completeness, including the ball and gap theorems, which allow us to juggle with equalities and inequalities, quantifier elimination, and exotic quantifiers, which extend the language of the theory. 

\subsection{The ``Ball Theorem''}
\label{subsec:ball}

The  original ``ball theorem'' is due to Vorob\cprime ev and Grigoriev~\cite{V84,GV89}. 
The following statement is weaker than the currently best known results, see~\cite[Theorem 3]{BR10}, but sufficient for all of our applications to the existential theory of the reals. 
A version stated for semialgebraic sets can be found in~\cite[Corollary 3.4]{SS17}, and Renegar~\cite[Proposition 1.3]{R92} stated a version which is useful at higher levels of the hierarchy.

\begin{theorem}[Vorob\cprime ev and Grigoriev]\label{thm:ball}
 If $f_i \in \Z[x_1, \ldots, x_n]$, $i \in [k]$ is a family of polynomials with coefficients of bitlength at most $L$ and total degree at most $d$, then every connected component of $\{x \in \R^n: f_i(x) \geq 0$ for all $i \in [k]\}$ intersects the ball of radius $2^{(L+ \log{k}) d^{cn}}$, for some constant $c > 0$ independent of $f$.
\end{theorem}

We want to point out that this theorem relies on the coefficients being integers.
If they are real numbers then no upper bound on the radius is known.
To see the difficulty to translate such a result to real coefficients consider the
linear equation $aX - bX = 1$. The solution is $X= \tfrac{1}{a-b}$, which
can be arbitrarily large.

The innocent-looking Theorem~\ref{thm:ball} has some far-reaching consequences. 
To start with, it can be used to write 
a decision algorithm for \ER\ that runs in exponential time~\cite{GV89}. 

The ball theorem also implies that solutions to computational problems in \ER\ can be bounded in size. As an example, consider the order type realizability problem~\ourref[Order Type Realizability]{p:ordertype}. One way to measure the size of a solution of a geometric problem like order type realizability is by working with {\em area}; to make this notion non-trivial---a realizable order type can be realized in any arbitrary small region---one typically adds some resolution requirement, say that the points must have at least unit distance. We can then define the {\em area} of a realization to be the area of a smallest bounding box; the area of a realizable instance of order type then is the smallest area of a realization in which every pair of points has distance at least one. The following result, stated for spread rather than area, can be found in~Goodman, Pollack, and Sturmfels~\cite[Theorem 2]{GPS90}.

\begin{theorem}[Goodman,Pollack, and Sturmfels]\label{thm:GPSorder}
  A realizable order type on $n$ points can be realized in area $2^{2^{O(n)}}$, and there are order types that require area $2^{2^{\Omega(n)}}$. 
\end{theorem}

The proof is conceptually simple: we can express the realizability of an order type on points $(p_i)_{i \in [n]}$ as the feasibility of a system of quadratic polynomials $f_i(p) \geq 0$, $i \in [k]$, where $k = O(n)$. We add $O(n^2)$ polynomial conditions of the form $\norm{p_i-p_j}^2 -1 \geq 0$ to enforce the minimum resolution. 
By Theorem~\ref{thm:ball}, there is a solution in which all variables have distance at most $R = 2^{2^{O(n)}}$ from the origin; in particular, this applies to the coordinates of the points $(p_i)_{i \in [n]}$, implying the area upper bound. 

The main obstacle to upper bounds like this is, occasionally, the encoding of the conditions in \ETR. A recent example is right-angle drawability~\cite{B20,S23b}. This issue becomes even more intricate if one aims at optimal bounds. This can require significant additional work, see, for example, the papers by McDiarmid and M\"{u}ller~\cite{McDM13} and Kang and M\"{u}ller~\cite{KM12}.

Goodman, Pollack, and Sturmfels~\cite{GPS90} also proved the corresponding lower bound, using a construction that lends itself to adaptation to other settings. Double-exponential lower bounds were reproduced in several early papers for various problems including rectilinear crossing number~\ourref[Graph(s)!Rectilinear Crossing Number]{p:rectcross} in~\cite{B91}, segment intersection graphs~\ourref[Segment Intersection Graph]{p:SEG} in~\cite{KM94}, and realizability of $4$-polytopes~\ourref[Polytope(s)!4-dimensional Polytope Realizability]{p:4DPolytope} in~\cite{RGZ95}. For most \ER-hardness reductions, double-exponential precision carries over, so these lower-bound results are nearly automatic, and are rarely stated explicitly these days, unless extra work is required.

Because the order type problem is universal, we cannot expect its realizations to be possible on a grid, but if we restrict ourselves to point sets in general position, so no three points are collinear, then there always is a solution on a grid. 

\begin{theorem}[Goodman,Pollack, and Sturmfels]\label{thm:GPDorderrational}
  If an order type is realizable by $n$ points in general position, then there is a realization on a grid of size $2^{2^{O(n)}} \times 2^{2^{O(n)}}$, and there are order types that require grids of size  $2^{2^{\Omega(n)}}\times 2^{2^{\Omega(n)}}$. 
\end{theorem}

The proof extends the proof of Theorem~\ref{thm:GPSorder} with a careful analysis on how far points can be perturbed without changing the order type. Kachiyan and Porkolab~\cite[Corollary 2.6]{KP97} observed that work by Basu, Pollack and Roy implies the general result that every open (actually, every full-dimensional) semialgebraic set always contains a rational point of at most exponential complexity. 
This automatically applies to any problem in \ER\ whose solution space is open, such as 
the simple stretchability~\ourref[Simple Stretchability]{p:Stretch} and the rectilinear crossing number~\ourref[Graph(s)!Rectilinear Crossing Number]{p:rectcross}. In the case of right-angle drawings with one bend, a similar upper bound can be obtained even though the realization space is not open~\cite[Theorem 15]{S23b}.

Tarski showed that his quantifier elimination algorithm works over an arbitrary real-closed field, implying that a formula can be satisfied over \R if and only of it can be satisfied over the algebraic numbers (which are real-closed). It follows that any non-empty semialgebraic set contains an algebraic point. Kachiyan and Porkolab~\cite[Proposition~2.2]{KP97} strengthened this result by showing that (non-empty) semialgebraic sets always contain algebraic points of at most exponential complexity; that is, there is an exponential bound on the size of a (Thom) encoding of such a point. This conclusion applies to 
most of the remaining problems, such as stretchability~\ourref[Stretchability]{p:Stretch} and the art gallery problem~\ourref[Art Gallery Problem]{p:ArtGallery}. 

\medskip

Theorem~\ref{thm:ball} can also be used to show that feasibility~\ourref[Polynomial(s)!feasibility]{p:feasibility} remains \ER-complete if restricted to a compact domain, say $[-1,1]^n$. As far as we know, this was first shown by Schaefer~\cite[Lemma 3.9]{S13}, and there is no result like this (for compact domains) for $\NP_{\R}$.

\begin{theorem}[Schaefer]\label{thm:feasiblecompact}
  Testing whether a polynomial $f \in \Z[x_1, \ldots, x_n]$ has a root in $[-1,1]^n$ is \ER-complete.
\end{theorem}

For open domains, like $(-1,1)^n$, the proof is much simpler, since $z \mapsto \frac{z}{1-z^2}$ is a surjective map from $(-1,1)$ to $\R$, but there cannot be a continuous surjective map from $[-1,1]$ to $\R$, since the image of such a map would be compact. An extension of this result to the second level of the hierarchy was obtained by D'Costa, Lefaucheux, Neumann, Ouaknine, and Worrellin~\cite[Lemma 10]{DCLNOW21} and Jungeblut, Kleist, and Miltzow~\cite{JKM22}.

We will look at further consequences of Theorem~\ref{thm:ball} in Section~\ref{sec:gap}, where we will see how it can be combined with gap theorems to replace equalities with strict inequalities. It can also be used to remove the exotic \HQ-quantifier, see Lemma~\ref{lem:Hremove}.

\subsection{Gap Theorems}\label{sec:gap}

Gap theorems establish lower bounds on the distance between two algebraic objects that have positive distance. 
One of the oldest is Cauchy's theorem which establishes upper and lower bounds on roots of a univariate polynomial~\cite{Y00}.
This can be generalized, under certain assumptions, to multivariate polynomials, as was done for example by Canny~\cite{C88,C88b}.
One can derive the multivariate separations via quantifier elimination from Cauchy's theorem, but there are also direct proofs. The following theorem follows from a strong separation result due to Jeronimo and Perruci~\cite{JP09}, the simple derivation can be found in~\cite[Corollary 3.6]{SS17}.

\begin{theorem}[Jeronimo, Perruci]\label{thm:polygap}
 If $f \in \Z[x_1, \ldots, x_n]$ is a polynomial with coefficients of bitlength at most $L$ and total degree at most $d$, 
 for which $\inf_{x \in \R^n} f(x) > 0$,
 then 
 \[\inf_{x \in \R^n} f(x) > 2^{-(L n^d+1)d^{n+1}} d^{-(n+1)d^{n+1}}.\] 
\end{theorem}

With this theorem, one can establish useful separation bounds for semialgebraic sets, e.g.\ the following bound from~\cite[Corollary 3.8]{SS17}. For a formula $\varphi$ we write $|\varphi|$ for its bitlength. 

\begin{lemma}[Schaefer, {\v{S}}tefankovi\v{c}]\label{lem:semiseparate}
  Suppose the distance between two semialgebraic sets $S = \{x \in \R^n: \varphi(x)\}$ and $T = \{x \in \R^n: \psi(x)\}$ 
  is positive. Then that distance is at least $2^{-2^{L+5}}$, assuming $L := |\varphi|+|\psi| \geq 5n$.
\end{lemma}

Combining this separation with Theorem~\ref{thm:ball} one can show that testing whether $f(x) > 0$ for all $x \in \R^n$ for a given polynomial $f$ is \VR-complete~\ourref[Polynomial(s)!Strict Positivity!Global]{p:GlobalPos}: Given $f$, if $f(x) = 0$ for
some $x \in \R^n$, then by Theorem~\ref{thm:ball}, there is an $x \in [-R,R]^n$, where $R = 2^{2^{O(L)}}$ and $L$ is the bitlength of $f$. Consider the sets $S = \{(x,y) \in \R^{n+1}: f(x) = y, x \in [-R,R]^n\}$ and $T = \{(x,0): x \in [-R,R]^n\}$.
If $f(x) = 0$ for some $x$, then the distance of $S$ and $T$ is $0$, otherwise, the distance is positive (since both sets are compact), and so by Lemma~\ref{lem:semiseparate} that distance is at least $2^{-2^{O(L)}}$. Here we use repeated squaring to express $R$ in the definition of $S$ and $T$, and we can also use it to express a lower bound $\delta > 0$ on the distance of $S$ and $T$ (if $f(x) > 0$) of bitlength at most $O(L)$. It follows that $f(x) = 0$ for some $x \in \R^n$ if and only if $|f(x)|< \delta$ for some $x \in \R^n$. It follows that strict global positivity~\ourref[Polynomial(s)!Strict Positivity!Global]{p:GlobalPos} is \VR-complete.

Since this idea allows us to replace equalities with strict inequalities, it also follows that \ETR\ remains \ER-complete even if 
equality and negation are not allowed. This result, proved in~\cite{SS17}, is an indication that the definition of \ER\ is quite robust.
It implies, as we mentioned earlier, that problems which can be defined without equality, like simple stretchability and segment intersection graphs have the same computational complexity as problems like feasibility and stretchability, which use equality.

A similar approach was used by B\"{u}rgisser and Cucker~\cite{BC09} to eliminate the exotic \HQ-quantifier, see Lemma~\ref{lem:Hremove} below.

We saw that $f(x) = 0$ can be equivalently replaced by $f(x) < \varepsilon$, where $\varepsilon$ is a doubly exponential small constant (depending on $f$). Another way of reading this is that finding an approximate solution to the feasibility problem, with a double exponential error, is as hard as the feasibility problem itself. This type of statement could be made for nearly every \ER-complete problem, though it is rarely made explicitly. Examples include drawings which are nearly right-angle drawings~\ourref[Graph(s)!Right Angle Crossing Drawability@Right-Angle Crossing Drawability]{p:racdraw}, and approximating the Hausdorff distance~\ourref[Semialgebraic Set(s)!Hausdorff distance]{p:Hausdorffdist} to within a double-exponential error.

\subsection{Quantifier Elimination}\label{sec:QE}

    Suppose we are given a two-dimensional region $S = \{x \in \R^2: \varphi(x)\}$ and we want to determine whether $S$ has diameter at most $1$. How hard is that?  Using the first-order theory of the reals, we can express the diameter condition as
    \begin{equation*}
        (\forall x, y \in \R^2)\ x \in S \wedge y \in S \rightarrow \norm{x-y} \leq 1.
    \end{equation*}
    Let us be a bit more honest and replace $\in S$ with $\varphi$, and eliminate the hidden square roots in $\norm{\cdot}$. That gives us
    \begin{equation}\label{eq:diameter}
        (\forall x, y \in \R^2)\ \varphi(x) \wedge \varphi(y) \rightarrow \norm{x-y}^2 \leq 1.    
    \end{equation}

    We can decide the truth of that sentence by using quantifier elimination. There are very sophisticated quantitative versions of quantifier elimination depending on the number of variables, degree of polynomials, and other parameters. For our purposes, a very simple version, which only eliminates a single quantifier is sufficient. This version follows, for example, from~\cite[Algorithm 14.5]{BPR06}, and that reference presents many more variants. As earlier, we write $|\Phi|$ for the bitlength of formula $\Phi$.
    
    \begin{lemma}\label{lem:quantelim}
        Let $\Phi(x,y)$ be a quantifier-free formula, with $x \in \R$ and $y \in \R^n$. In time $|\Phi|^{O(n)}$ one can construct a formula $\Psi(y)$, of size $|\Phi|^{O(n)}$, such that
        \[(\exists x\in \R)\ \Phi(x,y)\ \leftrightarrow \Psi(y).\]    
    \end{lemma}

    The lemma allows us to remove one quantifier at a time, and it can be used for both existential and universal quantifiers (using negation). Applying the lemma four times (twice for each point) to~\eqref{eq:diameter} yields, in polynomial time since $n$ is bounded, a variable-free formula whose truth can then be evaluated in polynomial time.
    So having diameter at most $1$ can be decided in polynomial time for semialgebraic sets in $\R^2$, since there is only a constant number of quantifiers that need to be eliminated. An analogous argument works for every fixed dimension $d$. If $d$ is part of the input, so not fixed, then the problem becomes \VR-complete~\ourref[Semialgebraic Set(s)!diameter]{p:diamsemi}. Eliminating quantifiers no longer works in polynomial time in this case, since we have $\Theta(d)$ many variables. Quantifier elimination still gives us an (exponential-time) decision procedure, of course. 

    Quantifier elimination can be used to show that a large number of problems are polynomial-time solvable in fixed dimension; this includes problems like convexity~\ourref[Semialgebraic Set(s)!convexity]{p:convexity}, radius~\ourref[Semialgebraic Set(s)!radius]{p:radiussemi}, Brouwer fixed point~\ourref[Polynomial(s)!Brouwer Fixed Point]{p:BrouwerFixedPoint}, and many variants of the Nash equilibrium problem, see Section~\ref{sec:Nash}.

\subsection{Replacing Exotic Quantifiers}\label{sec:repexotic}
    Sometimes problems are not captured well by traditional quantifiers. Let us have a closer look at the dimensionality problem of semialgebraic sets.
    A semialgebraic set $S \subseteq \R^d$ is {\em full-dimensional} if it contains an open subset, or, equivalently, an open ball. We can express this formally as
    \[(\exists x \in \R^d)(\exists \delta \in R)(\forall y \in \R^d)\ \norm{y-x} < \delta \rightarrow y \in S.\]
    So testing whether $S$ is full-dimensional lies in \EVR, the second level of the hierarchy. The condition can be expressed more succinctly using $\exists^*$, one of the {\em exotic quantifiers} introduced by Koiran~\cite{K99}, $(\exists^* x \in \R^d)\ \varphi(x)$ is defined as $(\exists x \in \R^d)(\exists \delta \in \R_{>0})(\forall y \in \R^d)\ \norm{y-x} < \delta \rightarrow \varphi(y)$. With that, our characterization of full-dimensionality of $S$ simplifies to
    \[(\exists^* x\in \R^d)\ x \in S.\]
    Koiran then went on to show that $\exists^*$ can be replaced with $\exists$ by adding some variables, also see~\cite[Section 8]{BC09}:
    \begin{lemma}[Koiran]
        Given a formula $\Phi(x,y)$ in prenex form with a fixed number of quantifier alternations, one can in polynomial time construct a formula $\Psi(x',y)$ such that 
        \[(\exists^*x \in \R^d)\ \Phi(x,y)\ \leftrightarrow (\exists x' \in \R^{d'})\ \Psi(x',y),\]
        $d'$ is polynomially bounded in $d$, and $\Psi(x',y)$ has the same pattern of quantifier alternations as $\Phi(x,y)$.
    \end{lemma}
    For the full-dimensionality example we can let $\Phi$ be the defining formula of $S = \{x \in \R^d: \varphi(x)\}$, so there are no additional quantifiers or variables. It follows that being full-dimensional lies in \ER, and it is easily seen to be \ER-hard, so the real dimension of a semialgebraic set~\ourref[Semialgebraic Set(s)!dimension]{p:dimension} is \ER-complete.

    The second exotic quantifier, $\forall^*$, is the negation of $\exists^*$, so it intuitively corresponds to saying ``for a dense set'', and, as the negation of $\exists^*$ can be replaced by $\forall$ using Koiran's lemma.
    
    Koiran's result significantly extends the expressive power of the existential theory of the reals and has been used for other examples including positivity~\ourref[Polynomial(s)!Positivity!Global]{p:GlobalPos}, continuity~\ourref[Arithmetic Circuit!Continuity]{p:ContinuityAC}, and image density~\ourref[Arithmetic Circuit!Image density]{p:imagedensity}, also see~\cite[Table 2]{BC09}.

    \smallskip

    The third exotic quantifier, \HQ, is a bit different. Whereas $\exists^*$ and $\forall^*$ are concerned with density, \HQ\ is concerned with limits, and tends to appear when suprema or infima are involved. 
    
    Define $(\HQ \varepsilon)$ as $(\exists \varepsilon' > 0)(\forall \varepsilon \in (0,\varepsilon'))$; \HQ\ can be read as
    ``for all sufficiently small $\varepsilon$''; it is introduced in 
    B\"{u}rgisser and Cucker~\cite{BC09}. We saw earlier that having diameter at most $1$ lies in \VR. What about requiring the diameter to be strictly less than $1$? It is not sufficient to say
    \begin{equation*}
        (\forall x, y \in \R^d)\ x \in S \wedge y \in S \rightarrow \norm{x-y} < 1,
    \end{equation*}
    since the diameter is the supremum of the distances of points in $S$. The $\HQ$-quantifier offers a solution:
    \begin{equation*}
        (\HQ\varepsilon > 0)(\forall x, y \in \R^d)\ x \in S \wedge y \in S \rightarrow \norm{x-y} \leq 1-\varepsilon,
    \end{equation*} 
    

    By a result of B\"{u}rgisser and Cucker~\cite[Theorem 9.2]{BC09}, a leading \HQ-quantifier can be eliminated. 

    \begin{lemma}[B\"{u}rgisser, Cucker]\label{lem:Hremove}
    Let $\Phi(\varepsilon)$ be a formula in prenex form. We can construct a  
    formula $\Psi$ in polynomial time such that 
    \[(\HQ\varepsilon > 0)\ \Phi(\varepsilon) \]
    is equivalent to $\Psi$, and $\Psi$ has the same pattern of quantifier alternations as $\Phi(\varepsilon)$.
    \end{lemma}

    In particular, if $\Phi$ is an existentially or universally quantified formula, then so is $\Psi$. (This special case was independently discovered by Schaefer, \v{S}tefankovi\v{c}~\cite{SS17}, already used in~\cite{S13}.) From this we can conclude that the diameter problem lies in \VR.
    
    Other computational problems that have been placed in \ER or \VR using the elimination of a leading \HQ-quantifier,
    including several properties of semialgebraic sets such as 
    unboundedess~\ourref[Semialgebraic Set(s)!unboundedness]{p:unbounded}, 
    a specific point belonging to the closure of a set~\ourref[Semialgebraic Set(s)!Zero in closure]{p:ZeroinClosure}, 
    linkage rigidity~\ourref[Linkage!Linkage Rigidity]{p:linkagerigid}, 
    a point not being an isolated zero of a polynomial~\ourref[Polynomial(s)!Isolated Zero]{p:ISO}, and
    the distance of semialgebraic sets~\ourref[Semialgebraic Set(s)!distance]{p:distancesemi}.
    An example from a different area is the angular resolution of a graph~\ourref[Graph(s)!Angular Resolution]{p:angularres}, which is defined with a supremum, that can be eliminated using the \HQ-quantifer.

\section{A History of the Existential Theory of the Reals}\label{sec:History}

The existential theory of the reals has roots in several areas of mathematics. 
In logic, it traces back to the axiomatization of the real numbers around the turn of the 19th century. 
One of the major early results in that tradition is Tarski's decision procedure 
for the first-order theory of the reals based on quantifier elimination. Tarski's work implies that the first-order theory of the reals is decidable~\cite{T48}. This result has been sharpened over the years, with seminal contributions by Collins, Renegar, Canny and many others (for a modern treatment, see~\cite{B17}), and decision procedures are implemented in modern solvers, such as Microsoft's Z3 and the redlog system~\cite{R?,R?b}. 

The logic roots branched in various ways, but there are two main branches along which, independently, the existential theory of the reals was discovered as a complexity class. One branch stems from Mn{\"{e}}v's universality theorem and Shor's presentation of it, and the other branch from the Blum-Shub-Smale model of real computation. These two traditions did not intermingle until quite recently. 

\subsection*{\ER\ in The BSS Tradition}

Let us first have a look at the discovery of \ER\ in the BSS-model, as $\BP(\NP_{\R}^0)$ or sometimes $\mathsf{NPR}$. 
The BSS-model model of computation was introduced by Blum, Shub, and Smale in~\cite{BSS89}, followed by the book~\cite{BCSS98}. 
This seminal paper defined the real complexity classes $\P_{\R}$ and $\NP_{\R}$, and asked whether $\P_{\R} \neq \NP_{\R}$, the real equivalent of the $\P \neq \NP$ question. 
The paper also establishes a foundational result, an ``analogue of Cook's theorem for $\R$''\index{Cook Levein theorem for ER@Cook-Levin theorem for \ER}, namely that the feasibility of a polynomial of degree at most $4$ (with real coefficients),  is complete for $\NP_{\R}$~\cite[Main Theorem, Section 6]{BSS89}. 
This result and the definition of $\NP_{\R}$ are the first two steps towards establishing $\BP(\NP_{\R}^0)$ as a complexity class. 
Two more steps are needed, the real Turing machines need to be restricted in two ways: 
they need to be constant-free (see below), denoted by the $0$ superscript in $\NP_{\R}^0$, and the input needs to be binary, denoted by the $\BP$ operator. 

The real Turing machines in the BSS-model are allowed real constants, making them quite powerful (every countable set can be encoded in a real constant, and a real Turing-machine can extract that information, so the halting problem, for example, lies in $\NP_{\R}$). 
In the {\em constant-free} (aka parameter-free) model, 
the only constants allowed are $0$ and $1$; this is equivalent to allowing the machine to use rational or even algebraic constants~\cite{ABKPM08}. 
One of the earliest papers we have found studying constant-free machines is by Cucker and Koiran~\cite[Section~7]{CK95}; they study a complexity class based on existentially quantified formulas (over the reals) but with the matrix only containing integers; 
however they do so only for matrices with addition (and subtraction), no multiplication and division (leading to so-called additive machines, which were well-studied at the time). 
There are other contemporary papers that looked at constant-free machines, including Koiran~\cite{K97b,K97c}, but the superscript $0$ notation, $\NP_{\R}^0$, first occurs in papers by Fournier, Koiran~\cite{FK99} and Chapuis and Koiran~\cite{CK99}. 
An explanation of the notation is given in the paper by Chapuis and Koiran~\cite{CK99}: for a real complexity class ${\cal C}$, the authors introduce the notation ${\cal C}^k$, where the real machines defining ${\cal C}$ are restricted to have at most $k$ constants.  

The earliest papers we can find that consider restricting the inputs of the real Turing machines to binary inputs are due to Koiran~\cite{K92, K94}, but neither paper introduces a name for the restriction. 
Probably the first paper to do so is by Cucker and Koiran~\cite[Definition 1]{CK95} which defines and names the {\em Boolean part} operator $\BP$ explicitly, describing it as capturing the ``the computational power of real Turing machines over binary inputs''. 
Starting with this paper, the Boolean part (or binary part) is studied in various papers, including~\cite{CM96,CG97,K97b,MM97}. 


Some papers, e.g. the paper by Cucker and Koiran~\cite{CK95} work with both Boolean parts and constant-free machines, but they do not combine them to define $\BP(\NP_{\R}^0)$. 
This seems to have first happened in the paper by B\"{u}rgisser and Cucker~\cite{BC06}, in which the authors consider both 
$\BP(\NP_{\C}^0)$ and $\BP(\NP_{\R}^0)$, and show that the feasibility~\ourref[Polynomial(s)!feasibility]{p:feasibility} and real dimension problem~\ourref[Semialgebraic Set(s)!dimension]{p:dimension} (based on Koiran~\cite{K99}) are complete for $\BP(\NP_{\R}^0)$, see their Proposition 8.5.

There are relatively few papers with completeness results in the BSS-tradition, even for the real complexity class $\NP_{\R}$, something commented on in the introduction to B\"{u}rgisser and Cucker~\cite{BC06}: ``these first completeness results where not followed by an avalanche of similar results''. 


This changed somewhat with a later paper by B\"{u}rgisser and Cucker~\cite{BC09}. 
Section~9 of that paper introduces the operator $\BPZ$ which restricts the real Turing machines to be both constant-free and the input to be in binary. The authors then define the ``constant-free Boolean part'' $\BPZ({\cal C})$ of a complexity class ${\cal C}$. 
(So $\BPZ({\NP_\R})$ and $ \BP(\NP_{\R}^0)$ become two different ways to denote the same class.)
With this operator, B\"{u}rgisser and Cucker define several complexity classes,
including $\PR = \BPZ(\P_{\R})$, $\mathsf{NPR} = \BPZ(\NP_{\R})$, the existential theory of the reals, and $\mathsf{coNPR} = \BPZ(\coNP_{\R})$, the universal theory of the reals, and they look at higher levels of the hierarchy as well. (The real hierarchy in the BSS-model was introduced by Cucker~\cite{C93}.)

They then argue that several known completeness results for $\NP_{\R}$, $\coNP_{\R}$ and even higher levels transfer to the discrete setting nearly automatically (see their Table~1), namely feasibility~\ourref[Polynomial(s)!feasibility]{p:feasibility}, first shown $\NP_{\R}$-complete in the original paper by Blum, Shub, and Smale~\cite{BSS89}, convexity, by Cucker and Rossell\'{o}~\cite{CR92}, and the real dimension problem, shown $\NP_{\R}$-complete by Koiran~\cite{K99} (using exotic quantifiers). Also in this category is 
an early result by Zhang~\cite{Z92} on neural networks~\ourref[Neural Networks!Training]{p:neuralnettrain}.  

B\"{u}rgisser and Cucker~\cite{BC09} also added several new problems (see Table~2 in their paper), including unboundedness of semialgebraic sets~\ourref[Semialgebraic Set(s)!unboundedness]{p:unbounded}, which is \ER-complete, and surjectivity of a function~\ourref[Polynomial(s)!Surjectivity]{p:surjectivity}, which is complete for $\VER$, or $\BPZ(\forall\exists)$ in the B\"{u}rgisser-Cucker notation. The main thrust of that paper was not actually the discrete setting, which may explain why it took more than ten years for researchers in the Mn{\"{e}}v/Shor-tradition to find these results, as its title suggested, the paper focused on the power of the exotic quantifiers \HQ{}, $\forall^*$ and $\exists^*$ introduced by Koiran~\cite{K99}. For example, ${\exists}^* x$ means ``for an open set of $x$''. B\"{u}rgisser and Cucker, building on the work of Koiran, were able to show that in many cases, and in all interesting cases at the first level, the exotic quantifiers $\forall^*$ and $\exists^*$ can be replaced with their non-exotic counterparts, and a fixed number of \HQ-quantifiers can be eliminated, see Section~\ref{sec:repexotic}. This shows that the classes $\PR$, $\mathsf{NPR}$, and $\mathsf{coNPR}$ are robust under modifications of the quantifiers, for more recent work extending this see~\cite{JJ23}. 

A comprehensive survey of (structural) complexity results in the BSS-model was written by Baartse and Meer~\cite{BM13}.

\subsection*{\ER\ in The  Mn{\"{e}}v/Shor Tradition} 

Perhaps the first paper to view the existential theory of the reals as a complexity class, without defining it explicitly, 
is Shor's presentation of Mn{\"{e}}v's universality theorem~\cite{M88,S91}. Mn{\"{e}}v showed that every semialgebraic set is ``equivalent'', in a technical sense, to the realization space of a stretchable arrangement of pseudolines (see~\ref{subsec:univ}). Mn{\"{e}}v's work was in algebraic geometry, and his notion of equivalence aimed at capturing algebraic similarity. Shor's paper recasts Mn{\"{e}}v's result with a computer science audience in mind; Mn{\"{e}}v's universality theorem is phrased as a reduction, showing how to translate a
system of equalities and inequalities into a stretchability problem, where equivalence is now at the decision level: the system of equalities and inequalities is solvable if and only of the pseudoline arrangement constructed by the reduction is stretchable. Shor's paper is also the first to show that \NP\ reduces to the stretchability problem, implying that $\NP \subseteq \ER$ (a result which is immediate if one starts with \ETR\ as the complete problem for \ER). 

Together with Kempe's 19th-century universality theorem on linkages~\cite{K77}, Mn{\"{e}}v's
result led to other universality theorems in algebraic geometry, but these often were not concerned with being effective reductions. Kempe's result can be made effective, implying that linkage realizability~\ourref[Linkage!Linkage Realizability]{p:linkagereal} is \ER-complete~\cite{KM02,A08,S13}), but this is not the case for all universality results, e.g.~\cite{S87,D03}.

Shortly after Shor's paper was published, Kratochv\'{\i}l and Matou\v{s}ek~\cite{KM94} showed that the segment intersection graph problem~\ourref[Segment Intersection Graph]{p:SEG} is as hard for the existential
theory of the reals, reducing from a system of strict inequalities (they cite a manuscript version of Shor's paper dated to 1988).  Kratochv\'{\i}l and Matou\v{s}ek
add two observations (based on work by Goodman, Pollack, and Sturmfels~\cite{GPS89}, which in turn is based on results of Grigor{\cprime}ev and Vorobjov~\cite{GV89} in algebraic geometry):  first, the segments realizing a particular segment intersection graph may require
double exponential precision to specify (in the number of segments), and double exponential precision is sufficient. 

A manuscript version of Kratochv\'{\i}l and Matou\v{s}ek~\cite{KM94} then influenced Bienstock~\cite{B91} who showed that the rectilinear crossing number~\ourref[Graph(s)!Rectilinear Crossing Number]{p:rectcross} is as hard as solving a system of strict inequalities, and also derives upper and lower bounds on the required precision of the vertex coordinates.

Other early papers influenced by the work of Shor, Kratochv\'{\i}l and Matou\v{s}ek include 
Tanenbaum, Goodrich, and Scheinerman~\cite{TGS95} on the point-halfspace order problem~\ourref[Point half space orders in Rd@Point-halfspace orders in $\R^d$]{p:pointhalfspaceorders}; 
Richter-Gebert and Ziegler~\cite{RGZ95} on the Steinitz problem; 
Davis, Gotts, and Cohn~\cite{DGC99} on topological inference with convexity~\ourref[Topological Inference in RCC8 with Convexity]{p:RCC8con}; 
Buss, Frandsen, and Shallit~\cite{BFS99} on various matrix problems (see Section~\ref{sec:MatandTen}), and  
Bose, Everett, Wismath~\cite{BEW03} on the arrangement graph problem~\ourref[Arrangement Graph]{p:arrangegraph}. 
The paper by Buss, Frandsen, and Shallit~\cite{BFS99} in particular comes pretty close to treating the existential theory of the reals (and other fields) as a complexity class.

Schaefer~\cite{S10} defined the existential theory of the reals as we defined it in this survey, as a computational complexity class based on \ETR, and introduced the notation \ER; this paper also mentioned the result (not published until 2017 by Schaefer and \v{S}tefankovi\v{c}~\cite{SS17}) that the definition of \ER\ is robust in the sense that if we restrict the language of \ETR\ to not allow equality and negation---so we can only express strict inequalities, then the resulting complexity class $\ER_{<}$ is the same as \ER. This implies that problems like simple stretchability, rectilinear crossing number and segment intersection graphs, which do not require equality, are as hard as problems like feasibility and stretchability, which do. 

In the following years, a majority of the research in the Mn{\"{e}}v/Shor tradition stayed in the areas of graph drawing and computational geometry, but some of the research branched off, looking at problems in analysis~\cite{OW14, S16b,OW17,SS17,SS18}, algebra~\cite{HL13,BK15,S16,BRS17,S17}, and other areas~\cite{HZ16,P16,P17}.

Inevitably, the Mn{\"{e}}v/Shor tradition also discovered the higher levels of the real polynomial hierarchy; this happened in a paper by Dobbins, Kleist, Miltzow, and Rz\polhk a\.{z}ewski~\cite{DKMR18,DKMR23}, who introduced $\EVR$ and $\VER$; several subsequent papers located problems beyond the first level of the hierarchy, including~\cite{BH21,DCLNOW21,JKM23}. Robustness results for all higher levels, similar to the ones known for the first level, were shown by Schaefer and \v{S}tefankovi\v{c}~\cite{SS23} extending work in~\cite{DCLNOW21,DKMR23}.

\subsection{Statistics}
\label{sec:Stats}

\begin{figure}[htbp]
    \centering
    \includegraphics[width =0.8\textwidth]{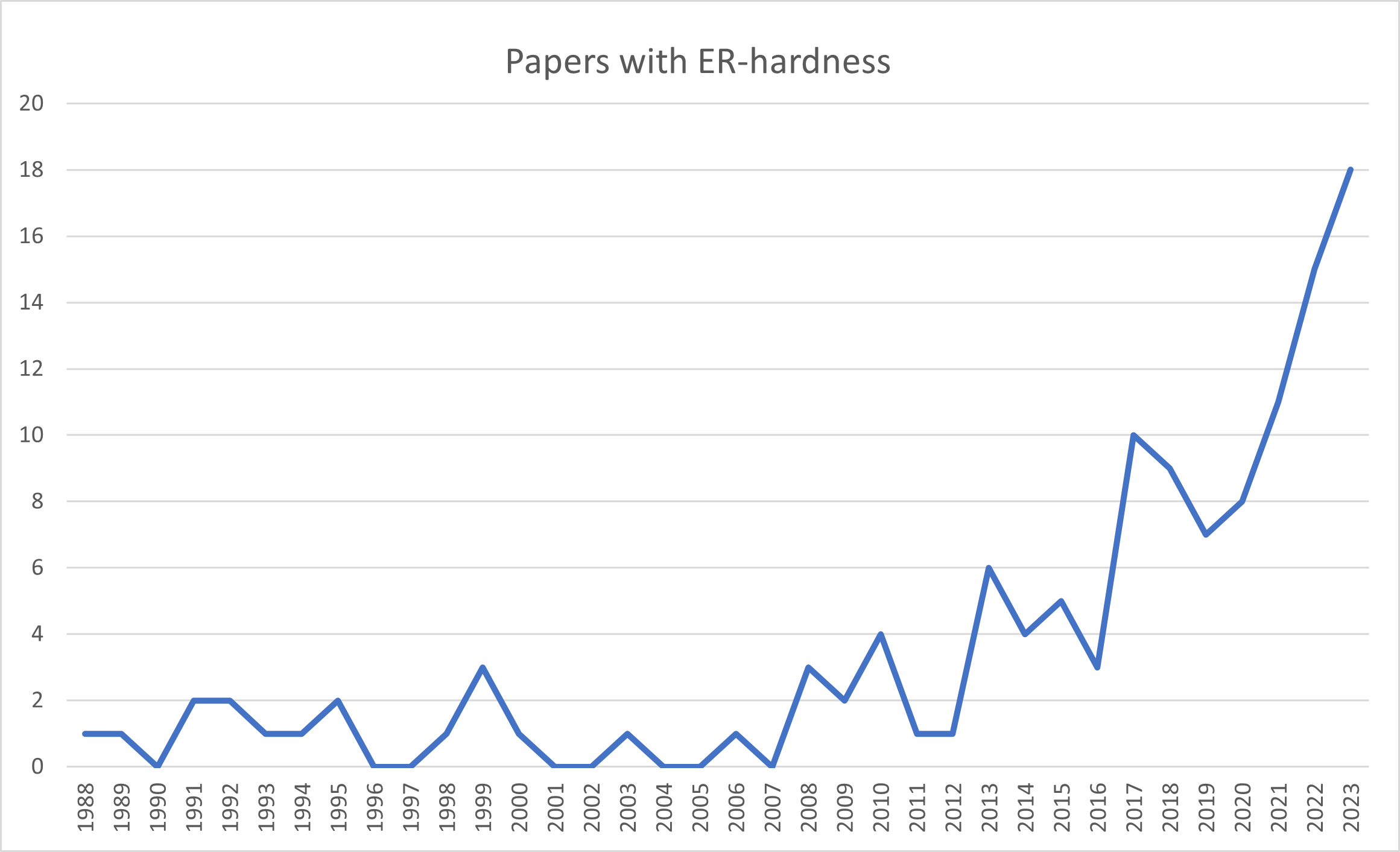}
    \caption{Number of papers with \ER-hardness results by year.}
    \label{fig:stats}
\end{figure}

Figure~\ref{fig:stats} shows the number of publications each year that (explicitly or implicitly) contain
\ER-hardness results. If a paper was published in multiple forms (conference/journal/preprint), we only counted the first version. Many of these papers contain multiple \ER-hardness results. 

Computational geometry is by far the most active area for \ER-complete problems, though we have to admit that the classification of problems is somewhat ambiguous.

\newpage
\part{A Tour of the Compendium}

We give a very brief preview of the compendium and some broader context.

Our compendium collects \ER-complete problems, but it also goes farther by including \ER-hard problems, as long as they belong to higher levels of the polynomial hierarchy beyond \ER, that is, problems that can be captured by the theory of the reals with a bounded number of quantifier alternations.

Within the entries, we will point out related universality results if we are aware of them. As we mentioned, there are many different notions of universality, depending on the equivalence relation that is used. We refer to Verkama~\cite{V23c} for a recent discussion on the notion of stable equivalence used by Richter-Gebert~\cite{RG96}.

We will not discuss problems complete for several other complexity classes that are related to \ER, but probably different. This includes $\NP^{\mathrm{SSQR}}$, $\P^{\mathrm{SSQR}}$, and $\ER\cap \VR$---which we already discussed earlier, variants of the BSS computation model, and classes such as \FIXP\ and \PPAD, which are important in the context of game theory. 

We also make no attempt to include problems complete for other existential theories, such as the existential theory of arbitrary rings and fields introduced in~\cite{BFS99}. The most interesting are perhaps the existential theory of $\operatorname{GF}(2)$, which yields \NP, \EQ, whose complexity is still unknown (possibly undecidable)~\cite{P09}, \EN, Hilbert's Tenth problem, which is undecidable by results of Davis, Robinson, and Matiyasevich~\cite{M06b}, and \EC\ which lies in the second level of the polynomial hierarchy by a result of Koiran's, assuming the generalized Riemann hypothesis~\cite{K97}.
However, if natural variants of \ER-hard problems turn out to be complete for other existential theories, we will mention this fact. For example, \EQ\ is relevant in the field of graph drawing, since it naturally occurs when looking at grid variants of graph drawing problems such as planar slope number, visibility graph, and right-angle crossing drawings, or in rational Nash Equilibria~\cite{BH22}.

The entries in the compendium are mostly self-contained, though some common notions are defined at the beginning of each  section. To keep the entries brief and focused on complexity aspects, we do not include definitions of some of the more intricate notions, relying on references in that case. We include cross-references between related entries. 

There are plenty of open problems, both within the entries, and in Section~\ref{sec:Candidates} on candidate problems. We have not carefully vetted all papers listed in this compendium, so we cannot guarantee correctness of results. Instead we opted to be inclusive to cover more ground, so we also include preprints; this allows the reader to follow up with the original sources. 
We will mention issues with papers if we are aware of them.

\section{Logic}
\label{sec:Tour-Logic}

The existential theory of the reals~\ourref[Existential Theory of the Reals]{p:ETR}, as the name suggests, is anchored in logic, as it is defined by the language \ETR, a fragment of the theory of the reals. As we have seen, \ETR\ is a powerful, and expressive language, which makes it useful for upper bounds, but inconvenient for reductions; we would prefer restricted versions of this language, and, as it turns our, \ER\ is quite robust when it comes to the exact definition and signature of the underlying logic. Variants include forcing variables to be distinct~\ourref[Distinct ETR@Distinct \ETR]{p:DistETR}, or ordered~\ourref[Existential Theory of Totally Ordered Real Variables]{p:ETRORV}, or allowing them to be approximate~\ourref[Approximate Existential Theory of the Reals]{p:epsETR}. Another family of variants looks is based on replacing multiplication with inversion, starting with \ETRINV~\ourref[ETRINV@\ETRINV]{p:ETRINV}, and including \PETRINV~\ourref[Planar ETRINV@\PETRINV]{p:PlanarETRINV} in which the variable constraint graph is planar, and other nonlinear constraints~\ourref[Continuous Constraint Satisfaction Problem]{p:CSSP}.  

These restrictions are useful as intermediate problem for reductions to show that other problems are \ER-complete.

A number of problems in this section also involve other formal systems such as the \textit{collinearity logic}~\ourref{p:CollLogic} 
the \textit{$\mathcal{LR}$-calculus} involving left/right orientations of triples of points in the plane~\ourref{p:LRCalculus},
the \textit{region-connection calculus}~\ourref{p:RCC8con}, 
or Birkhoff and von Neumann's \textit{quantum logic}~\cite{BvN36}~\ourref{p:QuantumStrong}, \ourref{p:QuantumWeak}.

\section{Algebra}
\label{sec:Tour-Algebra}

The section on algebra is split into three parts.
We deal with properties of polynomials and arithmetic circuits; matrices and tensors; as well as semialgebraic sets.

\begin{table}[htbp]
    \centering
    \caption{Properties of functions and their complexity.
    We use the abbreviation AC to indicate that the result is known for arithmetic circuits.}
    \vspace{0.2cm}
    \label{tab:polynomialProperties}
    \begin{tabular}{|l|c|p{7cm}|}
   \hline
   \multicolumn{1}{|c|}{Property}  & \multicolumn{1}{c|}{Complexity} & \multicolumn{1}{c|}{Source} \\ \hline
    Brouwer Fixed Point~\ourref{p:BrouwerFixedPoint} &\ER-complete & Schaefer, \v{S}tefankovi\v{c}~\cite{SS17}\\
    Continuity (AC)~\ourref{p:ContinuityAC} & \VR-complete & B\"{u}rgisser, Cucker~\cite[Corollary 9.4]{BC09}\\
    Continuity at a Point (AC)~\ourref{p:ContinuityatPointAC} & \VR-complete & B\"{u}rgisser, Cucker~\cite[Corollary 9.4]{BC09}\\
    Domain Density (AC)~\ourref{p:DomainDensityAC} & \VR-complete & B\"{u}rgisser, Cucker~\cite[Proposition 5.7, Corollary 9.4]{BC09}\\
    Feasibility \ourref{p:feasibility} & \ER-complete & folklore \\
    Fixed Point~\ourref{p:FixedPoint} & \ER-complete & Schaefer, \v{S}tefankovi\v{c}~\cite{SS17}\\
    (Strict) Global Positivity~\ourref{p:GlobalPos} & \VR-complete  & Ouaknine and Worrell~\cite[Theorem 7]{OW17}\\
    Image Density (AC)~\ourref{p:imagedensity} & \VER-complete & combining results from B\"{u}rgisser, Cucker~\cite[Proposition 5.7, Corollary 9.4]{BC09} and Junginger~\cite[Theorem 4.27]{JJ23}\\
    Injectivity (polynomial)~\ourref{p:injectivity} & \VR-complete & see entry\\
    Isolated Zero~\ourref{p:ISO} & \VR-complete & Koiran~\cite{K00}, Schaefer~\cite{S13} \\
   (Strict) Local Positivity~\ourref{p:LocalPos} & \ER-complete &  Schaefer, \v{S}tefankovi\v{c}~\cite{SS17}, also Ouaknine and Worrell~\cite[Theorem 7]{OW17}\\
    Nontrivial Homogeneous Zero~\ourref{p:HTN} & \ER-complete & Schaefer~\cite[Corollary 3.10]{S13}\\
    Polynomial Projection (Linear)~\ourref{p:PolyProj} & \ER-complete & Bl\"{a}ser, Rao, and Sarma~\cite{BRS17}\\
    Regularity (polynomial)~\ourref{p:regularity} &\VR-complete & Cucker, Rossell\'{o}~\cite{CR92}\\        Surjectivity~\ourref{p:surjectivity} & \VER-complete & Schaefer, \v{S}tefankovi\v{c}~\cite{SS23} \\
    Totality (AC)~\ourref{p:totality} & \VR-complete & 
    B\"{u}rgisser, Cucker~\cite[Proposition 4.1, Corollary 9.4]{BC09}\\
    \hline
    \end{tabular}
\end{table}

\subsection{Polynomials and Arithmetic Circuits.}
To us arithmetic circuits and polynomials are ways to encode functions $f:\R^n \rightarrow \R$.
Every arithmetic circuit can be expressed as a polynomial and every polynomial can be expressed as an arithmetic circuit.
The difference between the two is that certain functions can be expressed very efficiently using arithmetic circuits, but very lengthy using polynomials. 
The easiest example is the function $f(x,y) = (x+y)^{2^n}$ using general expressions,
we just showed how to encode it using roughly $\log n$ bits. 
We can express the function with arithmetic circuits using a linear number of bits, as can be seen in the drawing.

\begin{center}
    \includegraphics{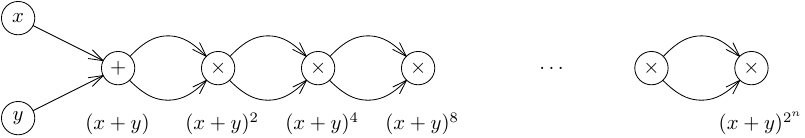}
\end{center}

Let us now consider the function~$f$ written down as a polynomial.
With $N = 2^n$, the function written as polynomial equals

\[f = \sum_{i=0}^{N} \,   \binom{N}{i} \, x^{i}y^{N-i}.\]

In this representation, 
we have an exponential number of coefficients, many of them having an exponential-size binary representation.
We are interested in properties like, does the function have a zero, can it be 
positive, is it convex, etc. 
Table~\ref{tab:polynomialProperties} contains an overview of studied properties and their complexity.

\begin{table}[tbp]
   \caption{Properties of Matrices and Tensors.
   }
   \label{tab:MatrixProp}
\begin{center}
   \begin{tabular}{|p{6cm}|c|p{6cm}|}
   \hline
   \multicolumn{1}{|c|}{Property}  & \multicolumn{1}{c|}{Complexity} & \multicolumn{1}{c|}{Source} \\ \hline
    Matrix Singularity~\ourref{p:MatSing}  & \ER-complete & Buss, Frandsen, Shallit~\cite{BFS99}  \\ 
    Minimum/Maximum Matrix Rank~\ourref{p:minmaxmatrank}  & \ER-complete & Buss, Frandsen, Shallit~\cite{BFS99}\\ 
    Minimum Rank of Linear Combination of Matrices~\ourref{p:minrankspanmat}  & \ER-complete & Bl\"{a}ser, Ikenmeyer, Lysikov, Pandey, and Schreyer~\cite[Theorem 2.3]{BILPS21}  \\ 
    Minimum Rank of Sign Pattern Matrix~\ourref{p:minranksignpatmat}  &  \ER-complete & Bhangale, Kopparte~\cite{BK15}  \\ 
    Monotone Matrix Rank~\ourref{p:monmatrank}  & \ER-complete & Lienkaemper~\cite{L23b} \\ 
    Non-negative Matrix Factorization~\ourref{p:nonnegmatfac}  & \ER-complete & Shitov~\cite{S16}  \\ 
    (Real) PSD Rank~\ourref{p:PSDrank}  &  \ER-complete & Shitov ~\cite[Theorem 3.4]{S23PSD}  \\ 
    (Real) CPSD Rank~\ourref{p:CPSDrank}  &   \ER-completen & Shitov~\cite[Theorem 3.10]{S23PSD} \\ 
    Phaseless Rank~\ourref{p:phaselessrank}  & \ER-complete & Shitov~\cite[Theorem 3.7]{S23PSD} \\ 
    Rank-Constrained Optimization~\ourref{p:rankconrtainedopti}  & \ER-complete & Bertsimas, Cory-Wright, and Pauphilet~\cite{BCWP22}  \\ 
    Tensor Rank~\ourref{p:tensorrank}  & \ER-complete & Schaefer, \v{S}tefankovi\v{c}~\cite{SS18} \\ 
    Tensor with Eigenvalue $0$~\ourref{p:tensorzero}  & \ER-complete & Schaefer~\cite[Corollary 3.10]{S13} \\ 
    Orbit Closure Containment~\ourref{p:orbitcontain}  & \ER-complete & Bl\"{a}ser, Ikenmeyer, Lysikov, Pandey, and Schreyer~\cite{BILPS21}\\ 
    
    \hline
   \end{tabular}
\end{center}
\end{table}

\subsection{Matrices and Tensors.}
Given an $n\times n$ matrix $M$, we can compute 
 a decomposition $M = A \cdot B$ ,
into a $n\times k$ matrix $A$ and an $k \times n$ matrix $B$ quickly,
if such a decomposition exists.
The underlying reason is a hidden linear structure.
The rank is the smallest
$k$ for which such a decomposition exists.

Can we also compute quickly closely related concepts of decomposition and rank?
The answer seems to be no in most cases.
The easiest example is arguably the non-negative rank of a matrix.
Here, we require that both that all entries of the input and the output are non-negative numbers.
Due to close inspection of the matrix

\[M = \begin{pmatrix}
1 & 1 & 0 & 0\\
0 & 1 & 1 & 0\\
0 & 0 & 1 & 1\\
1 & 0 & 0 & 1\\
\end{pmatrix},
\]
it was observed~\cite{thomas1974rank} that the non-negative rank can be different from the ordinary rank in 1974.
But it took until 2017, to note that a  non-negative decomposition might require irrational numbers~\cite{chistikov2017nonnegative}.
Shortly afterwards this result was improved to \ER-completeness by Shitov~\cite{S17}.
Thus, we see that the seemingly small requirement of non-negativity 
makes a big leap in terms of exploitable structure that we can use to find efficient algorithms.

Table~\ref{tab:MatrixProp} lists matrix and tensor properties for which we \ER-completeness are known.
Hillar and Lam~\cite{HL13} show that a fair number of problems related to tensors are \NP-hard, many of them may be candidates for \ER-completeness.

\subsection{Properties of Semialgebraic Sets.}

The study of properties of semialgebraic sets and their complexity began in the BSS-model, and, in a way, was a motivating influence. Early papers include Cucker and Rossell\'{o}~\cite{CR92}, settling boundedness, convexity, cardinality, and closure; and Koiran~\cite{K99} on the dimension problem. Later, many more problems were added by B\"{u}rgisser and Cucker~\cite{BC06,BC09}, and there are now additional results~\cite{SS17,JKM23,SS23}. Some problems remain intriguingly open, such as the complexity of deciding whether a semialgebraic set is closed. Table~\ref{tab:PropSemi} summarizes the known results.

\begin{table}[htbp]
   \caption{Properties of Semialgebraic Sets and their Complexity. Membership for closedness and compactness only applies to basic semialgebraic sets.
   }
   \label{tab:PropSemi}
\begin{center}
   \begin{tabular}{|l|c|p{7cm}|}
   \hline
   \multicolumn{1}{|c|}{Property}  & \multicolumn{1}{c|}{Complexity} & \multicolumn{1}{c|}{Source} \\ \hline
    Cardinality (Emptiness)~\ourref{p:cardinality}  & \VR-complete & Cucker, Rosell\'{o}~\cite{CR92} \\ 
    Convexity~\ourref{p:convexity} & \VR-complete & Cucker, Rosell\'{o}~\cite{CR92} \\ 
    Closedness~\ourref{p:closedness} & \VR-complete & B\"{u}rgisser, Cucker~\cite[Theorem 6.15, Corollary 9.4]{BC09}  \\
    Closure~\ourref{p:ZeroinClosure} & \ER-complete & Cucker, Rosell\'{o}~\cite{CR92} \\
    Compactness~\ourref{p:closedness} & \VR-complete & B\"{u}rgisser, Cucker~\cite[Theorem 6.15, Corollary 9.4]{BC09}  \\
    Density~\ourref{p:densitysemi} & \VR-complete & Koiran~\cite{K99} and B\"{u}rgisser, Cucker~\cite[Corollary 9.3]{BC09}, also see Schaefer, \v{S}tefankovi\v{c}~\cite{SS23} \\
    Diameter~\ourref{p:diamsemi} & \VR-complete &  Schaefer, \v{S}tefankovi\v{c}~\cite{SS23}\\ 
    Dimension~\ourref{p:dimension} & \ER-complete & Koiran~\cite{K99}, also see B\"{u}rgisser, Cucker~\cite[Corollary 9.4]{BC09} \\ 
    Distance~\ourref{p:distancesemi} & \ER-complete & Schaefer, \v{S}tefankovi\v{c}~\cite{SS17} \\
    Isolated Point~\ourref{p:isopoint} & \VR-complete & B\"{u}rgisser, Cucker~\cite[Corollary 9.4]{BC09} \\
    Hausdorff distance~\ourref{p:Hausdorffdist} & \VER-complete & Jungeblut, Kleist, Miltzow~\cite{JKM23},  also see Schaefer, {\v{S}tefankovi\v{c}}~\cite{SS23} \\
    Hyperplane Local Support~\ourref{p:LocSupp} & \ER-complete & B\"{u}rgisser, Cucker~\cite{BC09}, also see Schaefer, \v{S}tefankovi\v{c}~\cite{SS23} \\
    Local Dimension~\ourref{p:locdimension} & \ER-complete & B\"{u}rgisser and Cucker~\cite[Corollary 6.6,Corollary 9.4]{BC09} \\
    Radius~\ourref{p:radiussemi} & \VR-complete &  Schaefer, \v{S}tefankovi\v{c}~\cite{SS23}\\ 
    Unboundedness~\ourref{p:unbounded} & \ER-complete & Cucker, Rosell\'{o}~\cite{CR92}, also see B\"{u}rgisser, Cucker~\cite[Corollary 9.4]{BC09} \\ 
    \hline
   \end{tabular}
\end{center}
\end{table}

Other areas of algebra, such as group theory, are currently absent; Meer and Ziegler~\cite{MZ09} asked an interesting question in this context: is there a group whose word-problem is equivalent to $\NP_{\R}$? We can ask the same question for $\ER$ as well.

We want to mention that the term semialgebraic set is ambiguous in the literature. 
Many authors in algebraic geometry allow real coefficients of the defining polynomials, whereas in theoretical computer science it is more common to only allow integer coefficients.

\section{Matroids, Arrangements, Order Types, and Polytopes}

Mn{\"e}v's Universality Theorem~\cite{M88,RG99}, and its later complexity-theoretic interpretation by Shor~\cite{S91}, is one of the seminal result on which much of the more recent literature on \ER\ builds upon.

The theorem is usually stated in terms of realization spaces of \emph{oriented matroids}: Any semialgebraic set is \emph{stably equivalent} to the realization space of an oriented matroid. Just like matroids, oriented matroids can be defined in a variety of ways (\emph{cryptomorphisms}). They are intended to be a common abstraction of directed graphs, configurations of vectors, and arrangement of hyperplanes~\cite{BLVSWZ99}. One standard way of defining an oriented matroid of rank $d$ is via a \emph{chirotope}: a map from ordered $d$-tuples of elements of a ground set to the set $\{0,-1,+1\}$, that satisfy 
certain relations, coined the \emph{chirotope axioms}, derived from properties of determinants. In the computational geometry literature, rank-three chirotopes, or their isomorphism classes, are often referred to as \emph{order types}~\cite{GP83,AAK02}. The oriented matroid of rank three corresponding to a configuration of $n$ points in the plane is defined  by the order type that encodes the orientation of each ordered triple of points: clockwise, counterclockwise, or collinear. In general, in dimension $d$, the order type simply encodes the sign of the determinants defined by the ordered $d$-tuples of points. However, not all oriented matroids correspond to arrangements of points. The \emph{realization space} of an oriented matroid $M$ of rank $d$ on $n$ elements is the subset of $\R^{dn}$, every point of which corresponds to a set of $n$ points in $\R^d$ realizing $M$. Deciding whether this set is nonempty, hence whether $M$ is realizable in $\R^d$, is a canonical \ER-complete problem~\ourref{p:ormatroidreal}. Completeness actually holds for matroids as well~\ourref[Matroid!Real Realizability]{p:matroidreal}.

The \emph{Topological Representation Theorem} of Folkman and Lawrence~\cite{FL78} gives a convenient representation of \emph{any} oriented matroid, not only the representable ones: Oriented matroids of rank $d$ are in bijection with equivalence classes of arrangements of \emph{pseudo-hemispheres} of $\mathbb{S}^{d-1}$, whose boundaries are not geometric great circles, but behave combinatorially so. In rank three, for instance, oriented matroids are one-to-one with \emph{oriented pseudocircle arrangements} on the 2-sphere. Since Mn{\"e}v's Universality Theorem holds for fixed rank three, and applying some elementary simplifications, we obtain that the decision problem for the realizability of rank-three oriented matroid is essentially the pseudoline stretchability problem described earlier~\ourref[Stretchability]{p:Stretch}. Simple pseudoline arrangements are in fact one-to-one with \emph{uniform acyclic oriented matroids of rank 3}. We refer the interested reader to the standard textbook on oriented matroids~\cite{BLVSWZ99} for more details about the cryptomorphisms of oriented matroids and the definitions of the terms. Detailed and simplified proofs of the Topological Representation Theorem were given by Bokowski, Mock, Streinu~\cite{BMS01}, and Bokowski, King, Mock, Streinu~\cite{BKMS05}. An alternative treatment of order types corresponding to rank-three uniform oriented matroids was also given by Knuth~\cite{Knuth92}. The projective duality between sets of points and arrangements of hyperplanes directly yields that realizability of order types as point sets in the plane is also \ER-complete~\ourref[Order Type Realizability]{p:ordertype}.

Combinatorial abstractions of configurations of points or vectors have been defined in various ways in computational geometry. 
From the previous developments, it was shown that geometric realizability of many of them was \ER-hard to determine. These include Goodman and Pollack's allowable sequences~\ourref{p:allowableseq}~\cite{GP80}, radial systems~\ourref{p:radialsystem}, directional walks~\ourref{p:directwalk}, and convex geometries~\ourref{p:convexgeoreal} as defined by Edelman and Jamison~\cite{MR0815204}.

The Universality Theorem also applies to realization spaces of combinatorial types (face lattices) of polytopes~\cite{RGZ95,RG96,RG99,PT14,APT15}.
As a consequence, the problem of deciding whether a given combinatorial type of polytope is realizable is also a standard \ER-complete problem~\ourref[Polytope(s)!Polytope Realizability]{p:Polytope}.
A major achievement in the theory of \ER-completeness is the proof, due to Richter-Gebert~\cite{RG96}, that the problem is hard already in dimension 4~\ourref[Polytope(s)!4-dimensional Polytope Realizability]{p:4DPolytope}. 
The result also holds for simplicial and inscribed polytopes in arbitrary dimension~\ourref[Polytope(s)!4-dimensional Polytope Realizability]{p:ISPolytope}.
\emph{Delaunay triangulations} are projections of inscribed simplicial polytopes, and are therefore also hard to recognize in arbitrary dimension~\ourref[Delaunay Triangulation]{p:Delaunaytri}.

\section{Graph Drawing}

\begin{figure}[htbp]
    \centering
    \includegraphics[height=2in]{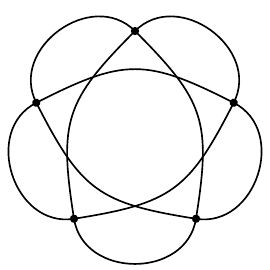}   
    \caption{A Lombardi drawing of $K_5$, each vertex has perfect angular resolution and edges are drawn as circular arcs. 
    }
    \label{fig:LombardiK5}
\end{figure}

Graph drawing forms the theoretical core of information visualization; it studies drawing styles of graphs---such as topological, polyline and straight-line drawings, and parameters that measure a drawing's readability, such as crossing number and rectilinear crossing number. Figure~\ref{fig:LombardiK5} illustrates a more recent drawing style, Lombardi drawings in which all vertices must have perfect angular resolution, that is, at each vertex, the angles between consecutive edge ends must be the same, and edges are realized by circular arcs~\ourref[Graph(s)!Lombardi Drawability with Rotation System@Lombardi-Drawability with Rotation System]{p:lombardidraw}.

Computational problems in graph drawing are often solvable in polynomial time if they are related to planarity and embeddability, or \NP-complete, typically when non-planarity comes into play. As soon as one replaces topological drawings, in which edges are represented as curves, with geometric drawings in which edges are straight-line or polygonal, the complexity often jumps from \NP\ to \ER. This applies to problems such as crossing number or pseudolinear crossing number versus rectilinear crossing number~\ourref[Graph(s)!Pseudolinear vs Rectilinear Crossing Number]{p:pseudorectcross}, or local crossing number versus rectilinear local crossing number~\ourref[Graph(s)!Rectilinear Local Crossing Number]{p:rectloccross}, weak realizability vs weak rectilinear realizability~\ourref[Graph(s)!Weak Rectilinear Realizability]{p:rectilinearreal}, simultaneous embedding vs simultaneous geometric embedding~\ourref[Graph(s)!Simultaneous Geometric Embedding]{p:SGE}, and thickness vs geometric thickness~\ourref[Graph(s)!Geometric Thickness of Multigraph]{p:geothickmulti}. 

\ER-hardness of straight-line realizations also arises from \textit{metric constraints}: graph realizability~\ourref[Graph(s)!Graph Realizability]{p:graphreal}, unit spanning ratio~\ourref[Graph(s)!Straight line Realizability with Unit Spanning Ratio]{p:unitspanratio}, 
from constraints on the \textit{geometric complexity} of the drawing: planar slope number~\ourref[Graph(s)!Planar Slope Number]{p:planarslopenumb}, line cover number~\ourref[Graph(s)!Line Cover Number in Rd@Line Cover Number in $\R^d$]{p:linecovernumb}, segment number~\ourref[Graph(s)!Segment Number]{p:segnumber}, and triangulation stretchability~\ourref[Stretchability!Triangulation]{p:TriStretch}, and from \textit{resolution constraints}: angular resolution~\ourref[Graph(s)!Angular Resolution]{p:angularres}, crossing resolution~\ourref[Graph(s)!Right Angle Crossing Drawability@Right-Angle Crossing Drawability]{p:racdraw}, and perfect resolution in a Lombardi drawing (with rotation system)~\ourref{p:lombardidraw}. The last result is one of the few graph drawing results for a drawing style in which edges are neither straight-edges, nor arbitrary curves, but circular arcs.

A tantalizing and fundamental open question in this area is the complexity of the partial drawing extensibility problem~\ourref[Graph(s)!Partial Drawing Extensibility (Open)]{p:partialdrawingext}: Is it \ER-complete to test whether a partial straight-line embedding of a graph can be extended to a straight-line embedding of the full grap? The problem is only known to be \ER-hard if the realization has to lie within a polygon~\ourref[Graph(s)!Graph in a Polygonal Domain]{p:GraphinPD}.

\section{Intersection Graphs}

Intersection graphs are graphs representing the intersection relation between objects. 
We refer to the book by McKee and McMorris for a general theory of such graphs~\cite{McKMcM99}. 
In computational geometry, \emph{geometric} intersection graphs are typically defined as follows: For a family
$\mathcal{F}$ of subsets of $\R^d$, a graph $G=(V,E)$ is an $\mathcal{F}$-intersection graph if there exists a map
$f:V\to\mathcal{F}$ such that $uv\in E$ if and only if $f(u)\cap f(v)\not=\emptyset$.
Perhaps the simplest family to consider is the family of intervals in $\R$, which yields the \emph{interval graphs}.
\emph{Disk} intersection graphs, defined as intersection graphs of disks in the plane, have been used in the context of wireless communications for modeling the network induced by antennas with bounded transmission ranges.
Other well-known families of geometric intersection graphs are \emph{segment intersection graphs}, intersection graphs of straight line segments in the plane, and \emph{string graphs}, intersection graphs of arbitrary curve segments ("strings") in the plane.

The \emph{recognition} problem for a class of graphs is the problem of deciding whether a given graph belongs to the class.
When defined as a language, the recognition problem is simply the set of graphs in the class, hence we can refer to the computational complexity
of the recognition problem for a class simply as the complexity of the class.

\begin{figure}[htbp]
    \centering
    \includegraphics[page = 4]{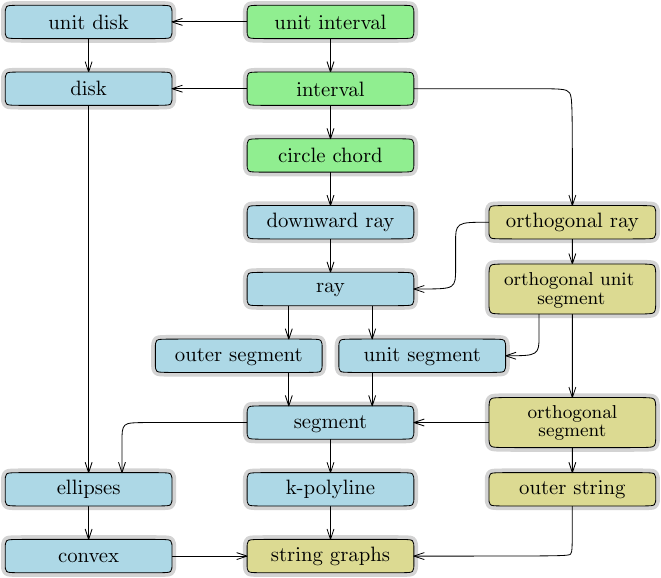}  
    \caption{The complexity of recognizing different geometric intersection graph classes. 
    Recognition of the green classes is in P, ochre is \NP-complete, blue is \ER-complete, and white is unknown.
   }
    \label{fig:Class-Overview}
\end{figure}

Early investigations of Kratochv{\'\i}l and Matou{\v{s}}ek~\cite{KM94} on the complexity of segment intersection graphs, and the finding that they
were \ER-complete, are one of the milestones of \ER-completeness theory~\ourref[Segment Intersection Graph]{p:SEG}. Since then, many classes have been proved \ER-complete.
Table~\ref{tab:Intersection} summarizes the complexities of some of the best-known geometric intersection  graphs. \ER-completeness holds in particular for convex sets in the plane~\ourref{p:convintersect}, unit disks~\ourref{p:unitdiskgraph}, and rays in the plane~\ourref{p:rayintersect}.
Figure~\ref{fig:Class-Overview} shows relationship between classes of geometric intersection graphs. 
More recently, there have also been \ER-completeness results on intersection graphs in hyperbolic geometry~\cite{BBDJ23, BFKS23}.
Matou{\v{s}}ek~\cite[Theorem 3.17]{M14} discusses the Oleinik–Petrovskii–Milnor–Thom theorem which allows one to give an $2^{O(n \log n)}$ upper bound on the number of non-isomorphic intersection graphs; see also Sauermann~\cite{Sa21} on this topic.

\emph{Geometric contact graphs} are defined similarly: 
A graph $G=(V,E)$ is an $\mathcal{F}$-contact graph for a family $\mathcal{F}$ of compact subsets of $\R^d$ if there exists a map $f:V\to\mathcal{F}$ such for any pair $u\not= v$ of vertices the sets $f(u)$ and $f(v)$ are interior-disjoint, and $uv\in E$ if and only if $f(u)\cap f(v)\not=\emptyset$. The Koebe-Andreev-Thurston Theorem states that planar graphs are exactly the contact graphs of disks. However, if we require that disks have unit radius, the recognition problem is known to be \NP-hard, but not known to be \ER-hard~\cite{BK96,HK01}~\ourref[Disk Graph!Unit Disk Contact Graph (Open)]{p:unitdiskcontact}.

\begin{table}[htbp]
   \caption{Classes of Geometric Intersection Graphs and their computational complexity.} 
   \label{tab:Intersection}
\begin{center}
   \begin{tabular}{|l|c|l|}
   \hline
   \multicolumn{1}{|c|}{Intersection graphs of} & \multicolumn{1}{c|}{Complexity} & \multicolumn{1}{c|}{Source} \\ \hline
    circle chords & \P & Gabor, Supowit, and Hsu~\cite{GSH89}, Spinrad~\cite{S94} \\  
    (unit) intervals & \P & Booth and Lueker~\cite{BL76} \\ 
    \hline
    outerstrings & \NP-complete & Kratochv{\'\i}l~\cite{K91} (see also Rok and Walczak~\cite{RW19}) \\ 
    strings & \NP-complete & Kratochv{\'\i}l~\cite{K91}, Schaefer and Sedgwick~\cite{SSS03} \\  
    \hline
    convex sets~\ourref{p:convintersect} & \ER-complete & Schaefer~\cite{S10} \\ 
    (unit) disks~\ourref{p:unitdiskgraph} & \ER-complete & McDiarmid and M{\"u}ller \cite{McDM13}\\  
    ellipses~\ourref{p:convintersect} & \ER-complete & Schaefer~\cite{S10} \\ 
    outer segments~\ourref{p:rayintersect} & \ER-complete & Cardinal et al.~\cite{CFMTV18} \\  
    $k$-polylines~\ourref{p:kpolySEG} & \ER-complete & Hoffmann et al.~\cite{HMWW24} \\
    (downward) rays~\ourref{p:rayintersect} & \ER-complete & Cardinal et al.~\cite{CFMTV18} \\  
    segments~\ourref{p:SEG} & \ER-complete & Kratochv{\'\i}l and Matou{\v{s}}ek~\cite{KM94} (see also~\cite{S10,M14})\\   
    3D segments~\ourref{p:3dSEG}  & \ER-complete & Evans, Rz\c{a}\.{z}ewski, Saeedi, Shin, and Wolff~\cite{ERSSW19} \\
    3D lines~\ourref{p:LIG}  & \ER-complete & Cardinal~\cite{C24} \\    
    unit balls~\ourref{p:unitballgraph} & \ER-complete & Kang and M{\"u}ller~\cite{KM12} \\
    unit segments~\ourref{p:uniSEG} & \ER-complete & Hoffmann et al.~\cite{HMWW24} \\  
    \hline
   \end{tabular}
\end{center}
\end{table}

\section{Visibility Graphs}

Visibility graphs are graphs whose vertices can be mapped to geometric objects so that two vertices are adjacent if and only if the two corresponding objects
are visible from each other, hence if there exists a ``line of sight'' that connects the two objects and does not hit any obstacle.
Typically, the objects are subsets of the plane, and lines of sight are straight line segments.
A generalized notion of visibility graphs was first suggested by Develin, Hartke, and Moulton~\cite{DHM03}, and we refer to O'Rourke~\cite{OR18} for a recent extensive bibliographical treatment.
An older survey on algorithmic aspects of visibility was written by Ghosh~\cite{G07}.

When the objects are points in the plane, and the obstacles are the points themselves, the graphs are called \emph{point visibility graphs}~\cite{KPW05, P08}.
The recognition problem for these graphs~\ourref[Visibility Graph!Points]{p:visibilitygraph} has been shown \ER-complete by Cardinal and Hoffmann~\cite{CH17}.

\emph{Polygon visibility graphs} are also the topic of an extended literature, and are closely related to the art gallery problem.
A graph on the vertex set $V$ is a polygon visibility graph if there exists a closed polygon whose vertices are in bijection with $V$,
and such that two vertices of $V$ are adjacent if and only if the line segment between the corresponding vertices of the polygon is completely contained in the polygon.
Lin and Skiena showed~\cite{LS95} that a realizing polygon can require exponential area. Hoffmann and Merckx~\cite{HM18} and proved that visibility graphs of polygons with holes are hard to recognize~\ourref[Visibility Graph!Polygon with Holes]{p:polygonholesvisibility}. Boomari, Ostovari, and Zarei~\cite{BOZ18} proved that if one is given two graphs on the same vertex set, it is \ER-complete to decide whether they correspond to the interior and exterior visibility graphs of a polygon~\ourref[Visibility Graph!Internal/External Polygon]{p:intextvisibility}. 
Still, the computational complexity of the recognition problem for simple polygon visibility graphs remains open, even for the variant in which the ordering of the vertices along the polygon is given. It is also not clear whether segment visibility graphs are \ER-complete~\ourref[Visibility Graph!Segment Visibility Graph (Open)]{p:segvisibilitygraph}.


O'Rourke and Streinu~\cite{OS98} introduced the bipartite {\em vertex-edge visibility graphs}
in which there is an edge between a vertex and a side of the polygon if the vertex can see some
interior points of the side. Similarly, the {\em edge-edge visibility graphs}
is a bipartite graph on pairs of sides of the polygon, and there is an edge between two sides if
there are interior points of the two sides that can see each other.
O'Rourke and Streinu~\cite{OS98} showed that from a vertex-edge visibility graph one can construct
the edge-edge visibility graph of a polygon in polynomial time and vice versa, so the
two recognition problems have the same complexity, and were both shown to lie in \NP~\cite{OS97}.


O'Rourke and Streinu~\cite{OS97} also defined \emph{polygon pseudo-visibility graphs}, in which, roughly, one relaxes straight-lines to pseudolines. 
Polygon pseudo-visibility graphs were shown to be distinct from visibility graphs by Streinu~\cite{S05}, and turn out to be recognizable in polynomial time, as showed recently by Ameer, Gibson-Lopez, Krohn, and Wang~\cite{AGKW22}.

\emph{Terrain visibility graphs} are defined as the visibility relations of vertices of an $x$-monotone polygonal line,
where the lines of sights must remain above the line~\cite{AGKSW20}. The complexity of the recognition problem for these graphs remains unknown as well. A related problem, on so-called triangulated irregular networks~\ourref[Visibility Graph!Triangulated Irregular Network]{p:tiangulatedirregular} was studied recently by Boomari, Ostovari, and Zarei~\cite{BOZ21}.


\begin{table}[t]
\label{tab:NE}
\caption{\ER-complete problems about the existence of Nash equilibria, adapted from Bil{\`o}, Hansen, and Mavronicolas~~\cite{BHM23}.}
\centering    
\begin{tabular}{|l|c|l|}
\hline
Problem & Condition & Source \\
\hline
NE in a ball~\ourref{p:NashinBall} & $\sigma_i(s_i)\leq u$ $\forall i$ and $s_i\in\Sigma_i$ & Schaefer, \v{S}tefankovi\v{c}~\cite{SS17} \\
Second NE~\ourref{p:NashSecond} & $\mathbf{\sigma}$ is not the only NE & Garg et al.~\cite{GMVY18} \\
\hline
NE with large payoffs~\ourref{p:NashBoundedPayoff} & $\mathsf{U}_i(\mathbf{\sigma})\geq u$ $\forall i$ & Garg et al.~\cite{GMVY18} \\
NE with small payoffs~\ourref{p:NashBoundedPayoff} & $\mathsf{U}_i(\mathbf{\sigma})\leq u$ $\forall i$ & \\
NE with large total payoff~\ourref{p:NashBoundedTotalPayoff} & $\sum_i \mathsf{U}_i(\mathbf{\sigma})\geq u$ & Bil{\`o}, Mavronicolas~\cite{BM21} \\
NE with small total payoff~\ourref{p:NashBoundedTotalPayoff} & $\sum_i \mathsf{U}_i(\mathbf{\sigma})\geq u$ & \\
\hline
NE with large supports~\ourref{p:NashBoundedSupp} & $|\mathsf{Supp}(\sigma_i)|\geq k$ $\forall i$ & Bil{\`o}, Mavronicolas~\cite{BM21} \\
NE with small supports~\ourref{p:NashBoundedSupp} & $|\mathsf{Supp}(\sigma_i)|\leq k$ $\forall i$ &  \\
NE with restricting supports~\ourref{p:NashRestrictedSupp}  & $T_i\subseteq\mathsf{Supp}(\sigma_i)$ $\forall i$ & Garg et al.~\cite{GMVY18} \\
NE with restricted supports~\ourref{p:NashRestrictedSupp} & $\mathsf{Supp}(\sigma_i)\subseteq T_i$ $\forall i$ & \\
\hline
Pareto-optimal NE~\ourref{p:ParetoNash}  & $\mathbf{\sigma}$ is Pareto-optimal & Berthelsen, Hansen~\cite{BH22} \\
Non-Pareto-optimal NE~\ourref{p:ParetoNash} & $\mathbf{\sigma}$ is not Pareto-optimal & \\
Strong NE~\ourref{p:strongNash} & $\mathbf{\sigma}$ is a strong NE & Berthelsen, Hansen~\cite{BH22} \\
Non-strong NE~\ourref{p:strongNash} & $\mathbf{\sigma}$ is not a strong NE & \\
\hline
\end{tabular}
\end{table}
\section{Nash Equilibria and Games}\label{sec:Nash}

The computational complexity of finding a Nash equilibrium of a game, the existence of which is guaranteed by the classical result of Nash, is the topic of an extensive and relatively recent literature. Daslakakis, Goldberg, and Papadimitriou~\cite{DGP09} and Cheng, Deng, and Teng~\cite{CDT09} proved that for two-player games, the problem is complete for the class $\mathsf{PPAD}$. Etessami and Yannakakis later proved that for $r\geq 3$ the problem is $\mathsf{FIXP}$-complete~\cite{EY10}. 

It is of interest to consider Nash equilibria in which some additional constraints are satisfied. For instance, does there exist a Nash equilibrium in which every payoff is bounded from above by some constant~\ourref{p:NashBoundedPayoff}, or in which the supports of the mixed strategies are restricted~\ourref{p:NashRestrictedSupp}?
For two players, many of these problems become $\NP$-complete~\cite{CS08}. For three players, it has been shown that many of them become $\ER$-complete. 
The first result in this vein is due to Schaefer and \v{S}tefankovi\v{c}~\cite{SS17}, who extended ~\cite{EY10} to show that $3$-player Nash equilibrium with additional constraints on the probabilities is \ER-complete~\ourref{p:NashinBall}.
Datta~\cite{D03} showed that the $3$-played Nash equilibrium problem is universal, but the reduction requires an exponential number of strategies. 

Table~\ref{tab:NE} lists a collection of problems about deciding the existence of Nash equilibria with some additional properties that are known to be \ER-complete~\cite{SS17,GMVY18,BM21,BH22,BHM23}.
In all cases, the number of players can be set to three. 
In most cases, except for Pareto-optimal and strong Nash equilibria, the game can be assumed to be a win-lose game, hence all payoffs are either 0 or 1. In many cases, variants of the problem in which the game is symmetric, and the sought Nash equilibria is constrained to be symmetric as well, are also \ER-complete.

A number of \ER-completeness results have been proved for problems on games in extensive form~\cite{H19,GPS20,HS20}, including for instance non-negativity of expected payoff for games with imperfect recall~\ourref{p:NonnegExt}, and the existence of constrained stationary Nash equilibria in acyclic games~\ourref{p:SNashAcyclic}. The existence of equilibria in Arrow-Debreu markets has also been shown \ER-complete by Garg, Mehta, Vazirani, and Yazdanbod~\cite{GMVY17}, \ourref{p:Market}.

\section{Machine Learning and Probabilistic Reasoning}

\begin{figure}[tbph]
    \centering
    \includegraphics{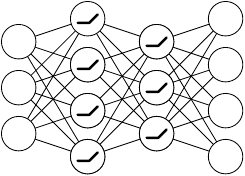}
    \caption{Artificial neural networks are currently among the prominent supervised learning models.}
    \label{fig:neural-networks}
\end{figure}

Suervised learning consists of developing systems that can solve certain tasks by leveraging example data instead of handcrafting a specific algorithm.
Given a model with the parameters $x\in \R^n$, and some training data, the \textit{training problem} consists of finding a good choice of the parameters using some optimization function indicating the fitting of the training data.
Artificial neural networks are popular such models.
Although the field of machine learning
stems from quite different paradigms and perspectives,
we can phrase the training problem as an computational problem
and study its computational complexity.
It turns out that even the simplest 
artificial neural networks are \ER-complete
to train to optimality~\ourref{p:neuralnettrain}.
As there is a broad range of models,  which typically work with real valued parameters, we expect that more training problems can be shown \ER-complete.
This section contains also a list of probabilistic reasoning problems that are \ER-complete~\ourref{p:satprobL}.
In probabilistic reasoning, we have a logical formula about probabilities of events, and the question is whether there exists consistent values for those probabilities.

\section{Markov Chains and Decision Processes}
A Markov Chain is a stochastic process where the state at time $t$ only depends on the state at time $t-1$.
Markov chains offer a rich language to study all types of stochastic phenomenon.
Markov decision processes are a bit more complex than Markov chains and are among the foundational models for the reinforcement learning problem and model checking.
Here, we have a controller that is allowed to perform an action in each time step. 
The next state depends probabilistically on the current state and the action taken by the controller, so that once we choose a fixed strategy of the controller the Markov decision process becomes a Markov Chain.
Typical questions related to those models are about values of state~\ourref{p:neverworseMDP} and~\ourref{p:equallybadMDP} or reachability~\ourref{p:reachMC} and ~\ourref{p:quantreachMC}.

\begin{figure}[tbp]
    \centering
\includegraphics{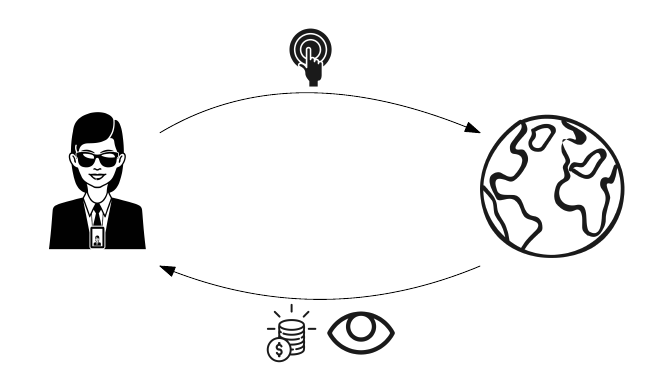}
    \caption{In a Markov Decision Process a controller can perform actions each time step. 
    In return the agent receives some observations about the state of world and a reward. There are many variants with the agent having more or less knowledge about the world, the reward function and the stochastic dynamics.}
    \label{fig:MarkovDecisionProcess}
\end{figure}

\section{Candidates}

As we performed the research for this survey, we came across many problems for which we do not know the complexity, but which are likely to be classifiable using \ER, \VR, or some higher level of the hierarchy. We have included these problems throughout the compendium, if they are variants of other listed problems, or otherwise in the final section containing candidate problems. 

Dobbins, Kleist, Miltzow, and Rz\polhk{a}\.{z}ewski~\cite[Section 5.2]{DKMR23} thoroughly investigate and classify the nature
of computational problems which may be complete for higher levels of the hierarchy. Some additional thoughts can be found in the conclusion section of~\cite{SS23}. 

\newpage
\part{Compendium}

\section{Logic}

\begin{quote}
Syntactically, a {\em (first-order) logic} is specified by its {\em signature}, consisting of the {\em predicate-} and {\em function-}symbols (including constants), together with its logical symbols, such as variables,  Boolean connectives, quantifiers and equality. There are then standard rules on how to form well-formed formulas in such a logic. Formulas without free variables are known as {\em sentences}. If all the quantifiers of a formula (or sentence) occur at the beginning, we say the formula is in {\em prenex} form. The quantifier-free part of a formula in prenex form is known as the {\em matrix} of the formula.

For a particular structure matching the signature of the logic, we can define whether a sentence is true or false in that structure, adding {\em semantics} to the logic. Most questions in this section will be concerned with the complexity of deciding truth in a particular structure, typically the real numbers. But one can also ask whether a set of sentences is {\em consistent}, that is, we can ask whether there is some structure that satisfies all the sentences; such a structure is often referred to as a {\em model}. 
\end{quote}

\begin{problemL}{Existential Theory of the Reals (\ETR)}
\label{p:ETR}
\mindex{Existential Theory of the Reals}
\index{ETR@\ETR|see {Existential Theory of the Reals}}
\given An existentially quantified sentence over the signature $(0,1,+,-,\cdot)$ with comparison operators $\{=, <, \leq, \geq, >\}$ and Boolean connectives $\vee, \wedge$ and $\neg$.
\question Is the sentence true over the real numbers?
\complexity \ER-complete by definition or by Mn\"{e}v~\cite{M88} and Shor~\cite{S91} via a reduction from simple stretchability~\ourref[Simple Stretchability]{p:Stretch}.
\comments Mn\"{e}v~\cite{M88}'s universality theorem can be seen as the earliest \ER-completeness result; it shows that (simple) stretchability can ``express'' semialgebraic sets. Shor's~\cite{S91} paper then draws the connection explicitly, he writes ``Mn\"{e}v [\dots] proved an even stronger theorem. When translated into complexity theory terms, his theorem implies that determining the stretchability of pseudoline arrangements is equivalent to the existential theory of the reals''. Schaefer~\cite{S10} and Schaefer, \v{S}tefankovi\v{c}~\cite{SS17} argue for the existential theory of the reals (\ETR) as the defining problem for the complexity class \ER. 
In the BSS-model, hardness of \ETR\ is the Cook-Levin theorem of that model, and one of the main results of the Blum-Shub-Smale paper~\cite{BSS89}, also see~\cite{C93}.
We want to point out that the encoding is not always defined in the same way~\cite{BPR06},
yet it is easy to show that all variants that we are aware of are polynomial time equivalent.
\universality Mn\"{e}v~\cite{M88}, also see Shor~\cite{S91}.
\also Existential theory of totally ordered real variables~\ourref{p:ETRORV}, feasibility~\ourref{p:feasibility}, simple stretchability~\ourref{p:Stretch}, approximate existential theory of the reals~\ourref{p:epsETR}, matroid realizability over the reals~\ourref{p:matroidreal}.
\end{problemL}

\begin{problemL}{Existential Theory of Totally Ordered Real Variables}
\label{p:ETRORV}
\mindex{Existential Theory of Totally Ordered Real Variables}
\given A set of conditions of the forms $x_i+x_j = x_k$ and $x_ix_j = x_k$.
\question Are there $1 = x_1 < x_2 < \cdots < x_n \in \R$ satisfying all conditions?
\complexity \ER-complete by Mn\"{e}v~\cite{M88} and Shor~\cite{S91} via a reduction from \ETR~\ourref[Existential Theory of the Reals]{p:ETR}.
\comments This useful intermediate problem was used by  Mn\"{e}v~\cite{M88} and named by Shor~\cite{S91}. It can be considered one of the first normal forms of \ETR; the ordering is essential in showing stretchability \ER-complete.
\universality Mn\"{e}v~\cite{M88}, also see Shor~\cite{S91}.
\also \ETR~\ourref{p:ETR}, stretchability~\ourref{p:Stretch}, 
continuous CSP~\ourref{p:CSSP}.
\end{problemL}

\begin{problemL}{Approximate Existential Theory of the Reals ($\varepsilon$-\ETR)}
\label{p:epsETR}
\mindex{Approximate Existential Theory of the Reals}
\given An existentially quantified sentence over the signature $(0,1,+,\cdot)$ with with comparison operators $\{<_{\varepsilon}, \leq_{\varepsilon}, \geq_{\varepsilon}, >_{\varepsilon}\}$ and logic connectives $\vee, \wedge$ and $\neg$, and $\varepsilon \in \Q_{\geq 0}$. Here $a <_{\varepsilon} b$ if $a < b + \varepsilon$, and similarly for the other comparison operators. 
\question Is the sentence true over the real numbers?
\complexity \ER-complete by Deligkas, Fearnley, Melissourgos, and Spirakis~\cite{DFMS22} via a reduction from feasibility~\ourref[Polynomial(s)!feasibility]{p:feasibility} for every fixed $\varepsilon \in \Q_{\geq 0}$..
\also \ETR~\ourref{p:ETR}, existential theory of totally ordered real variables~\ourref{p:ETRORV}, feasibility~\ourref{p:feasibility}.
\end{problemL}

\begin{problemL}{Distinct \ETR}
\label{p:DistETR}
\mindex{Distinct ETR@Distinct \ETR}
\given A set of conditions of the forms $x_i+x_j = x_k$ and $x_ix_j = x_k$.
\question Are there \emph{distinct} $x_1,\ldots,x_n \in \R$ satisfying all conditions?
\complexity \ER-completeness shown by Kim, de Mesmay, and Miltzow ~\cite{KdMM23} via a reduction from a variant of strict local positivity~\ourref[Polynomial(s)!Strict Positivity!Local]{p:LocalPos}.
The result also follows from the \ER-completeness of the existential theory of totally ordered real variables~\ourref[Existential Theory of Totally Ordered Real Variables]{p:ETRORV}, but that
result is more difficult to obtain. 
\comments This useful intermediate problem was used to show that matroid realizability over the reals~\ourref[Matroid!Real Realizability]{p:matroidreal} is \ER-complete.
\also \ETR~\ourref{p:ETR}, 
continuous CSP~\ourref{p:CSSP}, strict local positivity~\ourref{p:LocalPos}, matroid realizability over the reals~\ourref{p:matroidreal}.
\end{problemL}

\begin{problemL}{\ETRINV}
\label{p:ETRINV}
\mindex{ETRINV@\ETRINV}
\given A set of conditions/constraints of the forms $x_i+x_j = x_k$, $x_ix_j = 1$ and $x_i=1$.
\question Are there  $x_1,\ldots,x_n \in [1/2,2]$ satisfying all conditions?
\complexity \ER-completeness by Abrahamsen, Adamaszek, and Miltzow~\cite{AAM18} via a reduction from feasibility~\ourref[Polynomial(s)!feasibility]{p:feasibility}. .
\comments \ETRINV was introduced to show that the art gallery problem~\ourref[Art Gallery Problem]{p:ArtGallery} is \ER-complete. It has been refined in various ways, for example, see \PETRINV~\ourref[Planar ETRINV@\PETRINV]{p:PlanarETRINV}, \RGETRINV~\ourref[Range ETRINV@\RGETRINV]{p:RGETRINV} and the
continuous constraint satisfaction problem~\ourref[Continuous Constraint Satisfaction Problem]{p:CSSP}. 
The reduction has many small steps that are often only sketched, a detailed explanation can be found in~\cite{AM19}.
\also Art gallery problem~\ourref{p:ArtGallery}, \PETRINV~\ourref{p:PlanarETRINV}, \RGETRINV~\ourref{p:RGETRINV}, continuous constraint satisfaction problem~\ourref{p:CSSP}, feasibility~\ourref{p:feasibility}. 
\end{problemL}

\begin{problemL}{\PETRINV}
\label{p:PlanarETRINV}
\mindex{Planar ETRINV@\PETRINV}
\given A set of conditions/constraints of the forms $x+y = z$, $xy = 1$ and $x=1$, on the set of real variables $x_1,\ldots,x_n$.
Furthermore, the variable-constraint graph is planar.  
\question Are there  $x_1,\ldots,x_n \in \R$ satisfying all conditions?
\complexity \ER-completeness shown by Dobbins, Kleist, Miltzow, and Rz\polhk{a}\.{z}ewski~\cite{DKMR18}.
Later, Lubiw, Miltzow, and Mondal~\cite{LMM22} showed the same result for directed graphs.
\comments This intermediate problem was used to show \ER-completeness of  
Prescribed Area Extension~\ourref[Graph(s)!Prescribed Area Extension]{p:preareaext} and drawability of a graph in a polygonal domain~\ourref[Graph(s)!Graph in a Polygonal Domain]{p:GraphinPD}.
\also \ETRINV~\ourref{p:ETRINV}, graph in polygonal domain~\ourref{p:GraphinPD}.
\end{problemL}

\begin{problemL}{\RGETRINV}
\label{p:RGETRINV}
\mindex{Range ETRINV@\RGETRINV}
\given 
An instance~$\Phi$ of \ETRINV, a number $\delta = 2^{-\ell}$ for some natural number $\ell$ and 
for each variable $x_i$, an interval $I(x_i)$ of length at most $2\delta$.
\question Are there real numbers $x_1,\ldots,x_n \in [1/2,2]$ satisfying~$\Phi$ and with each $x_i\in I(x_i)$?
\complexity \ER-complete in the first Arxiv version of~\cite{AMS20}. 
\comments
In an updated version, the authors use Continuous Constraint Satisfaction Problems (CCSP) \ourref{p:CSSP}.
The problem is also difficult for $\delta = O(n^{-c})$, where $n$ is the number of variables and $c$ is an arbitrary constant. 
\also \ETRINV~\ourref{p:ETRINV}.
\end{problemL}

\begin{problemL}{Continuous Constraint Satisfaction Problems (CCSP)}
\label{p:CSSP}
\mindex{Continuous Constraint Satisfaction Problem}
\index{CCSP|see {Continuous Constraint Satisfaction Problem}}
\given Constraints of the form $x=\delta$, with $\delta \in \Q$, $x\geq 0$, $x+y=z$, and $f(x,y) = 0$ for a function $f: \R^2 \rightarrow \R$.
\question Is there a variable assignment that satisfies all constraints?
\complexity If the function $f$ is well-behaved and curved then the CCSP is \ER-complete by Miltzow and Schmiermann~\cite{MS24}. The problem remains \ER-complete if one adds range constraints, i.e., $x\in[\-\delta,\delta]$ or inequalities, i.e., $f(x,y)\geq 0$ and $g(x,y)\geq 0$, assuming $f$ and $g$ meet certain conditions~\cite{MS24}.
\comments CCSP generalizes many other \ER-complete results for technical problems such as \ETRINV~\ourref[ETRINV@\ETRINV]{p:ETRINV}. Continuous Constraint Satisfaction problems have also been studied by Jonsson and Thapper~\cite{JP14,JT18}. They showed \NP-hardness in case the set of relations contains a not essentially convex relation as well as a a primitive positively definable BNU (bounded, non-constant, and unary) relation.
\also \ETRINV~\ourref{p:ETRINV}.
\end{problemL}

\begin{problemL}{Collinearity Logic}
\label{p:CollLogic}
\mindex{Collinearity Logic}
\given A conjunctive logical formula using the ternary relation $\operatorname{COL}$ (collinear) and its negation. 
\question Are there points in the plane satisfying the constraints of the formula?
\complexity \ER-complete by Davis, Gotts, Cohn~\cite{DGC99} via a reduction from a restricted form of \ETR~\ourref[Existential Theory of the Reals]{p:ETR}.
\comments Collinearity logic is a special form of order type realizability~\ourref[Order Type Realizability]{p:ordertype}.  As an intermediate problem, Davis, Gotts, and Cohn~\cite{DGC99} work with a variant allowing the relation PAR (parallel), which is also \ER-complete. 
\openq Is collinearity logic universal? 
\sep\ Does the problem remain \ER-complete if each group of collinear points has fixed size? A careful analysis of the hardness proof for partial order type realizability~\ourref[Order Type Realizability]{p:ordertype} will probably show that the problem remains \ER-hard if each group of collinear points has size at most $6$ or $7$.
\also Order type realizability~\ourref{p:ordertype}, linear hypergraph realizability~\ourref[Linear Hypergraph Realizability]{p:linhypreal}.
\end{problemL}

\begin{problemL}{$\mathcal{LR}$-Calculus}
\label{p:LRCalculus}
\mindex{LR Calculus@$\mathcal{LR}$-Calculus}
\given A conjunctive logical formula using relations $\operatorname{left}(p,q,r)$ and $\operatorname{right}(p,q,r)$ for points in the Euclidean plane. 
\question Are there points in the plane satisfying the constraints of the formula? We say $\operatorname{left}(p,q,r)$ if $p$ and $q$ are distinct, and $r$ lies strictly to the left of the line from $p$ to $q$, and similarly for $\operatorname{right}(p,q,r)$.
\complexity \ER-complete by Mn\"{e}v~\cite{M88} since it is equivalent to simple order type realizability~\ourref[Simple Order Type Realizability]{p:ordertype}. Later, Lee~\cite{L13,L14} proved \ER-completeness for general logical formulas (not necessarily conjunctive) via a reduction from oriented matroid realizability~\ourref[Matroid!Oriented Matroid Realizability]{p:ordertype}, which is a slightly weaker result than Mn\"{e}v's. 
\comments The full $\mathcal{LR}$-calculus allows nine relations (e.g.\ see~\cite{SL15b}. For further \ER-completeness results on spatial calculi, see~\cite{L13,L14}.
\universality Mn\"{e}v~\cite{M88}.
\also  Order type realizability~\ourref{p:ordertype}.
\end{problemL}

\begin{problemL}{Cross-Product Expression Satisfiability}
\label{p:CrossProductExpr}
\mindex{Cross-Product Expression Satisfiability@Cross-Product Expression Satisfiability}
\given A cross-product term $t(x_1, \ldots, x_n)$; that is, a term that is either one of the variables, or of the form $(r \times s)$, where $r$ and $s$ are cross-product terms.
\question Are there vectors $v_1, \ldots, v_n \in \R^3$ such that $t(v_1, \ldots, v_n) = v_1$, where $\times$ is the cross-product over $\R^3$?
\complexity \ER-complete by Herrmann,  Sokoli, and Ziegler~\cite{HSZ13} via a reduction from Feasibility~\ourref[Polynomial(s)!feasibility]{p:feasibility}. 
\comments Herrmann,  Sokoli, and Ziegler show that the problem is equivalent to the existential theory of $F$ for any field $F \subseteq \R$, and that the problem remains \ER-complete over the real projective plane~\cite{HSZ13}. They also consider various variants that turn out to be equivalent to polynomial identity testing.
\end{problemL}

\begin{problemL}{Realizability of Euclidean Embeddings of Rankings (REER)}
\label{p:REER}
\mindex{Realizability of Euclidean Embeddings of Rankings}
\given A conjunctive logical formula of the forms $x_i \preceq_{\ell} x_j$ and $x_i \prec_{\ell} x_j$, 
where $i,j \in [n]$ and $\ell \in [d]$.
\question Are there points $x_i \in \R^d$, $i \in [n]$ and vectors $v_{\ell} \in \R^d$ such that all conditions are satisfied? Here $x_i \preceq_{\ell} x_j$  means that $v_{\ell}^T (x_i-x_j) \geq 0$, and  $x_i \prec_{\ell} x_j$  means that $v_{\ell}^T (x_i-x_j) > 0$.
\complexity \ER-complete by Schockaert and Lee~\cite{SL15b} for any fixed dimension $d \geq 2$ via a reduction from $\mathcal{LR}$-calculus.
\also $\mathcal{LR}$-calculus~\ourref{p:LRCalculus}
\end{problemL}

\begin{problemL}{Topological Inference in RCC8 with Convexity}
\label{p:RCC8con}
\mindex{Topological Inference in RCC8 with Convexity}
\given A {\em constraint network}, that is, a conjunctive logical formula using the predicate ``convex'', and the eight binary relations of the region-connection calculus (RCC8): DC (disconnected), EC (externally connected), EQ (equal), PO (partially overlapping), TPP (tangential proper part), TPPi (tangential proper part inverse), non-tangential proper part (NTTP), and non-tangential proper part inverse (NTTPi), see~\cite{DGC99} for definitions.
\question Are there regular regions in the plane satisfying the logical condition?  A region is {\em regular} if it is the closure of its interior. It does not have to be connected.
\complexity \ER-complete by Davis, Gotts, Cohn~\cite{DGC99} via a reduction from collinearity logic~\ourref[Collinearity Logic]{p:CollLogic}. The problem remains \ER-complete if relations are restricted to EC, PP and ``convex'''~\cite{DGC99}, even in (fixed) higher dimension~\cite[Appendix A]{SL15}. It also remains complete if the set of relations is PO, DC and ``convex'' and the relation between any two regions has to be specified (a {\em fully specified contraint network})~\cite{S10}; this result is obtained via a reduction from intersection graphs of convex sets~\ourref[Convex Set(s)!Intersection Graph]{p:convintersect}.
\comments Membership in \ER\ is not trivial, since regions only have to be regular. 
\openq Does the topological inference problem in RCC8 with convexity remain in \ER, if we added the predicate ``connected''~\cite[Remark 4]{S10}?
\also Convex set intersection graph~\ourref{p:convintersect}.
\end{problemL}

\begin{problemL}{Quantum Logic Strong Satisfiability}
\label{p:QuantumStrong}
\mindex{Quantum Logic Strong Satisfiability}
\given A propositional formula $\phi(x)$, with $x=(x_1,x_2,\ldots ,x_n)$.
\question Is the formula $\phi(x)$ {\em strongly satisfiable} over the modular ortholattice $L(\R^d)$ for some constant $d\geq 3$, i.e., does there exist an assignment of values $a_i\in L(\R^d)$ to the $x_i$ such that $\phi (a)=\mathbf{1}$ when interpreted in $L(\R^d)$?
\complexity \ER-complete by \cite{HZ16}.
\comments The lattice $L(\R^d)$ is the lattice of \emph{linear subspaces of $\R^d$}, with the following three operations, where $U$ and $V$ are two such linear subspaces:
\begin{description}
\item[meet:] $U\wedge V = U\cap V$,
\item[complement:] $\neg U = \{x\in\R^d : \langle x, u\rangle = 0\ \forall u\in U \}$, the orthogonal subspace,
\item[join:] $U\vee V = U + V$, where $+$ denotes the Minkowski sum. Hence the join $U\vee V$ is simply the linear subspace spanned by $U$ and $V$.
\end{description}
In addition, we define the constants $\mathbf{0} = \{ (0,0,\ldots ,0) \}$ (the origin) and $\mathbf{1} = \R^d$ (the whole space).
The lattices $L(\R^d)$ are \emph{modular ortholattices}. They form the basis of the so-called \emph{quantum logic} proposed by Birkhoff and von Neumann in 1936~\cite{BvN36}. The problem studied by Hermann and Ziegler~\cite{HZ16} should not be confused with another quantum satisfiability problem referred to as QSAT. For $d=2$, the problem is \NP-complete. 
\also Quantum Logic Weak Satisfiability~\ourref{p:QuantumWeak}.
\end{problemL}

\begin{problemL}{Quantum Logic Weak Satisfiability}
\label{p:QuantumWeak}
\mindex{Quantum Logic Weak Satisfiability}
\given A propositional formula $\phi(x)$, with $x=(x_1,x_2,\ldots ,x_n)$.
\question Is the formula $\phi(x)$ {\em weakly satisfiable} over $L(\R^d)$ for some fixed constant $d\geq 3$, i.e., does there exist an assignment of values $a_i\in L(\R^d)$ to the $x_i$ such that $\phi (a)\not= \mathbf{0}$ when interpreted in $L(\R^d)$?
\complexity \ER-complete by \cite{HZ16}.
\comments See quantum logic strong satisfiability~\ourref{p:QuantumStrong} for definitions. For $d=2$, the problem is \NP-complete. 
\also Quantum Logic Strong Satisfiability~\ourref{p:QuantumStrong}.
\end{problemL}

\begin{problemL}{Arithmetic Data Exchange Realizability}
\label{p:arithdataex}
\mindex{Arithmetic Data Exchange Realizability}
\given Two {\em (database) schemas}, $S$ and $T$ (finite relational signatures over real arithmetic), an {\em arithmetic schema mapping}, $(S,T, \Sigma_{st}, \Sigma_{t})$, where $\Sigma_{st}$ and $\Sigma_{t}$ are finite sets of sentences over $S\cup T$ and $T$, respectively (with $\Sigma_{st}$ encoding translation constraints, and $\Sigma_{t}$ encoding target constraints), and an {\em instance} $I$ of $S$. The target constraints $\Sigma_t$ are assumed to be {\em weakly acyclic}. (For detailed definitions, see~\cite{tCKO13}.)
\question Is there an instance $J$ of $T$ such that $(I,J)$ satisfies both $\Sigma_{st}$ and $\Sigma_{t}$?
\complexity \ER-complete by ten Cate, Kolaitis, and Othman~\cite[Theorem 4.10]{tCKO13} via a reduction from rectilinear crossing number~\ourref[Graph(s)!Rectilinear Crossing Number]{p:rectcross}.
\comments Ten Cate, Kolaitis, and Othman also show that the {\em certain answer variant} of the data exchange problem is \VR-complete for conjunctive queries. 
\also Rectilinear crossing number~\ourref{p:rectcross}.

\end{problemL}

\section{Algebra}\label{sec:A}

\begin{quote}
Polynomials in this section are {\em multivariate} unless explicitly stated. Polynomials can be represented in various different ways; our default representation is the {\em explicit (or sparse)} representation as a sum of monomial terms with integer coefficients. A denser representation of polynomials can be achieved through an {\em arithmetic (or algebraic) circuit}, which is a directed graph with vertices ({\em gates}) that represent variables or the mathematical operators $+,-,\cdot$. The difference between the explicit and the circuit representation corresponds to the difference between Boolean formulas and Boolean circuits in logic: circuits can reuse already computed values. Circuits can be extended with additional operators, such as the sign-function or division, which allows them to represent non-polynomial functions~\cite{BC09}. 

In this section we will occasionally encounter \EC, the existential theory of the complex numbers. 
\end{quote}

\subsection{Polynomials and Arithmetical circuits}

\begin{problemA}{Feasibility}  
\label{p:feasibility}
\mindex{Polynomial(s)!feasibility}
\index{feasibility|see {Polynomial(s)}}
\given  A family $f = (f_1, \ldots, f_k): \R^n \rightarrow \R^k$ of polynomials $f_i \in \Z[x_1, \ldots, x_n]$. 
\question Is there an $x \in \R^n$ such that $f_i(x) = 0$ for all $i \in [k]$? 
\complexity \ER-complete, even if the total degree of all polynomials is at most $2$ or if $k = 1$ and the single polynomial is non-negative and has total degree at most $4$~\cite{BSS89, BFS99, BCSS98,SS17}. 
Reductions are from \ETR~\ourref[Existential Theory of the Reals]{p:ETR}. Also remains \ER-complete if the domain is a compact set like the unit ball or $[-1,1]^n$ and each polynomial has total degree at most $2$~\cite[Lemma 3.9]{S13}. For a fixed number $n$ of variables, the feasibility problem can be decided in polynomial time (using quantifier elimination). 
If $k = 1$ and the degree is at most $3$, the problem is solvable in polynomial time by Triesch~\cite{T90}.
\comments In the BSS-model this result traces back to~\cite{BSS89,BCSS98}. Buss, Frandsen, and Shallit~\cite{BFS99} define the existential theory $ETh(F)$ of a field $F$ (as a decision problem), and show that the feasibility problem over $F$ is complete for $ETh(F)$. 
In the real case this is true even for $k = 1$. 
Arguably, the \ER-completeness result is implicit in the normal forms devised by Mn\"{e}v~\cite{M88} and Shor~\cite{S91}. 
Triesch's result was proved in the BSS-model~\cite{T90}, but carries over. 
\openq Does the problem remain \ER-complete for $n = 1$ and $k=1$ if the polynomial is represented by a straight-line program? See the candidate problem feasibility of a univariate polynomial. 
\also \ETR~\ourref{p:ETR}, feasibility of univariate polynomial~\ourref[Polynomial(s)!Feasibility of Univariate Polynomial (Open)]{p:UniFEAS}, existential theory of invertible matrices~\ourref{p:ETInvMat}.
\end{problemA}

\begin{problemA}{(Strict) Local Positivity}  
\label{p:LocalPos}
\mindex{Polynomial(s)!Positivity!Local}\mindex{Polynomial(s)!Strict Positivity!Local}
\given A family $f = (f_1, \ldots, f_k): \R^n \rightarrow \R^k$ of polynomials $f_i \in \Z[x_1, \ldots, x_n]$. We say that $f$ is {\em (strictly) locally positive} if $f(x) \geq 0$ ($f(x) > 0$) for some $x \in \R^n$. 
\question Is $f$ (strictly) locally positive?
\complexity \ER-complete for family of polynomials of degree $2$, by Schaefer, \v{S}tefankovi\v{c}~\cite{SS17}, or a single polynomial of degree $8$ by Ouaknine and Worrell~\cite[Theorem 7]{OW17} via a reduction 
from Feasibility~\ourref[Polynomial(s)!feasibility]{p:feasibility}. The reductions work for both strict and non-strict versions. 
The result also remains true for bounded domains~\cite{OW17,SS23}. (Ouaknine and Worrell only prove the bounded case and write that it is ``straightforward'' to adapt to the case $\R^n$; a proof of that case can be found in~\cite{SS23} which also lowers the degree to $4$.)
Kiefer and Tang~\cite{KT24} give an independent proof for a single multivariate polynomial of degree $6$ on a bounded domain.
\comments The reductions do not (and cannot) maintain strong forms of universality (such as stable equivalence), since feasibility describes closed sets and positivity describes open sets. Strict local positivity is called \SINEQ\index{SINEQ@\SINEQ} by Kratochv\'{\i}l and Matou\v{s}ek~\cite{KM94}.
Cucker and Rossell\'{o}~\cite{CR92} discuss this problem under the name $SPOS_d$.
\also Feasibility~\ourref{p:feasibility}, global positivity~\ourref{p:GlobalPos}, positivity of linear recurrence~\ourref{p:poslinrecur}.
\end{problemA}

\begin{problemA}{(Strict) Global Positivity}   
\label{p:GlobalPos}
\mindex{Polynomial(s)!Positivity!Global}\mindex{Polynomial(s)!Strict Positivity!Global}
\index{Positivity|see {Polynomial(s)}}
\given  A polynomial $f: \R^n \rightarrow \R$ in $\Z[x_1, \ldots, x_n]$. If $f(x) \geq 0$ ($f(x) > 0$) for all $x \in \R^n$ we say that $f$ is {\em (strictly) globally positive}. 
\question Is $f$ (strictly) globally positive?
\complexity \VR-complete by Ouaknine and Worrell~\cite[Theorem 7]{OW17} via a reduction from Feasibility~\ourref[Polynomial(s)!feasibility]{p:feasibility}. Remains hard for bounded domain and total degree at most $8$.
\comments Strict global positivity is what is typically meant by a function being positive. Since $f$ is strictly globally positive (globally positive) if and only if $-f$ is not locally positive (not strictly locally positive), \VR-completeness follows from Ouaknine and Worrell~\cite[Theorem 7]{OW17} showing that (strict) local positivity is \ER-complete. In the BSS-model, Cucker and Rossell\'{o}~\cite{CR92} called global positivity {\em Hilbert's 17th problem}\index{Hilbert's 17th problem}, and classified its complement using the exotic quantifier $\exists^*$ later introduced by Koiran~\cite{K99,K00}. 
\also Feasibility~\ourref{p:feasibility}, local positivity~\ourref{p:LocalPos}.
\end{problemA}

\begin{problemA}{Fixed Point}   
\label{p:FixedPoint}
\mindex{Polynomial(s)!Fixed Point}
\index{Fixed Point|see {Polynomials(s)}}
\given A family $f = (f_1, \ldots, f_n): \R^n \rightarrow \R^n$ of polynomials $f_i \in \Z[x_1, \ldots, x_n]$. 
\question Does $f$ have a {\em fixed point}, that is an $x \in \R^n$ with $f(x) = x$?
\complexity \ER-complete by Schaefer, \v{S}tefankovi\v{c}~\cite{SS17} via a (trivial) reduction from Feasibility~\ourref[Polynomial(s)!feasibility]{p:feasibility}. For fixed dimension $n$, the problem can be decided in polynomial time (using quantifier elimination).
\comments The reduction is trivial. 
\also Feasibility~\ourref{p:feasibility}, Brouwer Fixed Point~\ourref{p:BrouwerFixedPoint}.
\end{problemA}

\begin{problemA}{Brouwer Fixed Point}    
\label{p:BrouwerFixedPoint}
\mindex{Polynomial(s)!Brouwer Fixed Point}
\index{Brouwer Fixed Point|see {Polynomial(s)}}
\given A family $f = (f_1, \ldots, f_n): [-1,1]^n \rightarrow [-1,1]^n$ of polynomials in $\Z[x_1, \ldots, x_n]$, $q \in \Q^n$, $\delta \in \Q_{\geq 0}$.
\question Is there a {\em fixed point} of $f$ within distance $\delta$ of $q$, that is an $x \in [-1,1]^n$ with $f(x) = x$ and $\norm{x-q}_{\infty}< \delta$?
\complexity \ER-complete by Schaefer, \v{S}tefankovi\v{c}~\cite{SS17} via a reduction from Fixed Point~\ourref[Polynomial(s)!Fixed Point]{p:FixedPoint}. The problem remains \ER-complete for other natural compact domains and norms. For fixed dimension $n$, the problem can be decided in polynomial time (using quantifier elimination). 
\comments Without the restriction $\norm{x-q}_{\infty}< \delta$, there always is a fixed point, by Brouwer's fixed-point theorem. Ettessami and Yannakakis~\cite{EY10} introduced the complexity class \FIXP\ to capture the complexity of these types of problems. 
\also Fixed point~\ourref{p:FixedPoint}.
\end{problemA}

\begin{problemA}{Nontrivial Homogeneous Zero (\HTN)}
\label{p:HTN}
\mindex{Polynomial(s)!Nontrivial Homogenous Zero}
\index{HTN@\HTN|see {Polynomial(s),Nontrivial Homogenous Zero}}
\given A family $f = (f_1, \ldots, f_k): \R^n \rightarrow \R^k$ of homogeneous polynomials $f_i \in \Z[x_1, \ldots, x_n]$. 
\question Is there a {\em non-trivial zero}, that is, an $x \in \R^k$ with $f(x) = 0$ and $x\neq 0$?
\complexity \ER-complete by Schaefer~\cite[Corollary 3.10]{S13} from feasibility~\ourref[Polynomial(s)!feasibility]{p:feasibility} with compact domain. Remains \ER-complete even if the total degree of each polynomial is $2$.
\comments This problem was introduced as \HTN, abbreviating Hilbert's Homogeneous Nullstellensatz\index{Hilbert's Homogenous Nullstellensatz}, by Koiran and shown \NP-hard~\cite[Theorem 6]{K00}. The proof in~\cite{S13} yields polynomials of degree $4$, but that is unnecessary: the term $\sum_{i=1}^n x_i^2$ in the proof does not need to be square, which implies that $y_0^4$ and $x_0^4$ can be replaced by $y_0^2$ and $x_0^2$, yielding quadratic polynomials. This problem is equivalent to a $3$-dimensional tensor having eigenvalue $0$. The problem is called Quadratic Feasibility in~\cite{HL13} which also contains a sketch of a hardness proof which has issues~\cite[Example 2.5]{SS18}.
\also Feasibility~\ourref{p:feasibility}, \ISO~\ourref{p:ISO}, 
tensor with eigenvalue $0$~\ourref{p:tensorzero}.  
\end{problemA}

\begin{problemA}{Isolated Zero (\ISO)}
\label{p:ISO}
\mindex{Polynomial(s)!Isolated Zero}
\index{Isolated Zero|see {Polynomial(s)}}
\index{ISO@\ISO}
\given A family $f = (f_1, \ldots, f_k): \R^n \rightarrow \R^k$ of polynomials $f_i \in \Z[x_1, \ldots, x_n]$. 
\question Does $f$ have an {\em isolated zero}, that is, an $x \in \R^k$ such that $f(x) = 0$ and $f(x') \neq 0$ in an open neighborhood of $x$?
\complexity \VR-complete, by Koiran~\cite{K00}, Schaefer~\cite{S13}  via a reduction from \HTN~\ourref[Polynomial(s)!Nontrivial Homogenous Zero]{p:HTN} due to Koiran. 
Membership is shown in Schaefer,  \v{S}tefankovi\v{c}~\cite{SS17}. 
Remains \ER-complete even if the total degree of each polynomial is $4$.
\comments Koiran~\cite{K00} introduced the problem \ISO\ as the question whether the algebraic variety
$\{x\in \R^n: f(x) = 0\}$ contains an isolated point, and observed that \HTN~\ourref[Polynomial(s)!Nontrivial Homogenous Zero]{p:HTN} reduces to the complement of \ISO; since he also showed that \HTN~\ourref[Polynomial(s)!Nontrivial Homogenous Zero]{p:HTN} is \NP-hard, this implied \coNP-hardness for \ISO. 
\also \HTN~\ourref{p:HTN}, isolated point~\ourref{p:isopoint}, semialgebraic set with isolated point~\ourref{p:semiisolated}.
\end{problemA}

\begin{problemA}{Injectivity (polynomial)}
\label{p:injectivity}
\mindex{Polynomial(s)!Injectivity}
\index{Injectivity|see {Polynomial(s)}}
\given A family $f = (f_1, \ldots, f_n): \R^n \rightarrow \R^n$ of polynomials $f_i \in \Z[x_1, \ldots, x_n]$.
\question Is $f$ injective on $\R^n$?
\complexity \VR-complete (unpublished, see comments). 
 For functions represented by arithmetic circuits, this was shown by B\"{u}rgisser, Cucker~\cite[Corollary 9.4]{BC09}.
\comments Start with the feasibility problem for a polynomial $g \in \Z[x_1, \ldots, x_{n-1}]$. 
Then $g$ is feasible if and only if $f(x_1, \ldots, x_n) := (x_1, \ldots, x_{n-1}, x_n g(x_1, \ldots, x_{n-1}))$ is not injective. 
Starting with feasibility on a bounded or open domain implies hardness for testing injectivity on these domains as well. 
\also Surjectivity (second level)~\ourref{p:surjectivity}.  
\end{problemA}

\begin{problemA}{Regularity (polynomial)}
\label{p:regularity}
\mindex{Polynomial(s)!Regularity}
\index{Regularity|see {Polynomial(s)}}
\given A polynomial $f \in \Z[x_1, \ldots, x_n]$. 
\question Is $f$ {\em regular}? That is, there is no common real zero of $f$ and its partial derivatives. 
\complexity \VR-complete by Cucker, Rossell\'{o}~\cite{CR92} via a reduction from (non)-feasibility~\ourref[Polynomial(s)!feasibility]{p:feasibility}.
\comments A polynomial $g \in \Z[x_1, \ldots, x_{n}]$ is feasible if and only if $f = g^2$ is not regular. 
\also Feasibility~\ourref{p:feasibility}.  
\end{problemA}

\begin{problemA}{Totality (arithmetic circuit)}
\label{p:totality}
\mindex{Arithmetic Circuit!Totality}
\given An arithmetic circuit with division defining a (partial) function $f: \R^n \rightarrow \R$. The value of the circuit is not {\em defined} if it requires division by zero. 
\question Is $f$ total on $\R^n$, that is, is $f$ defined for every $x \in \R^n$? 
\complexity \VR-complete by B\"{u}rgisser, Cucker~\cite[Proposition 4.1, Corollary 9.4]{BC09} via a reduction from Feasibility~\ourref[Polynomial(s)!feasibility]{p:feasibility}. 
\comments The problem is hard for a single division gate, since we can let $f = 1/g$, where $g$ is a polynomial. Then $f$ is total if and only if $g$ is not feasible. Without division gate, the problem is trivial. 
\also Feasibility~\ourref{p:feasibility}, injectivity~\ourref[Polynomial(s)!Injectivity]{p:injectivity}, surjectivity (second level)~\ourref{p:surjectivity}.  
\end{problemA}

\begin{problemA}{Continuity (arithmetic circuit)}
\label{p:ContinuityAC}
\mindex{Arithmetic Circuit!Continuity}
\given An arithmetic circuit, division-free but with sign function $\sgn$, defining a function $f: \R^n \rightarrow \R$. 
\question Is $f$ continuous on $\R^n$?
\complexity \VR-complete by B\"{u}rgisser, Cucker~\cite[Corollary 9.4]{BC09} via a reduction from Feasibility~\ourref[Polynomial(s)!feasibility]{p:feasibility}. 
\comments A polynomial $g$ is feasible if and only if $f = \sgn(g^2)$ is not continuous, showing hardness of continuity. The trickier part is showing membership in \VR, since expressing continuity seems to require a quantifier alternation; B\"{u}rgisser and Cucker solve this problem via exotic quantifiers, which can be eliminated in this case. 
\also Feasibility~\ourref{p:feasibility}, continuity at a point (arithmetic circuit)~\ourref[Arithmetic Circuit!Continuity at a Point]{p:ContinuityatPointAC}.  
\end{problemA}

\begin{problemA}{Continuity at a Point (arithmetic circuit)}
\label{p:ContinuityatPointAC}
\mindex{Arithmetic Circuit!Continuity at a Point}
\given An arithmetic circuit, division-free but with sign function $\sgn$, defining a function $f: \R^n \rightarrow \R$, a point $x \in Q^n$. 
\question Is $f$ continuous at $x$?
\complexity \VR-complete by B\"{u}rgisser, Cucker~\cite[Corollary 9.4]{BC09} via a reduction from unboundedness of a semialgebraic set~\ourref[Semialgebraic Set(s)!unboundedness]{p:unbounded}. 
\comments A polynomial $g$ is feasible if and only if $f = \sgn(g^2)$ is not continuous, showing hardness of continuity. The trickier part is showing membership in \VR, since expressing continuity seems to require a quantifier alternation; B\"{u}rgisser and Cucker solve this problem via exotic quantifiers, which can be eliminated in this case. 
\also Continuity (arithmetic circuit)~\ourref[Arithmetic Circuit!Continuity]{p:ContinuityAC}, unboundedness of semialgebraic set~\ourref{p:unbounded}.  
\end{problemA}

\begin{problemA}{Domain Density (arithmetic circuit)}
\label{p:DomainDensityAC}
\mindex{Arithmetic Circuit!Domain density}
\given An arithmetic circuit (with division gate and sign function), defining a (partial) function $f: \R^n \rightarrow \R$.
\question Is the domain of $f$ dense in $\R^n$?
\complexity \VR-complete, by B\"{u}rgisser, Cucker~\cite[Proposition 5.7, Corollary 9.4]{BC09}.
\comments  In the Zariski topology, B\"{u}rgisser, Cucker~\cite[Proposition 5.7, Corollary 9.4]{BC09} show that the problem is \ER-complete. 
\also Image density (second level)~\ourref{p:imagedensity}, density of semialgebraic set~\ourref{p:radiussemi}.  
\end{problemA}

\begin{problemA}{Polynomial Projection (Linear)}
\label{p:PolyProj}
\mindex{Polynomial(s)!Projection (Linear)}
\given Two polynomials $f,g:\R^n\to\R$ in $\Z[x_1,x_2,\ldots,x_n]$.
\question Is there a matrix $A$ such that $f(x)=g(Ax)$?
\complexity \ER-complete by Bl\"{a}ser, Rao, and Sarma~\cite{BRS17} via a reduction from the symmetric tensor rank problem~\ourref[Tensor!Rank]{p:tensorrank} .
\comments Over an arbitrary integral domain, the problem is complete for the existential theory of that domain~\cite{BRS17}. If $A$ is a diagonal, invertible matrix, then the polynomial projection problem can be solved in randomized polynomial time~\cite{BRS17}. The general (affine) polynomial projection problem asks whether there is a matrix $A$ and a vector $b$ such that $f(x) = g(Ax+b)$. The special case $A = I$ is known as polynomial equivalence under shifts~\ourref[Polynomial Equivalence under Shifts (Open)]{p:polyshifts}. 
\openq Is the affine polynomial projection problem \ER-hard? This version is known to be \NP-hard by Kayal~\cite{K12}.
\also Polynomial equivalence under shifts~\ourref{p:polyshifts}.    
\end{problemA}

\subsection{Matrices and Tensors}\label{sec:MatandTen}

\begin{quote}
A $d$-dimension (real) tensor $T$ is an element in $\R^{n_1\times \cdots \times n_d}$. The tensor has rank at most $k$ if it can be written as $\sum_{i=1}^k v_1 \otimes \cdots \otimes v_k$, where $v_i \in \R^{n_i}$ for $1 \leq i \leq d$, and $\otimes$ is the outer product of two vectors. Matrices are then the special case of $2$-dimensional tensors. There are other notions of tensor- and matrix-rank, some of which we discuss below.
\end{quote}

\begin{problemA}{Matrix Singularity}
\label{p:MatSing}
\mindex{Matrix!Singularity}
\given A square matrix $M(x_1, \ldots, x_k)$ with variable entries $x_1, \ldots, x_k$ and entries in $\Q$. 
\question Is $M(x_1, \ldots, x_k)$ singular, that is, does $M(x_1, \ldots, x_k)$ have determinant $0$ for some $(x_1, \ldots, x_k) \in~\R^k$?
\complexity \ER-complete by Buss, Frandsen, Shallit~\cite{BFS99} via a reduction from Feasibility~\ourref[Polynomial(s)!feasibility]{p:feasibility}.
\comments Buss, Frandsen, Shallit~\cite{BFS99} show that the singularity problem over any commutative ring $R$ is complete for the existential theory of $R$. The reduction is based on Valiant's universality proof for the determinant~\cite{V79}. Non-singularity, that is, testing whether $M(x_1, \ldots, x_k)$ is non-singular for some $x_1, \ldots, x_k$ lies in randomized polynomial time, so is unlikely to be \ER-complete~\cite{BFS99}. Brijder, Geerts, and Van den Bussche~\cite{BGvdBW19b} introduce a formal matrix language (MATLANG) in which matrix singularity is expressible, implying that expression evaluation in that language is \ER-complete.
\also Existential theory of invertible matrices~\ourref{p:ETInvMat}, constrained matrix invertibility problem~\ourref{p:constmatinv}, feasibility~\ourref{p:feasibility}.
\end{problemA}

\begin{problemA}{Minimum/Maximum Matrix Rank}
\label{p:minmaxmatrank}
\mindex{Matrix!Rank}
\given A square matrix $M(x_1, \ldots, x_k)$ variable entries $x_1, \ldots, x_k$ and entries in $\Q$, integer $k$. 
\question Is the rank of $M(x_1, \ldots, x_k)$ at most/at least $k$ for some $(x_1, \ldots, x_k) \in \R^k$?
\complexity \ER-complete by Buss, Frandsen, Shallit~\cite{BFS99} via a reduction from Feasibility~\ourref[Polynomial(s)!feasibility]{p:feasibility}.
\comments Buss, Frandsen, Shallit~\cite{BFS99} show that the minimum and maximum matrix rank problems over any field $F$ is complete for the existential theory of $F$, even if the matrix entries lie in $\{0,1\}$. A minimum rank problem for matrices with affine entries can be used to prove \ER-completeness of tensor rank~\ourref{p:tensorrank}, see~\cite{SS18}.
\openq Buss, Frandsen, Shallit~\cite{BFS99} show that approximating the minimum rank of a matrix to within a factor of $1+\varepsilon$ is \NP-hard for sufficiently small $\varepsilon$, is it also \ER-hard? \sep\ How hard is the minimum matrix rank problem if each variable occurs only once in the matrix? This is the well-known minimum rank matrix completion problem, the maximum rank version of this problem can be solved in randomized polynomial time~\cite{BFS99}. Related to this is the problem of computing the minimum rank of a sign pattern matrix~\ourref[Sign Pattern Matrix!Minimum Rank]{p:minranksignpatmat}.  \sep\ Is the problem hard for constant $k$, say $k = 4$? 
\also Feasibility~\ourref{p:feasibility}, tensor rank~\ourref{p:tensorrank}, minimum rank of sign pattern matrix~\ourref{p:minranksignpatmat}, minimum graph rank~\ourref{p:mingraphrank}, low rank approximation (matrix)~\ourref{p:lowrankapp}.
\end{problemA}

\begin{problemA}{Minimum Rank of Linear Combination of Matrices}
\label{p:minrankspanmat}
\mindex{Matrix!Minimum Rank of Linear Combination}
\given A family of matrices $M_1, M_2,\ldots ,M_k \in \Q^{m\times n}$ and an integer $r$.
\question Is the rank of $\sum_i \lambda_iM_i$ at most $r$ for some nonzero combination $(\lambda_1, \ldots, \lambda_k) \in \R^k$?
\complexity \ER-complete by Bl\"{a}ser, Ikenmeyer, Lysikov, Pandey, and Schreyer~\cite[Theorem 2.3]{BILPS21} via a reduction from \HTN~\ourref[Polynomial(s)!Nontrivial Homogenous Zero]{p:HTN} even for $r = 1$.
\comments There are interesting connections to various notions of tensor rank, including slice rank and border rank, see~\cite{BILPS21}. The result is used to prove \ER-hardness of the orbit closure containment problem~\ourref[Tensor!Orbit Closure Containment]{p:orbitcontain}.
\openq There is an affine version of this problem, in which we require that $\lambda_1 = 1$; this version is known to be \NP-hard~\cite{BIJL18}, but is it \ER-complete? 
\also Minimum matrix rank~\ourref{p:minmaxmatrank}, tensor rank~\ourref{p:tensorrank}, orbit closure containment~\ourref{p:orbitcontain}.
\end{problemA}

\begin{problemA}{Minimum Rank of Sign Pattern Matrix}
\label{p:minranksignpatmat}
\mindex{Sign Pattern Matrix!Minimum Rank}
\given A {\em generalized sign pattern matrix}, that is, $M \in \{+,-,0\}^{n \times n}$, integer $k$.
\question Is there a matrix $A \in \R^{n\times n}$ of rank at most $k$ so that $\sgn(A) = M$, where $\sgn$ is the sign function applied to each entry?
\complexity \ER-complete by Bhangale, Kopparte~\cite{BK15} via a reduction from simple stretchability~\ourref[Simple Stretchability]{p:Stretch}, even for $k = 3$.
\comments For $k = 2$ the problem is solvable in polynomial time~\cite{BK15}. The maximum rank variant is equivalent to a matching problem and can be solved in polynomial time~\cite{BK15}.
\openq Generalized in the name refers to allowing the entry $0$; in a {\em sign pattern matrix} zero entries are not allowed; in this case Bhangale, Kopparte~\cite{BK15} can still show that the problem is \NP-hard, but \ER-completeness is open.
\also Minimum/maximum matrix rank~\ourref{p:minmaxmatrank}, feasibility~\ourref{p:feasibility}.
\end{problemA}

\begin{problemA}{Monotone Matrix Rank}
\label{p:monmatrank}
\mindex{Matrix!Monotone Rank}
\given A matrix $M \in \Q^{m \times n}$, integer $k$. 
\question Is the {\em monotone rank} of $M$ at most $k$, that is, is there a matrix $A$ of rank at most $k$, and monotone functions $f_1, \ldots, f_n$ such that $M_{ij} = f_j(A_{ij})$, where $i \in [m], j \in [n]$?
\complexity \ER-complete for $k = 2$ by Lienkaemper~\cite{L23b} via a reduction from simple allowable sequence~\ourref[Simple Allowable Sequence]{p:allowableseq}.
\also Minimum/maximum matrix rank~\ourref{p:minmaxmatrank}, minimum rank of sign pattern matrix~\ourref{p:minranksignpatmat}, (simple) allowable sequence~\ourref{p:allowableseq}.
\end{problemA}

\begin{problemA}{Minimum Graph Rank}
\label{p:mingraphrank}
\mindex{Minimum Graph Rank}
\given A graph $G$ on $n$ vertices, an integer $k\geq 0$.
\question Is there a real symmetric matrix $A \in \R^{n \times n}$ of rank at most $k$ so that $v_iv_j \in E(G)$ if and only if $A_{i,j} \neq 0$, for $1 \leq i \neq j \leq n$?
\complexity \ER-complete even for $k = 3$ by Grace, Hall, Jensen and Lawrence~\cite[Section 4]{L23} via a reduction from feasibility~\ourref[Polynomial(s)!feasibility]{p:feasibility}. 
\comments It is also claimed that over an arbitrary infinite field instead of $\R$, the problem (with $A$ being Hermitian instead of symmetric) is equivalent to the existential theory of the field.
\also Minimum matrix rank~\ourref{p:minmaxmatrank}. 
\end{problemA}

\begin{problemA}{Non-negative Matrix Factorization (NMF)}
\label{p:nonnegmatfac}
\mindex{Matrix!Non-negative Factorization}
\given A square matrix $M \in \Q^{n \times n}$ with all entries non-negative and a natural number $k\in \N$. 
\question Are there $n\times k$ and $k\times n$ matrices $A,B$ such that $M = AB$ and all entries of $A$ and $B$ are non-negative?
\complexity \ER-complete by Shitov~\cite{S16} via a reduction from feasibility~\ourref[Polynomial(s)!feasibility]{p:feasibility}.
\comments Vavasis had earlier shown NP-hardness~\cite{V09}. The smallest $k$ for which $M$ has a non-negative factorization equals the non-negative rank of $M$, that is the smallest number of rank-$1$ matrices with non-negative entries, whose sum is $M$~\cite[Lemma 2.1]{CR93}, so the non-negative matrix rank problem\mindex{Matrix!Non-negative Rank}\index{Non-negative matrix rank|see {Matrix}} is also \ER-complete.
Rank problems for matrices with affine entries can be used to prove \ER-completeness of tensor rank~\ourref[Tensor!Rank]{p:tensorrank}, see~\cite{SS18}. 
\universality Shitov~\cite{S16} also shows universality.
\openq Is the problem already hard for some fixed $k$, say four?
\also Nested polytopes~\ourref{p:nestpolytopes}, Hadamard decomposition~\ourref{p:Hadamard}.
\end{problemA}

\begin{problemA}{(Real) PSD Rank}
\label{p:PSDrank}
\mindex{Matrix!(Real) PSD Rank}
\given A matrix $A \in \Q^{m\times n}$, a natural number $r\in \N$.
\question Is $\operatorname{psd-rank}(A)\leq r$? That is, are there Hermitian
positive semidefinite $r \times r$ matrices $(B_1,  \ldots , B_m)$ and $(C_1, \ldots ,C_n)$ with $A(i,j) = \trace(B_iC_j )$, for all $i , j$? If the $B_i$ and $C_j$ are restricted to be real, we obtain the \textit{real psd rank} of $A$. 
\complexity Shitov established \ER-completeness~\cite[Theorem 3.4]{S23PSD} of the (complex) PSD rank.
He also established \ER-completeness~\cite{shitov2017complexity} of the real PSD rank. The problem remains hard in case the nominator and denominator of each entry are given in unary. 
\comments The abbreviation psd stands for positive semi-definite. 
The literature often does not distinguish between real and complex psd rank. 
PSD Rank can be solved in polynomial time for any fixed value of $r$.
\also CPSD Rank~\ourref{p:CPSDrank}
\end{problemA}

\begin{problemA}{(Real) CPSD Rank}
\label{p:CPSDrank}
\mindex{Matrix!CPSD Rank}
\given A matrix $A \in \Q^{n\times n}$, a natural number $r\in \N$
\question Is $\operatorname{cpsd-rank}(A) \leq r$? That is, are there Hermitian psd matrices $(P_1, \ldots , P_n)$ of size $r \times r$ such that $A(i,j) = \trace P_i P_j$ for all $i,j$? We say $A$ is \textit{completely positive semi-definite}. If the $P_i$ are restricted to be real matrices, we obtain the \textit{real CPSD rank} of $A$. 
\complexity Shitov established \ER-completeness~\cite[Theorem 3.10]{S23PSD} of both the cpsd and the real cpsd rank.
\comments The problem is even hard in case the nominator and denominator of each entry are given in unary. 
For any fixed integer $r$, there is a polynomial time algorithm to determine
whether the cpsd rank or real cpsd rank of a given matrix are at most $r$~\cite[Theorem 3.14]{S23PSD}.
\also PSD Rank~\ourref{p:PSDrank}.
\end{problemA}

\begin{problemA}{Phaseless Rank}
\label{p:phaselessrank}
\mindex{Matrix!Phaseless Rank}
\given A matrix $A \in \Q^{m\times n}$, a natural number $r\in \N$
\question Does $A$ have {\em phaseless rank} at most $r$, that is, is there a matrix $\varepsilon \in \C^{m \times n}$ with $|\varepsilon_{ij}| = 1$ such that $A \odot \varepsilon$ has rank at most $r$, where $\odot$ denotes the element-wise matrix product?
\complexity Shitov established \ER-completeness~\cite[Theorem 3.7]{S23PSD}.
\comments The problem is hard even if the numerator and denominator of each entry are given in unary. 
\end{problemA}

\begin{problemA}{Rank-Constrained Optimization}
\label{p:rankconrtainedopti}
\mindex{Matrix!Rank-Constrained Optimization}
\given Two matrices $A,B\in\R^{\ell\times n}$, a natural number $k$.
\question Is there an $n\times n$ matrix $X$ in the positive semidefinite cone $S^n_+$ such that $AX=B$ and $\operatorname{rank}(X)\leq k$?
\complexity \ER-complete by Bertsimas, Cory-Wright, and Pauphilet~\cite{BCWP22} via a reduction from
linkage realizability~\ourref[Linkage!Linkage Realizability]{p:linkagereal}. 
\comments  In Bertsimas et al.~\cite{BCWP22}, the general problem of minimizing an expression of the form $\langle C,X\rangle + \lambda\cdot\operatorname{rank}(X)$ subject to $AX=B$, $\operatorname{rank}(X)\leq k$ and $X\in\mathcal{K}$ for some proper cone $\mathcal{K}$ is considered.
This problem is already \ER-hard for $\mathcal{K}=S^n_+$, $k=2$, $\lambda = 1$ and $C=\mathbf{0}$ by reduction from unit linkage realizability.
\also Linkage realizability~\ourref{p:linkagereal}. 
\end{problemA}

\begin{problemA}{Tensor Rank}
\label{p:tensorrank}
\mindex{Tensor!Rank}
\given A $3$-dimensional tensor $T = (t_{ijk}) \in \Q^{d_1\times d_2\times d_3}$, a natural number $k$.
\question Is the {\em (tensor) rank} of $T$ at most $k$; that is, are there at most $k$ rank-$1$ tensors $T_1, \ldots, T_k$ such 
that $T = \sum_{i \in [k]} T_i$? A tensor has {\em rank-$1$} if it can be written as $u \otimes v \otimes w$ for real vectors $u,v,w$, where $\otimes$ is the outer (Kronecker) product of two vectors.
\complexity \ER-complete by Schaefer, \v{S}tefankovi\v{c}~\cite{SS18} via a reduction from a restricted version of the minimum matrix rank problem~\ourref{p:minmaxmatrank}, and Shitov~\cite{S16b} via a reduction from feasibility~\ourref[Polynomial(s)!feasibility]{p:feasibility}. 
The problem remains hard for symmetric tensors~\cite{S16b}.
\comments Tensor rank was famously shown \NP-hard by {H\r{a}stad}~\cite{H90}. 
Shitov~\cite{S16b} showed that over an integral domain, the (symmetric) tensor rank problem is as hard as the existential theory of that domain. 
For a related notion, called {\em border rank}, whose complexity is still open, see~\cite{BIJL18}.
\universality There is a universality result~\cite{SS24} (which fixes the version in~\cite{SS18}).
\openq Does the problem remain hard for fixed $k$?
\also Minimum matrix rank~\ourref{p:minmaxmatrank}.
\end{problemA}

\begin{problemA}{Tensor with Eigenvalue $0$}
\label{p:tensorzero}
\mindex{Tensor!Eigenvalue 0}
\given A $3$-dimensional tensor $T = (t_{ijk}) \in \Q^{n\times n\times n}$.
\question Does $T$ have $0$ as an {\em eigenvalue}; that is, is there a non-zero vector $v \in R^n$ for which $\sum_{i,j=1}^{n} t_{i,j,k}v_iv_j = 0$ for all $k\in [n]$?
\complexity \ER-complete by Schaefer~\cite[Corollary 3.10]{S13}, also see Schaefer, \v{S}tefankovi\v{c}~\cite[Example 2.5]{SS18}. 
\comments There also is a proof by Hillar and Lim~\cite{HL13}, but it appears to have a gap, see the discussion in~\cite[Example 2.5]{SS18}. The problem is equivalent to the nontrivial homogenous zero problem (\HTN)~\ourref[Polynomial(s)!Nontrivial Homogenous Zero]{p:HTN}. 
We should note that there are other, non-equivalent, definitions of tensor eigenvalues.  
\openq Is the eigenvalue problem \ER-complete for symmetric tensors (with eigenvalues in $\Q[d]$)? Hillar and Lam~\cite[Theorem 9.3]{HL13} show that this problem is \NP-hard. 
\also Tensor rank~\ourref{p:tensorrank} , \HTN~\ourref{p:HTN}, tensor eigenvector approximation~\ourref{p:tensoreigenapp}, tensor singular value~\ourref{p:tensorsingular}.
\end{problemA}

\begin{problemA}{Orbit Closure Containment}
\label{p:orbitcontain}
\mindex{Tensor!Orbit Closure Containment}
\given Two $3$-dimensional tensors $T, T' \in \Q^{n\times n\times n}$.
\question Is the orbit closure of $T$ contained in the orbit closure of $T'$ under the $\mathsf{GL}_n\times \mathsf{GL}_n\times \mathsf{GL}_n$ action in $\R$?
\complexity \ER-complete by Bl\"{a}ser, Ikenmeyer, Lysikov, Pandey, and Schreyer~\cite{BILPS21} via a reduction from minimum rank of linear combination of matrices~\ourref[Matrix!Minimum Rank of Linear Combination]{p:minrankspanmat}.
\comments The orbit of $T$ under $\mathsf{G}_n$ is the set $\{ gT : g\in \mathsf{G}_n\}$, and we consider the closure of these orbits in the Zariski topology. We refer to Bl\"{a}ser et al.~\cite{BILPS21} for detailed definitions and motivations.
\also Minimum rank of linear combination of matrices~\ourref{p:minrankspanmat}, tensor rank~\ourref{p:tensorrank}, minimum matrix rank~\ourref{p:minmaxmatrank}.
\end{problemA}

\subsection{Semialgebraic and Algebraic Sets}\label{sec:SAS}

\begin{quote}
For polynomials and their representations, see Section~\ref{sec:A}.
An {\em algebraic set} is the zero-set of a family of polynomials: $S = \{x\in R^n: f_i(x) = 0$ for all $i \in [k]\}$ and is specified  by its defining polynomials, $f_i$, $i \in [k]$. A {\em basic semialgebraic set} is a set of the form $S = \{x: \in \R^n: f_i(x) = 0$ for all $i \in [k], g_i(x) > 0$ for all $i \in [m]\}$. A {\em semialgebraic set} is the finite union of basic semialgebraic sets. 
We will see two ways of specifying a semialgebraic set: via a formula $\varphi$ as
$S = \{x \in \R^n: \varphi(x)\}$, where $\varphi$ is a quantifier-free formula in the theory of the reals with free variables $x$; and via an arithmetic circuit $C$, where $S = \{x: \in \R^n: C(x) = 1\}$, where $x$ are the variables of the circuit $C$. Unless we say otherwise, we assume a semialgebraic set is specified using the formula $\varphi$.
\end{quote}

\begin{problemA}{Cardinality of Semialgebraic Set}
\label{p:cardinality}
\mindex{Semialgebraic Set(s)!cardinality}
\given A semialgebraic set $S \subseteq \R^n$, a natural number $k\geq 1$ or $k = \infty$.
\question Does $S$ contain fewer than $k$ elements? 
\complexity \VR-complete by Cucker and Rossell\'{o}~\cite{CR92} via a reduction from feasibility~\ourref[Polynomial(s)!feasibility]{p:feasibility} for any fixed $k$, including $k = \infty$. 
\comments Special (named) cases are $k = 1$, the {\em emptiness}\mindex{Semialgebraic Set(s)!emptiness} and $k = \infty$, the {\em finiteness}\mindex{Finiteness of Semialgebraic Set} problem. Cucker and Rossell\'{o}~\cite{CR92} work in the BSS-model, but their proof carries over to \VR. If $k$ is not fixed, the problem is complete for the counting version of \ER, see~\cite[Proposition 8.5(ii)]{BC06}.
\also Dimension of semialgebraic set~\ourref{p:dimension}, feasibility~\ourref{p:feasibility}. 
\end{problemA}

\begin{problemA}{Dimension of Semialgebraic Set}
\label{p:dimension}
\mindex{Semialgebraic Set(s)!dimension}
\given A semialgebraic set $S \subseteq \R^n$, an integer $d\geq 0$.
\question Does $S$ have (real) dimension at least $d$; that is, is there a projection $\pi: \R^n\rightarrow \R^d$ (dropping $n-d$ coordinates), such that $\pi(S)$ contains a non-empty interior? 
\complexity \ER-complete by Koiran~\cite{K99} via reduction from feasibility~\ourref[Polynomial(s)!feasibility]{p:feasibility}, even for any fixed $d \geq 0$. 
\comments For $d = 0$ the problem degenerates to the non-emptiness problem of a semialgebraic set~\ourref{p:cardinality}.
Cucker and Rossell\'{o}~\cite{CR92} looked at the cases $d = n-1$, which they call the hypersurface problem\index{Hypersurface|see {Semialgebraic Set(s),dimension}}, and $d = n$, which they call the NEI (nonempty interior) problem\index{Nonempty Interior|see {Semialgebraic Set(s),dimension}}. They showed that both problems can be described using the exotic quantifier $\exists^*$ (not formally introduced until Koiran's papers~\cite{K99,K00}).

The \ER-hardness reduction is straight-forward, the non-trivial part is showing that the problem lies in \ER. Koiran~\cite{K99} first shows this for the BSS-model, but also explains how to translate the result into the discrete setting, corresponding to \ER. Koiran had earlier shown that the {\em complex} dimension of a semialgebraic set is \EC-complete (again for the BSS-model)~\cite{K97}. For his result Koiran introduces the exotic quantifier $\exists^*$ and shows how to eliminate it. 
\also Local dimension of semialgebraic set~\ourref{p:locdimension}, feasibility~\ourref{p:feasibility}. 
\end{problemA}

\begin{problemA}{Local Dimension of Semialgebraic Set}
\label{p:locdimension}
\mindex{Semialgebraic Set(s)!local dimension}
\given A semialgebraic set $S \subseteq \R^n$, $x \in \Q^n$, an integer $d\geq 0$.
\question Does $S$ have local dimension at least $d$ at $x$; that is, is there a sufficiently small neighborhood $U$ of $x$ such that the dimension of $S \cap U$ is at least $d$? 
\complexity \ER-complete by B\"{u}rgisser and Cucker~\cite[Corollary 6.6,Corollary 9.4]{BC09} even for $d = 1$ via a reduction from testing whether $0$ lies in the closure of a semialgebraic set~\ourref[Semialgebraic Set(s)!Zero in closure]{p:ZeroinClosure}.
\comments \ER-membership uses the exotic quantifier \HQ{}, which can be eliminated.
\also Dimension of semialgebraic set~\ourref{p:dimension}. 
\end{problemA}

\begin{problemA}{Convexity of Semialgebraic Set}
\label{p:convexity}
\mindex{Semialgebraic Set(s)!convexity}
\given A semialgebraic set $S \subseteq \R^n$, a natural number $n \geq 2$.
\question Is $S$ {\em convex}, that is, does the line segment $st$ belong to $S$ for every two points $s, t \in S$? 
\complexity \VR-complete by Cucker and Rossell\'{o}~\cite{CR92}. Also, see Schaefer, \v{S}tefankovi\v{c}~\cite{SS23}.
\comments The proof by Cucker and Rossell\'{o}~\cite{CR92} is phrased in the BSS-model, but translates easily to the \VR-completeness setting, as shown in~\cite{SS23}.
\also Star-shapedness of semialgebraic set~\ourref{p:starshaped}, convexity through translation~\ourref{p:convtrans}.
\end{problemA}

\begin{problemA}{Unboundedness of Semialgebraic Set}
\label{p:unbounded}
\mindex{Semialgebraic Set(s)!unboundedness}
\given A semialgebraic set $S \subseteq \R^n$.
\question Is $S$ {\em unbounded}, that is, is it true that for every $r > 0$ there is an $x \in S$ with $\norm{x} > r$?
\complexity \ER-complete by Cucker, Rossell\'{o}~\cite{CR92}, also see
B\"{u}rgisser, Cucker~\cite[Proposition 6.4,Corollary 9.4]{BC09}, and Schaefer, \v{S}tefankovi\v{c}~\cite{SS23}.
\comments Membership in \ER\ is the non-trivial part here, since the definition uses a universal quantifier. Cucker, Rossell\'{o}~\cite{CR92} work in the BSS-model, but consider a restricted variant which corresponds to \ER, and obtain membership using an upper bound from real algebraic geometry. B\"{u}rgisser and Cucker~\cite{BC09} work with an exotic quantifier and show how to eliminate it; Schaefer and \v{S}tefankovi\v{c}~\cite{SS23} give a reduction from  feasibility over a bounded domain. 
\end{problemA}

\begin{problemA}{Radius of Semialgebraic Set}
\label{p:radiussemi}
\mindex{Semialgebraic Set(s)!radius}
\index{Radius|see {Semialgebraic Set(s)}}
\given A semialgebraic set $S \subseteq \R^n$, $d\in \Q_{\geq 0}$.
\question Does $S$ have radius at most $d$; that is, is there a point $c \in \R^n$ such that $\norm{c-s}\leq d$ for every $s \in S$?
\complexity \VR-complete by Schaefer, \v{S}tefankovi\v{c}~\cite{SS23} via reduction from Feasibility~\ourref[Polynomial(s)!feasibility]{p:feasibility} over a bounded domain.
\comments The \ER-hardness reduction is straight-forward, 
the difficulty lies in showing that the problem lies in \VR\ by eliminating the quantification over the center. 
\also Diameter of semialgebraic set~\ourref{p:diamsemi}.
\end{problemA}

\begin{problemA}{Diameter of Semialgebraic Set}
\label{p:diamsemi}
\mindex{Semialgebraic Set(s)!diameter}
\index{Diameter|see {Semialgebraic Set(s)}}
\given A semialgebraic set $S \subseteq \R^n$, $d\in \Q_{\geq 0}$.
\question Does $S$ have diameter at most $d$; that is, do every two points in $S$ have distance at most $d$ from each other?
\complexity \VR-complete by Schaefer, \v{S}tefankovi\v{c}~\cite{SS23} via reduction from Feasibility~\ourref[Polynomial(s)!feasibility]{p:feasibility} over a bounded domain.
\comments The strict version, in which we require the diameter to be strictly less than $d$ is also \VR-complete; the only issue is \VR-membership, but that can be handled using the exotic \HQ-quantifier as we saw in Section~\ref{sec:repexotic}. 
\also Radius of semialgebraic set~\ourref{p:radiussemi}.
\end{problemA}

\begin{problemA}{Density of Semialgebraic Set}
\label{p:densitysemi}
\mindex{Semialgebraic Set(s)!density}
\given A semialgebraic set $S \subseteq \R^n$.
\question Is $S$ dense in $\R^n$? A set $S \subseteq T$ is {\em dense} in $T$ if for every $t \in T$ and every $\varepsilon > 0$ there is an $s \in S$ such that $\norm{s-t} < \varepsilon$. 
\complexity In \VR\ by Koiran~\cite{K99}. \VR-complete by B\"{u}rgisser, Cucker~\cite[Proposition 5.2,Corollary 9.3]{BC09}, and Schaefer, \v{S}tefankovi\v{c}~\cite{SS23} via reduction from strict positivity over a bounded domain.
\comments The tricky part here is membership in \VR, that follows using Koiran's method~\cite{K99}.  B\"{u}rgisser, Cucker~\cite{BC09} showed the problem complete for semialgebraic sets defined by arithmetic circuits, Schaefer, \v{S}tefankovi\v{c}~\cite{SS23} strengthened the result to semialgebraic sets defined by formulas. 

The directed Hausdorff distance of $S$ and $T$~\ourref[Semialgebraic Set(s)!Hausdorff distance]{p:Hausdorffdist} is $0$ if and only if $S \cap T$ is dense in $T$, so replacing $\R^n$ with another semialgebraic set leads to a \VER-complete problem.  
\openq Does the problem remain hard for algebraic sets?
\also Distance of semialgebraic sets~\ourref{p:distancesemi}, (directed) Hausdorff distance of semialgebraic sets (second level)~\ourref{p:Hausdorffdist}, zero in closure of semialgebraic set~\ourref{p:ZeroinClosure}, Image density (second level)~\ourref{p:imagedensity}, Domain density (arithmetic circuit)~\ourref{p:DomainDensityAC}.
\end{problemA}

\begin{problemA}{Zero in Closure of Semialgebraic Set}
\label{p:ZeroinClosure}
\mindex{Semialgebraic Set(s)!Zero in closure}
\given A semialgebraic set $S \subseteq \R^n$.
\question Does $0$ belong to the closure of $S$? That is, are there $s_n \in S$ such that $\norm{s_i}\rightarrow 0$ for $n \rightarrow \infty$?
\complexity \ER-complete by Cucker, Rossell\'{o}~\cite{CR92}.
\comments Membership is the interesting result here, it requires an upper bound from real algebraic geometry. Cucker and Rossell\'{o}~\cite{CR92} call this the {\em (Euclidean) adherence}\mindex{Semialgebraic Set(s)!Adherence} problem.  B\"{u}rgisser, Cucker~\cite[Proposition 5.1,Corollary 9.4]{BC09} study both the Euclidean and Zariski adherence problem.
\openq How hard is this problem in the Zariski topology~\cite[Table 2]{BC09}?
\also Distance of semialgebraic sets~\ourref{p:distancesemi}.
\end{problemA}

\begin{problemA}{Isolated Point of Semialgebraic Set}
\label{p:isopoint}
\mindex{Semialgebraic Set(s)!isolated point}
\given A semialgebraic set $S \subseteq \R^n$.
\question Is $0$ an isolated point in $S$, that is, does $S$ not contain other points arbitrarily close to $0$?
\complexity \VR-complete by B\"{u}rgisser, Cucker~\cite[Corollary 6.7, Corollary 9.4]{BC09} via a reduction from  testing whether $0$ lies in the closure of a semialgebraic set~\ourref[Semialgebraic Set(s)!Zero in closure]{p:ZeroinClosure}. 
\comments The problem remains hard for algebraic sets, see the isolated zero problem~\ourref[Polynomial(s)!Isolated Zero]{p:ISO}. The complexity of testing whether a semialgebraic set contains an isolated point is open~\ourref[Semialgebraic Set(s)!isolated point existence (Open)]{p:semiisolated}. 
\also Isolated zero~\ourref{p:ISO}, semialgebraic set with isolated point~\ourref{p:semiisolated}, zero in closure of semialgebraic set~\ourref{p:ZeroinClosure}. 
\end{problemA}

\begin{problemA}{Closedness of Basic Semialgebraic Set}
\label{p:closedness}
\mindex{Basic Semialgebraic Set!closedness}
\index{Closedness|see {Basic Semialgebraic Set}}
\given Basic Semialgebraic set $B \subseteq \R^n$, a natural number $n \geq 2$.
\question Is $B$ {\em closed}, that is, does it contain all its limit points? 
\complexity \VR-complete by B\"{u}rgisser, Cucker~\cite[Lemma 6.20,Corollary 9.4]{BC09} via a reduction from Feasibility~\ourref[Polynomial(s)!feasibility]{p:feasibility}.
\comments Hardness is easy: let $B = \{(x,y) \in R^{n+1}: f(x) = 0, y > 0\}$, then $f$ is not feasible if and only if the basic semialgebraic set $B$ is closed; membership in \VR\ is much trickier, and is open for semialgebraic sets in general\index{Semialgebraic Set(s)!closedness (Open)}.
The problem remains \VR-hard for compactness instead of closedness\index{Semialgebraic Set(s)!compactness (Open)}. 
\openq Does the problem lie in \VR\ for semialgebraic sets? 
\also Zero in Closure of semialgebraic sets~\ourref{p:ZeroinClosure}. 
\end{problemA}

\begin{problemA}{Distance of Semialgebraic Sets}
\label{p:distancesemi}
\mindex{Semialgebraic Set(s)!distance}
\given Two semialgebraic sets $S, T \subseteq \R^n$.
\question Do $S$ and $T$ have distance $0$, that is, are there $s_n \in S$ and $t_n \in T$ such that $\norm{s_i-t_t}\rightarrow 0$ for $n \rightarrow \infty$?
\complexity \ER-complete by Schaefer, \v{S}tefankovi\v{c}~\cite{SS17} via a reduction from feasibility~\ourref[Polynomial(s)!feasibility]{p:feasibility}.
\comments \ER-hardness is easy, more interesting is membership in \ER, since there is a universal quantifier which has to be eliminated. The special case in which $T$ consists of a single point amounts to determining whether a point lies in the closure of a semialgebraic set~\ourref[Semialgebraic Set(s)!Zero in closure]{p:ZeroinClosure}.
\also Zero in closure of semialgebraic set~\ourref{p:ZeroinClosure}, Hausdorff distance~\ourref{p:Hausdorffdist}.
\end{problemA}

\begin{problemA}{Hyperplane Local Support of Semialgebraic Set}
\label{p:LocSupp}
\mindex{Semialgebraic Set(s)!local support by hyperplane}
\given A semialgebraic set $S \subseteq \R^n$, hyperplane $H$.
\question Does $H$ locally support $S$; that is, some open subset of $H$ belongs to $S$ and, close to that subset, all of $S$ lies on the same side of $H$?
\complexity \ER-complete by B\"{u}rgisser, Cucker~\cite{BC09} and Schaefer, \v{S}tefankovi\v{c}~\cite{SS23} via a reduction from feasibility~\ourref[Polynomial(s)!feasibility]{p:feasibility}.
\comments Hardness was shown by B\"{u}rgisser and Cucker~\cite{BC09} for arithmetic circuits for a (potentially) larger class. Schaefer and \v{S}tefankovi\v{c}~\cite{SS23} consider semialgebraic sets defined by formulas. 
\also Distance of semialgebraic sets~\ourref{p:distancesemi}.
\end{problemA}

\section{Computational Geometry}

\subsection{Realizability of Configurations}

\begin{quote}
A rank-$d$ \emph{chirotope} is a function $\chi: \{1,2,\ldots ,n\}^d: \rightarrow \{-1,0,1\}$ with the following properties:
(i) exchanging two components $i$ and $j$ changes the sign: $\chi (\tau_{ij}(\lambda)) = -\chi(\lambda)$ , hence it is an \emph{alternating sign map}; (ii) the function $|\chi|$ is the characteristic function of independent sets of a \emph{matroid}; (iii) the following \emph{Grassmann-Pl\"ucker relation} holds:
for every $\lambda\in [n]^{d-2}$ and $a,b,c,d\in [n]\setminus\lambda$, the set 
\[
\{ 
\chi(\lambda , a,b)\cdot \chi(\lambda ,c,d), -\chi(\lambda ,a,c)\cdot \chi(\lambda ,b,d), \chi(\lambda ,a,d)\cdot \chi(\lambda ,b,c)
\}
\]
either contains $\{ -1,+1\}$ or equals $\{ 0\}$.
We usually identify a chirotope $\chi$ with it opposite $-\chi$.
A chirotope defines an oriented matroid of rank $d$. In rank three, chirotopes are often referred to as \emph{order types}. 
it can be checked that for a given set of $n$ vectors $\{ v_1,v_2,\ldots ,v_n\}$ in $\R^d$, the function $\chi$ that returns 
$\sgn (\det (v_{\lambda_1}, v_{\lambda_2}, \ldots , v_{\lambda_d}))$ satisfies the above properties. This function indicates
the orientation of the simplex formed by any $d$-tuple of vectors. In the plane, we can associate to every point $p\in\R^2$ the vector $(p_1,p_2,1)$. Then the rank-three chirotope given by the determinant indicates whether a given ordered triple of points
forms a clockwise, counter-clockwise, or flat triangle. 
It is not true that every rank-three chirotope corresponds to a set of points in the plane, in that sense. Deciding this is one of the fundamental \ER-complete problem.
\end{quote}

\begin{problemCG}{(Simple) Order Type Realizability}
\label{p:ordertype}
\mindex{Order Type Realizability}\mindex{Simple Order Type Realizability}
\given An order type $\chi: [n]^3: \rightarrow \{-1,0,1\}$.
\question Are there points $p_1, \ldots, p_n$ in the plane realizing this order type, hence such that $p_k$ lies to the left of $p_ip_j$ if $\chi(p_i,p_j,p_k) = -1$, to the right if $\chi(p_i,p_j,p_k) = 1$, and the three points are collinear otherwise?
\complexity \ER-complete by Mn\"{e}v~\cite{M88, S91} even if the order type is {\em simple} (aka {\em uniform}), that is, does not take on value $0$.
\comments This is one of the original \ER-complete problems, dual to simple stretchability~\ourref[Simple Stretchability]{p:Stretch}, and a special case of (rank-$3$) oriented matroid realizability\index{Matroid!Oriented Matroid Realizability}. For a detailed proof of \ER-completeness, see Matou\v{s}ek~\cite{M14}. In spatial reasoning, the order type is captured by the $\mathcal{LR}$-calculus (left/right calculus,~\ourref[LR Calculus@$\mathcal{LR}$-Calculus]{p:LRCalculus}), which is therefore \ER-complete~\cite{WL10}. There is a partial variant, in which $\chi$ is only required to be defined on subsets of $[n]$ of fixed size and fixed overlap; that version remains \ER-complete~\cite{S21b}, which is useful in establishing results for problems with fixed parameters, such as segment intersection graphs of bounded degree~\ourref[Segment Intersection Graph]{p:SEG}, pseudo-segment stretchability~\ourref[Stretchability!Pseudo-segments]{p:SegStretch}, and the rectilinear local crossing number~\ourref[Graph(s)!Rectilinear Local Crossing Number]{p:rectloccross}. 
\universality Mn\"{e}v~\cite{M88}, also see Shor~\cite{S91}. 
\also Directional walk~\ourref{p:directwalk}, collinearity logic~\ourref{p:CollLogic}, $\mathcal{LR}$-calculus~\ourref{p:LRCalculus}, pseudo-segment stretchability~\ourref[Stretchability!Pseudo-segments]{p:SegStretch}, rectilinear local crossing number~\ourref{p:rectloccross}.
\end{problemCG}

\begin{problemCG}{Oriented Matroid Realizability Over the Reals}
\label{p:ormatroidreal}
\mindex{Matroid!Real Realizability}
\given A rank-$d$ oriented matroid, for some fixed $d$, given for example as a chirotope $\chi:[n]^d\to \{-1,0,1\}$. 
\question Can this oriented matroid be realized over the reals, hence are there vectors $v_1,\ldots,v_n \in \R^d$ such that $\chi (\lambda) = \sgn(\det(v_{\lambda_1}, v_{\lambda_2}, \ldots, v_{\lambda_d}))$?
\complexity \ER-complete from order type realizability~\ourref[Order Type Realizability]{p:ordertype}.
\also Order type realizability~\ourref{p:ordertype}. 
\end{problemCG}

\begin{problemCG}{Matroid Realizability Over the Reals}
\label{p:matroidreal}
\mindex{Matroid!Real Realizability}
\given A family $F\subset 2^{[n]}$ and a dimension $d\in \N$. 
\question Are there vectors $v_1,\ldots,v_n \in \R^d$ such that for every set $s\in 2^{[n]}$, we have $s\in F$ if and only if the vectors $\{v_i : i \in S \}$ are linearly independent?
\complexity \ER-complete by Kim, Mesmay, and Miltzow~\cite{KdMM23}; the paper contains two reductions: one from order type realizability~\ourref[Order Type Realizability]{p:ordertype}, the other from distinct \ETR~\ourref[Distinct ETR@Distinct \ETR]{p:DistETR}.  
\comments The proof follows from a careful reading of the \ER-completeness proofs of order type realizability~\ourref[Order Type Realizability]{p:ordertype}~\cite{M14}, but
the result was first explicitly stated by Kim, Mesmay, and Miltzow~\cite{KdMM23}
in an arguably simpler and self-contained way.
Matroids can be represented in various ways, affecting the complexity status of the realizability problem.
Typically matroids are given by a circuit or black-box model. Compared to that,
the explicit representation via independent sets may appear wasteful. However,
representing the problem explicitly an only make it easier, but it is still \ER-hard. Moreover,
matroid realizability is \ER-complete for every fixed $d\geq 3$, so at most $O(n^d)$ sets have to be specified.

Sturmfels also studied the problem of matroid realizability over the rational numbers and showed similar equivalences with finding rational solutions to Diophantine equations~\cite{S87}.
However Sturmfels' reduction is not polynomial time, so does not establish an \EQ-completeness result.
\also Order type realizability~\ourref{p:ordertype}, distinct \ETR~\ourref{p:DistETR}. 
\end{problemCG}

\begin{problemCG}{(Simple) Allowable Sequence}
\label{p:allowableseq}
\mindex{Allowable Sequence}
\mindex{Simple Allowable Sequence}
\given A sequence $\pi_1, \ldots, \pi_k$ of permutations of $[n]$.  
\question Is the given sequence {\em allowable}, that is, is there a set of $n$ points in the plane so that the permutations $\pi_1, \ldots, \pi_k$ are obtained, in that order, by projecting the points onto a line rotating by a half turn around some point of the plane?
\complexity \ER-complete by Hoffmann and Merckx~\cite{HM18} even if the sequence is {\em simple}, that is, every two consecutive permutations differ by transposition of two adjacent elements via a reduction from order type realizability~\ourref[Order Type Realizability]{p:ordertype}. 
\comments Allowable sequences are one of the oldest tools in the computational geometers tool set, first introduced in the 19th century, see~\cite{GP80}.
\universality Hoffmann and Merckx~\cite{HM18}.
\also Order type realizability~\ourref{p:ordertype}, visibility graph of polygon with holes~\ourref{p:polygonholesvisibility}.
\end{problemCG}

\begin{problemCG}{Radial System Realizability}
\label{p:radialsystem}
\mindex{Radial System!Realizability}
\given An (abstract) radial system, that is a family $(\pi_i)_{i \in [n]}$ such that $\pi_i$ is a permutation of $[n]-\{i\}$, for each $i$.
\question Are there points $p_1, \ldots, p_n$ in the plane so that for each $i$, the points seen from $p_i$ occur in radial clockwise order as described by $\pi_i$?
\complexity \ER-complete by Cardinal and Kusters~\cite{CK15} via a reduction from order type realizability~\ourref[Order Type Realizability]{p:ordertype}.
\comments There is a partial variant, in which the radial system only needs to be specified for point sets of bounded size and overlap, see~\cite{S21b}; the partial variant can be used to establish hardness for problems with fixed parameters, such as existence of simultaneous geometric embeddings for a fixed number of graphs, and straight-line realizability of bounded-degree graphs with rotation system~\ourref[Graph(s)!Straight-line Realizability of Graph with Rotation System]{p:straightrotation}.
\also Order type realizability~\ourref{p:ordertype}, straight-line realizability of graph with rotation system~\ourref{p:straightrotation}, simultaneous geometric embedding~\ourref{p:SGE}.
\end{problemCG}

\begin{problemCG}{Directional Walk}
\label{p:directwalk}
\mindex{Directional Walk}
\given A walk $W$ on $n$ vertices, and a direction left/right for every triple of consecutive points along $W$. 
\question Are there $n$ points in the plane so that for all triples $(p,q,r)$ of consecutive points along the walk, $r$ lies to the left or right or $pq$ as specified and distinct vertices are mapped to distinct locations?
\complexity \ER-complete~\cite{S21b} via a reduction from a partial version of order type realizability~\ourref[Order Type Realizability]{p:ordertype}. Remains \ER-complete even if each edge occurs at most once on the walk~\cite{KR23}.  
\comments The directional walk problem can be viewed as a sparsified version of order type realizability, or a special case of the left/right calculus~\cite{WL10}. 
\openq Does the problem remain \ER-complete if the realizing walk has to be crossing-free? 
\sep\ The proof in~\cite{S21b} implies that the problem remains \ER-complete even if each vertex occurs only a finite number of times along the walk. Is the directional walk problem \ER-complete even if each vertex occurs at most twice (once being trivial)?
\also Order type realizability~\ourref{p:ordertype}.
\end{problemCG}

\begin{problemCG}{Finite Convex Geometry Realizability}
\label{p:convexgeoreal}
\mindex{Finite Convex Geometry Realizability}
\given A {\em finite convex geometry} $(V, \mathcal{C})$, that is a set $\mathcal{C}\subseteq 2^{V}$ that contains $\emptyset$, $V$, is closed under intersection, and for every $C \in \mathcal{C}\setminus \{V\}$ there is an $v \in V$ such that $C \cup \{v\} \in \mathcal{C}$. 
\question Are there $n$ points in $\R^d$ which realize the given finite convex geometry? Letting $\conv(C)$ denote the convex hull of the points in $C$, a point set $P$ {\em realizes} the finite convex geometry $(P, \{C \subseteq P: \conv(C) \cap P = C\})$.   
\complexity \ER-complete by Adaricheva and Wild~\cite{AW10,HM18} for $d = 2$, via a reduction from order type realizability~\ourref[Order Type Realizability]{p:ordertype}, and by Hoffmann and Merckx~\cite{HM18} for $d \geq 2$ via a reduction from allowable sequence~\ourref[Allowable Sequence]{p:allowableseq}.  
\also Allowable sequence~\ourref{p:allowableseq}, order type realizability~\ourref{p:ordertype}, 
convex neural codes~\ourref{p:convneuralcode}. 
\end{problemCG}

\begin{problemCG}{Free Space Matrix Realizability}
\label{p:freespacemat}
\mindex{Free Space Matrix!Realizability}
\given A matrix $M_{\varepsilon} \in \{0,1\}^{n \times n}$, $\varepsilon \in \Q_{>0}$. $M_{\varepsilon}$ is the {\em free space matrix} for two polygonal curves $P = (p_1, \ldots, p_n)$ and $Q = (q_1, \ldots, q_m)$ in $\R^d$, if $M_{\varepsilon}$ has a $1$ in cell $(i,j)$ if and only if $\norm{p_i-q_j} < \varepsilon$.
\question Are there two polygonal curves $P$ and $Q$ so that $M_{\varepsilon}$ is the free space matrix for $P$ and $Q$?
\complexity \ER-complete for every $d \geq 2$ by Akitaya, Buchin, Mirzanezhad, Ryvkin, and Wenk~\cite[Theorem 14]{ABMRW23} via a reduction from oriented hyperplane realizability~\ourref[Oriented Hyperplane Realizability in Rd@Oriented Hyperplane Realizability in $\R^d$]{p:orientedhyperplanereal}.
\comments In $\R^1$, the problem can be decided in polynomial time~\cite[Theorem 16]{ABMRW23}.
\also Free space diagram realizability~\ourref{p:freespacediag}, oriented hyperplane realizability~\ourref{p:orientedhyperplanereal}.
\end{problemCG}

\subsection{Arrangements and Stretchability}

\begin{quote}
Arrangements of vectors or points can be dualized into arrangements of hyperplanes. From the topological representation theorem~\cite{FL78,BKMS05}, oriented matroids can always be realized as arrangements of pseudo-hyperplanes. In the planar case, we obtain \emph{pseudoline arrangements}. In a projective setting, a pseudoline arrangement is defined on a two-sphere, but a pseudoline arrangement can also be defined in the plane as a collection of $x$-monotone curves that pairwise intersect exactly once. 
For our purpose, the only relevant information about a pseudoline arrangement is the underlying combinatorics, or equivalence class under homeomorphisms of the plane. A convenient way to specify this information is in the form of a \emph{wiring diagram}, or by specifying, for each pseudoline, the left-to-right order of its intersection points with the other pseudolines.
\end{quote}

\begin{problemCG}{(Pseudoline) Stretchability}  
\label{p:Stretch}
\mindex{Stretchability!Pseudolines}\mindex{Stretchability}
\mindex{Simple Stretchability}
\mindex{Hyperbolic Simple Stretchability}
\given A pseudoline arrangement.
\question Is the arrangement {\em stretchable}, that is, is it isomorphic to a straight-line arrangement? 
\complexity \ER-complete, by Mn\"{e}v~\cite{M88}, and Shor~\cite{S91} via a reduction from the existential theory of totally ordered real variables~\ourref[Existential Theory of Totally Ordered Real Variables]{p:ETRORV}. The problem remains \ER-complete if the given arrangement is {\em simple}, that is, at most two pseudolines intersect in each point~\cite{SS17}. Since simple stretchability in the plane is equivalent to simple stretchability in the hyperbolic plane, hyperbolic simple stretchability is also \ER-complete~\cite{B22,BBDJ23}.
\comments Stretchability is the original \ER-complete problem in the Mn\"{e}v/Shor tradition. Mn\"{e}v~\cite{M88} showed that pseudoline stretchability is universal, and Shor~\cite{S91} recast the result from a computer scientists point of view. A nice introductory presentation of the result is due to Matou\v{s}ek~\cite{M14}. Shor~\cite{S91} also explicitly shows that pseudo-line stretchability is \NP-hard.
\universality Both pseudoline stretchability and simple pseudoline stretchability are universal, as shown by Mn\"{e}v~\cite{M88}, also see Shor~\cite{S91}. 
\also Existential theory of totally ordered real variables~\ourref[Existential Theory of Totally Ordered Real Variables]{p:ETRORV}, pseudo-segment stretchability~\ourref{p:SegStretch}.
\end{problemCG}

\begin{problemCG}{Pseudo-segment Stretchability}  
\label{p:SegStretch}
\mindex{Stretchability!Pseudo-segments}
\given An arrangement of \emph{pseudo-segments}: $x$-monotone curves such that each pair intersects exactly once.
\question Is the arrangement {\em stretchable}, that is, is it isomorphic to an arrangement of straight-line segments? 
\complexity \ER-complete by Mn\"{e}v~\cite{M88}, and Shor~\cite{S91}, since pseudoline stretchability~\ourref[Stretchability]{p:Stretch} can be viewed as a special case. The problem remains \ER-complete if each pseudo-segment is involved in at most $72$ crossings and the arrangement is simple, by Schaefer~\cite{S21b} via a reduction from a partial version of order type realizability~\ourref[Order Type Realizability]{p:ordertype}. F{\"{o}}rster, Kindermann, Miltzow, Parada, Terziadis, and Vogtenhuber~\cite{FKMPTV24} show that the problem remains \ER-hard even if the intersection graph of the arrangement has chromatic number at most $30$.
\comments This result can be used to show that segment intersection graphs~\ourref{p:SEG} of degree at most $216$ are \ER-complete to recognize~\cite{S21b} and that the geometric thickness of multigraphs~\ourref[Graph(s)!Geometric Thickness of Multigraph]{p:geothickmulti} is \ER-complete~\cite{FKMPTV24}.
\openq What is the smallest number of crossings on a pseudo-segment for which the pseudo-segment stretchability remains \ER-complete? \sep\ Does the problem remain \ER-complete if the intersection graph of the arrangement has chromatic number at most $2$?~\cite{FKMPTV24}.
\also  Order type realizability~\ourref{p:ordertype}, geometric thickness of multigraphs~\ourref{p:geothickmulti}, segment intersection graph~\ourref{p:SEG}.
\end{problemCG}

\begin{problemCG}{Arrangement Graph}
\label{p:arrangegraph}
\mindex{Arrangement Graph}
\given A graph $G$.
\question Is $G$ an arrangement graph? The vertices of an {\em arrangement graph} are intersection points of an arrangement $L$ of lines in the plane, and two vertices are adjacent if and only if they lie on the same line of $L$ and the line segment between them does not intersect any of the lines in $L$.
\complexity \ER-complete, by Bose, Everett, Wismath~\cite{BEW03} via a reduction from simple stretchability~\ourref[Simple Stretchability]{p:Stretch}.
\also Segment number~\ourref{p:segnumber}.
\end{problemCG}

\begin{problemCG}{(Pseudocircle) Circularizability}
\label{p:pseudocirccirc}
\mindex{Pseudocircle Circularizability}
\given An \emph{arrangement} of \emph{pseudocircles}, that is, a collection of closed curves, any two of which are either disjoint or intersect exactly twice.
\question Is the arrangement {\em circularizable}, that is, is there a homeomorphism of the plane taking it to an arrangement of circles? 
\complexity \ER-complete by Kang and M\"uller~\cite{KM14} (see also Felsner and Scheucher~\cite{FS20}) via a reduction from simple stretchability~\ourref[Simple Stretchability]{p:Stretch}. The problem remains \ER-hard when restricted to arrangements of great pseudocircles~\cite{FS20}. 
\also Simple stretchability~\ourref{p:Stretch}.
\end{problemCG}

\begin{problemCG}{Oriented Hyperplane Realizability in $\R^d$}
\label{p:orientedhyperplanereal}
\mindex{Oriented Hyperplane Realizability in Rd@Oriented Hyperplane Realizability in $\R^d$}
\given A set $S \subseteq \{-,+\}^n$ with $\{(-,\ldots,-), (+,\ldots,+)\}\subseteq S$.
\question Are there $n$ oriented hyperplanes in $\R^d$ such that for every vector $s \in S$ there is a point in $\R^d$ whose characteristic vector with respect to the hyperplane arrangement is $s$? 
\complexity \ER-complete  by Tanenbaum, Goodrich, and Scheinerman~\cite{TGS95}, Basri, Felzenszwalb, Girshick, Jacobs, and Klivans~\cite{BFGJK09}, and Kang and M\"uller~\cite{KM12} for every $d \geq 2$ via a reduction from simple stretchability~\ourref[Simple Stretchability]{p:Stretch}. 
\comments Tanenbaum, Goodrich, and Scheinerman~\cite{TGS95} cast the problem slightly differently, as that of recognizing point-halfspace orders~\ourref[Point half space orders in Rd@Point-halfspace orders in $\R^d$]{p:pointhalfspaceorders}. For $d = 2$ this is very close to the simple stretchability problem~\ourref[Simple Stretchability]{p:Stretch}, but with the additional condition that $\{(-,\ldots,-), (+,\ldots,+)\}\subseteq S$. Kang and M\"uller~\cite{KM12} also show that specifying a realizing hyperplane arrangement may require $2^{\Theta(n)}$ bits. Basri, Felzenszwalb, Girshick, Jacobs, and Klivans~\cite{BFGJK09} introduce oriented hyperplane realizability under the name matrix visibility\index{Matrix Visibility} problem and show it is \ER-hard (they only claim \NP-hardness, but reduce from oriented matroid realizability~\ourref[Matroid!Oriented Matroid Realizability]{p:ordertype}). For $d = 2$, McDiarmid and M\"{u}ller study this as the oriented line arrangement problem\index{Oriented Line arrangement}.
\also Point-halfspace orders~\ourref{p:pointhalfspaceorders}, dot-product graph~\ourref{p:dotprodgraph}, unit ball graph~\ourref{p:unitballgraph}, Euclidean preference~\ourref{p:Euclideanpref}, oriented matroid realizability~\ourref{p:ordertype}. 
\end{problemCG}

\begin{problemCG}{Point-halfspace orders in $\R^d$}
\label{p:pointhalfspaceorders}
\mindex{Point half space orders in Rd@Point-halfspace orders in $\R^d$}
\given A bipartite poset $(B,T,\leq)$.
\question Are there injective maps $f$ from $B$ to $\R^2$ and $g$ from $T$ to the set of halfplanes in $\R^2$ such that $f(x)\in g(y)$ if and only if $x\leq y$? 
\complexity \ER-complete  by Tanenbaum, Goodrich, and Scheinerman~\cite{TGS95} via a reduction from simple stretchability~\ourref[Simple Stretchability]{p:Stretch}. Also \ER-complete in any fixed higher dimension $d$.
\comments This is a different but essentially equivalent encoding of the oriented hyperplane realizability problem~\ourref[Oriented Hyperplane Realizability in Rd@Oriented Hyperplane Realizability in $\R^d$]{p:orientedhyperplanereal}. 
\also Oriented hyperplane realizability~\ourref{p:orientedhyperplanereal}. 
\end{problemCG}

\begin{problemCG}{Circle Orders}
\label{p:circleorders}
\mindex{Circles orders}
\given A poset $P = (S,\leq)$.
\question Is $P$ the containment order of a set of closed disks in the plane, hence does there exist an injective map $f$ from $S$ to the set of closed disks in $\R^2$ such that $f(x)\subseteq f(y)$ if and only if $x\leq y$? 
\complexity \ER-complete by Cardinal~\cite{C24} via a reduction from pseudocircle circularizability~\ourref[Pseudocircle Circularizability]{p:pseudocirccirc}. Also \ER-complete if $P$ is a bipartite poset.
\also Point-halfspace orders~\ourref{p:pointhalfspaceorders}.
\end{problemCG}

\subsection{Covering and Packing}

\begin{problemCG}{Art Gallery Problem}
\label{p:ArtGallery}
\mindex{Art Gallery Problem}
\given A simple polygon $P$ with corners in $\Q^2$ and a number $k \in \N$.
\question Does there exist a set $G\subset P$ of $k$ guards such that every point inside the polygon is visible from~$G$? Two points $a$, $b$ are said to be visible to each other in case that the line segment $ab$ is completely contained in the polygon~$P$. The polygon is considered to be closed.
\complexity \ER-complete, by Abrahamsen, Adamaszek, Miltzow~\cite{AAM18,AAM22}, and Stade~\cite{S23c}
\comments  Also known as the minimum star cover problem~\cite{OR87}\index{Minimum Star Cover}. The original \ER-completeness proof in~\cite{AAM22} is fairly involved, but Stade found a shorter proof in 2022~\cite{S23}.
It is known that irrational coordinates can already be necessary for two and three guardable galleries~\cite{AAM17, MM22}. 
There is a series of papers that try to find practical algorithms for the art gallery problem.
Interestingly some of them come with provable performance guarantees under some mild assumptions~\cite{HM21}.
\universality The art gallery problem admits topological universality~\cite{BEMMSW22,BEMMSW23,ST22}.
\openq Does the problem lie in \ER\ for semialgebraic art galleries, even in $\R^2$?
\also Polygon convex cover~\ourref{p:polygonconvcov}, star-shapedness of semialgebraic set~\ourref{p:starshaped}. 
\end{problemCG}

\begin{problemCG}{Boundary-Guarding}
\label{p:boundaryguard}
\mindex{Boundary Guarding@Boundary-Guarding}
\given A simple polygon $P$ with corners in $\Q^2$ and a number $k \in \N$.
\question Does there exist a set $G\subset \partial P$ of $k$ guards such that every point inside the polygon is visible by at least one guard in~$G$?
\complexity \ER-complete, by Stade~\cite{S23c} via a reduction from \RGETRINV~\ourref[Range ETRINV@\RGETRINV]{p:RGETRINV}.
\comments The boundary guarding problems differs from the art gallery problem~\ourref[Art Gallery Problem]{p:ArtGallery} in that the guards must be on the boundary.
\openq It is open if boundary-boundary guarding is \ER-complete; in the boundary-boundary variant, all guards must be placed on the boundary, and only the boundary has to be guarded~\cite{S23c}. This is similar to the terrain guarding problem, which turns out to be \NP-complete~\cite{KK11,FHK16}.
\also Art gallery problem~\ourref{p:ArtGallery}, \ETRINV~\ourref{p:ETRINV}.
\end{problemCG}

\begin{problemCG}{Geometric Packing}
\label{p:geopack}
\mindex{Geometric Packing}
\given A simple polygon $C$, the {\em container}, and polygonal pieces $p_1,\ldots,p_m$, all polygons have corners in $\Q^2$.
\question Is it possible to move (translate and rotate) all the pieces so they lie in the container without overlapping?
\complexity This problem is \ER-complete, by Abrahamsen, Miltzow, and Seiferth~\cite{AMS20}, even
if the container is a square and all pieces are convex. The reduction is from \ETRINV~\ourref[ETRINV@\ETRINV]{p:ETRINV}. 
\comments The problem is in \NP\ in case that only translations are allowed. 
The proof from~\cite{AMS20} required that either the pieces or the container had a partially curved boundary. This condition was removed in a paper by Miltzow and Schmiermann~\cite{MS24} about continuous constraint satisfaction problems~\ourref[Continuous Constraint Satisfaction Problem]{p:CSSP}, which implies that geometric packing remains \ER-complete even if all the pieces are convex and polygonal, a simplified proof of that result can also be found in~\cite{AMS24}.
\openq Does the problem remain \ER-hard if the pieces are restricted to a fixed number of shapes (one, for example)~\cite{S21b}? 
\sep\
Is the problem fixed-parameter tractable in the number $m$ of pieces?
\sep\ How hard is the packing problem if the pieces to be packed are unit disks? Hardness may be easier to show for polygons with holes~\cite{A24b}.
\end{problemCG}

\begin{problemCG}{Polygon Convex Cover}
\label{p:polygonconvcov}
\mindex{Polygon!Convex Cover}
\given A simple polygon, a natural number $k \in \N$.
\question Can the polygon be written as the union of at most $k$ convex polygons?
\complexity \ER-complete by Abrahamsen~\cite{A22} via a reduction from \RGETRINV~\ourref[Range ETRINV@\RGETRINV]{p:RGETRINV}. The problem remains \ER-complete even if all the convex polygons have to be triangles.
\comments Also known as the {\em minimum convex cover problem}\index{Minimum Convex Cover}, it was previously known to be \NP-hard~\cite{CR94} and to lie in \ER~\cite{OR82}. Replacing convex sets with star-shaped sets one obtains the art gallery problem~\ourref[Art Gallery Problem]{p:ArtGallery}. 
\openq Abrahamsen~\cite{A24b} asks whether the polygon cover problem remains \ER-hard if a set of convex polygons is given and the original polygon has to be covered by rotated and translated copies of the convex polygons.   
\also \ETRINV~\ourref{p:ETRINV}, \RGETRINV~\ourref{p:RGETRINV}, art gallery problem~\ourref{p:ArtGallery}.
\end{problemCG}

\subsection{Geometric Representations of Graphs and Hypergraphs}

\begin{problemCG}{Unit Disk Graph}
\label{p:unitdiskgraph}
\mindex{Disk Graph!Unit Disk Graph}
\given A graph $G$.
\question Is $G$ the intersection graph of unit disks in the plane?
\complexity \ER-complete, by McDiarmid, M\"{u}ller~\cite{McDM10,McDM13} via a reduction from oriented hyperplane realizability~\ourref[Oriented Hyperplane Realizability in Rd@Oriented Hyperplane Realizability in $\R^d$]{p:orientedhyperplanereal}. The problem remains \ER-complete for unit balls in $d$-dimensional space~\cite{KM12} again via a reduction from oriented hyperplane realizability~\ourref[Oriented Hyperplane Realizability in Rd@Oriented Hyperplane Realizability in $\R^d$]{p:orientedhyperplanereal}.
\comments This problem was earlier shown to be \NP-hard in~\cite{BK98}.
\openq Does the problem remain hard for graphs of bounded degree~\cite{S21b}? 
\also Disk graph~\ourref{p:diskgraph}, hyperbolic unit disk graph~\ourref{p:hypunitdiskgraph}, unit ball graph~\ourref{p:unitballgraph}, unit disk contact graph~\ourref{p:unitdiskcontact}, oriented hyperplane realizability~\ourref{p:orientedhyperplanereal}.
\end{problemCG}

\begin{problemCG}{Disk Graph}
\label{p:diskgraph}
\mindex{Disk Graph}
\given A graph $G$.
\question Is $G$ the intersection graph of disks in the plane?
\complexity \ER-complete, by McDiarmid, M\"{u}ller~\cite{McDM13} via a reduction from oriented hyperplane realizability~\ourref[Oriented Hyperplane Realizability in Rd@Oriented Hyperplane Realizability in $\R^d$]{p:orientedhyperplanereal}. 
\openq Does the problem remain \ER-complete for balls in $d$-dimensional space~\cite{McDM13}? \sep\ Does the problem remain hard for graphs of bounded degree?
\also Unit disk graph~\ourref{p:unitdiskgraph}, unit ball graph~\ourref{p:unitballgraph}.
\end{problemCG}

\begin{problemCG}{Hyperbolic Unit Disk Graph}
\label{p:hypunitdiskgraph}
\mindex{Disk Graph!Hyperbolic Unit Disk Graph}
\given A graph $G$. 
\question Is $G$ the intersection graph of unit disks in the hyperbolic plane?
\complexity \ER-complete, by Bieker, Bl{\"{a}}sius, Dohse, and Jungeblut~\cite{BBDJ23} via a reduction from hyperbolic simple stretchability~\ourref[Hyperbolic Simple Stretchability]{p:Stretch}.
\comments The proof uses the result that every unit disk graph is also a hyperbolic unit disk graph~\cite{BFKS23}.  
\also Unit disk graph~\ourref{p:unitdiskgraph}, hyperbolic simple stretchability (under Stretchability~\ourref{p:Stretch}).
\end{problemCG}

\begin{problemCG}{Unit Ball Graph in $\R^d$}
\label{p:unitballgraph}
\mindex{Unit Ball Graph in Rd@Unit Ball Graph in $\R^d$}
\given A graph $G$.
\question Is $G$ the intersection graph of unit balls in $\R^d$?
\complexity \ER-complete for every fixed $d \geq 2$ by Kang, M\"{u}ller~\cite{KM12} via a reduction from oriented hyperplane realizability~\ourref[Oriented Hyperplane Realizability in Rd@Oriented Hyperplane Realizability in $\R^d$]{p:orientedhyperplanereal}. This generalizes the result for unit disk graphs~\ourref{p:unitdiskgraph}, the case $d = 2$, by McDiarmid and M\"{u}ller~\cite{McDM10}. 
\comments Kang and  M\"{u}ller~\cite{KM12} call this the {\em $d$-sphericity} problem\index{sphericity@$d$-sphericity}, and they establish exponential lower and upper bounds on the centers of a family of balls representing $G$. Fiduccia, Scheinerman, Trenk and Zito~\cite{FSTZ98} asked whether the problem lies in \NP. For $d = 1$ we get the {\em unit interval problem}, which is solvable in linear time~\cite{LO93}.
\openq Does recognizing unit ball graphs remain \ER-complete in $d$-dimensional hyperbolic space~\cite{BBDJ23}?
\also Unit disk graph~\ourref{p:unitdiskgraph}, oriented hyperplane realizability~\ourref{p:orientedhyperplanereal}, dot-product graph~\ourref{p:dotprodgraph}.
\end{problemCG}

\begin{problemCG}{Dot-Product Graph in $\R^d$}
\label{p:dotprodgraph}
\mindex{Dot Product Graph in Rd@Dot-Product Graph in $\R^d$}
\given A graph $G = (V,E)$.
\question Are there vectors $v_1, \ldots, v_n \in \R^d$ such that $ij \in E$ if and only if $v_i^Tv_j \geq 1$ for all $ 1\leq i,j \leq n$?
\complexity \ER-complete, by Kang, M\"{u}ller~\cite{KM12} for every fixed $d \geq 2$ via a reduction from oriented hyperplane realizability~\ourref[Oriented Hyperplane Realizability in Rd@Oriented Hyperplane Realizability in $\R^d$]{p:orientedhyperplanereal}. This generalizes the result for unit disk graphs~\ourref{p:unitdiskgraph}, the case $d = 2$, by McDiarmid and M\"{u}ller~\cite{McDM10}. 
\comments The problem was introduced by Fiduccia, Scheinerman, Trenk and Zito~\cite{FSTZ98}. The smallest $d$ for which $G$ has a dot-product representation is the {\em dot-product dimension} of the graph. Kang and M\"uller~\cite{KM12} also establish exponential upper and lower bounds on the precision of the vectors~$v_i$. 
\also Oriented hyperplane realizability~\ourref{p:orientedhyperplanereal}, unit ball graph~\ourref{p:unitballgraph}, unit disk graph~\ourref{p:unitdiskgraph}, representing graph with ortogonal vectors~\ourref{p:graphorthogonalvectorsR3}. 
\end{problemCG}

\begin{problemCG}{Segment Intersection Graph (\SEG)}
\label{p:SEG}
\mindex{Segment Intersection Graph}
\index{SEG@\SEG|see {Segment Intersection Graph}}
\given A graph $G$.
\question Is $G$ the intersection graph of line segments in the plane?
\complexity \ER-complete, by Kratochv{\'\i}l and Matou{\v{s}}ek~\cite{KM94}, reduction from simple stretchability~\ourref[Simple Stretchability]{p:Stretch}. 
The problem remains \ER-hard for graphs of bounded degree~\cite{S21b} via a reduction from pseudo-segment stretchability~\ourref[Stretchability!Pseudo-segments]{p:SegStretch}.
\comments A slightly simplified \ER-hardness proof can be found in~\cite{S10} and, in full detail, in~\cite{M14}. Kratochv{\'\i}l and Matou{\v{s}}ek~\cite{KM94} show that there are graphs on $n$ vertices which require segment representations of size $2^{2^{\Omega(n^{0.5})}}$, and that there are always representations of size at most $2^{2^{O(n)}}$.
\openq The problem is known to be \NP-hard for graphs of girth at least $k$~\cite{KP07}, does it remain \ER-hard?
\also Unit segment intersection graph~\ourref{p:uniSEG}, 3D segment intersection graph~\ourref{p:3dSEG}, planar slope number~\ourref{p:planarslopenumb}, pseudo-segment stretchability~\ourref{p:SegStretch}, $k$-polyline intersection graph~\ourref{p:kpolySEG}, incidence graph of geometric $3$-configuration~\ourref{p:geoconfig}, linear hypergraph realizability~\ourref[Linear Hypergraph Realizability]{p:linhypreal}.
\end{problemCG}

\begin{problemCG}{Unit Segment Intersection Graph}
\label{p:uniSEG}
\mindex{Segment Intersection Graph!Unit Segment Intersection Graph}
\given A graph $G$.
\question Is $G$ the intersection graph of unit line segments in the plane?
\complexity \ER-complete, by Hoffmann, Miltzow, Weber, and Wulf~\cite{HMWW24}, reduction from simple stretchability~\ourref[Simple Stretchability]{p:Stretch}. 
\also Segment intersection graph~\ourref{p:SEG}.
\end{problemCG}

\begin{problemCG}{$k$-Polyline Intersection Graph}
\label{p:kpolySEG}
\mindex{Segment Intersection Graph!$k$-Polyline Intersection Graph}
\given A graph $G$.
\question Is $G$ the intersection graph of $k$-polylines in the plane? A {\em $k$-polyline} is a polygonal chain consisting of $k-1$ line segments.
\complexity \ER-complete, by Hoffmann, Miltzow, Weber, and Wulf~\cite{HMWW24} for any fixed $k$ via a reduction from simple stretchability~\ourref[Simple Stretchability]{p:Stretch}. 
\also Segment intersection graph~\ourref{p:SEG}.
\end{problemCG}

\begin{problemCG}{(Downward) Ray Intersection Graph}
\label{p:rayintersect}
\mindex{Ray Intersection Graph}
\mindex{Ray Intersection Graph!Downward Ray Intersection Graph}
\given A graph $G$.
\question Is $G$ the intersection graph of (downward) rays in the plane?
\complexity The recognition of both ray and downward ray intersection graphs is \ER-complete by Cardinal, Felsner, Miltzow, Tompkins and Vogtenhuber~\cite{CFMTV18} via a reduction from pseudoline stretchability~\ourref[Stretchability]{p:Stretch}.
\comments Downward ray intersection graphs are equivalent to grounded segment graphs~\cite{CFMTV18}. The paper~\cite{CFMTV18} also shows \ER-completeness of outer segment graph recognition. Outer segments have one endpoint on a common circle and the other endpoint in the interior of the circle.
\also Segment intersection graph~\ourref{p:SEG}.
\end{problemCG}

\begin{problemCG}{3D Segment Intersection Graph}
\label{p:3dSEG}
\mindex{Segment Intersection Graph!3D Segment Intersection Graph}
\index{3D Segment Intersection Graph|see {Segment Intersection Graph}}
\given A graph $G$.
\question Is $G$ the intersection graph of line segments in $\R^3$?
\complexity \ER-complete, by Evans, Rz\c{a}\.{z}ewski, Saeedi, Shin, and Wolff~\cite{ERSSW19} via a reduction from simple stretchability~\ourref[Simple Stretchability]{p:Stretch}.
\comments The authors remark that the problem remains \ER-hard in any fixed dimension.
\also Segment intersection graph~\ourref{p:SEG}.
\end{problemCG}

\begin{problemCG}{3D Line Intersection Graph}
\label{p:LIG}
\mindex{Line Intersection Graph}
\given A graph $G$.
\question Is $G$ the intersection graph of lines in $\R^3$?
\complexity \ER-complete, by Cardinal~\cite{C24} via a reduction from matroid realizability over the reals~\ourref{p:matroidreal}.
\also Segment intersection graph~\ourref{p:SEG}, 3D segment intersection graph~\ourref{p:3dSEG}.
\end{problemCG}

\begin{problemCG}{Convex Set Intersection Graph}
\label{p:convintersect}
\mindex{Convex Set Intersection Graph, see {Convex Set(s)}}
\mindex{Convex Set(s)!Intersection Graph}
\given A graph $G$.
\question Is $G$ the intersection graph of a family of convex sets in the plane?
\complexity \ER-complete, by Schaefer~\cite{S10} by a reduction from simple stretchability~\ourref[Simple Stretchability]{p:Stretch}. Remains \ER-complete if the graph has bounded degree~\cite{S21b}.
\comments While not explicitly claimed in~\cite{S10}, \ER-hardness also holds for any family of convex sets that can arbitrarily closely approximate any line segment; this includes ellipses\index{Ellipses Intersection Graph}, triangles and convex polygons. It was shown by M\"{u}ller, van Leeuwen, van Leeuwen~\cite{MvLvL13} that the intersection graph recognition
problem for translates (or homothets) of any fixed convex polygon can be solved on an exponential-size grid.
This implies that the corresponding recognition problems lie in \NP.
\openq Bieker, Bl{\"{a}}sius, Dohse, and Jungeblut~\cite{BBDJ23} ask whether the problem remains \ER-complete in the hyperbolic plane.
\also Segment intersection graph~\ourref{p:SEG}, topological inference in RCC8 with convexity~\ourref{p:RCC8con}.
\end{problemCG}

\begin{problemCG}{Nerves of Convex Set}
\label{p:nervesconv}
\mindex{Convex Set(s)!Nerves}
\given A simplicial complex $S$ of dimension $k$.
\question Is $S$ the $k$-skeleton of the nerve of $c$-dimensional convex sets in $\R^d$?
\complexity \ER-complete for many parameter combinations with $k\geq 2$, in particular
$k \geq d$ and $c = d-1$ or $c=d$, by Schnider and Weber~\cite{SW23}.
\comments This defines a family of problems. This family is parameterized by
 the dimension of the convex sets $c$,
the ambient dimension $d$ containing the convex sets
and the dimension $k$ of the dimension of the input simplicial complex.
Note that for $k=1$, this is a question about graphs.
For example, $(k,c,d)=(1,1,1)$ asks if the underlying graph is an interval graph, 
which is known to be polynomial time solvable~\cite{BL76}.
Another illustrative example is $(k,c,d)=(1,1,2)$. Here, we ask if the underlying graph is the intersection graph
of segments, which is known to be \ER-complete~\cite{KM94}.
The hardness results stem from the hardness results of graphs
combined with a lifting lemma due to Tancer~\cite{Tan10}. 
Some other parameter combinations are always possible to realize (specifically $d,c\geq 2k+1$~\cite{W67})
and thus the computational problem becomes trivial.

Given a family $A$ of sets, we say the nerve of $A$ is a simplicial complex that has one vertex for each member of $A$, and any subset  $B\subseteq A$ forms a simplex if their intersection is non-empty.

\openq All remaining parameter combinations are still open.
\also Segment intersection graph~\ourref{p:SEG}, convex set intersection graph~\ourref{p:convintersect}, 
convex neural codes~\ourref{p:convneuralcode},
finite convex geometry realizability~\ourref{p:convexgeoreal}.
\end{problemCG}

\begin{problemCG}{Hypergraph Induced by Halfspaces and Points}
\label{p:inducedhyp}
\mindex{Hypergraph Induced by Halfspaces and Points}
\given A hypergraph $H=(V,E)$, an integer $d \geq 0$.
\question Is there a set of halfspaces and points in $\R^d$ such that the induced hypergraph equals $H$? The hypergraph has one vertex per point, and a hyperedge consists of all points belonging to the same halfspace. 
\complexity \ER-complete by Tanenbaum, Goodrich, Scheinerman~\cite{TGS95} via a reduction from pseudoline stretchability~\ourref[Stretchability]{p:Stretch} for every fixed $d \geq 2$. The problem remains \ER-complete if halfspaces are replaced by balls\index{Hypergraphs Induced by Balls and Points}, ellipsoids\index{Hypergraphs Induced by Ellipsoids and Points} and, so-called, bi-curved, difference-separable, and computable convex sets as shown by Bertschinger, El Maalouly, Kleist, Miltzow,
and Weber~\cite{BEMKMW23b}.
\openq Does the problem remain \ER-complete for any other natural geometric objects? The authors of~\cite{BEMKMW23b} note that they are not aware of any such examples.
\also Stretchability~\ourref[Stretchability]{p:Stretch}, point-halfspace orders~\ourref{p:pointhalfspaceorders}, oriented hyperplane realizability~\ourref{p:orientedhyperplanereal}.
\end{problemCG}

\begin{problemCG}{Linear Hypergraph Realizability}
\label{p:linhypreal}
\mindex{Linear Hypergraph Realizability}
\given A {\em linear hypergraph} $H$, that is, a hypergraph in which every two edges share at most one point. 
\question Is there a {\em straight-line realization} of $H$; that is, can every hyperedge be drawn as a straight-line segment containing all vertices of the hyperedge and no other vertices of $H$? The segments are allowed to cross. 
\complexity \ER-complete, by Dobler, Kobourov, Lenhart, Mchedlidze, N\"{o}llenburg, and Symvonis~\cite{DKLMNS24} via a reduction from matroid realizability over the reals~\ourref[Matroid!Real Realizability]{p:matroidreal}
\comments Linear hypergraphs of degree at most $2$ are always realizable in the sense defined above. The {\em rank} of a hypergraph is the maximum size of a hyperedge. The straight-line realizability problem becomes non-trivial already for hypergraphs of rank and degree at most $3$~\cite{DKLMNS24}; this corresponds to incidence graphs of geometric $3$-configurations~\ourref[Incidence Graph of Geometric 3 Configuration@Incidence Graph of Geometric $3$-Configuration]{p:geoconfig}
\openq Dobler, Kobourov, Lenhart, Mchedlidze, N\"{o}llenburg, and Symvonis ask whether the problem remains \ER-hard if the hypergraph has bounded rank, of even rank $3$. The answer for bounded rank is likely yes, see the comments on a bounded version of collinearity logic~\ourref[Collinearity Logic]{p:CollLogic}.
\also Matroid realizability over the reals~\ourref{p:matroidreal}, collinearity logic~\ourref[Collinearity Logic]{p:CollLogic}, segment intersection graph~\ourref{p:SEG}, incidence graph of geometric $3$-configuration~\ourref{p:geoconfig}.
\end{problemCG}

\begin{problemCG}{(Point) Visibility Graph}
\label{p:visibilitygraph}
\mindex{Visibility Graph!Points}
\index{Point Visibility Graph|see {Visibility Graph}}
\given A graph $G$.
\question Is $G$ the visibility graph of a set of points in the plane? The visibility graph of a set of points in the plane has one vertex per point, and two vertices are adjacent if and only if the line segment between the two points does not contain any other point of the set.
\complexity \ER-complete, by Cardinal and Hoffmann~\cite{CH17}.
\comments If the points in the realization have to lie on a grid, the problem becomes \EQ-complete~\cite{CH17}. 
\openq The complexity of recognizing visibility graphs of polygons is open. In the visibility graph of a polygon, two vertices of the polygon are adjacent if and only if the line segment between them is contained in the polygon.
\also Visibility graph of polygon with holes~\ourref{p:polygonholesvisibility}, unit spanning ratio~\ourref{p:unitspanratio}, segment visibility graph~\ourref{p:segvisibilitygraph}, obstacle number~\ourref{p:obstaclenumb}, hidden set in polygon~\ourref{p:hiddenpoly}.
\end{problemCG}

\begin{problemCG}{Visibility Graph of Polygon with Holes}
\label{p:polygonholesvisibility}
\mindex{Visibility Graph!Polygon with Holes}
\index{Polygon!Visibility Graph of Polygon with Holes|see {Visibility Graph}}
\index{Visibility Graph of Polygon with Holes|see {Visibility Graph}}
\given A graph $G$.
\question Is there a polygon with holes such that $G$ is the visibility graph of the vertices of $G$; that is, the vertices of $G$ are the corners of the polygon and its holes, and there is an edge between two vertices if the segment connecting the two vertices lies inside the polygon?
\complexity \ER-complete by Hoffmann and Merckx~\cite{HM18} via a reduction from allowable sequence~\ourref[Allowable Sequence]{p:allowableseq} and Boomari, Ostovari, and Zarei~\cite{BOZ18} from simple stretchability~\ourref[Simple Stretchability]{p:Stretch}. 
\comments For convex polygons with a single hole the visibility graphs are circular arc graphs and therefore recognizable in polynomial time~\cite{CE95}.
\openq What is the complexity of recognizing visibility graphs of polygons without holes\index{Visibility Graph!Polygon (Open)}? Boomari, Ostovari, and Zarei~\cite{BOZ18} conjecture that the problem remains \ER-complete. 
\also Allowable sequence~\ourref{p:allowableseq}, point visibility graph~\ourref{p:visibilitygraph}, internal/external polygon visibility graph~\ourref{p:intextvisibility}. 
\end{problemCG}

\begin{problemCG}{Internal/External Polygon Visibility Graph}
\label{p:intextvisibility}
\mindex{Visibility Graph!Internal/External Polygon}
\index{Internal/External Polygon Visibility Graph|see {Visibility Graph}}
\given A graph $G = G_{\mathit{int}} \cup G_{\mathit{ext}}$.
\question Is there a simple polygon such that $G_{\mathit{int}}$ is the visibility graph of the vertices of $G$ inside the polygon, and $G_{\mathit{ext}}$ is the visibility graph of the vertices of $G$ outside the polygon? Two vertices can see each other if the line segment connecting them does not cut through the side of the polygon.
\complexity \ER-complete by Boomari, Ostovari, and Zarei~\cite{BOZ18} from simple stretchability~\ourref[Simple Stretchability]{p:Stretch}. The problem remains \ER-complete even if the blocking vertex for each pair of vertices which cannot see each other is specified (see~\cite{BOZ18} for definition of blocking vertex).  
\openq What is the complexity of recognizing polygon visibility graphs\index{Visibility Graph!Polygon (Open)}, that is, if just $G_{\mathit{int}}$ is given? 
\also Visibility graph of polygon with holes~\ourref{p:polygonholesvisibility}, point visibility graph~\ourref{p:visibilitygraph}. 
\end{problemCG}

\begin{problemCG}{Triangulated Irregular Network Visibility Graph}
\label{p:tiangulatedirregular}
\mindex{Visibility Graph!Triangulated Irregular Network}
\index{Triangulated Irregular Network Visibility Graph| see {Visibility Graph}}
\given A graph $G$, a plane triangulation $T$. 
\question Is there a triangulated irregular network (TIN) with visibility graph $G$ and $xy$-projection isomorphic to $T$? The {\em triangulated irregular network (TIN)} assigns coordinates in $\R^3$ to the vertices of $T$. Two vertices in the TIN can {\em see} each other if the straight-line segment connecting them lies above the TIN. 
\complexity \ER-complete by Boomari, Ostovari, and Zarei~\cite{BOZ21} from triangulation stretchability~\ourref[Stretchability!Triangulation]{p:TriStretch}. 
\also Triangulation stretchability~\ourref{p:TriStretch}, boundary guarding~\ourref{p:boundaryguard}. 
\end{problemCG}

\begin{problemCG}{Generalized Transmission Graph}
\label{p:transgraph}
\index{Generalized Transmission Graph, see {Graph(s)}}
\mindex{Graph(s)!Generalized Transmission Graph}
\given A directed graph $G = (V,E)$, a family $\mathcal{C}$ of geometric objects (e.g.\ lines in $\R^2$).
\question Are there objects $C(v) \in \mathcal{C}$ and points $p(v) \in C(v)$ for every vertex $v \in V$, such that $p(u) \in C(v)$ if and only if $uv \in E$ for all $u,v \in V$?
\complexity \ER-complete for $\mathcal{C}$ being lines in $\R^2$ by Klost and Mulzer~\cite{KM18} via a reduction from simple stretchability~\ourref[Simple Stretchability]{p:Stretch}. The problem is also \ER-complete for $\mathcal{C}$ being a restricted type of circular sectors instead of lines~\cite{KM18}.  
\openq For \ER-completeness of circular sectors, Klost and Mulzer~\cite{KM18} require a somewhat technical condition (they call wide-spread); they conjecture that that condition can be removed.
\sep\ What about other natural families of geometric objects? Are there general conditions on $\mathcal{C}$ that would guarantee $\ER$-hardness?
\also Segment intersection graph~\ourref{p:SEG}.
\end{problemCG}

\subsection{Graph Drawing}

\begin{quote}
Graphs in this section are simple (no multiple edges, no loops) and undirected, unless we say otherwise. A {\em (topological) drawing} of a graph assigns a different location to each vertex of the graph, typically in $\R^2$, and connects the endpoints of each edge by a (Jordan) curve. We assume that there are no vertices in the interior of edges, edges intersect finitely often, and no more than two edges cross in a point. In a {\em geometric/straight-line} drawing, all edges are represented by straight-line segments. We refer to crossing-free drawings as {\em embeddings} or {\em planar}.

A {\em rotation} at a vertex is a cyclic permutation of the edges incident on the vertex (meant to represent the order in which the edges leave the vertex in a drawing or embedding). A {\em rotation system} assigns a rotation to each vertex of a graph. 

We will occasionally encounter \EQ\, the existential theory of the rationals.
\end{quote}

\begin{problemCG}{Planar Slope Number}
\label{p:planarslopenumb}
\mindex{Graph(s)!Planar Slope Number}
\index{Planar Slope Number|see {Graph(s)}}
\given A graph $G$, an integer $k \geq 0$.
\question Is there a crossing-free straight-line drawing of $G$ in which the edges are drawn using at most $k$ different slopes?
\complexity \ER-complete, by Hoffmann~\cite{H17}, reduction from pseudoline stretchability~\ourref[Stretchability]{p:Stretch}. For fixed $k$, the problem lies in \NP~\cite{H17} using a result from~\cite{KM94} which shows that intersection graphs of segments with a fixed number of slopes can be recognized in \NP. 
\comments A drawing realizing the planar slope number of the graph may not be possible on a grid, and it is\EQ-complete to decide whether a grid realization exists, and there is a $2^{2^{\Omega(n^{1/3})}}$ lower bound on the grid size~\cite{H17}.  
\openq How hard is it to determine the (non-planar) slope number\index{Graph(s)!Slope Number (Open)} of a graph~\cite{H17}? 
\also Upward planar slope number~\ourref{p:upplanarslopenumb}, upward outerplanar slope number~\ourref[Graph(s)!Upward Outerplanar Slope Number]{p:upouterplanarslopenumb}, segment intersection graph~\ourref{p:SEG}, line cover number~\ourref{p:linecovernumb}, segment number~\ourref{p:segnumber}, faithful drawability of edge-colored graph~\ourref{p:faithfullcolor}. 
\end{problemCG}

\begin{problemCG}{Upward Planar Slope Number}
\label{p:upplanarslopenumb}
\mindex{Graph(s)!Upward Planar Slope Number}
\index{Upward Planar Slope Number|see {Graph(s)}}
\given A directed graph $G$, an integer $k \geq 0$. A (straight-line) drawing of $G$ is {\em upward} if $u$ lies below $v$ for every directed edge $uv \in E(G)$.
\question Is there an upward planar straight-line drawing of $G$ in which the edges are drawn using at most $k$ different slopes?
\complexity \ER-complete, by Quapil~\cite{Q21}, reduction from pseudoline stretchability~\ourref[Stretchability]{p:Stretch}. 
\comments A drawing realizing the planar slope number of the graph may not be possible on a grid, and it is\EQ-complete to decide whether a grid realization exists, and there is a $2^{2^{\Omega(n^{1/3})}}$ lower bound on the grid size~\cite{H17}.  
\openq Does the problem turn \NP-complete for fixed $k$, as is the case for planar slope number~\ourref[Graph(s)!Planar Slope Number]{p:planarslopenumb}? It is known to be \NP-hard in that case~\cite{KZ23}. \sep\ Is the problem \EQ-complete if the vertices have to lie on a grid?
\also Planar slope number~\ourref[Graph(s)!Planar Slope Number]{p:planarslopenumb}, upward outerplanar slope number~\ourref[Graph(s)!Upward Outerplanar Slope Number]{p:upouterplanarslopenumb}. 
\end{problemCG}

\begin{problemCG}{Line Cover Number in $\R^d$}
\label{p:linecovernumb}
\mindex{Graph(s)!Line Cover Number in Rd@Line Cover Number in $\R^d$}
\index{Line Cover Number|see {Graph(s)}}
\given A graph $G$, an integer $k \geq 0$
\question Are there $k$ lines in $\R^d$ whose union contains a crossing-free straight-line drawing of $G$? We say the lines {\em cover} $G$.
\complexity \ER-complete by Chaplick, Fleszar, Lipp, Ravsky, Verbitsky, and Wolff~\cite{CFLRVW23} for $d \in \{2,3\}$.
\comments Chaplick \textit{et al.}~\cite{CFLRVW23} define the {\em affine cover number}\index{Graph(s)!Affine Cover Number in Rd@Affine Cover Number in $\R^d$}
as the smallest number of $\ell$-dimensional planes in $\R^d$ that cover a straight-line drawing of $G$. Since the affine cover number of a graph is the same for $d > 3$ and $d = 3$, it is sufficient to study line cover numbers in $\R^2$ and line and plane cover numbers in $\R^3$. 
\openq Chaplick \textit{et al.}~\cite{CFLRVW23} show that the plane cover number is \NP-hard in $\R^3$. Is it \ER-complete? Is it \ER-complete for other values of $(\ell,d)$? 
\also Planar slope number~\ourref{p:planarslopenumb}, circle cover number under spherical cover number~\ourref{p:sphericalcovernumb}, minimum line number~\ourref{p:minlinenumb}, planar stick number of drawing~\ourref{p:planarsticknumb}.
\end{problemCG}

\begin{problemCG}{Segment Number}
\label{p:segnumber}
\mindex{Graph(s)!Segment Number}
\given A graph $G$, an integer $k \geq 0$.
\question Is there a crossing-free straight-line drawing of $G$ which is the union of $k$ straight-line segments? The smallest such $k$ is known as the {\em (planar) segment number}.
\complexity \ER-complete by Okamoto, Ravsky, and Wolff~\cite{ORW19} via a reduction from arrangement graph recognition~\ourref[Arrangement Graph]{p:arrangegraph} even if the maximum degree of the graph is $4$ and by Erickson~\cite{E19} via a reduction from simple stretchability~\ourref[Simple Stretchability]{p:Stretch} if the graph is $4$-regular and given with a fixed embedding which has to be realized. Several variants are \ER-complete as well, including the case that polyline drawings are allowed ({\em bend segment number}\index{Graph(s)!bend segment number}), that crossings are allowed ({\em crossing segment number}\index{Graph(s)!crossing segment number}), or that the straight-line drawing lies in $\R^3$ ({\em 3D segment number}\index{Graph(s)!3D segment number})~\cite{ORW19}.
\comments Erickson's result is phrased as straightening a closed curve, see the isotopic polygon problem~\ourref[Polygon!Isotopic]{p:isopolygon}. The arc number~\ourref[Graph(s)!Arc Number (Open)]{p:arcnumb} was introduced by Schulz~\cite{S15} as a variation of the segment number. 
\also Arrangement graph~\ourref{p:arrangegraph}, arc number~\ourref{p:arcnumb}, minimum line number~\ourref{p:minlinenumb}, isotopic polygon~\ourref{p:isopolygon}, planar stick number of drawing~\ourref{p:planarsticknumb}.
\end{problemCG}

\begin{problemCG}{Isotopic Polygon}
\label{p:isopolygon}
\mindex{Polygon!Isotopic}
\index{Isotopic Polygon|see {Polygon}}
\given A closed curve $\gamma$, that is, a $4$-regular plane graph, an integer $k \geq 0$.
\question Is there a polygon with $k$ vertices isotopic to $\gamma$?
\complexity \ER-complete by Erickson~\cite{E19} via a reduction from simple stretchability~\ourref[Simple Stretchability]{p:Stretch}.
\comments The isotopic polygon problem is a special case of the
planar stick number of drawing~\ourref[Graph(s)!Planar Stick Number of Drawing (Open)]{p:planarsticknumb}, and can be viewed as the fixed embedding variant of the segment number~\ourref[Graph(s)!Segment Number]{p:segnumber}, and the 
\universality Universal~\cite{E19}.
\also Segment number~\ourref{p:segnumber}, planar stick number of drawing~\ourref{p:planarsticknumb}.
\end{problemCG}

\begin{problemCG}{Planar Stick Number of Drawing}
\label{p:planarsticknumb}
\mindex{Graph(s)!Planar Stick Number of Drawing}
\index{Planar Stick Number of Drawing (Open)|see {Graph(s)}}
\given A graph $G$ with topological drawing $D$ (not necessarily crossing-free), an integer $k \geq 0$.
\question Is there a poly-line drawing of $G$ isomorphic to $D$ with at most $k$ segments in total? 
\complexity \ER-complete via a reduction from isotopic polygon~\ourref[Polygon!Isotopic]{p:isopolygon}.
\comments This problem was suggested by the work of Khandhawit, Pongtanapaisan, and Wasun~\cite{KPW23} who introduce the stick number of a spatial diagram of a graph; since this notion requires knot equivalence, there is currently no chance to relate it to \ER, but the simplified version presented here should still be of interest, as the underlying computational problem. Erickson~\cite{E24} observed that hardness follows from the isotopic polygon problem~\ourref[Polygon!Isotopic]{p:isopolygon}, even if $G = K_3$: start with the closed curve $\gamma$ from that problem and view it as a topological drawing of $K_3$.
\also Segment number~\ourref{p:segnumber}, isotopic polygon~\ourref{p:isopolygon}, line cover number~\ourref{p:linecovernumb}, minimum line number~\ourref{p:minlinenumb}. 
\end{problemCG}

\begin{problemCG}{Triangulation Stretchability}  
\label{p:TriStretch}
\mindex{Stretchability!Triangulation}
\index{Triangulation Stretchability|see {Stretchability}}
\given A plane triangulation $T$, and a collection of paths in $T$.
\question Is there a straight-line drawing of $T$ in which each path in the collection lies on a line?
\complexity \ER-complete by Boomari, Ostovari, and Zarei~\cite{BOZ21} from simple stretchability~\ourref[Simple Stretchability]{p:Stretch}. 
\also Triangulated irregular network visibility graph~\ourref{p:tiangulatedirregular}, simple stretchability~\ourref{p:Stretch}. 
\end{problemCG}

\begin{problemCG}{Graph in a Polygonal Domain}
\label{p:GraphinPD}
\mindex{Graph(s)!Graph in a Polygonal Domain}
\index{Graph in a Polygonal Domain|see {Graph(s)}}
\given A graph $G$ and a polygon $P$, the {\em domain}.
Some of the vertices of $G$ are specified to be vertices of~$P$.
\question Does there exists a straight line drawing $D$ of $G$ such that
$D$ lies inside the domain $P$? 
\complexity \ER-complete~\cite{LMM22} via a reduction from \PETRINV~\ourref[Planar ETRINV@\PETRINV]{p:PlanarETRINV}.
\comments The problem is closely related to partial drawing extensibility (PDE)~\ourref[Graph(s)!Partial Drawing Extensibility (Open)]{p:partialdrawingext}.
In the PDE problem, we are given a partial straight-line drawing of a 
graph, and we are asked whether this drawing can be extended to a straight-line drawing of the full graph~\cite{P06}.
This differs from the graph in a polygonal domain problem in that in PDE we are 
not allowed to place vertices on existing edges, whereas in the current problem we are allowed to place vertices on the already existing polygon.
\openq Patrignani~\cite{P06} showed that the partial drawing extensibility problem~\ourref[Graph(s)!Partial Drawing Extensibility (Open)]{p:partialdrawingext} is \NP-hard, is it \ER-complete? 
\also Partial drawing extensibility~\ourref{p:partialdrawingext}.
\end{problemCG}

\begin{problemCG}{Drawing Graph on Segments with Obstacles}
\label{p:drawgraphsegobstacles}
\mindex{Graph(s)!Drawing Graph on Segments with Obstacles}
\index{Drawing Graph on Segments with Obstacles|see {Graph(s)}}
\given A plane graph $G$, a line-segment $\ell_v$ (open or closed) for every $v \in V(G)$, and a set $S$ of line-segments, all in the plane. 
\question Is there a straight-line drawing of $G$, isomorphic to the given embedding, in which every vertex $v$ lies on $\ell_v$, and edges do not cross the line-segments in $S$?
\complexity \ER-complete by Bieker~\cite{B20} via a reduction from a variant of \PETRINV~\ourref[Planar ETRINV@\PETRINV]{p:PlanarETRINV}.
\comments Since open line segments can be points, this problem can be turned into a restricted version of the partial drawing extensibility problem~\ourref[Graph(s)!Partial Drawing Extensibility (Open)]{p:partialdrawingext}, whose complexity is open.  
\openq Segments are allowed to be both open and closed. Does the problem remain \ER-hard if all segments are open, or all segments are closed? \sep\ The proof requires segments with four different slopes, does the problem become easier with fewer slopes? \sep\ Does the problem remain \ER-hard if $S$ is empty? \sep\ Does the problem remain \ER-hard if the embedding of the graph is not fixed?
\also Graph in a polygonal domain~\ourref{p:GraphinPD}, partial drawing extensibility~\ourref{p:partialdrawingext}, pseudo-segment stretchability~\ourref{p:SegStretch}.
\end{problemCG}

\begin{problemCG}{Rectilinear Crossing Number}
\label{p:rectcross}
\mindex{Graph(s)!Rectilinear Crossing Number}
\index{Rectilinear Crossing Number|see {Graph(s)}}
\given A graph $G$, an integer $k \geq 0$.
\question Does $G$ have a straight-line drawing with at most $k$ crossings? The smallest such $k$ is the {\em rectilinear crossing number} of $G$.
\complexity \ER-complete. \ER-hardness was shown by Bienstock~\cite{B91} via a reduction from simple stretchability~\ourref[Simple Stretchability]{p:Stretch} and Dean~\cite{D02} showed that the problem lies in \ER. 
\comments For $k \leq 2$ it is known that the rectilinear crossing number equals the standard crossing number~\cite{BD93}, so the problem can be decided in polynomial time for those values. (The case $k = 3$ is also claimed, but only sketched, in the paper.)
\universality Bienstock's reduction yields algebraic universality~\cite{B91}.
\openq Does the problem remain \ER-complete for graphs of bounded degree?
\sep\ Does the problem remain \ER-complete for fixed $k$? What is the complexity of the problem for $k = 4$?
\sep\ Does the problem remain \ER-complete if each edge consists of $t$ segments, for some fixed $t$. What about $t = 2$? This question is based on Bienstock~\cite{B91} who named the corresponding crossing number the {\em $t$-polygonal crossing number}, and pointed out that realizing an optimal $t$-polygonal crossing number drawing may require exponential precision for every fixed $t$.
\also Rectilinear local crossing number~\ourref{p:rectloccross}, pseudolinear vs rectilinear crossing number~\ourref{p:pseudorectcross}, segment intersection graph~\ourref{p:SEG}, maximum rectilinear crossing number~\ourref{p:maxrectcross}.
\end{problemCG}

\begin{problemCG}{Pseudolinear vs Rectilinear Crossing Number}
\label{p:pseudorectcross}
\mindex{Graph(s)!Pseudolinear vs Rectilinear Crossing Number}
\index{Pseudolinear vs Rectilinear Crossing Number|see {Graph(s)}}
\given A graph $G$.
\question If $G$ has a pseudolinear drawing with at most $k$ crossings, does $G$ have a straight-line drawing with $k$ crossings, for every $k$? In other words, does the pseudolinear crossing number of $G$ equal its rectilinear crossing number~\ourref[Graph(s)!Rectilinear Crossing Number]{p:rectcross}? A {\em pseudolinear drawing} of a graph is a pseudoline arrangement in which every edge is drawn as part of a pseudoline.
\complexity \ER-complete by Hern\'{a}ndez-V\'{e}lez, Lea\~{n}os, and Salazar via a reduction from pseudoline stretchability~\ourref[Stretchability]{p:Stretch}.
\also Rectilinear crossing number~\ourref{p:rectcross}, stretchability~\ourref[Stretchability]{p:Stretch}.
\end{problemCG}

\begin{problemCG}{Rectilinear Local Crossing Number}
\label{p:rectloccross}
\mindex{Graph(s)!Rectilinear Local Crossing Number}
\index{Rectilinear Local Crossing Number|see {Graph(s)}}
\given A graph $G$, an integer $k \geq 0$.
\question Does $G$ have a straight-line drawing with at most $k$ crossings {\em per edge}? The smallest such $k$ is the {\em rectilinear local crossing number} of $G$.
\complexity \ER-complete~\cite{S21}. Remains \ER-complete for fixed $k \geq 867$~\cite{S21}.
\comments \ER-complete~\cite{S21} via a reduction from simple pseudo-segment stretchability~\ourref[Stretchability!Pseudo-segments]{p:SegStretch}. \NP-complete for $k = 1$~\cite{S14}, and \NP-hard for every fixed $k \geq 2$~\cite{S21}.
\universality Reduction yields algebraic universality~\cite{S21}.
\openq Does the problem remain \ER-complete for graphs of bounded degree? \sep\ What is the largest $k$ for which the problem lies in \NP? What is the smallest $k$ for which the problem is \ER-complete?
\also Rectilinear crossing number~\ourref{p:rectcross}, segment intersection graph~\ourref{p:SEG}, pseudo-segment stretchability~\ourref{p:SegStretch}.
\end{problemCG}

\begin{problemCG}{Geometric Partial $1$-Planarity}
\label{p:geopartial1planarity}
\mindex{Graph(s)!Geometric Partial 1 Planarity@Geometric Partial $1$-Planarity}
\index{Geometric Partial 1 Planarity@Geometric Partial $1$-Planarity|see {Graph(s)}}
\given A graph $G = (V,E)$, a subset of edges $F \subseteq E$.
\question Is there a straight-line drawing of $G$ in which every edge of $F$ has at most one crossing?
\complexity \ER-complete by Schaefer~\cite{S14} via a reduction from pseudoline stretchability~\ourref[Stretchability]{p:Stretch}. 
\comments The special case of geometric $1$-planarity in which $F = E$ is only \NP-complete, a folklore result~\cite{S14}. 
\openq What is the complexity of geometric partial planarity~\ourref[Graph(s)!Geometric Partial Planarity]{p:geopartialplanarity}, the variant of the problem in which edges in $F$ have to be crossing-free? 
\also Geometric partial planarity~\ourref{p:geopartialplanarity}, rectilinear local crossing number~\ourref{p:rectloccross}. 
\end{problemCG}

\begin{problemCG}{Weak Rectilinear Realizability}
\label{p:rectilinearreal}
\mindex{Graph(s)!Weak Rectilinear Realizability}
\mindex{Graph(s)!Strict Rectilinear Realizability}
\given A graph $G = (V,E)$, a set $R \subseteq \binom{E}{2}$.
\question Is there a straight-line drawing of $G$ such that $\{e,f\} \in R$ for every two edges $e$ and $f$ that cross in the drawing?
\complexity \ER-complete, even for complete graphs by Kyn\v{c}l~\cite{K11} via a reduction from simple stretchability~\ourref[Simple Stretchability]{p:Stretch}. Remains \ER-complete in the {\em strict} case, in which $R$ contains exactly the pairs of edges that cross in the drawing, even if $G$ is the complete graph.
\comments The strict case for a matching is the segment intersection graph problem~\ourref[Segment Intersection Graph]{p:SEG}, which was known to be \ER-complete~\cite{KM94}; the fact that the problem remains hard for complete graphs is novel here~\cite{K11}. For topological drawings, the problem in \NP-complete~\cite{K91b,SSS03}.
\also segment intersection graph~\ourref{p:SEG}, straight-line realizability of graph with rotation system~\ourref{p:straightrotation}.
\end{problemCG}

\begin{problemCG}{Straight-line Realizability of Graph with Rotation System}
\label{p:straightrotation}
\mindex{Graph(s)!Straight-line Realizability of Graph with Rotation System}
\given A graph $G = (V,E)$ with {\em rotation system} $\pi$, that is, a function that assigns to every vertex $v \in V$ a cyclic permutation of the edges incident on $v$.
\question Is there a straight-line drawing of $G$ such that the cyclic permutation of the edges incident to each vertex $v$ agrees with $\pi(v)$? 
\complexity \ER-complete for complete graphs~\cite{K11} via a reduction from strict rectilinear realizability~\ourref[Graph(s)!Strict Rectilinear Realizability]{p:rectilinearreal}. Remains \ER-complete for graphs of bounded degree~\cite{S21b} via a reduction from the partial version of radial system realizability~\ourref[Radial System!Realizability]{p:radialsystem}~\cite{S21b}.
\comments For complete graphs, the pairs of intersecting edges completely determine the rotation system of a straight-line drawing and vice versa; this implies that for complete graphs, the problem is equivalent to the strict rectilinear realizability problem. 
\openq How hard is the problem for cubic graphs? \sep\ Is it \NP-complete to test whether a graph with rotation system has a good (topological) drawing? For complete graphs, this can be decided in polynomial time~\cite{K20}. 
\also Weak rectilinear realizability~\ourref{p:rectilinearreal}, simultaneous geometric embedding~\ourref{p:SEG}, radial system realizability (partial version)~\ourref{p:radialsystem}.
\end{problemCG}

\begin{problemCG}{Simultaneous Geometric Embedding (\SGE)}
\label{p:SGE}
\mindex{Simultaneous Geometric Embedding, see {Graph(s)}}
\mindex{Graph(s)!Simultaneous Geometric Embedding}
\index{SGE@\SGE|see {Graph(s), Simultaneous Geometric Embedding}}
\given A family $(G_i)_{i \in [k]}$ of graphs on a shared vertex set.
\question Is there a straight-line drawing of $G = \bigcup_{i \in [k]} G_i$ in which each $G_i$ by itself is planar?
\complexity \ER-complete by Cardinal and Kusters~\cite{CK15}. Remains \ER-complete even if $k$ is fixed and all graphs are paths~\cite{S21b} and in the case that there are no {\em public} edges (edges belonging to more than one graph), the {\em empty sunflower case}~\cite{KR23} even for fixed $k=31$~\cite{FKMPTV24}. 
A weak hardness result for \SGE\ was first presented in~\cite{EBGJPSS08}.
\comments Explicitly shown \ER-complete in~\cite{CK15} via a reduction from simple stretchability~\ourref[Simple Stretchability]{p:Stretch}; the result can also be proved by reduction from strict rectilinear realizability of complete graphs~\ourref[Graph(s)!Strict Rectilinear Realizability]{p:rectilinearreal}~\cite{K11}, but the reduction from simple stretchability leads to stronger precision bounds: a simultaneous geometric embedding may require a grid of size $2^{2^{\Omega(n)}}$, where $n = V(G)$. The reductions for fixed $k$ start with pseudo-segment stretchability~\ourref[Stretchability!Pseudo-segments]{p:SegStretch}. 
\openq Does the empty sunflower case remain \ER-hard if all graphs are trees? \sep\ Does the problem remain \ER-complete for $k = 2$~\cite{C15}?
\also Strict rectilinear realizability~\ourref{p:rectilinearreal}, geometric thickness of multigraphs~\ourref{p:geothickmulti}, geometric thickness~\ourref{p:geothick}, pseudo-segment stretchability~\ourref{p:SegStretch}.
\end{problemCG}

\begin{problemCG}{Geometric Thickness of Multigraph}
\label{p:geothickmulti}
\mindex{Graph(s)!Geometric Thickness of Multigraph}
\given A multigraph $G$, a natural number $k \in \N$.
\question Does $G$ have a straight-line drawing in which the edges can be colored with $k$ colors in such a way that no crossing in the drawing involves two edges of the same color?
\complexity \ER-complete by F{\"{o}}rster, Kindermann, Miltzow, Parada, Terziadis, and Vogtenhuber~\cite{FKMPTV24} for every fixed $k \geq 57$ via a reduction from pseudo-segment stretchability~\ourref[Stretchability!Pseudo-segments]{p:SegStretch}. The problem remains \ER-complete even if $G$ has fixed local crossing number~\cite{FKMPTV24}.
\comments Geometric thickness was introduced by Kainen~\cite{K73}. It has similarities to the simultaneous geometric embedding problem~\ourref[Graph(s)!Simultaneous Geometric Embedding]{p:SGE}, which is \ER-complete. For a brief discussion, see~\cite[Remark 7]{S21b}. The complexity of the geometric thickness of (simple) graphs~\ourref[Graph(s)!Geometric Thickness (Open)]{p:geothick} remains open. 
\also Geometric thickness~\ourref{p:geothick}, simultaneous geometric embedding~\ourref{p:SGE}.
\end{problemCG}

\begin{problemCG}{Right-Angle Crossing Drawability}
\label{p:racdraw}
\mindex{Graph(s)!Right Angle Crossing Drawability@Right-Angle Crossing Drawability}
\index{Right-Angle Crossing Drawability|see {Graph(s)}}
\index{RAC Drawability|see {Graph(s),Right-Angle Crossing Drawability}}
\given A graph $G$.
\question Does $G$ have a straight-line drawing in which all crossings are right-angle crossings? Such drawings are known as right-angle crossing drawings, or {\em RAC-drawings}, for short. 
\complexity \ER-complete by Schaefer~\cite{S23b} via a reduction from feasibility~\ourref[Polynomial(s)!feasibility]{p:feasibility} (using straight-line drawability with junctions as an intermediate problem). Remains \ER-complete even if the topological embedding of the graph is given and each edge is involved in at most eleven crossings, and crossing angles are doubly exponentially close to being right-angle~\cite{S23b}.
\comments Determining whether $G$ has a RAC-drawing on the integer grid is \EQ-complete~\cite{S23b}. The {\em resolution} at a crossing is the smaller of the two angles incident to a crossing; the {\em crossing resolution} of a straight-line drawing is the largest resolution at any crossing, so RAC-drawings have maximal crossing resolution. The maximal resolution at vertices is known as the angular resolution~\ourref[Graph(s)!Angular Resolution]{p:angularres}.
\openq In a RAC$_k$ drawing each edge is allowed to have up to $k$ bends. Are RAC$_1$- and RAC$_2$-drawability \ER-hard? \sep\ Does RAC-drawbility remain \ER-hard for graphs of bounded degree? \sep\ Does RAC-drawability remain \ER-hard for $1$-planar drawing, that is, drawings with at most one crossing per edge? See geometric partial $1$-planarity~\ourref[Graph(s)!Geometric Partial 1 Planarity@Geometric Partial $1$-Planarity]{p:geopartial1planarity}. 
\sep\ Does RAC-drawability remain \ER-complete if all right-angle crossings have to occur between axis-parallel edges? This variant is introduced as apRAC-drawability in~\cite{ABKKPU23} who showed that the problem is \NP-hard.
\also Angular resolution~\ourref{p:angularres}, geometric partial $1$-planarity~\ourref{p:geopartial1planarity}. 
\end{problemCG}

\begin{problemCG}{Lombardi-Drawability With Rotation System}
\label{p:lombardidraw}
\mindex{Graph(s)!Lombardi Drawability with Rotation System@Lombardi-Drawability with Rotation System}
\index{Lombardi drawability (Open)|see {Graph(s)!Lombardi-Drawability with Rotation System}}
\index{Lombardi drawability with Rotation System|see {Graph(s)}}
\given A graph $G=(V,E)$ with {\em rotation system} $\pi$, that is, a function that assigns to every vertex $v \in V$ a cyclic permutation of the edges incident on $v$.
\question Does $G$ have a Lombardi drawing in the plane respecting $\pi$? In a {\em Lombardi drawing} all edges must be drawn as circular arcs (crossings are allowed), and every vertex must have {\em perfect angular resolution}, that is, all angles between consecutive ends at the vertex must be the same. The drawing {\em respects} the rotation system if the cyclic permutation of the edges incident to each vertex $v$ agrees with $\pi(v)$? 
\complexity \ER-complete by Jungeblut~\cite{J24} via a reduction from simple stretchability~\ourref[Simple Stretchability]{p:Stretch}.
\comments 
\openq How hard is Lombardi drawability if no rotation system is given~\cite[Open Problem 1]{J24}? \sep\
How hard is Lombardi drawability if we require the drawing to be crossing-free (with or without rotation system)n~\cite[Open Problem 2]{J24}?
\sep\ Does Lombardi drawability remain hard for graphs of bounded degree?
\also Straight-line realizability of graph with rotation system~\ourref{p:straightrotation}, angular resolution~\ourref[Graph(s)!Angular Resolution]{p:angularres}, arc number~\ourref{p:arcnumb}.
\end{problemCG}

\begin{problemCG}{Angular Resolution}
\label{p:angularres}
\mindex{Graph(s)!Angular Resolution}
\index{Angular Resolution|see {Graph(s)}}
\given A graph $G$, a natural number $k \geq 2$.
\question Does $G$ have a straight-line drawing with {\em angular resolution} at least $\pi/(2k)$, that is, a drawing in which all angles between incident edges are at least $\pi/(2k)$? The {\em angular resolution} of $G$ is the supremum of the angular resolution of all straight-line drawings of $G$. 
\complexity \ER-complete by Schaefer~\cite{S23} via a reduction from feasibility~\ourref[Polynomial(s)!feasibility]{p:feasibility} (using straight-line drawability with junctions as an intermediate problem). Remains \ER-complete even if a crossing-free topological embedding of the graph is given~\cite{S23}.
\comments There is a subtle difference between a graph having angular resolution at least $\alpha$ and there being a drawing of the graph with angular resolution at least $\alpha$, since angular resolution of a graph is defined as a supremum, as observed in~\cite{FHHKLSWW93}. Both versions are \ER-complete.  
\openq Testing for angular resolution at least $\pi/4$ is \NP-complete. Is the angular resolution problem \ER-complete for $\pi/3$~\cite{S23}? \sep\ Does the problem remain \ER-complete if $G$ is planar, but the angular-resolution drawing does not have to be~\cite{S23}?
\also RAC-drawability~\ourref{p:racdraw}.
\end{problemCG}

\begin{problemCG}{Prescribed Area Extension}
\label{p:preareaext}
\mindex{Graph(s)!Prescribed Area Extension}
\index{Prescribed Area Extension|see {Graph(s)}}
\given A plane graph $G$, an area assignment for $G$, where an {\em area assignment} for $G$ assigns a positive real number to every face of $G$; locations for some of the vertices of $G$ are fixed in the plane. 
\question Does $G$ have a straight-line drawing realizing the area assignment, that is, each face having the area prescribed by the area assignment, with the fixed vertices at their prescribed location? 
\complexity \ER-complete by Dobbins, Kleist, Miltzow and Rz\polhk a\.{z}ewski~\cite{DKMR23} via a reduction from \PETRINV~\ourref[Planar ETRINV@\PETRINV]{p:PlanarETRINV}.
\openq Does the problem remain \ER-hard without fixing vertex locations?  
\also Prescribed area~\ourref{p:prearea}, area universality~\ourref{p:areauniversal}, area universality for triples with partial area assignment~\ourref{p:areauniversaltriples}.  
\end{problemCG}

\begin{problemCG}{Straight-line Realizability with Unit Spanning Ratio}
\label{p:unitspanratio}
\mindex{Graph(s)!Straight line Realizability with Unit Spanning Ratio}
\index{Straight-line Realizability with Unit Spanning Ratio, see {Graph(s)}}
\given A graph $G$.
\question Is there a {\em proper} straight-line drawing of $G$ in the plane with spanning ratio one? Here proper means that no two vertices are drawn on the same point, and no point corresponding to a vertex lies on a segment representing an edge, apart from the two endpoints. The {\em spanning ratio} of a drawing is the maximum ratio, over all pairs of vertices, of the length of a shortest path in the drawing and the Euclidean distance. 
\complexity \ER-complete, by Aichholzer, Borrazzo, Bose, Cardinal, Frati, Morin, Vogtenhuber~\cite{ABBCFMV22} via a reduction from the point visibility graph problem~\ourref[Visibility Graph!Points]{p:visibilitygraph}.
\comments Aichholzer et al.~\cite{ABBCFMV22} show that a graph has a proper drawing with spanning ratio $1$ if and only if it is the visibility graph of a set of points in the plane~\ourref[Visibility Graph!Points]{p:visibilitygraph}. They also show that the problem becomes linear-time solvable if the drawing has to be planar (crossing-free). 
\also Point visibility graph~\ourref{p:visibilitygraph}.
\end{problemCG}

\begin{problemCG}{Graph Realizability}
\label{p:graphreal}
\mindex{Graph(s)!Graph Realizability}
\index{Graph Realizability|see {Graph(s)}}
\given A graph $G$ with weight function $\ell: E(G) \rightarrow \Q_{\geq 0}$.
\question Is there a straight-line drawing of $G$ in which every edge $e$ of $G$ has length $\ell(e)$?  
\complexity \ER-complete by Schaefer~\cite{S13} via a reduction from pseudoline stretchability~\ourref[Stretchability]{p:Stretch}. The problem remains \ER-complete even if $\ell(e) = 1$ for all $e \in E(G)$.
\comments Graph drawings are assumed to be non-degenerate in that vertices do not overlap with other vertices or edges; relaxing that condition leads to the linkage realizability problem~\ourref{p:linkagereal}. 
The problem was first shown \NP-hard by Saxe~\cite{S79} and Yemini~\cite{Y79}; Eades and Wormald~\cite{EW90} showed \NP-hardness if the drawing has to be crossing-free (and all edges have the same length), and that result was strengthened by Cabello, Demaine, and Rote~\cite{CDR07}. The problem is known under different names to different communities, e.g.\ as the Euclidean distance matrix completion problem\index{Euclidean Distance Matrix Completion}~\cite{L09}, and it has many variants and applications~\cite{LLMM14}. A related problem is the reconstruction problem: given a multiset of distances (of the form $\sqrt{q}$ for $q \in \Q$) are there $n$ points in $\R^d$ such that the multiset contains exactly their pairwise distances. This problem turns out to be \NP-complete~\cite{LSS03}.
\openq It is open whether the problem remains \ER-complete for embedded graphs.
(A claimed \ER-hardness proof~\cite{ADDELS16} may not be complete. We were informed by one of the authors
that the PhD thesis of Abel~\cite{PhDAbel2016} contains more details, `` but is still lacking some detail and edits''.) 
\also Unit distance graph~\ourref{p:unitdist}, linkage realizability~\ourref{p:linkagereal}, graph flattenability~\ourref{p:graphflatten}. 
\end{problemCG}

\begin{problemCG}{Graph with Long and Short Edges}
\label{p:graphlongshort}
\mindex{Graph(s)!Graph with long and short edges}
\index{Graph with long and short edges|see {Graph(s)}}
\index{Dichotomous ordinal graph|see {Graph(s)}}
\given A graph $G = (V,E)$ with a partition of the edges $E = E_s \cup E_{\ell}$ into {\em short} and {\em long} edges. 
\question Is there a straight-line drawing of $G$ in which exactly the short edges have length at most $1$?   
\complexity \ER-complete by Angelini, Cornelsen, Haase, Katsanou, Montecchiani, Symvonis~\cite{ACHKMS24} via a unit disk graph~\ourref[Disk Graph!Unit Disk Graph]{p:unitdiskgraph}.
\comments Graphs with long and short edges were introduced in~\cite{ACHKMS24} under the name {\em dichotomous ordinal graphs}. The same paper suggests a reduction from Euclidean preference~\ourref[Euclidean Preference]{p:Euclideanpref}, but there also is a direct reduction from unit disk graphs~\ourref[Disk Graph!Unit Disk Graph]{p:unitdiskgraph}: $G = (V,E)$ is a unit disk graph if and only if the graph $H$ on vertex set $V$ can be realized with short edges $E$ and long edges $\binom{V}{2}-E$. 
\openq Does the problem remain \ER-complete if the realization has to be crossing-free? If the short edges have to induce a plane graph, then, as~\cite{ACHKMS24} point out, the problem is \NP-hard~\cite{AKPT16}. 
\sep\ Does the problem remain \ER-complete if the underlying graph is a complete bipartite graph? That is the case of most interest in~\cite{ACHKMS24}.
\also Unit disk graph~\ourref{p:unitdiskgraph}, Euclidean preference~\ourref{p:Euclideanpref}, graph realizability~\ourref{p:graphreal}. 
\end{problemCG}

\begin{problemCG}{(Strict) Unit Distance Graph}
\label{p:unitdist}
\mindex{Graph(s)!Unit Distance Graph}\mindex{Graph(s)!Strict Unit Distance Graph}
\given A graph $G$.
\question Is $G$ a (strict) subgraph of $E_2 = (\R^2, \{uv \in (\R^2)^2: |u-v| = 1\})$?
\complexity Recognition of both unit distance graphs and strict unit distance graphs is \ER-complete by Schaefer~\cite{S13} via a reduction from graph realizability~\ourref[Graph(s)!Graph Realizability]{p:graphreal}.
\comments By definition, different vertices of $G$ correspond to distinct points in $E_2$; if we relax this condition, we get the degenerate unit distance graph problem~\cite{HKP11}, which we discuss as part of the linkage realizability problem~\ourref{p:linkagereal}. Erd\"{o}s, Haray and Tutte~\cite{EHT65} introduced the {\em Euclidean dimension}\index{Graph(s)!Euclidean Dimension}\index{Euclidean dimension|see {Graph(s)}} of a graph, which is the smallest $d$ for which $G$ is a subgraph of $E_d = (\R^d, \{uv \in (\R^d)^2: |u-v| = 1\})$; so testing whether a graph has dimension at most $2$ is already \ER-complete.
\universality Maehara~\cite{M91} showed that the distances that occur in rigid unit distance graphs are exactly the algebraic numbers. 
\openq Is the Euclidean dimension \ER-complete for every fixed $d$?
\also Linkage realizability~\ourref{p:linkagereal}, unit disk contact graph~\ourref{p:unitdiskcontact}, graph realizability~\ourref{p:graphreal}. 

\end{problemCG}

\begin{problemCG}{Linkage Realizability}
\label{p:linkagereal}
\mindex{Linkage!Linkage Realizability}
\given A graph $G$ with a function $\ell: E(G) \rightarrow \Q_{\geq 0}$.
\question Can we assign a point $p_u \in \R^d$ to every $u \in V(G)$ such that $|f(u)-f(v)| = \ell(uv)$ for every edge $uv$?
\complexity \ER-complete by Abbott~\cite{A08} via a reduction from Feasibility~\ourref[Polynomial(s)!feasibility]{p:feasibility} for every $d\geq 2$, The problem remains \ER-complete even for unit linkages , that is if $\ell(e) = 1$ for all $e \in E(G)$, in $\R^2$ by Schaefer~\cite{S13} via a reduction from pseudoline stretchability~\ourref[Stretchability]{p:Stretch}. 
\comments The main difference to graph realizability~\ourref[Graph(s)!Graph Realizability]{p:graphreal} is that vertices can overlap with other vertices and edges. In some definitions, linkages also come with points with pre-assigned locations. Unit linkages are called {\em degenerate unit distance graphs} by Horvat, Kratochv\'{\i}l, and Pisanski~\cite{HKP11} who showed that the problem is \NP-hard.
\universality The universality of linkage realizability is Kempe's universality theorem~\cite{KM02, A08}.
\openq Is realizability of unit linkages \ER-complete in arbitrary dimension? There is a gadget, due to Saxe~\cite{S79}, that increases the dimension by one, but it is not a unit linkage.
\also Unit distance graph~\ourref{p:unitdist}, graph realizability~\ourref{p:graphreal}, linkage rigidity~\ourref{p:linkagerigid}. 
\end{problemCG}

\begin{problemCG}{Linkage Rigidity}
\label{p:linkagerigid}
\mindex{Linkage!Linkage Rigidity}
\given A graph $G$ with a function $\ell: E(G) \rightarrow \Q_{\geq 0}$, and a {\em configuration}, that is, $p_u \in \Q^d$ for every $u \in V(G)$ such that $|p_u-p_v| = \ell(uv)$ for every edge $uv$.
\question Is the configuration rigid, that is, do small changes to all $p_u$ so that the length conditions are still satisfied lead to the same configuration modulo rigid motions of the plane?
\complexity \VR-complete by Abbott~\cite{A08}, Schaefer~\cite{S13} for $d \geq 2$ via a reduction from \ISO~\ourref[Polynomial(s)!Isolated Zero]{p:ISO}. Abbott~\cite{A08} shows that \ISO\ reduces to linkage rigidity for $d = 2$, and $d-1$-dimensional linkage rigidity reduces to $d$-dimensional linkage rigidity. Together with the \ER-completeness of \ISO, this implies \ER-completeness of $d$-dimensional linkage rigidity for every fixed $d \geq 2$. Membership of linkage rigidity in \VR\ is shown in~\cite{S13}.
\comments The main difference to graph realizability~\ourref[Graph(s)!Graph Realizability]{p:graphreal} is that vertices can overlap with other vertices and edges. 
\also Unit distance graph~\ourref{p:unitdist}, linkage realizability~\ourref{p:linkagereal}, graph realizability~\ourref{p:graphreal}. 
\end{problemCG}

\subsection{Polytopes and Simplicial Complexes}

\begin{quote}
A polytope is the convex hull of a finite point set in $\R^d$.
The {\em face lattice} of a polytope is the lattice of its faces: the {\em join}
of two faces is the smallest face containing both faces, and the {\em meet}
is the intersection of the faces. Given a lattice, that is, a partial order with
join and meet operations, one can ask whether it is {\em polytopal}, namely
whether it is the face lattice of a polytope. 
We refer to the face lattice of $P$ as its \emph{combinatorial type}, and to $P$ itself as a
\emph{realization} of this combinatorial type. 
The realizability question is known as the {\em algorithmic Steinitz problem}, due to a famous theorem of Steinitz.
Let the {\em graph of a polytope} be the graph formed by the vertices and edges of the
polytope. The graph is also referred to as the \emph{$1$-skeleton}, and the $k$-skeleton for $k\geq 1$ is the 
complex of all faces of dimension up to $k$.
Steinitz theorem characterizes the graphs of $3$-dimensional polytopes:
they are precisely the simple, planar, $3$-connected graphs~\cite[Section 4]{Z95}, hence easy to
recognize. For $3$-polytopes the facial lattice is determined by the graph of the polytope,
so the algorithmic Steinitz problem for $3$-polytopes is tractable.

Note that the complete combinatorial type of a polytope can be deduced from its set of $(n-1)$-dimensional faces, or \emph{facets}: the lower-dimensional faces are simply obtained by intersecting subsets of facets. The input to the problems can therefore be restricted to be a list of facets.

We obtain an interesting variant if we restrict the problem to \emph{simple} polytopes, the skeletons of which are regular graphs. In that case the facial lattice is determined by the graph of the polytope 
(Blind and Mani with an easy proof by Kalai, see~\cite[Section~3.4]{Z95} and~\cite[Problem 15]{KP03}).

A polytope is said to be \emph{simplicial} if all its facets are simplices. 

\emph{Subdivisions} are closely related to polytopes. Given a set of points, a
subdivision is a covering of their convex hull by polytopes, or \emph{cells}, the vertices of which are
points of the set, such that any two intersect in a common face.
If all cells are simplices, then the subdivision is called a \emph{triangulation}.
A \emph{Delaunay subdivision} is such that the cells are defined by an empty sphere condition:
A cell $C$ is part of the Delaunay subdivision of a set of points in $\R^d$ if and only if there exists a
$(d-1)$-sphere passing through the vertices of $C$ but containing no other point of the set in its interior.
For generic point sets, such cells are always simplices, and we talk of \emph{Delaunay triangulations}.

An \emph{abstract simplicial $k$-complex} is a collection of sets of size at most $k$ that is closed under taking subsets. More generally, an \emph{(abstract) polyhedral complex} is a collection of (combinatorial types of) polytopes, such that every face of every polytope is also part of the collection, and any two polytopes in the collection intersect in a common face. 
\end{quote}

\begin{problemCG}{Polytope Realizability}
\label{p:Polytope}
\mindex{Polytope(s)!Polytope Realizability}
\index{Polytope Realizability|see {Polytope(s)}}
\given A combinatorial type, given as a list of facets. 
\question Does there exist a polytope of this combinatorial type? 
\complexity \ER-complete, by Mn\"{e}v~\cite{M88} via a reduction from existential theory of totally ordered real variables~\ourref[Existential Theory of Totally Ordered Real Variables]{p:ETRORV}.
\comments This problem is also known as the \emph{algorithmic Steinitz problem}. The problem is known to be hard in particular for $n$-dimensional polytopes with 
$n+4$ vertices. The input is then a collection of subsets of size at least $n$ of the $n+4$ vertices.
\universality Mn\"{e}v~\cite{M88}, also see Richter-Gebert~\cite{RG99}.
\also 4-dimensional Polytope Realizability~\ourref{p:4DPolytope}, Simplicial Polytope Realizability~\ourref{p:SPolytope}, Inscribed Simplicial Polytope Realizability~\ourref{p:ISPolytope}, treetope graph~\ourref{p:treetopegraph}, existential theory of totally ordered real variables~\ourref{p:ETRORV}.
\end{problemCG}

\begin{problemCG}{4-dimensional Polytope Realizability}
\label{p:4DPolytope}
\mindex{Polytope(s)!4-dimensional Polytope Realizability}
\index{4-dimensional Polytope Realizability|see {Polytope(s)}}
\given A combinatorial type, given as a list of facets. 
\question Does there exist a 4-dimensional polytope of this combinatorial type? 
\complexity \ER-complete, by Richer-Gebert and Ziegler~\cite{RGZ95} and Richter-Gebert~\cite{RG96} via a reduction from existential theory of totally ordered real variables~\ourref[Existential Theory of Totally Ordered Real Variables]{p:ETRORV}. 
\universality Richter-Gebert and Ziegler~\cite{RGZ95}, and Richter-Gebert~\cite{RG96,RG99}.
\openq The complexity of the problem for simplicial polytopes of constant dimension $d\geq 4$ is open, see Richter-Gebert~\cite{RG96}. The complexity of the related problem in dimension 4 in which we are only given the graph ($1$-skeleton) of the polytope is also open.
\also Polytope Realizability~\ourref{p:Polytope}.
\end{problemCG}

\begin{problemCG}{Simplicial Polytope Realizability}
\label{p:SPolytope}
\mindex{Polytope(s)!Simplicial Polytope Realizability}
\index{Simplicial Polytope Realizability|see {Polytope(s)}}
\given A combinatorial type, given as a list of facets. 
\question Does there exist simplicial polytope of this combinatorial type? 
\complexity \ER-complete, by Mn{\"e}v~\cite{M88}, and Adiprasito and Padrol~\cite{AP17} even in even dimension via a reduction from oriented matroid realizability~\ourref[Matroid!Oriented Matroid Realizability]{p:ormatroidreal}.
\universality Adiprasito and Padrol~\cite{AP17}.
\comments Universality of simplicial polytopes was announced by Mn{\"e}v~\cite{M88}, but with an incomplete proof. Adiprasito and Padrol prove the universality result for \emph{neighborly polytopes}, the $d$-dimensional polytopes whose $\lfloor d/2\rfloor$-skeleton is complete. Neighborly polytopes in even dimension are simplicial, so this implies the current result, but we can further assume neighborliness. 
The problem is known to be hard in particular for $n$-dimensional polytopes with 
$n+4$ vertices. 
\openq Encouraged by the results of Richter-Gebert~\cite{RG96} and Bokowski and Guedes de Oliveira~\cite{BG90}, Adiprasito and Padrol make the ``daring conjecture'' that the result holds in dimension $4$. 
\also Inscribed Simplicial Polytope Realizability~\ourref{p:ISPolytope}.
\end{problemCG}

\begin{problemCG}{Inscribed Simplicial Polytope Realizability}
\label{p:ISPolytope}
\mindex{Polytope(s)!Inscribed Simplicial Polytope Realizability}
\index{Inscribed Simplplical Polytope Realizability|see {Polytope(s)}}
\given A combinatorial type, given as a list of facets. 
\question Does there exist simplicial polytope of this combinatorial type, all vertices of which lie on a sphere? 
\complexity \ER-complete, by Adiprasito, Padrol, and Theran~\cite{APT15}.
\universality Adiprasito, Padrol, and Theran~\cite{APT15}.
\comments Delaunay subdivisions are obtained by stereographic projections of inscribed polytopes.
The problem is known to be hard in particular for $n$-dimensional polytopes with 
$n+4$ vertices. 
\openq The complexity of the problem for inscribed polytopes (not necessarily simplicial) of constant dimension $d\geq 4$ is open as well. In dimension $3$, a result by Rivin implies that inscribed polytopes can be recognized in polynomial time~\cite{R96}, although it is not clear how to efficiently produce a realization.
\also Delaunay Triangulation~\ourref{p:Delaunaytri}.
\end{problemCG}

\begin{problemCG}{Nested Polytopes}
\label{p:nestpolytopes}
\mindex{Polytope(s)!Nested Polytopes}
\index{Nested Polytopes|see {Polytope(s)}}
\given Two polytopes $A\subseteq B \subseteq \R^d$ and a number $k\in \N$. The polytope $A$ is described by vertices and the polytope $B$ is described by its facets.
\question Does there exist a polytope $X$ on at most $k$ vertices with $A\subseteq X \subseteq B$?
\complexity \ER-complete, by Dobbins, Holmsen and Miltzow~\cite{DHM19} via a direct reduction from \ETRINV~\ourref[ETRINV@\ETRINV]{p:ETRINV}.
\universality Universality follows from~\cite{DHM19}.
\comments  Vavasis~\cite{V09} 
showed that there is a reduction from non-negative matrix factorization~\ourref[Matrix!Non-negative Factorization]{p:nonnegmatfac} to nested polytope,
for the special case $k=d+1$.
(In the case $k=d-1$ the intermediate polytope is a simplex, and thus they call this problem the intermediate simplex problem.
\index{Intermediate Simplex Problem|see {Polytope(s)}})

Shitov showed in turn \ER-completeness and universality for non-negative matrix factorization and thus also implied the nested polytope result~\cite{S16}.
\also \ETRINV~\ourref{p:ETRINV}.
\end{problemCG}

\begin{problemCG}{Delaunay Triangulation}
\label{p:Delaunaytri}
\mindex{Pointset!Delaunay Triangulation}
\index{Delaunay Triangulation|see {Pointset}}
\given A collection $F$ of subsets of size $n-4$ of $[n]$.
\question Does there exist a $(n-5)$-dimensional Delaunay triangulation on $n$ vertices, such that the vertices of the cells of the triangulations are exactly the subsets in $F$?
\complexity \ER-complete, by Adiprasito, Padrol, and Theran~\cite{APT15} via a reduction from Inscribed Simplicial Polytope Realizability~\ourref[Inscribed Simplicial Polytope Realizability]{p:ISPolytope}.
\universality Adiprasito, Padrol, and Theran~\cite{APT15}.
\comments The result is directly reduced from the complexity of Inscribed Simplicial Polytope Realizability~\ourref[Inscribed Simplicial Polytope Realizability]{p:ISPolytope}: An inscribed polytope in dimension $d$ is mapped to a Delaunay subdivision in dimension $d-1$ via a stereographic projection.
\openq The complexity of the problem in constant dimension $d\geq 3$ is open. In dimension two, it is known from a result by Rivin that Delaunay triangulations can be recognized in polynomial time~\cite{R96}, although it is not clear how to efficiently produce a realization. Adiprasito, Padrol, and Theran conjecture that universality already holds for $3$-dimensional Delaunay subdivisions~\cite{APT15}. Padrol and Theran~\cite{PT14} give examples of Delaunay triangulations with disconnected realization spaces.
\also Inscribed Simplicial Polytope Realizability~\ourref{p:ISPolytope}, Gabriel graph~\ourref{p:Gabrielgraph}. 
\end{problemCG}

\begin{problemCG}{Simplicial Complex}
\label{p:simplicialcomplex}
\mindex{Simplicial Complex}
\given An abstract simplicial $k$-complex $C$ and a natural number $d \in \N$.
\question Does $C$ have a geometric embedding in $\R^d$?
\complexity \ER-complete for all $d\geq 3$ and $k\in\{d-1,d\}$, by Abrahamsen, Kleist, and Miltzow~\cite{AKM23} via a reduction from stretchability~\ourref[Stretchability]{p:Stretch}.
\comments For piecewise linear embedding, the problem has been shown to be \NP-hard even in dimension 3, see~\cite{MTW11, MRST20}.
\openq For which combinations of $k$ and $d$ is the problem \ER-complete?
An interesting case is when $k=2$ and $d=4$.
\also Polyhedral complex~\ourref{p:polyhedralcomplex}.
\end{problemCG}

\begin{problemCG}{Polyhedral Complex}
\label{p:polyhedralcomplex}
\mindex{Polyhedral Complex}
\given An abstract polyhedral complex $C$ composed of quadrangles and triangles.
\question Does $C$ have a geometric embedding in $\R^3$ in which the image of every polygon is a convex polygon, and their relative interiors are disjoint?
\complexity \ER-complete, by Brehm (unpublished), via a reduction from collinearity~\ourref[Collinearity Logic]{p:CollLogic}.
\comments The result has been announced and presented, but never published formally. It is hinted at by Ziegler~\cite{Z08}. Wilson~\cite{W12} also showed that the same result holds for polyhedral complexes composed of convex three-dimensional polyhedra.
\also Collinearity logic~\ourref{p:CollLogic}, simplicial complex~\ourref{p:simplicialcomplex}.
\end{problemCG}

\section{Game Theory}

\subsection{Games in Strategic Form}

\begin{quote}
We briefly recall the definitions related to games given in \emph{strategic form}.
An $r$-player game $\Gamma$ in strategic form is a pair $\Gamma =(\Sigma_i, \mathsf{U}_i)$, for $i\in 1,2,\ldots ,r$, where $\Sigma_i$ is the set of strategies available to the $i$th player, and $\mathsf{U}_i$ is the payoff function of player $i$, of the form $\mathsf{U}_i:\Sigma_1\times\Sigma_2\times\ldots\times\Sigma_r\to\R$.

A \emph{mixed strategy} for the $i$th player, as opposed to a \emph{pure strategy} $s\in\Sigma_i$, is a probability distribution $\sigma_i$ on the pure strategies in $\Sigma_i$. Let $\mathbf{\sigma}=(\sigma_i)_{i=1,2,\ldots,r}$ be a collection of mixed strategies for each of the $r$ players. The expected payoff of the $i$th player under these probability distributions is:
\[
\mathsf{U}_i(\mathbf{\sigma}) = \sum_{s\in\prod_{i=1}^r \Sigma_i}
\mathsf{U}_i(s) \cdot
\prod_{i=1}^r \sigma_i(s_i)\cdot
\]

A Nash equilibrium is a collection of mixed strategies such that no player can improve her expected payoff by deviating from her current mixed strategy. Hence it consists of a collection of mixed strategies $\mathbf{\sigma}$ such that for every $i\in\{1,2,\ldots ,r\}$ and every $\mathbf{\sigma}'$ obtained by replacing $\sigma_i$ by another mixed strategy $\sigma'_i$, we have $\mathsf{U}_i(\mathbf{\sigma}) \geq \mathsf{U}_i(\mathbf{\sigma}')$. The classical result of Nash is that such an equilibrium exists for any game.
A Nash equilibrium is said to be \emph{pure} if it consists only of pure strategies. 
\end{quote}

\begin{problemGT}{Nash Equilibrium in a Ball}
\label{p:NashinBall}
\mindex{Nash Equilibrium!In a Ball}
\given A $3$-player game $\Gamma$ in strategic form, $u\in\Q$. 
\question Does $\Gamma$ have a Nash equilibrium $\mathbf{\sigma}$ such that $\sigma_i(s)\leq u$ for all $i=1,2,3$ and all $s\in\Sigma_i$?
\complexity \ER-complete by Schaefer, \v{S}tefankovi\v{c}~\cite{SS17} via a reduction from Brouwer Fixed Point~\ourref[Polynomial(s)!Brouwer Fixed Point]{p:BrouwerFixedPoint} due to Ettessami and Yannakakis~\cite{EY10}.
\comments The result holds for a fixed $u=1/2$~\cite{SS17}.
\also Brouwer Fixed Point~\ourref{p:BrouwerFixedPoint}.
\end{problemGT}

\begin{problemGT}{Nash Equilibrium with Bounded Payoffs}
\label{p:NashBoundedPayoff}
\mindex{Nash Equilibrium!Bounded Payoffs}
\given A $3$-player game $\Gamma$ in strategic form, $u\in\Q$. 
\question Does $\Gamma$ have a Nash equilibrium $\mathbf{\sigma}$ such that $\mathsf{U}_i(\mathbf{\sigma})\geq u$ for all $i\in\{ 1,2,3\}$?
\complexity \ER-complete by Garg, Mehta, Vazirani, Yazdanbod~\cite{GMVY18} via a reduction from Nash Equilibrium in a ball~\ourref{p:NashinBall}.
\comments The problem in which we require that $\mathsf{U}_i(\mathbf{\sigma})\leq u$ instead is also \ER-complete. Both results hold for symmetric win-lose games as well, as shown by Bil{\`o}, Hansen, Arnsfelt and Mavronicolas~\cite{BHM23}. Berthelsen and Hansen~\cite{BH22} also show that for zero-sum games, deciding the existence of a Nash equilibrium in which $\mathsf{U}_i(\mathbf{\sigma})=0$ for all $i\in\{ 1,2,3\}$  is \ER-complete. The proofs of the latter are related to those in Hansen~\cite{H19}.
\end{problemGT}

\begin{problemGT}{Nash Equilibrium with Restricted Support}
\label{p:NashRestrictedSupp}
\mindex{Nash Equilibrium!Restricted Support}
\given A $3$-player game $\Gamma$ in strategic form, a subset $T_i\subseteq\Sigma_i$ for each $i=1,2,3$. 
\question Does $\Gamma$ have a Nash equilibrium $\mathbf{\sigma}$ such that the set of strategies $s\in\Sigma_i$ such that $\sigma_i(s)>0$ is contained in $T_i$ for every $i\in\{1,2,3\}$?
\complexity \ER-complete by Garg, Mehta, Vazirani, Yazdanbod~\cite{GMVY18} via a reduction from Nash Equilibrium in a ball~\ourref{p:NashinBall}.
\comments The problem in which we require that the set contains $T_i$ instead is also \ER-complete~\cite{GMVY18,BM21}. The two problems are respectively called \textit{Superset} and \textit{Subset} in~\cite{GMVY18}.
Both results hold for win-lose games as well~\cite{BHM23}.
\end{problemGT}

\begin{problemGT}{Second Nash Equilibrium}
\label{p:NashSecond}
\mindex{Nash Equilibrium!Second}
\index{Second Nash Equilibrium|see {Nash Equilibrium}}
\given A $3$-player game $\Gamma$ in strategic form. 
\question Does $\Gamma$ have at least two distinct Nash equilibria? 
\complexity \ER-complete by Garg, Mehta, Vazirani, Yazdanbod~\cite{GMVY18} via a reduction from Nash Equilibrium with Bounded Payoffs~\ourref{p:NashBoundedPayoff}. The problem is referred to as \textit{NonUnique} in ~\cite{GMVY18}.
\end{problemGT}

\begin{problemGT}{Nash Equilibrium with Bounded Total Payoff}\mindex{Nash Equilibrium!Bounded Total Payoff}
\label{p:NashBoundedTotalPayoff}
\given A $3$-player game $\Gamma$ in strategic form, $u\in\Q$. 
\question Does $\Gamma$ have a Nash equilibrium $\mathbf{\sigma}$ such that $\sum_{i\in\{1,2,3\}}\mathsf{U}_i(\mathbf{\sigma})\geq u$?
\complexity \ER-complete by Bil{\`o}, Mavronicolas~\cite{BM21} via a reduction from Nash Equilibrium with Restricted Support~\ourref{p:NashRestrictedSupp}.
\comments The problem in which we require that $\sum_i\mathsf{U}_i(\mathbf{\sigma})\leq u$ instead is also \ER-complete.
An alternative hardness proof is also given by Berthelsen and Hansen~\cite{BH22}.
Both results hold for symmetric win-lose games as well~\cite{BHM23}. 
\end{problemGT}

\begin{problemGT}{Nash Equilibrium with Bounded Support}
\label{p:NashBoundedSupp}
\mindex{Nash Equilibrium!Bounded Support}
\given A $3$-player game $\Gamma$ in strategic form, a natural number $k\in\N$. 
\question Does $\Gamma$ have a Nash equilibrium $\mathbf{\sigma}$ such that the number of nonzero components of $\sigma_i$ is at least $k$, for every $i\in\{1,2,3\}$?
\complexity \ER-complete by Bil{\`o}, Mavronicolas~\cite{BM21} via a reduction from Nash Equilibrium with Restricted Support~\ourref{p:NashRestrictedSupp}.
\comments The problem in which we require that the support has size at most $k$ is also \ER-complete. Both results hold for symmetric win-lose games as well~\cite{BHM23}.
\end{problemGT}

\begin{problemGT}{Minmax Value}
\label{p:minmaxGame}
\mindex{Game in Strategic Form!Minmax Value}
\index{Minmax Value|see {Game in Strategic Form!}}
\given A 3-player game $\Gamma$ in strategic form.
\question Is the minmax value of player 1 at most 0? Here the minmax value of player $i$ is the minimum utility $\min_{\sigma} \max_{s_i} \mathsf{U}_i (\sigma \setminus s_i)$ of a best reply strategy, where $\sigma \setminus s_i$ is the collection of strategies obtained by replacing $\sigma_i$ by $s_i$ in $\sigma$.
\complexity \ER-complete by Hansen~\cite{H19} via a reduction from a variant of~\ourref{p:feasibility} similar to that of~\cite[Lemma 3.9]{S13}, where the solution of a system of quadratic equations is promised to lie in a polyhedron contained in the standard simplex.
\comments The problem remains \ER-complete under various promises, see Hansen~\cite{H19} for details.
\end{problemGT}

\begin{problemGT}{Trembling Hand Perfect Equilibrium}
\label{p:thpNash}
\mindex{Nash Equilibrium!Trembling Hand Perfect}
\given A 3-player game $\Gamma$ in strategic form, and a pure Nash equilibrium $\mathbf{\sigma}$ of $\Gamma$.
\question Is $\mathbf{\sigma}$ a trembling hand perfect equilibrium? 
A collection of strategies $\sigma$ is a trembling hand perfect equilibrium if and only if it is the limit of a sequence of $\varepsilon$-perfect equilibria with $\varepsilon\to 0^+$, where an $\varepsilon$-perfect equilibrium is a Nash equilibrium that is fully mixed, and in which only pure strategies that are best replies get probability more than $\varepsilon$, see~\cite{S75,M78}.
\complexity \ER-complete by Hansen~\cite{H19} via a reduction from Minmax value~\ourref{p:minmaxGame}.
\end{problemGT}

\begin{problemGT}{Proper Equilibrium}
\label{p:pNash}
\mindex{Nash Equilibrium!Proper}
\given A 3-player game $\Gamma$ in strategic form, and a pure Nash equilibrium $\mathbf{\sigma}$ of $\Gamma$.
\question Is $\mathbf{\sigma}$ a proper equilibrium? Here $\sigma$ is a proper equilibrium if it is the limit point of a sequence of $\varepsilon$-proper equilibria with $\varepsilon\to 0^+$, where an $\varepsilon$-proper equilibrium is a fully mixed equilibrium such that given two pure strategies $k$ and $\ell$ of the same player, if $k$ is a worse reply against $\sigma$ than $\ell$, then $\sigma$ must assign a probability to $k$ that is at most $\varepsilon$ times the probability it assigns to $\ell$, see~\cite{S75}.
\complexity \ER-complete by Hansen~\cite{H19} via a reduction from Minmax value~\ourref{p:minmaxGame}.
\end{problemGT}

\begin{problemGT}{Strategy Sets Closed under Rational Behavior}
\label{p:curbGame}
\mindex{Game in Strategic Form!Strategy Sets Closed under Rational Behavior}
\given A 3-player game $\Gamma$ in strategic form, and a collection of subsets $S_i\subseteq\Sigma_i$, $i=1,2,3$.
\question Is the collection $\{ S_i\}$ closed under rational behavior (CURB)? Here sets of pure strategies
$\{ S_i\}_{i=1}^m$ of an $m$-player game is CURB if and only if for all pure strategies $s$ of Player $i$ such that $s$ is a best reply to a distribution $S_1\times S_2 \times\ldots\times S_{i-1} \times S_{i+1} \times\ldots\times S_m$,
we have that $s\in S_i$.
\complexity \ER-complete by Hansen~\cite{H19} via a reduction from Minmax value~\ourref{p:minmaxGame}.
\end{problemGT}


\begin{problemGT}{Pareto Optimal Nash Equilibrium}
\label{p:ParetoNash}
\mindex{Nash Equilibrium!Pareto Optimal}
\index{Pareto Optimal Nash Equilibrium|see {Nash Equilibrium}}
\given A $3$-player game $\Gamma$ in strategic form.
\question Does $\Gamma$ have a  Pareto optimal Nash equilibrium? A collection $\mathbf{\sigma}$ of mixed strategies is Pareto optimal if there does not exist any $\mathbf{\sigma}'$ such that $\mathsf{U}_i(\mathbf{\sigma}')\geq \mathsf{U}_i(\mathbf{\sigma})$ for all $i\in\{1,2,3\}$, and $\mathsf{U}_i(\mathbf{\sigma}')> \mathsf{U}_i(\mathbf{\sigma})$ for at least one $i\in\{1,2,3\}$. 
\complexity \ER-complete by Berthelsen, Hansen~\cite{BH22} via a reduction from the variant of Nash Equilibrium with Bounded Payoffs~\ourref{p:NashBoundedPayoff}, in which we require $\mathsf{U}_i(\mathbf{\sigma})=0$. 
\comments The result also holds for zero-sum games. The problem of deciding whether there exists a non-Pareto optimal Nash equilibrium is \ER-hard for symmetric win-lose games as well~\cite{BM21,BHM23}. 
\openq It is open whether the result also holds for win-lose games~\cite{BHM23}.
\end{problemGT}

\begin{problemGT}{Strong Nash Equilibrium}
\label{p:strongNash}
\mindex{Nash Equilibrium!Strong}
\index{Strong Nash Equilibrium|see {Nash Equilibrium}}
\given A $3$-player game $\Gamma$ in strategic form.
\question Does $\Gamma$ have a strong Nash equilibrium? A collection $\mathbf{\sigma}$ of mixed strategies is a strong Nash equilibrium if no coalition of players can together increase their payoff, hence if there is no nonempty subset $B\subseteq\{1,2,3\}$ such that all players in $B$ can increase their payoff by simultaneously deviating from their strategies in $\mathbf{\sigma}$. 
\complexity \ER-complete by Berthelsen, Hansen~\cite{BH22} via a reduction from the variant of Nash Equilibrium with Bounded Payoffs~\ourref{p:NashBoundedPayoff} in which we require $\mathsf{U}_i(\mathbf{\sigma})=0$.
\comments The result also holds for zero-sum games. The problem of deciding whether there exists a non-strong Nash equilibrium is \ER-hard for symmetric win-lose games as well~\cite{BHM23}.
\openq It is open whether the result also holds for win-lose games~\cite{BHM23}. 
\end{problemGT}

\begin{problemGT}{Nonsymmetric Nash Equilibrium}
\label{p:nonsymNash}
\mindex{Nash Equilibrium!Nonsymmetric}
\index{Nonsymmetric Nash Equilibrium|see {Nash Equilibrium}}
\given A symmetric $3$-player game $\Gamma$ in strategic form.
\question Does $\Gamma$ have a nonsymmetric Nash equilibrium? 
\complexity \ER-complete by Berthelsen, Hansen~\cite{BH22} via a reduction from the variant of Nash Equilibrium with Bounded Payoffs~\ourref{p:NashBoundedPayoff} in which we require $\mathsf{U}_i(\mathbf{\sigma})=0$.
\end{problemGT}

\begin{problemGT}{Irrational Nash Equilibrium}
\label{p:irrationalNash}
\mindex{Nash Equilibrium!Irrational}
\index{Irrational Nash Equilibrium|see {Nash Equilibrium}}
\given A $3$-player game $\Gamma$ in strategic form.
\question Does $\Gamma$ have a Nash equilibrium $\mathbf{\sigma}$ in which $\sigma_i(s)\not\in\Q$ for some $i\in\{1,2,3\}$ and some $s\in\Sigma_i$?
\complexity \ER-hard by Berthelsen, Hansen~\cite{BH22} via a reduction from the variant of Nash Equilibrium with Bounded Payoffs~\ourref{p:NashBoundedPayoff} in which we require $\mathsf{U}_i(\mathbf{\sigma})=0$.
\comments The result also holds for zero-sum games~\cite{BH22}, and for win-lose games~\cite{BHM23}.
The problem of deciding whether there exists a rational Nash equilibrium, in which $\sigma_i(s)\in\Q$ for all $i\in\{1,2,3\}$ and all $s\in\Sigma_i$ is shown to be \EQ-hard~\cite{BH22}. It is open whether the problem is even decidable.
\end{problemGT}

\subsection{Games in Extensive Form}

\begin{quote}
Games in \emph{extensive form} are defined by a rooted tree with utility vectors at the leaves. Nodes of the tree are controlled either by one of the player, or by a so-called chance player, in which case we associate a probability distribution to the outgoing edges. In games in extensive form with \emph{imperfect information}, strategies can only be based on partial information, or \emph{signals} regarding the moves of the opponent.
A behavioral strategy is a mixed strategy based only on the received signals.
In games with \emph{imperfect recall}, not only strategies can only be based on partial information, but players can forget about the sequence of signals or actions.
The definitions extend to games that are defined on directed acyclic graphs.
We refer to Hansen~\cite{H19} and Gimbert, Paul, Srivathsan~\cite{GPS20} for detailed definitions.
\end{quote}

\begin{problemGT}{Trembling Hand Perfect Equilibrium for Games in Extensive Form}
\label{p:NashTremblingExt}
\mindex{Nash Equilibrium!Game in Extensive Form!Trembling Hand Perfect}
\index{Trembling Hand Perfect Equilibrium for Game in Extensive Form|see {Nash Equilibrium, Game in Extensive Form}}
\mindex{Nash Equilibrium!Game in Extensive Form!Quasi Perfect}
\index{Quasi Perfect Equilibrium for Game in Extensive Form|see {Nash Equilibrium, Game in Extensive Form}}
\mindex{Nash Equilibrium!Game in Extensive Form!Sequential}
\index{Sequential Equilibrium for Game in Extensive Form|see {Nash Equilibrium, Game in Extensive Form}}
\given A 3-player game $\Gamma$ in extensive form, and a pure Nash equilibrium $\mathbf{\sigma}$ of $\Gamma$.
\question Is $\mathbf{\sigma}$ a trembling hand perfect equilibrium? 
\complexity \ER-complete by Hansen~\cite{H19} via a reduction from Minmax value~\ourref{p:minmaxGame}.
\comments We refer~\cite{S75,vD84} for the definition of trembling hand perfect equilibria for games in extensive form. Hansen~\cite{H19} also gives \ER-completeness results for other properties of pure Nash equilibria of games in extensive form, in particular that of belonging to a \emph{sequential equilibrium}~\cite{KW82} and being a \emph{quasi-perfect equilibrium}~\cite{vD84}. 
\also Trembling Hand Perfect Equilibrium~\ourref{p:thpNash}.
\end{problemGT}

\begin{problemGT}{Non-negative Expected Payoff for Games in Extensive Form}
\label{p:NonnegExt}
\mindex{Game in Extensive Form!Non-negative Expected Payoff}
\index{Non-negative Expected Payoff for Games in Extensive Form|see {Game in Extensive Form}}
\given A one-player game in extensive form, with imperfect recall.
\question Is the maximum expected payoff, over all behavioral strategies, non-negative?
\complexity \ER-complete by Gimbert, Paul, Srivathsan~\cite{GPS20}, via a reduction from feasibility~\ourref[Polynomial(s)!feasibility]{p:feasibility}. The result also holds for deciding whether the maximum payoff is strictly positive.
\also Feasibility~\ourref{p:feasibility}.
\end{problemGT}

\begin{problemGT}{Nonnegative Maxmin Payoff for Games in Extensive Form}
\label{p:NonnegMaxminExt}
\mindex{Game in Extensive Form!Non-negative Maxmin Payoff}
\index{Non-negative Maxmin Payoff for Games in Extensive Form|see {Game in Extensive Form}}
\given A zero-sum two-player game in extensive form, with imperfect recall.
\question Is the maxmin payoff, over all behavioral strategies, non-negative?
\complexity Both \ER-hard and \VR-hard, and belongs to \EVR by Gimbert, Paul, Srivathsan~\cite{GPS20}. Reductions are from Non-negative Expected Payoff for Games in Extensive form~\ourref{p:NonnegExt}.
\also Non-negative Expected Payoff for Games in Extensive form~\ourref{p:NonnegExt}.
\end{problemGT}

\begin{problemGT}{Stationary Nash Equilibrium in Acyclic Games}
\label{p:SNashAcyclic}
\mindex{Nash Equilibrium!Game in Extensive Form!Stationary Nash Equilibrium in Acyclic Games}
\given An acyclic $m$-player recursive game with payoff demands $L\in\R^m$.
\question Does there exist a stationary Nash equilibrium satisfying all the demands? A stationary equilibrium is one involving stationary strategies, that only depend on the current node of the graph.
\complexity \ER-complete by Hansen and S\o lvsten~\cite{HS20}. The reduction is from feasibility~\ourref[Polynomial(s)!feasibility]{p:feasibility} with homogeneous quadratic polynomials over a simplex.
\comments \ER-completeness holds even for $7$-player games with non-negative rewards.
\end{problemGT}

\subsection{Fair Division Problems}

\begin{problemGT}{Square-Cut Pizza Sharing}
\label{p:pizzasquarecut}
\mindex{Pizza Sharing!Square-Cut}
\index{Square-Cut Pizza Sharing|see {Pizza Sharing}}
\given A collection of $n$ pizzas, where a {\em pizza} is a mass distribution (measure) on $[0,1]^2$ that can be computed for polygonal subsets using arithmetic circuits. 
\question Can all pizzas be halved simultaneously by a {\em square cut}, that is, a single cut which is a polygonal arc consisting of horizontal and vertical segments, and such that the square cut makes at most $n-2$ turns? 
\complexity \ER-complete by Deligkas, Fearnley, and Melissourgos~\cite{DFM20,DFM22} via a reduction from consensus halving~\ourref[Consensus Halving]{p:consensushalving}. Remains \ER-complete even if each pizza consists of at most $6$ square or triangular pieces of mass (not necessarily uniform), and each point belongs to at most $3$ pizzas~\cite{DFM20,DFM22}. 
\comments The \ER-hardness result is sharp in the sense that $n$ pizzas can always be halved simultaneously with a square-cut making at most $n-1$ turns~\cite{KRPS16}.
\also Line-cut pizza sharing~\ourref{p:pizzalinecut}, consensus halving~\ourref{p:consensushalving}. 
\end{problemGT}

\begin{problemGT}{Line-Cut Pizza Sharing}
\label{p:pizzalinecut}
\mindex{Pizza Sharing!Line-Cut}
\index{Line-Cut Pizza Sharing|see {Pizza Sharing}}
\given A collection of $2n$ pizzas, where a {\em pizza} is a mass distribution (measure) on $[0,1]^2$ that can be computed for polygonal subsets using arithmetic circuits. 
\question Can all pizzas be halved simultaneously by making $k$ line cuts (and assembling the resulting pieces appropriately)? 
\complexity \ER-complete by Schnider~\cite{S21d} via a reduction from consensus halving~\ourref[Consensus Halving]{p:consensushalving} with $k = n-1$  
\comments The \ER-hardness result is sharp in the sense that $2n$ pizzas can always be halved simultaneously with at most $n$ line cuts~\cite{HK20}.
\also Square-cut pizza sharing~\ourref{p:pizzasquarecut}, consensus halving~\ourref{p:consensushalving}. 
\end{problemGT}

\begin{problemGT}{Consensus Halving}
\label{p:consensushalving}
\mindex{Consensus Halving}
\given A family of $n$ arithmetic circuits computing functions $F_i: [0,1] \rightarrow \R$, $i \in [n]$, a natural number $k \in \N$. The functions $F_i$ are used to valuate intervals and unions of intervals by defining $F_i([a,b]) = F_i(b)-F_i(a)$, and $F_i(A \cup B) = F_i(A) + F_i(B)$ for disjoint subsets $A$ and $B$ of $[0,1]$. 
\question Is there a {\em $k$-cut}, that is, $k$ points $0 \leq x_1 \leq \cdots \leq x_k \leq 1$, with the following property: if we split $[0,1]$ into
two sets $A_+ = \bigcup_{0 \leq 2i \leq k} [x_{2i}, x_{2i+1}]$ and $A_- = \bigcup_{0 \leq 2i+1 \leq k} [x_{2i+1}, x_{2i+2}]$, where we let $x_0 = 0$ and $x_{k+1} = 1$, then $F_i(A_+) = F_i(A_-)$ for all $i \in [n]$? That is all agents, the holders of the valuation functions, valuate the two ``halves'' $A_+$ and $A_-$ equally.
\complexity \ER-complete by Deligkas, Fearnley, Melissourgos, and Spirakis~\cite{DFMS22} via a reduction from feasibility~\ourref[Polynomial(s)!feasibility]{p:feasibility} over a compact domain. The problem is hard for $k = n-1$ and always solvable (using the Borsuk-Ulam Theorem) for $k = n$.
\openq Does the problem remain hard if the $F_i$ are given as polynomials? 
\also Square-cut pizza sharing~\ourref{p:pizzasquarecut}, line-cut pizza sharing~\ourref{p:pizzalinecut}. 
\end{problemGT}

\subsection{Markets}

\begin{problemGT}{Equilibrium in Exchange Markets}
\label{p:Market}
\mindex{Equilibrium in Exchange Markets}
\given An Arrow-Debreu exchange market, consisting of a set $\mathcal{G}$ of $g$ divisible goods, a set $\mathcal{A}$ of agents, a matrix $W$ such that $W_{ij}$ is the amount of good $j$ with agent $i$, and a non-negative piecewise-linear concave utility function $U_i:\R^g_+\to\R_+$.
\question Is there an equilibrium in the market, that is, $p\in\R_+^g$ and $x_i\in\R^g$ for all $i\in\mathcal{A}$, such that
\begin{eqnarray*}
   \forall i\in\mathcal{A}: x_i & = & \mathrm{arg}\max_y \{U_i(y) : \sum_j y_jp_j\leq\sum_j W_{ij}p_j \} \\
\forall j\in\mathcal{G}: \sum_i x_{ij} & = & \sum_i W_{ij}?
\end{eqnarray*}
\complexity \ER-complete by Garg, Mehta, Vazirani, and Yazdanbod~\cite{GMVY17} via a reduction from feasibility~\ourref[Polynomial(s)!feasibility]{p:feasibility}.
The result holds even under the assumption that the utility functions are Leontief utility functions, a special subclass of piecewise-linear concave functions, see~\cite{GMVY17} for definitions.
\end{problemGT}

\section{Machine Learning and Probabilistic Reasoning}

\begin{quote}
Artificial neural networks come in  many variants. For the purpose of this section,
we only define fully-connected, feed-forward, ReLU networks.
Such a network realizes a parameterized function $N_p: \R^{n} \rightarrow \R^m$ which maps an input $x\in \R^n$ to some output $N_p(x) \in \R^m$ given some parameters $p$ (weights and biases).
Given a set of data points $(x,y)$, the goal is to find the right $p$ such that $N_p$ predicts the $y$-values of the data points to within a given threshold.
The function $N_p$ is the composition of two types of functions.
The first type is a simple affine transformation $x \mapsto Ax + b$.
The second type is an activation function; examples include the identity function and the rectified linear unit activation function \ReLU. 
(The name is more complicated than the function itself.)
\ReLU maps $x \in \R^n$ to $y \in \R^n$, where  $y_i = \max(x_i,0)$.
An artificial neural network can be described as a repeated composition
of affine functions and \ReLU's.
We refer the interested reader to the comprehensive book by Goodfellow, Bengio and Aaron on  Deep learning~\cite{goodfellow2016deep}, for an in-depth introduction of the great variety neural network architectures and training methods.

\bigskip

When we want to reason about probabilities in a formal sense, we can define 
formal languages $\mathcal{L}$ which can express probabilities of events,
where events can again be combined using elementary variables, conditional probabilities and conjunction of events, comparison of probabilities, independence statements, confirmation statements and potentially many more. 
There is an illustrative example that shows us that irrational numbers might be needed to find a solution.
We try to find probabilities such that $ \Pr(A \cap B ) = 1 - \Pr(A \cap B)$ and 
$\Pr(A \cap B ) = \Pr(B) \cdot \Pr(B)$ 
implies that $\Pr(B) = \sqrt{1/2}$.
\end{quote}

\begin{problemML}{Training Neural Networks}
\label{p:neuralnettrain}
\mindex{Neural Networks!Training}
\index{Training Neural Network|see {Neural Networks}}
\given A set of data points, a neural network architecture, a cost function, a threshold
\question Do there exist biases and weights such that the total error is below the threshold?
\complexity First Abrahamsen, Kleist and Miltzow~\cite{AKM21} showed \ER-completeness for arbitrary network architectures, two layers, and the identity activation function.
Bertschinger, Hertrich, Jungeblut, Miltzow, and Weber~\cite{BHJMW22} strengthened the result to fully-connected neural networks with only two input, two output neurons and the ReLU activation function.
In case that we allow that some weights are given in the input than the problem was shown to be \ER-complete already in 1992 by Zhang~\cite{Z92}.
\comments The \ER-hardness results were extended in part to other activation functions involving non-algebraic functions~\cite{HHKV23}.
In that case the problems might not be even decidable.
To make the problems with non-algebraic activation functions easier, the neural network problem was studied in a regime where incorrect answers are allowed~\cite{BDC23}.
\also Reachability in neural network~\ourref{p:neuralnetreach}.
\end{problemML}

\begin{problemML}{Reachability in Neural Networks}
\label{p:neuralnetreach}
\mindex{Neural Networks!Reachability}
\index{Reachability in Neural Network|see {Neural Networks}}
\given A feedforward neural network $N$ with weights and biases, seen as a function from $N:\R^n\to\R^m$, and two convex polytopes $A\subseteq \R^n$ and $B\subseteq \R^m$, described by linear inequalities. 
\question Do there exist points $x \in A$ and $y \in B$ such that $N(x) = y$?
\complexity The complexity of the problem depends on the type of activation functions that are allowed, see Wurm~\cite{W23b}. 
The problem becomes \ER-hard for many common non-linear activation functions.
\comments Wurm~\cite{W23b} shows a close relation to constraint satisfaction problems (CSPs).
Different computational complexities of CSPs translate directly to the corresponding complexity of the neural network reachability problem.
\also Training neural network~\ourref{p:neuralnettrain}.
\end{problemML}

\begin{problemML}{Satisfiability of Probabilistic Languages \L{}}
\label{p:satprobL}
\mindex{Satisfiability of Probabilistic Languages}
\given A formula $\varphi$ in the probabilistic language \L .
\question Does there exists an assignment of probabilities that makes $\varphi$ true?
\complexity Moss\'{e}, Ibeling, and Icard~\cite{MII22} showed that the problem is \ER-complete for the languages
$\L^{\textrm{cond}}_{\textrm{prob}}$, $\L^{\textrm{poly}}_{\textrm{prob}}$, $\L^{\textrm{cond}}_{\textrm{causal}}$, and $\L^{\textrm{poly}}_{\textrm{causal}}$.
Ibeling, Icard, Mierzewski, and Mossé~\cite{IIMM23} showed \ER-completeness
for the languages 
$\L^{\textrm{ind}}$, $\L^{\textrm{confirm}}$, $\L^{\textrm{cond}}$, and 
$\L^{\textrm{poly}}$
The problem is known to be \NP-complete for the languages 
$\L^{\textrm{comb}}_{\textrm{prob}}$, 
$\L^{\textrm{lin}}_{\textrm{prob}}$, 
$\L^{\textrm{comp}}_{\textrm{causal}}$, and
or $\L^{\textrm{lin}}_{\textrm{causal}}$, see also Fagin~\cite{F86}.
The reductions are from \ETRINV~\ourref[ETRINV@\ETRINV]{p:ETRINV}.
\comments The results indicate that the difference between 
\NP-completeness and \ER-completeness is caused by 
allowing conditional probability statements 
or the multiplication of probabilities.
The satisfiability problem for probabilistic languages that are allowed to contain summation symbols, e.g.\ abbreviating a sum like $1+2+3 +...+ n$ as $\sum_{i=1}^{n}i$, is complete for the class \succER, a real version of \NEXP~\cite{vdZBL23}.
\also \ETRINV~\ourref{p:ETRINV}.
\end{problemML}

\begin{problemML}{Gaussian Conditional Independence Inference Problem}
\label{p:gaussianindependence}
\mindex{Gaussian Conditional Independence Inference Problem}
\given A Conditional Independence Inference Rule $\varphi$. See~\cite{studeny2006probabilistic} for definitions.
\question Is the rule $\varphi$ valid for all regular Gaussians?
\complexity Boege (see~\cite[Theorem 5.38]{B22b} and \cite[Theorem 11]{B22c}) showed that the problem is \VR-complete.
\comments The proof is via a reduction from feasibility~\ourref[Polynomial(s)!feasibility]{p:feasibility}, using Shor's normal and an encdoing via geometric von Staudt gadgets.
\also Feasibility~\ourref{p:feasibility}.
\end{problemML}

\section{Markov Chains and Decision Processes}

\begin{quote}
We recommend the book by Sutton and Barto on reinforcement learning~\cite{sutton2018reinforcement} for the definition of Markov Decision Processes (MDPs) and their variants, to which a full chapter is dedicated, and which does an excellent job of illustrating the notion in-depth.
\end{quote}

\begin{problemMDP}{Realizability of polynomial parametric Markov Decision Processes}
\label{p:realpolyMC}
\mindex{Markov Decision Process!Realizability of polynomial parametric MDP}
\index{Realizability of polynomial parametric MDP|see {Markov Decision Process}}
\given A polynomial, parametric Markov Decision Process with at least two states.
\question Is the polynomial parametric Markov Decision Process realizable?
\complexity The problem is first claimed to be \ER-complete by Junges, Katoen, P\'{e}rez, and Winkler~\cite{JKPW21} via a reduction from local positivity over a bounded domain~\ourref[Polynomial(s)!Positivity!Local]{p:LocalPos}.
They in turn attribute this to earlier work by
Lanotte, Maggiolo{-}Schettini, Troina~\cite{LMT07}.
\comments Polynomial and parametric in the context of Markov Decision Processes mean that the transition probabilities are polynomial functions $f(x)$ instead of real numbers.
The question is whether there exists some $x$ such that $f$ describes a well-defined Markov Decision Process.
\end{problemMDP}

\begin{problemMDP}{Never Worse States in Graph Parametric MDP}
\label{p:neverworseMDP}
\mindex{Markov Decision Process!Never Worse States in Graph Parametric MDP}
\index{Never Worse States in Graph Parametric MDP|see {Markov Decision Process}}
\given A (non-weighted) parametric Markov Decision Process $\mathcal{M}$ with graph $G(\mathcal{M}) = (V,E)$, together with $v\in V$ and $W\subseteq V$.
\question Is $v$ never worse than $W$?
\complexity Engelen, P\'{e}rez, and Rao showed that the problem is \VR-complete~\cite{EPR23} via a reduction from (the complement of) strict local positivity~\ourref[Polynomial(s)!Strict Positivity!Local]{p:LocalPos}.
\also Equally bad states in graph parameteric MDP~\ourref{p:equallybadMDP}, strict local positivity~\ourref{p:LocalPos}.
\end{problemMDP}

\begin{problemMDP}{Equally Bad States in Graph Parametric MDP}
\label{p:equallybadMDP}
\mindex{Markov Decision Process!Equally Bad States in Graph Parametric MDP}
\index{Equally Bad States in Graph Parametric MDP|see {Markov Decision Process}}
\given A (non-weighted) parametric Markov Decision Process $\mathcal{M}$ with graph $G(\mathcal{M}) = (V,E)$, together with $v,w\in V$.
\question Is $v$ never worse than $w$ and vice versa?
\complexity Engelen, P\'{e}rez, and Rao showed that the problem is \VR-complete~\cite{EPR23} via a reduction from~\ourref[Markov Decision Process!Never Worse States in Graph Parametric MDP]{p:neverworseMDP}.
\also Never worse states in graph parameteric MDP~\ourref{p:neverworseMDP}.
\end{problemMDP}

\begin{problemMDP}{Total Variation Distance in MDP}
\label{p:totvardistMDP}
\mindex{Markov Decision Process!Total Variation Distance}
\index{Total Variation Distance in MDP|see {Markov Decision Process}}
\given A (labeled) Markov Decision Process and two initial distributions $\mu$ and $\nu$.
\question Is there a (memoryless) strategy such that total variation distance in the induced Markov~Chain equals $0$?
\complexity Kiefer and Tang showed that the problem is \ER-complete~\cite{KT20} via a reduction from non-negative matrix factorization~\ourref[Matrix!Non-negative Factorization]{p:nonnegmatfac}. Remains \ER-complete for total variation distance less than~$1$~\cite{KT20}.
\comments The problem is polynomial time decidable if we ask if the total variation distance can be made larger than $0$.
For the related notion of probabilistic bisimilarity all studied cases are either \NP-complete or in \P~\cite{KT20}.
\openq If we ask for total variation distance equal to one then it is known that the problem is \NP-hard and in \ER~\cite{KT20}. 
So it remains to settle this case.
\also Non-negative matrix factorization~\ourref{p:nonnegmatfac}.
\end{problemMDP}

\begin{problemMDP}{Reachability for Augmented Interval Markov Chain}
\label{p:reachMC}
\mindex{Markov Chain!Reachability for Augmented Interval MC}
\index{Reachability for Augmented Interval MC|see {Markov Chain}}
\given An augmented interval Markov Chain $\mathcal{M}$, an initial vertex $s$, a target vertex $t$, and a threshold $\tau\in [0,1]$.
\question Does there exists an Markov Chain $M\in [\mathcal{M}]$ refining $\mathcal{M}$ such that 
the probability to go from $s$ to $t$ is at least/most $\tau$?
\complexity Chonev showed that the problem is \ER-complete~\cite{C19}. 
\comments The problem is also \ER-complete, if we consider the approximate version.
\also Quantitative reachabilty for parametric acyclic Markov chain~\ourref{p:quantreachMC}.
\end{problemMDP}

\begin{problemMDP}{Model Checking for Interval Markov Chains w.r.t.\ Unambiguous Automata} \label{p:MCforMCwrtUA}
\mindex{Markov Decision Process!{Model Checking for Interval Markov Chains with respect to Unambiguous Automata}}
\index{Model Checking for Interval Markov Chains with respect to Unambiguous Automata|see {Markov Chain}}
\given An interval Markov chain and a regular language $E$.
\question Does there exists values for the undetermined transitions of the Markov chain such that $E$ is satisfied?
\complexity The problem was shown to be $\NP_\R$-complete by Benedikt, Lenhardt, and  Worrell~\cite{BLW13b}. 
The same proof easily also shows \ER-completeness.
\comments 
They reduce from Local Positivity~\ourref{p:LocalPos} on a bounded domain with a single multivariate polynomial of a specific form.
\end{problemMDP}

\begin{problemMDP}{Quantitative Reachability for Parametric Acyclic Markov Chains}
\label{p:quantreachMC}
\mindex{Markov Chain!Quantitative Reachability for Parametric Acyclic MC}
\index{Quantitative Reachability for Parametric Acyclic MC|see {Markov Chain}}
\given A parametric \textit{acyclic} Markov Chain and a target state $T$.
\question Does there exists a well-defined and graph preserving valuation of the probability transitions such that the probability to reach $T$ is at least/most $1/2$?
\complexity The problem was shown to be \ER-complete by Junges, Katoen, P\'{e}rez, and Winkler~\cite{JKPW21}.
\comments The problem is very reminiscent to the Reachability problem for Augmented Interval Markov Chains. 
Clearly, the \ER-completeness also carries over to Markov Decision Processes.
The authors also use the term \textit{parameter synthesis} in this context to describe the question of finding parameters for some property to hold.
\also Reachabilty for augmented interval Markov chain~\ourref{p:reachMC}.
\end{problemMDP}

\begin{problemMDP}{Memoryless Distance Minimisation Problem for Markov Decision Processes}
\label{p:memorylessDMPforMDPs}
\mindex{Markov Chain!Memoryless Distance Minimisation Problem for MDP}
\index{Memoryless Distance Minimisation Problem for Markov Decision Processes|see {Markov Decision Process}}
\given A Markov Decision Process with two initial states, and a threshold.
\question Does there exists a memoryless strategy such that the distance between the two initial states is at most the threshold?
\complexity The problem was shown to be \ER-complete by Kiefer and Tang~\cite{KT24}.
\comments They reduce from strict local Positivity \ourref{p:LocalPos}, with bounded domain and a single polynomial of degree $6$.
\end{problemMDP}

\section{Miscellaneous}

\begin{quote}
In this section, we collect problems that we could not fit into any
of the other categories.
\end{quote}

\begin{problemMisc}{Euclidean Preference}
\label{p:Euclideanpref}
\mindex{Euclidean Preference}
\given A finite set $A$, and a {\em profile} $V$, a collection of strict orders $\succ_v$ over $A$, that is, each $\succ_v$ is total, irreflexive, antisymmetric and transitive. The orders are also called {\em preference relations} or {\em votes}. 
\question Is there a function $x: V\cup A \rightarrow \R^d$ such that $a \succ b$ implies that $\norm{x(v)-x(a)}_2 < \norm{x(v)-x(b)}_2$
for all $v \in V$ and $a,b \in A$? In that case the profile $V$ is known as {\em $d$-Euclidean}.
\complexity \ER-complete for any fixed $d \geq 2$ by Peters~\cite{P17} via a reduction from oriented hyperplane realizability~\ourref[Oriented Hyperplane Realizability in Rd@Oriented Hyperplane Realizability in $\R^d$]{p:orientedhyperplanereal}. Remains \ER-complete if all orders are {\em dichotomous}, that is, there is no triple $a,b,c$ with $a \succ_v b \succ_v c$ for any $v \in V$~\cite{P16,P17}.
\comments For the Manhattan and maximum metrics the problem lies in \NP, so it is unlikely to be \ER-hard in those cases~\cite{P17}; \NP-hardness seems to be open for those metrics. 
\also Oriented hyperplane realizability~\ourref{p:orientedhyperplanereal}.
\end{problemMisc}

\begin{problemMisc}{(Ultimate) Positivity of Linear Recurrence}
\label{p:poslinrecur}
\mindex{Linear Recurrence!Positivity}\index{Positivity of Linear Recurrence|see {Linear Recurrence}}
\mindex{Linear Recurrence!Ultimate Positivity}\index{Ultimate Positivity of Linear Recurrence|see {Linear Recurrence}}
\given Coefficients $a_1, \ldots, a_k \in \Q^k$ and initial values $u_0, \ldots, u_{k-1} \in \Q^k$. 
With these numbers we define the sequence $u_n = \sum_{i=1}^k a_i u_{n-i}$ for $n \geq k$.
\question Is the sequence $(u_n)_{n \geq 0}$ {\em (ultimately) positive}; that is, is $u_n > 0$ for (almost) all $n$?
\complexity Both positivity and ultimate positivity are \VR-hard by Ouaknine and Worrell~\cite{OW17} via a reduction from strict local positivity (of a polynomial)~\ourref[Polynomial(s)!Strict Positivity!Local]{p:LocalPos}. For simple recurrences (the characteristic polynomial has no multiple roots), the problems can be decided in $\coNP^{\ER}$~\cite{OW14,OW17}. 
\comments These problems are related to Skolem's problem which asks whether $u_n = 0$ for some $n$. Decidability of Skolem's problem is a famously open question. 
\openq Does positivity (or ultimate positivity) lie in \VR? \sep\ What is the complexity of the robust version of the problem (in which each $a_i$ is given with a small neighborhood), as studied in~\cite{V23}?
\also Strict local positivity~\ourref{p:LocalPos}, zeroness of convolution recursive sequence~\ourref{p:zerconvolution}. 
\end{problemMisc}

\begin{problemMisc}{Linear Update Loop Termination}
\label{p:linearloop}
\mindex{Linear Update Loop Termination}
\given Boolean formula $\varphi(x)$ over polynomial inequalities in $x \in \R^n$, matrix $A \in \R^{n\times n}$, vector $b \in \R^n$.
\question Does the program $\mathrm{\mathbf{while}}\ \varphi(x)\ \mathrm{\mathbf{do}}\ x = Ax+b$ terminate for every initial value $x \in \R^n$?
\complexity \VR-complete by Frohn, Hark, Giesl~\cite[Theorem 7.5]{FHG23} via a reduction from feasibility~\ourref[Polynomial(s)!feasibility]{p:feasibility}. Remains \VR-complete if $b = 0$, but becomes \coNP-complete if $\varphi$ consists of linear polynomial conditions only~\cite[Theorem 7.3]{FHG23}. 
\comments The non-trivial part in this result is \VR-membership, the \VR-hardness is immediate with program $\mathrm{\mathbf{while}}\ \varphi(x)\ \mathrm{\mathbf{do}}\ x = x$. For earlier work on related problems, see~\cite{COW15}.
\openq What is the complexity of the problem if instead we ask whether the program terminates for a specific initial value $x \in \Q^n$? For specific types of conditions $\varphi(x)$ this was called the Extended Orbit Problem\index{Extended Orbit Problem (Open)} in~\cite{COW15} and shown to lie in \ER. Another problem that has similarities is the compact escape problem~\ourref[Compact Escape Problem]{p:compactescape}.
\also (Ultimate) positivity of linear recurrence~\ourref{p:poslinrecur}, compact escape problem~\ourref{p:compactescape}, feasibility~\ourref{p:feasibility}.
\end{problemMisc}

\newpage
\section{Higher Levels of the Real Hierarchy}
\label{sec:HigherLevels}

\subsection{Logic}

\begin{problemL}{$Q_1\cdots Q_k$-Fragment of the Theory of the Reals}
\label{p:quantifiedR}
\mindex{Q1Qk fragment of the theory of the reals@$Q_1\cdots Q_k$-Fragment of the Theory of the Reals}
\given A sentence over the signature $(0,1,+,\cdot)$ with alternating quantifier blocks $Q_1, \ldots, Q_k$, with $Q_i \in \{\exists,\forall\}$ comparison operators $\{=, < \leq, \geq, >\}$ and logic connectives $\vee, \wedge$ and $\neg$. 
\question Is the sentence true if interpreted over the reals?
\complexity Complete for $Q_1\cdots Q_k \R$ by B\"{u}rgisser and Cucker~\cite{BC09}. The problem remains complete for
$Q_1\cdots Q_k \R$ even if the matrix is of the form $f(x) > 0$ or $f(x) \geq 0$, for a polynomial $f$, and, under certain conditions, for $f(x) = 0$, and if the quantifiers are restricted to $(-1,1)$ or $[-1,1]$, as shown by Schaefer and \v{S}tefankovi\v{c}~\cite[Tables 1 and 2]{SS23}. There are also some robustness results, mostly on the first and second level, that allow the use of exotic quantifiers, $\exists^*, \forall^*$  and \HQ, see~\cite{BC09,JJ23}.
\also \ETR~\ourref{p:ETR}. 
\end{problemL}

\subsection{Algebra}

\begin{problemA}{Compact Escape Problem}
\label{p:compactescape}
\mindex{Compact Escape Problem}
\given A compact semialgebraic set $K \subseteq \R^n$, a matrix $A \in \Q^{n\times n}$.
\question For all $x \in K$ is there a $k$ such that $A^kx \not\in K$? 
\complexity \VER-complete by D'Costa, Lefaucheux, Neumann, Ouaknine, and Worrell~\cite{DCLNOW21} via a reduction from the $\forall\exists$-fragment of the theory of the reals~\ourref[Q1Qk fragment of the theory of the reals@$Q_1\cdots Q_k$-Fragment of the Theory of the Reals]{p:quantifiedR}.
\comments  D'Costa, Lefaucheux, Neumann, Ouaknine, and Worrell~\cite{DCLNOW21} show that the problem is hard for $\forall\exists_{<}\R$, hardness then follows, since $\VER = \forall\exists_{<}\R$ by Schaefer, \v{S}tefankovi\v{c}~\cite{SS23}.
\also $Q_1\cdots Q_k$-fragment of the theory of the reals~\ourref{p:quantifiedR}, linear update loop termination~\ourref{p:linearloop}.
\end{problemA}

\begin{problemA}{Hausdorff Distance}
\label{p:Hausdorffdist}
\mindex{Hausdorff distance|see {Semialgebraic Set(s)}}
\mindex{Semialgebraic Set(s)!Hausdorff distance}
\given Two semialgebraic sets $A, B \subseteq \R^n$, $d \in \Q$. The {\em Hausdorff distance} is defined as $d_H(A,B) := \max(\sup_{a \in A} \inf_{b \in B} \norm{a-b}, \sup_{b \in B} \inf_{a \in A} \norm{a-b})$. 
\question Do $A$ and $B$ have Hausdorff distance at most $d$, that is,
is $d_H(A,B) \leq d$? 
\complexity \VER-complete by Jungeblut, Kleist, Miltzow~\cite{JKM23} and Schaefer, \v{S}tefankovi\v{c}~\cite{SS23} via reductions from the $\forall\exists$-fragment of the theory of the reals~\ourref[Q1Qk fragment of the theory of the reals@$Q_1\cdots Q_k$-Fragment of the Theory of the Reals]{p:quantifiedR}. The problem remains hard for $d = 0$~\cite{JKM23,SS23} or $d$ double-exponentially small~\cite{JKM23}. 
\comments  Jungeblut, Kleist, Miltzow~\cite{JKM23} show that the problem is complete for $\forall\exists_<\R$, Schaefer, \v{S}tefankovi\v{c}~\cite{SS23} show the full result, based on B\"{u}rgisser, Cucker~\cite{BC09} who called the $d = 0$ case {\em Euclidean Relative Denseness}.\index{Euclidean Relative Denseness}. The directed Hausdorff distance, $\sup_{a \in A} \inf_{b \in B} \norm{a-b}$ is also \VER-complete~\cite{JKM23, SS23}. The proofs are 
\openq Jungeblut 
asks whether the problem becomes \EVER-complete if we ask whether there is a translation $t$ in $\R^n$ such that $d_H(t(A),B) \leq d$; this is a restricted case of the  Gromov-Hausdorff distance~\ourref[Semialgebraic Set(s)!Gromov-Hausdorff Distance (Open)]{p:Gromovdistance}.
\also Distance of semialgebraic sets~\ourref{p:distancesemi}, Gromov-Hausdorff distance~\ourref{p:Gromovdistance}, Fr\'{e}chet distance of surfaces~\ourref{p:frechetdist}. 
\end{problemA}

\begin{problemA}{Surjectivity}
\label{p:surjectivity}
\mindex{Polynomial(s)!Surjectivity}
\given A polynomial $f$ in $\Z[x_1, \ldots, x_n]$, domain $A$, codomain $B$. 
\question Is $f$ surjective as a function from $A$ to $B$?
\complexity \VER-complete by Schaefer, \v{S}tefankovi\v{c}~\cite{SS23} extending work by B\"{u}rgisser and Cucker~\cite[Proposition 4.1]{BC09}. Hardness can be obtained via a reduction from a constrained version of the $Q_1\cdots Q_k$-fragment of the theory of the reals~\ourref[Q1Qk fragment of the theory of the reals@$Q_1\cdots Q_k$-Fragment of the Theory of the Reals]{p:quantifiedR}. The problem remains \VER-complete if $A$ and $B$ are Cartesian products of $\R$ and $[-1,1]$ and the total degree of $f$ is bounded. 
\comments  B\"{u}rgisser and Cucker~\cite{BC09} showed that surjectivity of rational functions given by arithmetic circuits is \VER-complete, Schaefer and \v{S}tefankovi\v{c}~\cite{SS23} proved the result for polynomials.
\openq Does the problem remain \VER-complete if both $A$ and $B$ are of the form $\R^n$? 
\also Image density (arithmetic circuit)~\ourref{p:imagedensity}.
\end{problemA}

\begin{problemA}{Image Density (arithmetic circuit)}
\label{p:imagedensity}
\mindex{Arithmetic Circuit!Image density}
\given An arithmetic circuit (with sign function), defining a function $f: \R^n \rightarrow \R$.
\question Is the range of $f$ dense in $\R^n$?
\complexity \VER-complete, combining results from B\"{u}rgisser, Cucker~\cite[Proposition 5.7, Corollary 9.4]{BC09} and Junginger~\cite[Theorem 4.27]{JJ23}. 
\comments B\"{u}rgisser and Cucker show that this problem is hard for a complexity class involving exotic quantifiers; Junginger shows that that class equals \VER. In the Zariski topology, B\"{u}rgisser, Cucker~\cite[Proposition 5.7, Corollary 9.4]{BC09} show that the problem is \ER-complete. 
\also Surjectivity (second level)~\ourref{p:surjectivity}, domain density (arithmetic circuit)~\ourref[Arithmetic Circuit!Domain density]{p:DomainDensityAC}, density of semialgebraic set~\ourref{p:radiussemi}.  
\end{problemA}

\subsection{Computational Geometry}

\begin{problemCG}{Area Universality For Triples with Partial Area Assignment}
\label{p:areauniversaltriples}
\mindex{Graph(s)!Area Universality For Triples with Partial Area Assignment}
\index{{Area Universality For Triples with Partial Area Assignment}|see {Graph(s)}}
\given Vertices $V$, a set of vertex triples $F \subseteq \binom{V}{3}$, and a {\em partial area assignment}, that is, a function $\alpha': F' \rightarrow \R_{>0}$, where $F' \subseteq F$.
\question An {\em area assignment} assigns a positive real number to every triple in $F$. Can every area assignment $\alpha$ that extends $\alpha'$, that is, agrees with it on $F'$, be realized by placing the vertices of $V$ in the plane so that every vertex triple in $F$ forms a triangle with the area prescribed by $\alpha$? 
\complexity \VER-complete by Dobbins, Kleist, Miltzow, and Rz\polhk{a}\.{z}ewski~\cite{DKMR23} via a reduction from a constrained form of the $\forall\exists$-fragment of the theory of the reals~\ourref[Q1Qk fragment of the theory of the reals@$Q_1\cdots Q_k$-Fragment of the Theory of the Reals]{p:quantifiedR}. The problem remains hard if the size of $F$ is linear in the size of $V$
\comments This problem is a step towards settling the complexity of the area universality problem~\ourref[Graph(s)!Area Universality (Open)]{p:areauniversal}. Dobbins, Kleist, Miltzow, and Rz\polhk{a}\.{z}ewski~\cite{DKMR23} suggest studying triangulations in a next step.
\also Area Universality~\ourref{p:areauniversal}, prescribed area extension~\ourref{p:preareaext}.  
\end{problemCG}

\newpage
\section{Candidates}
\label{sec:Candidates}

\subsection{Logic}

\begin{problemLO}{Universality of Unambiguous Context-Free Grammar (UUCFG)}
\label{p:universalunambig}
\mindex{Grammar!Universality of Unambiguous Context-Free Grammar (Open)}
\index{Universality of Unambiguous Context-Free Grammar (Open)|see {Grammar}}
\given A context-free grammar $G$ over alphabet $\Sigma$ which is {\em unambiguous}, that is, every word generated by the grammar has a unique derivation tree.
\question Is the grammar {\em universal}, that is, is $L(G) = \Sigma^*$?
\complexity This problem was shown to lie in \VR\ by Clemente~\cite{C20} using another interesting intermediate problem, zeroness of convolution recursive sequences~\ourref[Sequence!Zeroness of Convolution Recursive Sequence (Open)]{p:zerconvolution}, which also lies in \VR. Clemente also mentions a more direct proof of membership in Stanis{\l}aw Purga{\l}'s master's thesis. Finally, he shows that a slightly generalized problem, the coin-flip measure comparison problem\index{coin-flip measure comparison}, is as hard as \SSQR\ problem\index{sum of square roots}.
\also Positivity of linear recurrence~\ourref{p:poslinrecur}, zeroness of convolution recursive sequence~\ourref{p:zerconvolution}.
\end{problemLO}

\begin{problemLO}{Existential Theory of Invertible Matrices}
\label{p:ETInvMat}
\mindex{Matrix!Existential Theory of Invertible Matrices (Open)}
\index{Existential Theory of Invertible Matrices (Open)|see {Matrix}}
\given A conjunction of conditions of the form $AX_i = X_jB$, where the $X_k$ are (matrix) variables and $A,B \in \Q^{n \times n}$.
\question Are there invertible matrices $X_k \in \R^{n \times n}$ satisfying all conditions?
\complexity In \ER. 
\comments The problem was introduced by Dubut~\cite{D20} to upper-bound the complexity of the bisimilarity of diagrams (itself a possible candidate for \ER-completeness?). Without the invertibility contraints, the problem turns into a linear program becoming solvable in polynomial time. It may be easier to show the problem \ER-complete if one is given access to the inverses of the $X_i$, but that's not the case in the original problem.  
\also Matrix singularity~\ourref{p:MatSing}, constrained matrix invertibility problem~\ourref{p:constmatinv}, feasibility~\ourref{p:feasibility}.
\end{problemLO}

\subsection{Algebra}

\begin{problemAO}{Feasibility of Univariate Polynomial}  
\label{p:UniFEAS}
\mindex{Polynomial(s)!Feasibility of Univariate Polynomial (Open)}
\given A univariate polynomial $f \in \Z[x]$ defined by a straight-line program or arithmetic circuit.
\question Does $f$ have a root, that is, is there an $x$ such that $f(x) = 0$?
\complexity Known to be \NP-hard (for a history of proofs, see Perrucci, Sabia~\cite{PS07}), and in \ER. Is it \ER-complete?
\comments This candidate is a bit surprising; the known \NP-hardness proofs are quite intricate. Does the problem lie in \NP? 
If one asks whether $f$ has a {\em rational} root in a given interval, then the problem is decidable using Eisenstein's criterion, see for example~\cite[Lemma 5.22]{DGVvG99}; that makes it somewhat less likely that the rational version of the problem is \EQ-hard. See~\cite{BJ24} for complexity results on counting the number of real roots.
\openq What is the computational complexity of determining whether $f$ has a rational root in some given interval? 
\also Feasibility~\ourref{p:feasibility}.
\end{problemAO}

\begin{problemAO}{Polynomial Equivalence under Shifts}
\label{p:polyshifts}
\mindex{Polynomial Equivalence under Shifts (Open)|see {Polynomial(s)}}
\mindex{Polynomial(s)!Equivalence under Shifts (Open)}
\given Two polynomials $f,g:\R^n\to\R$ in $\Z[x_1,x_2,\ldots,x_n]$.
\question Is there a vector $b$ such that $f(x)=g(x+b)$?
\complexity In \ER.  
\comments This problem is discussed in Chillara, Grichener, and Shpilka~\cite{CGS23}, but the authors switch to a minimality version: given a polynomial $f$, is there a vector $b$ such that $f(x+b)$ has fewer monomials? For this version the authors show that over an integral domain which is not a field, the minimality version is as hard as the existential theory of the domain. Over $\Z$ this implies that the problem is undecidable, but the reduction does not apply to $\R$ (or $\Q$ or $\C$). 
\also Polynomial projection (linear)~\ourref{p:PolyProj}.    
\end{problemAO}

\begin{problemAO}{Semialgebraic Set with Isolated Point}
\label{p:semiisolated}
\mindex{Semialgebraic Set(s)!isolated point existence (Open)}
\given A semialgebraic set $S \subseteq \R^n$.
\question Does $S$ contain an isolated point; that is, is there an $x \in S$ such that $S$ contains no other points close to $x$?
\complexity In \EVR\ and known to be \VR-hard by B\"{u}rgisser, Cucker~\cite[Corollary 6.8, Corollary 9.4]{BC09}, but not known to be \EVR-complete.
\comments Testing whether a specific point belonging to $S$ is isolated, is \VR-complete, see the Isolated zero problem~\ourref[Polynomial(s)!Isolated Zero]{p:ISO}. 
\also Isolated point of semialgebraic set~\ourref{p:isopoint}, isolated zero~\ourref{p:ISO}. 
\end{problemAO}

\begin{problemAO}{Star-Shapedness of Semialgebraic Set}
\label{p:starshaped}
\mindex{Semialgebraic Set(s)!star-shapedness (Open)}
\given A semialgebraic set $S \subseteq \R^n$.
\question Is $S$ {\em star-shaped}? That is, is there a point in $S$ which can {\em see} all points in $S$ in the sense that the line segment connecting the points belongs to $S$?
\complexity In \EVR\ by definition.
\comments Suggested in Schaefer, \v{S}tefankovi\v{c}~\cite{SS23}.
\also Convexity of semialgebraic set~\ourref{p:convexity}. 
\end{problemAO}

\begin{problemAO}{Vapnik-\v{C}ervonenkis Dimension of Family of Semialgebraic Sets}
\label{p:semiVC}
\mindex{Semialgebraic Set(s)!Vapnik-\v{C}ervonenkis Dimension (Open)}
\given A family of semialgebraic sets ${\cal F} = \{\{x \in \R^n: \varphi(x,y)\}: y \in \R^m\}$, integer $k$.
\question Does ${\cal F}$ have Vapnik-\v{C}ervonenkis Dimension at least $k$, that is, are there $k$ points $P$ in $\R^n$ such that for every subset $P' \subseteq P$ of those points, there is an $S \in {\cal F}$ such that $P' = S \cap P$?
\complexity In \EVER; probably not complete for that class, since the universal quantifier is discrete, not real. 
\comments Suggested in Schaefer, \v{S}tefankovi\v{c}~\cite{SS23}. Over discrete sets this problem is complete for the third level of the polynomial-time hierarchy~\cite{S99}.
\end{problemAO}

\begin{problemAO}{Gromov-Hausdorff Distance in Euclidean Space}
\label{p:Gromovdistance}
\mindex{Semialgebraic Set(s)!Gromov-Hausdorff Distance (Open)}
\given Two semialgebraic sets $A, B \subseteq \R^n$, $d \in \Q$. The {\em Hausdorff distance} of two sets $A$ and $B$ is $d_H(A,B) := \max(\sup_{a \in A} \inf_{b \in B} \norm{a-b}, \sup_{b \in B} \inf_{a \in A} \norm{a-b})$. The {\em Gromov-Hausdorff distance, $d_{GH}(A,B)$, of $A$ and $B$ in Euclidean space} is the infimum of $d_H(f(A),g(B))$ over all isometries (distance-preserving functions) $f,g$ from $\R^n$ to $\R^n$.
\question Do $A$ and $B$ have Gromov-Hausdorff distance at most $d$, that is,
is $d_{GH}(A,B) \leq d$? 
\complexity In \EVER\ by definition, and \VER-hardness follows from the \VER-hardness of the Hausdorff distance~\ourref[Semialgebraic Set(s)!Hausdorff distance]{p:Hausdorffdist}.
\comments The Gromov-Hausdorff distance has a more general definition than the one given here; the fact that for Euclidean spaces it is equivalent to the definition of $d_{GH}$ above, was shown by M{\'{e}}moli~\cite{M08}. To reduce the Hausdorff distance to the Gromov-Hausdorff distance, we can add a set of isolated points to the two sets that forces the isometries to be the identity.
\also Hausdorff distance~\ourref{p:Hausdorffdist}.
\end{problemAO}

\begin{problemAO}{Constrained Matrix Invertibility Problem}
\label{p:constmatinv}
\mindex{Matrix!Constrained Invertibility (Open)}
\given Two matrices $A,B \in \{0,*\}^{n \times n}$.
\question Can the $*$-entries in $A$ and $B$ be replaced by real entries so that $AB = I$, where $I$ is the $n$-dimensional identity matrix?
\complexity Over finite fields, the problem is known to be \NP-complete, by Bjerkevik, Botnan, and Kerber~\cite{BBK20}.
\comments This problem arises in the study of interleaving and stability of persistence modules~\cite{BBK20}, there may be further relevant open problems in that area.
\also Matrix singularity~\ourref{p:MatSing}, existential theory of invertible matrices~\ourref{p:ETInvMat}.
\end{problemAO}

\begin{problemAO}{Matrix $\ell^p$-Low Rank Approximation}
\label{p:lowrankapp}
\mindex{Matrix!$\ell^p$-Low Rank Approximation (Open)}
\given A matrix $A \in \Q^{m\times n}$, an integer $k \geq 0$, $\varepsilon \in \Q_{>0}$.
\question Is there a matrix $B \in \R^{m\times n}$ such that $\norm{A-B}_p < \varepsilon$ and $\operatorname{rank}(B) \leq k$? Here, $\norm{}_p$ denotes the {\em $\ell^p$-norm}, which is $(\sum_{i\in [m],j\in [n]} |A_{i,j}-B_{i,j}|^p)^{1/p}$ for $p \geq 1$, and $|\{(i,j) \in [m]\times [n]: A_{i,j} \neq B_{i,j}$ for $p = 0$.
\complexity In \ER, and known to be \NP-hard for $p \in \{0,1\}$ even for $k=1$~\cite{GV18}.
\also Maximum/minimum matrix rank~\ourref{p:minmaxmatrank}. 
\end{problemAO}

\begin{problemAO}{Hadamard Decomposition}
\label{p:Hadamard}
\mindex{Matrix!Hadamard Decomposition}
\given A matrix $M \in \Q^{m \times n}$, natural numbers $k_1, k_2 \in \N$. 
\question Are there matrices $M_1, M_2$ of rank $k_1$ and $k_2$ such that
$M = M_1 \odot M_2$, where $\odot$ is the component-wise product, known as the {\em Hadamard product}, of two matrices?
\complexity In \ER as observed by Ciaperoni, Gionis, and Mannila~\cite{CGM24} who asked the question. 
\also non-negative matrix factorization~\ourref{p:nonnegmatfac}.
\end{problemAO}

\begin{problemAO}{Tensor Eigenvector Approximation}
\label{p:tensoreigenapp}
\mindex{Tensor!Eigenvector approximation (Open)}
\given A $3$-dimensonal tensor $T = (t_{ijk}) \in \Q^{d\times d\times d}$, $\lambda \in \Q$, $\epsilon \in \Q_{>0}$.
\question Is there an $x \in \R^d$ within distance $\epsilon$ of an eigenvector $x'$ of $T$ for eigenvalue $\lambda$; that is, $\sum_{i,j} t_{i,j,k} x'_i x'_j = \lambda x'_k$, and $\norm{x-x'} < \epsilon$?
\complexity Known to be \NP-hard by Hillar, Lim~\cite{HL13} and in \ER\ by definition.  Is it \ER-complete?
\also Tensor with eigenvalue $0$~\ourref{p:tensorzero}, tensor rank~\ourref{p:tensorrank}.  
\end{problemAO}

\begin{problemAO}{Tensor Singular Value}
\label{p:tensorsingular}
\mindex{Tensor!Singular value (Open)}
\given A $3$-dimensional tensor $T = (t_{ijk}) \in \Q^{d\times d\times d}$, $\sigma \in \Q_{>0}$.
\question Is $\sigma$ a singular value of $T$? (See Hillar, Lim~\cite[Section 6]{HL13} for relevant definitions.)
\complexity Known to be \NP-hard even for $\sigma = 0$, by Hillar, Lim~\cite[Theorem 6.2]{HL13} and in \ER\ by definition.  Is it \ER-complete?
\comments It is also open whether testing if $\sigma$ is an upper bound on the spectral norm\index{Tensor!spectral norm} of $T$ is \ER-complete; it is known to be \NP-hard by Hillar, Lim~\cite[Theorem 1.10]{HL13}.
\also Tensor with eigenvalue $0$~\ourref{p:tensorzero}. 
\end{problemAO}

\begin{problemAO}{Zeroness of Convolution Recursive Sequence}
\label{p:zerconvolution}
\mindex{Sequence!Zeroness of Convolution Recursive Sequence (Open)}
\index{Zeroness of Convolution Recursive Sequence (Open)|see {Sequence}}
\given A convolution recursive sequence $f$. A sequence $f:\N\rightarrow \R$ is {\em convolution recursive} if there is a family $g = (g_1, \ldots, g_k): \R^k \rightarrow \R^k$ of polynomials $g_i \in \Q[x_1, \ldots, x_k]$ and $k$ sequences of real numbers $f_1 = f, f_2, \ldots, f_k$ such that $\sigma f_i = g_i(f_1, \ldots, f_k)$, where the polynomials $g_i$ are interpreted as follows: addition is addition of sequences and multiplication is convolution of sequences; the operator $\sigma$ shifts the sequence by one position. 
\question Is $f(n) = 0$ for all $n$?
\complexity This problem lies in \VR\ by Clemente~\cite{C20} who uses it to show that universality of an unambiguous context-free grammar~\ourref[Grammar!Universality of Unambiguous Context-Free Grammar (Open)]{p:universalunambig} lies in \VR. 
\also Universality of unambiguous context-free grammar~\ourref{p:universalunambig}, positivity of linear recurrence~\ourref{p:poslinrecur}.
\end{problemAO}

\subsection{Computational Geometry}

\begin{problemCGO}{Geometric Thickness}
\label{p:geothick}
\mindex{Graph(s)!Geometric Thickness (Open)}
\given A graph $G$, a natural number $k\in \N$.
\question Does $G$ have a straight-line drawing in which the edges can be colored with $k$ colors in such a way that no crossing in the drawing involves two edges of the same color?
\complexity Known to be \NP-hard even for $k=2$~\cite{DGM16}, but no better upper bound than \ER\ is known, including the case $k=2$. For multigraphs, and $k=30$ the problem is \ER-complete~\cite{FKMPTV24}.
\comments Geometric thickness was introduced by Kainen~\cite{K73}. It has similarities to the simultaneous geometric embedding problem~\ourref[Graph(s)!Simultaneous Geometric Embedding]{p:SGE}, which is \ER-complete. For a brief discussion, see~\cite[Remark 7]{S21b}. 
\also Geometric thickness of multigraphs~\ourref{p:geothickmulti}, simultaneous geometric embedding~\ourref{p:SGE}. 
\end{problemCGO}

\begin{problemCGO}{Colin de Verdi\`{e}re Invariant}
\label{p:ColindeVerdiere}
\mindex{Graph!Colin de Verdi\`{e}re invariant (Open)}
\index{Colin de Verdiere invariant open@Colin de Verdi\`{e}re invariant (Open)|see {Graph}}
\given A graph $G = (V,E)$ and a number $k\in \N$.
\question Is the Colin de Verdi\`{e}re invariant $\mu(G)$ at least $k$? For a definition see the comment section below.
\complexity Lies in \ER\ as shown by Kaluza and Tancer~\cite{KT19}.
\comments The graph parameter was introduced by Colin de Verdi\`{e}re~\cite[French]{de1990nouvel}
and is connected to well-studied graph classes. 
For example $\mu(G) \leq 2$ if and only if $G$ is an outerplanar graph and $\mu(G) \leq 3$ if and only if $G$ is a planar graph~\cite{van1999colin}.
The parameter $\mu(G)$ is the largest corank
of a symmetric matrix $M \in \R^{n\times n}$ such that
\begin{itemize}
    \item $ij\in E$ implies $M_{ij} <0$ and $M_{ij} =0$ otherwise;
    \item $M$ has exactly one negative eigenvalue, of multiplicity 1;
    \item There is a matrix $X$ such that $MX = 0$ and $X$ has only zero 
    entries on the diagonal and at the non-zero entries of $M$.
\end{itemize}
For fixed $k$ the invariant $\mu$ can be computed in polynomial-time, since it is minor-closed.
\also minimum graph rank~\ourref{p:mingraphrank}, minimium matrix rank~\ourref{p:minmaxmatrank}.
\end{problemCGO}

\begin{problemCGO}{Unit Disk Contact Graph}
\label{p:unitdiskcontact}
\mindex{Graph(s)!Unit Disk Contact Graph (Open)}
\given A graph $G$.
\question Is $G$ the {\em contact graph} of a collection of disjoint (open) unit disks in the plane, where two disks form a contact if they share exactly one point on their boundaries?
\complexity Known to be \NP-hard~\cite{BK96,KNP15} even if $G$ is outerplanar. 
\comments Unit disk contact graphs are planar, and by Koebe's theorem all planar graphs are contact graphs of disks. There also is a weak version, which allows contacts between disks whose corresponding vertices are not adjacent; weak unit disk contact graphs thus are the subgraphs of unit disk contact graphs. 
\also Ply number~\ourref{p:plynumb}, kissing number~\ourref{p:kissingnumb}.  
\end{problemCGO}

\begin{problemCGO}{Ply Number}
\label{p:plynumb}
\mindex{Graph(s)!Ply number (Open)}
\given A graph $G$, an integer $k \geq 0$.
\question Is there a straight-line drawing of $G$ with the following property: If at each vertex we draw an open disk with radius half the length of the longest edge incident to the vertex, then no more than $k$ disks intersect at any point in the plane? The smallest such $k$ is the {\em ply number} of $G$.
\complexity The problem is \NP-hard, since having ply number at most $1$ is equivalent to being a unit disk contact graph~\ourref[Disk Graph!Unit Disk Contact Graph (Open)]{p:unitdiskcontact}, see~\cite{DGDHKKLMSY15}, and that problem is \NP-hard~\cite{KNP15}.
\comments The ply number of a graph was introduced as a new metric for measuring the quality of graph drawings in~\cite{DGDHKKLMSY15}.
\also Unit distance graphs~\ourref{p:unitdist}, unit disk contact graphs~\ourref{p:unitdiskcontact}. 
\end{problemCGO}

\begin{problemCGO}{Free Space Diagram Realizability}
\label{p:freespacediag}
\mindex{Free Space Diagram Realizability (Open)}
\given A {\em free space diagram}, that is, a rectangle $[0,\ell_1]\times [0,\ell_2]$ in $\R^2$ which is subdivided into a grid of rectangular cells; each cell may be associated with an ellipse (which is cropped to the area of the cell), and a parameter $\varepsilon \in \Q_{\geq 0}$. 
\question Are there two polygonal curves $P_1,P_2$ of length $\ell_1$ and $\ell_2$ such that the union of the croppsed ellipses in the free space diagram corresponds to $\{(x,y): \norm{P_1(x)-P_2(y)}_2 \leq \varepsilon\}$? Here, $P(z)$ denotes the parameterization of polygon $P$ by length.
\complexity In \ER.
\comments The paper~\cite{ABMRW23} shows that planar linkage realizability~\ourref[Linkage!Linkage Realizability]{p:linkagereal} reduces to free space diagram realizability, but an earlier claim that planar linkage realizability is \ER-complete may have been erroneous, leaving the complexity of the free space diagram realizability problem in limbo as well.
\also Free space matrix realizability~\ourref{p:freespacemat}, linkage realizability~\ourref{p:linkagereal}, Fr\'{e}chet distance of surfaces~\ourref{p:frechetdist}.
\end{problemCGO}

\begin{problemCGO}{Maximum Rectilinear Crossing Number}
\label{p:maxrectcross}
\mindex{Graph(s)!Maximum Rectilinear Crossing Number (Open)}
\index{Maximum Rectilinear Crossing Number (Open)| see {Graph(s)}}
\given A graph $G$, an integer $k \geq 0$.
\question Does $G$ have a straight-line drawing with at least $k$ crossings? The largest such $k$ is the {\em maximum rectilinear crossing number} of $G$.
\complexity In \ER. Bald, Johnson, and Liu~\cite{BJL16} showed that the problem is \NP-hard. 
\openq What is the complexity of the maximum rectilinear local crossing number\index{Graph(s)!Maximum Rectilinear Local Crossing Number (Open)}?
\also Rectilinear crossing number~\ourref{p:rectcross}, rectilinear local crossing number~\ourref{p:rectloccross}.
\end{problemCGO}

\begin{problemCGO}{Partial Drawing Extensibility}
\label{p:partialdrawingext}
\mindex{Graph(s)!Partial Drawing Extensibility (Open)}
\index{Partial Drawing Extensibility (Open)|see {Graph(s)}}
\given A graph $G$, prescribed locations for a subset of the vertices of $G$.
\question Is there a planar straight-line drawing of $G$ respecting the prescribed vertex locations? 
\complexity Known to be \NP-hard by Patrignani~\cite{P06}, in \ER\ by definition. 
\comments Closely related to the graph in a polygonal domain problem~\ourref[Graph(s)!Graph in a Polygonal Domain]{p:GraphinPD}. The graph on segments with obstacles problem~\ourref[Graph(s)!Drawing Graph on Segments with Obstacles]{p:drawgraphsegobstacles} can be considered a restricted variant.  
\also Graph in a polygonal domain~\ourref{p:GraphinPD}, graph on segments with obstacles~\ourref{p:drawgraphsegobstacles}.
\end{problemCGO}

\begin{problemCGO}{Geometric Partial Planarity}
\label{p:geopartialplanarity}
\mindex{Graph(s)!Geometric Partial Planarity (Open)}
\index{Geometric Partial Planarity (Open)|see {Graph(s)}}
\given A graph $G = (V,E)$, a subset of edges $F \subseteq E$.
\question Is there a straight-line drawing of $G$ in which all edges in $F$ are crossing-free?
\comments If we replace ``are crossing-free'' with ``have at most one crossing'', then the problem is \ER-complete, see geometric partial $1$-planarity~\ourref[Graph(s)!Geometric Partial 1 Planarity@Geometric Partial $1$-Planarity]{p:geopartial1planarity}.
\also Geometric partial $1$-planarity~\ourref{p:geopartialplanarity}, rectilinear local crossing number~\ourref{p:rectloccross}. 
\end{problemCGO}

\begin{problemCGO}{Convex Neural Codes}
\label{p:convneuralcode}
\mindex{Convex Neural Codes|see {Convex Set(s)}}
\mindex{Convex Set(s)!Neural Codes}
\given A \emph{combinatorial neural code} $\mathcal{C}$: a set of subsets of $[n]$.
\question Does there exist a collection of convex sets $X_1,X_2,\ldots ,X_n$ in $R^d$, for some fixed constant $d\geq 2$, 
such that $\mathcal{C} = \{ I(x) : x\in\R^d \}$, where $I(x)=\{i\in [n] : x\in X_i\}$ is the \emph{intersection pattern} of $x$?
\complexity \ER-hard even for $d=2$, from~\cite{KLR23}, and decidable if $d=2$, from~\cite{BJ23}. 
\openq Does the problem lie in \ER\ for $d=2$? 
Is the problem decidable for $d=3$?
\also Convex set intersection graph~\ourref{p:convintersect}, nerves of convex sets~\ourref{p:nervesconv}, finite convex geometry realizability~\ourref{p:convexgeoreal}.
\end{problemCGO}

\begin{problemCGO}{Convex Clustered Planarity}
\label{p:conclusteredplanarity}
\mindex{Graph(s)!Convex Clustered Planarity (Open)}
\index{Convex Clustered Planarity (Open)|see {Graph(s)}}
\given A graph $G = (V,E)$, a hierarchy ${\cal C} \subseteq 2^V$. In a {\em hierarchy} if any two sets, {\em clusters}, intersect, one most be a subset of the other.
\question Is there a straight-line drawing of $G$ such that for every $C \in {\cal C}$ the convex hull of the vertices in $C$ contains no vertices not in $C$, and any edge that crosses the boundary of the convex hull of $C$ may do so at most once? This is a {\em clustered convex} straight-line embedding of $G$.
\comments This is a variant of the clustered planarity problem, which is now known to be solvable in \P~\cite{FT22}. The traditional variant does not allow vertices on the boundary of a cluster. This may not be the most likely candidate for \ER-hardness, is there a stretchability result like F\'{a}ry's theorem instead?
\also Simultaneous geometric embedding~\ourref[Graph(s)!Simultaneous Geometric Embedding]{p:SGE}, convex set intersection graph~\ourref{p:convintersect}.

\end{problemCGO}

\begin{problemCGO}{Arc Number}
\label{p:arcnumb}
\mindex{Graph(s)!Arc Number (Open)}
\given A graph $G$, an integer $k \geq 0$.
\question Is there a crossing-free circular-arc drawing of $G$ which is the union of $k$ circular arcs? The smallest such $k$ is known as the {\em arc number}.
\comments The arc number was introduced by Schulz~\cite{S15} as a variation of the segment number~\ourref[Graph(s)!Segment Number]{p:segnumber}. 
\also Segment number~\ourref{p:segnumber}, Lombardi drawability~\ourref{p:lombardidraw}, spherical cover number~\ourref{p:sphericalcovernumb}. 
\end{problemCGO}

\begin{problemCGO}{Spherical Cover Number}
\label{p:sphericalcovernumb}
\mindex{Graph(s)!Spherical Cover Number (Open)}
\index{Spherical Cover Number|see {Graph(s)}}
\index{Graph(s)!Circle Cover Number (Open), see {Spherical Cover Number (Open)}}
\index{Circle Cover Number (Open), see Graph(s),Spherical Cover Number (Open)}
\given A graph $G$, an integer $k \geq 0$.
\question Are there $k$ spheres in $\R^d$ whose union contains a crossing-free circular-arc drawing of $G$? We say the spheres {\em cover} $G$.
\complexity Lies in \ER.
\comments Introduced by Kryven, Ravsky, and Wolff~\cite{KRW19}, the case $d = 2$ is also known as the {\em circle cover number}\index{Circle Cover Number (Open), see Graph(s),Spherical Cover Number (Open)}.
\openq 
\end{problemCGO}

\begin{problemCGO}{Minimum Line Number}
\label{p:minlinenumb}
\mindex{Graph(s)!Minimum Line Number (Open)}
\index{Minimum Line Number|see {Graph(s)}}
\given A graph $G$, an integer $k \geq 0$.
\question Is there a straight-line drawing of $G$ with minimum segment number, and line-cover number at most $k$? (See entries for segment number~\ourref[Graph(s)!Segment Number]{p:segnumber} and line-cover number~\ourref[Graph(s)!Line Cover Number in Rd@Line Cover Number in $\R^d$]{p:linecovernumb} for definitions.)
\complexity In \ER.
\comments This problem is very similar to both segment number and line cover number, but differs from both~\cite{KRW19}. It was introduced in~\cite{DMNW13}.
\also Segment number~\ourref{p:segnumber}, line cover number~\ourref{p:linecovernumb}, planar stick number of drawing~\ourref{p:planarsticknumb}. 
\end{problemCGO}

\begin{problemCGO}{Upward Outerplanar Slope Number}
\label{p:upouterplanarslopenumb}
\mindex{Graph(s)!Upward Outerplanar Slope Number}
\index{Upward Outerplanar Slope Number|see {Graph(s)}}
\given A directed graph $G$, an integer $k \geq 0$. A (straight-line) drawing of $G$ is {\em upward} if $u$ lies below $v$ for every directed edge $uv \in E(G)$.
\question Is there an upward, outerplanar straight-line drawing of $G$ in which the edges are drawn using at most $k$ different slopes? 
\complexity \NP-hard by Klawitter and Zink~\cite{KZ23} for $k = 3$.
\comments A drawing realizing the planar slope number of the graph may not be possible on a grid, and it is\EQ-complete to decide whether a grid realization exists, and there is a $2^{2^{\Omega(n^{1/3})}}$ lower bound on the grid size~\cite{H17}.  
\openq How hard is it to determine the (non-planar) slope number\index{Graph(s)!Slope Number (Open)} of a graph~\cite{H17}? 
\also Planar slope number~\ourref{p:planarslopenumb}, upward planar slope number~\ourref{p:upplanarslopenumb}.
\end{problemCGO}

\begin{problemCGO}{Incidence Graph of Geometric Configuration}
\label{p:geoconfig}
\mindex{Incidence Graph of Geometric 3 Configuration@Incidence Graph of Geometric $3$-Configuration}
\given A bipartite cubic graph $G = (L \cup P, E)$.
\question Is $G$ the incidence (Levi) graph of a geometric $3$-configuration, that is, are there points for $P$, and lines for $L$ such that $E$ has an edge $p\ell$ if and only if $p$ is incident to $\ell$?
\complexity In \ER.
\comments Asked by Pisanski on mathoverflow~\cite{P10}. This is a special case of the linear hypergraph realizability problem~\ourref[Linear Hypergraph Realizability]{p:linhypreal}.
\also Segment intersection graph~\ourref{p:SEG}, pseudo-segment stretchability~\ourref{p:SegStretch}, linear hypergraph realizability~\ourref{p:linhypreal}.
\end{problemCGO}

\begin{problemCGO}{Representing Graph with Orthogonal Vectors in $\R^3$}
\label{p:graphorthogonalvectorsR3}
\mindex{Graph(s)!Representing Graph with Orthogonal Vectors in $\R^3$ (Open)}
\index{Graph(s)!Representing Graph with Orthogonal Vectors in $\R^3$ (Open)|see {Graph(s)}}
\given A graph $G$ on $n$ vertices.
\question Can we assign $n$ distinct vectors in $\R^3$ to the vertices in $G$ so that any two vectors belonging to adjacent vertices are orthogonal to each other? (No restrictions on non-adjacent vertices.)  
\complexity Arends, Ouaknine, Jo\"{e}l and Wampler~\cite{AOW11} show that the problem lies in \ER. 
\comments Arends, Ouaknine, Jo\"{e}l and Wampler name this problem the ``embeddability problem''. They arrive at it in their search for small Kochen-Specker vector systems, which play a role in quantum computing. They consider additional restrictions on $G$ which are natural in this context, but which are all of an \NP\ nature, so the problem, as presented here seems to capture the computational core of the problem. 
\also Dot-product graph~\ourref{p:dotprodgraph},  unit ball graph~\ourref{p:unitballgraph}. 

\end{problemCGO}

\begin{problemCGO}{Graph Flattenability}
\label{p:graphflatten}
\mindex{Graph(s)!Flattenability (Open)}
\index{Flattenability|see {Graph(s)}}
\given A graph $G$, integer $d \geq 0$.
\question For every straight-line embedding of $G$ in $\R^{d'}$, with arbitrary $d' \geq d$, is there a straight-line embedding in $\R^d$ with the same distances between all pairs of vertices?
\complexity The problem lies in \VER~\cite{BC07} (the authors observe that it is equivalent to testing the equality of two semialgebraic sets\index{Semialgebraic Set(s)!equality}). For fixed $d$, graph flattenability is minor-closed and can therefore be solved in polynomial time; for hardness results it will therefore be essential that $d$ is part of the input. Lists of excluded minors are known for $d\leq 3$~\cite{BC07}.
\comments The problem was introduced under the name $d$-realizability by Belk and Connelly~\cite{BC07}, but it had been studied earlier. 
\openq Is the problem \NP- or \coNP-hard? 
\also Graph realizability~\ourref{p:graphreal}, area universality~\ourref{p:areauniversal}, unit distance graph~\ourref{p:unitdist}. 
\end{problemCGO}

\begin{problemCGO}{Faithful Drawability of Edge-Colored Graph}
\label{p:faithfullcolor}
\mindex{Graph(s)!Faithful Drawability of Edge-Colored Graph (Open)}
\index{Graph(s)!Faithful Drawability of Edge-Colored Graph (Open)|see {Graph(s)}}
\given A graph $G$ and a proper edge-coloring of $G$, that is, any two adjacent edges differ in color.
\question Is there a {\em faithful} drawing of $G$, that is, a drawing of $G$ in which all edges of the same color lie on concurrent lines, and no two edges lie on the same line?
\complexity In \ER, but otherwise open even for $4$-regular graphs.
\comments Richter shows that faithful drawability can be recognized in polynomial time for cubic graphs and conjectures that the problem is \ER-complete for $4$=regular graphs~\cite[Conjecture 9]{R11}. By moving the common points of the bundles of concurrent lines to infinity, we obtain a drawing in which all edges of the same color are parallel; this seems to suggest a connection to the planar slope number~\ourref[Graph(s)!Planar Slope Number]{p:planarslopenumb}.
\also Planar slope number~\ourref{p:planarslopenumb}.
\end{problemCGO}

\begin{problemCGO}{Kissing Number}
\label{p:kissingnumb}
\mindex{Semialgebraic Set(s)!kissing number (Open)}
\given A compact semialgebraic set $S \subseteq \R^n$, an integer $k \geq 0$..
\question Are there $k$ non-overlapping congruent copies of $S$ each touching $S$? The largest such $k$ is the {\em kissing number} of $S$.
\complexity Open. 
\comments For many natural geometric sets, such as polygons or spheres, there are known upper bounds on the kissing number, which places the problem in \ER, so solvers can, in principle, be used to find the kissing number in those instances. A similar problem, the largest number of pairwise touching non-overlapping congruent copies of cylinders, was tackled using a similar, but more sophisticated approach, see~\cite{BLR15}.  
\also Unit disk contact graph~\ourref{p:unitdiskcontact}.
\end{problemCGO}

\begin{problemCGO}{Euclidean Minimum Spanning Tree}
\label{p:euclideanMST}
\mindex{Graph(s)!Euclidean Minimum Spanning Tree (Open)}
\index{Graph(s)!Euclidean Minimum Spanning Tree (Open)|see {Graph(s)}}
\index{Pointset!Euclidean Minimum Spanning Tree (Open)|see {Graph(s)}}
\given A tree $T$.
\question Is $T$ the Euclidean minimum spanning tree of a set of points in the plane? In the Euclidean minimum spanning tree the weights of the edges are the distances between the points in the plane. 
\complexity In \ER\ and known to be \NP-hard by Eades and Whitesides~\cite{EW96b}. 
\comments The maximum degree of $T$ has to be $6$, since every tree of maximum degree $5$ is a Euclidean Minimum Spanning Tree; the construction showing this requires a double-exponential grid; as far as we know it is open whether this can be improved~\cite{MS92}.
\also Nearest neighbor graph~\ourref{p:NNgraph}, Gabriel graph~\ourref{p:Gabrielgraph}.
\end{problemCGO}

\begin{problemCGO}{Nearest Neighbor Graph}
\label{p:NNgraph}
\mindex{Graph(s)!Nearest Neighbor Graph (Open)}
\index{Nearest Neighbor Graph (Open)|see {Graph(s)}}
\index{Pointset!Nearest Neighbor Graph (Open)|see {Graph(s)}}
\given A graph $G$.
\question Is $G$ the nearest neighbor graph of a set of points in the plane? The {\em nearest-neighbor graph} of a point set has a vertex for each point, and there is an edge between two vertices if they are nearest neighbors to each other. Here $p$ is a {\em nearest neighbor} of $q$ if there is no point closer to $p$ than $q$.
\complexity In \ER\ and known to be \NP-hard by Eades and Whitesides~\cite{EW96}. 
\comments There are several variants of this problem, the variant we presented here is called the {\em mutual} nearest neighbor graph in~\cite{EW96}.
\also Euclidean minimum spanning tree~\ourref{p:euclideanMST}, Gabriel graph~\ourref{p:Gabrielgraph}.
\end{problemCGO}

\begin{problemCGO}{Gabriel Graph}
\label{p:Gabrielgraph}
\mindex{Graph(s)!Gabriel Graph (Open)}
\index{Gabriel Graph (Open)|see {Graph(s)}}
\index{Pointset!Gabriel Graph (Open)|see {Graph(s)}}
\given A graph $G$.
\question Is $G$ a Gabriel graph in the plane? The {\em Gabriel graph} of a point set has a vertex for each point, and there is an edge between two vertices $pq$ if the disk with diameter $pq$ contains no other vertices of the graph. 
\complexity In \ER\ and known to be \NP-hard by Eades and Whitesides~\cite{EW96}. 
\comments The realization of a Gabriel graph contains the Euclidean minimum spanning tree~\ourref[Graph(s)!Euclidean Minimum Spanning Tree (Open)]{p:euclideanMST} and the nearest neighborhood graph~\ourref[Graph(s)!Nearest Neighbor Graph (Open)]{p:NNgraph} of the point set, and is itself a subset of the Delaunay triangulation~\ourref[Delaunay Triangulation]{p:Delaunaytri}.
\also Euclidean minimum spanning tree~\ourref{p:euclideanMST}, nearest neighbor graph~\ourref{p:NNgraph}, Delaunay triangulation~\ourref{p:Delaunaytri}.
\end{problemCGO}

\begin{problemCGO}{Hidden Set in Polygon}
\label{p:hiddenpoly}
\mindex{Polygon!Hidden Set}
\index{Hidden Set in Polygon|see {Polygon}}
\given A simple polygon $P$, natural number $k \geq 0$.
\question Are there $k$ points inside the region enclosed by $P$ which cannot pairwise see each other?
\complexity Known to be \NP-hard by Eidenbenz~\cite{E02} and to lie in \ER~\cite{F24}.
\comments As with visibility graphs one can consider variants in which the hidden points must lie on the boundary or the polygon has holes. There are also variants based on terrains rather than polygons. (This problem is also sometimes called the \textit{bird nesting box problem}. The idea is that you want to place as many bird nesting boxes as possible. However, if two bird nesting boxes see each other they may be avoided by birds as many bird species avoid other birds.)
\also Visibility graphs~\ourref{p:visibilitygraph}.
\end{problemCGO}

\begin{problemCGO}{Watchman Route Problem}
\label{p:watchmanroute}
\mindex{Polygon!Watchman Route Problem (Open)}
\index{Watchman Route Problem (Open)|see {Polygon}}
\given A polygon with holes, an integer $k \geq 0$.
\question Is there a polygonal curve of length at most $k$ such that every point of the polygon can be seen from some point on the polygonal curve?
\complexity In \ER, as shown by Volk~\cite{V23b}. 
\NP-hardness is shown by Dumitrescu and T\'{o}th~\cite{DT12}, who explain why the earlier \NP-hardness proof by Chin and Ntafos~\cite{CN88}, in the paper that introduced the problem, does not work. The problem is polynomial-time solvable for simple polygons, see~\cite{DT12} for more background.
\comments The definition of the problem only gives \EVR-membership, showing the problem lies in \ER\ requires additional analysis. Volk~\cite{V23b} also considers the variant in which the polygonal curve has at most $k$ bends. 
\openq What is the complexity of the problem in unbounded dimension? In that case, \EVR is the best upper bound.  
\also Lawn moving problem~\ourref{p:lawnmoving} art gallery problem~\ourref{p:ArtGallery}, visibility graph of polygon with holes~\ourref{p:polygonholesvisibility}.  
\end{problemCGO}

\begin{problemCGO}{Lawn Mowing Problem}
\label{p:lawnmoving}
\mindex{Polygon!Lawn Moving Problem (Open)}
\index{Lawn Moving Problem (Open)|see {Polygon}}
\given A polygon with holes, an integer $k \geq 0$.
\question Is there a polygonal curve of length at most $k$ such that every point of the polygon has distance at most $1$ from the polygonal curve?
\complexity In \EVR, as shown by Volk~\cite{V23b} and \NP-hard, as shown by Arkin, Fekete, and Mitchell~\cite{AFM00}.
\comments As an intermediate problem, Volk~\cite{V23b} works with the variant in which the polygonal curve has at most $k$ bends. Arkin, Fekete, and Mitchell~\cite{AFM00} also showed \NP-hardness of the {\em milling problem}\index{Milling Problem}, in which the mower may not cross the boundary of the polygon with holes. 
\openq Does the problem lie in \ER? For the similar  
\also Watchman route problem~\ourref{p:watchmanroute}, art gallery problem~\ourref{p:ArtGallery}.
\end{problemCGO}

\begin{problemCGO}{Sweeping Points}
\label{p:sweepingpoints}
\mindex{Pointset!Sweeping Points (Open)}
\index{Sweeping Points (Open)|see {Pointset}}
\given Finite set $P \subseteq Q^2$, $\ell \in \Q_{\geq 0}$.
\question Can all the points be swept to the origin using sweeps of total length at most $\ell$? In a single {\em sweep}, an infinite line is moved in parallel from one location to another location (the distance swept is the contribution of that sweep to the total length). Points encountered along the sweep get moved with the line. 
\comments Dumitrescu and Jiang~\cite{DJ11} attribute this problem to  {Pawe\l \ \.{Z}yli\'{n}ski}; they study approximation algorithms, and ask whether it can be solved in polynomial time. 
\openq Can the shortest length solution always be achieved in a polynomial number of sweeps~\cite{DJ11}? 
\end{problemCGO}

\begin{problemCGO}{Hypercube}
\label{p:hypercube}
\mindex{Pointset!Hypercube Problem (Open)}
\index{Hypercube Problem (Open)|see {Pointset}}
\given Finite set of points $P \subseteq [0,1]^d$, a natural number $k\in \N$.
\question Is there a pointset $Q\subseteq [0,1]^d$  of
size $k$ such that $P \subset \conv(Q)$, where $\conv(Q)$ is the convex hull of $Q$?
\complexity The hypercube problem is \NP-hard via a reduction from 3SAT, and lies in \ER as shown by Kiefer and Wacther~\cite{KW14}. Kiefer and Wachter also observe that for fixed $d$, the problem lies in \P. 
\comments The problem remains \NP-hard if $P$ is required to contain the origin, the {\em restricted hypercube problem}. Kiefer and Wachter introduced this problem to isolate the geometric difficulty at the root of the
probabilitic automaton minimisation problem~\ourref[Automata!Probabilistic Automaton Minimisation]{p:probautomin}.
The hypercube problem is reminiscent of the nested polytope problem~\ourref[Polytope(s)!Nested Polytopes]{p:nestpolytopes} as defined by~\cite{DHM19}.
\openq Kiefer and Wachter~\cite{KW14} observe that for $d = 2$ the set $Q$ can always be assumed to be rational, and ask whether this remains true for higher dimensions.
\also Probabilitic automaton minimisation~\ourref{p:probautomin}, nested polytope problem~\ourref{p:nestpolytopes}.
\end{problemCGO}

\begin{problemCGO}{$k$-Median of Polygonal Curves}
\label{p:mediancurve}
\mindex{Median of Polygonal Curves (Open)@$k$-Median of Polygonal Curves (Open)}
\given A family of polygonal curves $P_1, \ldots, P_n$ in $\R^d$, all of the same length, $c \in \Q$.
\question Is there a polygonal curve $P$ of length $k$ such that $\sum_{i \in [n]} d_F(P, P_i) \leq c$? Here, $d_F$ denotes the Fr\'{e}chet distance, see references for a definition.
\complexity In \ER, as shown by Rohde~\cite{R22}, and \NP-hard by Buchin, Driemel, and Struijs~\cite{BDS20}.
\comments The Fr\'{e}chet distance between two polygonal curves can be computed efficiently~\cite{AG95}.
\also Fr\'{e}chet distance of surfaces~\ourref{p:frechetdist}.
\end{problemCGO}

\begin{problemCGO}{Fr\'{e}chet Distance of Surfaces}
\label{p:frechetdist}
\mindex{Fr\'{e}chet Distance of Surfaces (Open)}
\given Two homeomorphic surfaces $R$ and $S$ together with immersions $\varphi_R: R \rightarrow \R^3$, and $\varphi_S: S \Rightarrow \R^3$, $\delta \in \Q_{\geq 0}$.
\question Is the Fr\'{e}chet distance $\delta_F(R,S) \leq \delta$? Here $\delta_F(R,S)$ is the infimum over all homeomorphisms $f: R\rightarrow S$ of $\sup_{x \in R} \norm{\varphi_R(x) - \varphi_S(f(x))}_2$.
\complexity One has to assume that $R$, $S$, $\varphi_R$ and $\varphi_S$ have reasonable effective representations for this problem to fall into the domain of \ER. There are various relevant results, e.g.\ Godau's proof that the problem is \NP-hard~\cite{G99}, and Nayyeri and Xu's result that the problem lies in \ER\ if one of the surfaces is a triangle~\cite{NX18}.
\also Hausdorff distance~\ourref{p:Hausdorffdist}, $k$-Median of Polygonal Curves~\ourref{p:mediancurve}. 
\end{problemCGO}

\begin{problemCGO}{Compatible Triangulations of Polygons with Holes}
\label{p:comptriagpolygon}
\mindex{Polygon!Compatible Triangulations of Polygons with Holes (Open)}
\index{Compatible Triangulations of Polygons with Holes (Open)|see {Polygon}}
\given Two vertex-labeled polygons $P$ and $Q$ with holes, an integer $k \geq 0$.
\question Can $P$ and $Q$ be extended to vertex-labeled triangulations which are compatible by adding at most $k$ points? Here two triangulations are {\em compatible} if they have the same rotation system, that is, the cyclic permutations of neighbors at each vertex is the same for both triangulations. 
\complexity In \ER\ and \NP-hard, as shown by Lubiw and Mondal~\cite{LM20}, who also ask whether the problem is \ER-hard.
\comments The additional points are known as {\em Steiner points}; the given polygons have to be compatible to start with for the problem to be solvable.
\openq Lubiw and Mondal~\cite{LM20} ask whether the problem remains \NP-hard for simple polygons (no holes). 
\end{problemCGO}

\begin{problemCGO}{Convexity Through Translation}
\label{p:convtrans}
\mindex{Pointset!Convexity Through Translation (Open)}
\given A family of point sets in $\Q^2$.
\question Can the point sets be translated so that the resulting points are in convex position, that is, lie on the convex hull of the union of the translated point sets? 
\complexity In \ER\ and \NP-hard, as shown by Hoffmann, Kusters, Rote, Saumell, and Silveira~\cite{HKRSS13}.
\comments If the point sets can be scaled as well as translated, the problem is not even known to be \NP-hard~\cite{HKRSS13}.
\also Convexity of semialgebraic set~\ourref{p:convexity}.
\end{problemCGO}

\begin{problemCGO}{Prescribed Area}
\label{p:prearea}
\mindex{Graph(s)!Prescribed Area (Open)}
\index{Prescribed Area (Open)|see {Graph(s)}}
\given A plane graph $G$, an area assignment for $G$, where an {\em area assignment} for $G$ assigns a positive real number to every face of $G$. 
\question Does $G$ have a straight-line drawing realizing the area assignment, that is, each face having the area prescribed by the area assignment? 
\complexity In \ER. If some of the vertex locations can be fixed, the problem is \ER-complete, as shown by Dobbins, Kleist, Miltzow and Rz\polhk a\.{z}ewski~\cite{DKMR23}.
\comments Dobbins, Kleist, Miltzow and Rz\polhk a\.{z}ewski~\cite{DKMR23} conjecture that this problem is \ER-complete. 
\also Prescribed area extension~\ourref{p:preareaext}, area universality~\ourref{p:areauniversal}, area universality for triples with partial area assignment~\ourref{p:areauniversaltriples}.  
\end{problemCGO}

\begin{problemCGO}{Area Universality}
\label{p:areauniversal}
\mindex{Graph(s)!Area Universality (Open)}
\index{Area Universality (Open)|see {Graph(s)}}
\given A plane graph $G$.
\question An {\em area assignment} for $G$ assigns a positive real number to every face of $G$. Can every area assignment be realized by a straight-line embedding of $G$? In that realization, every face must have the area assigned to it. 
\complexity Dobbins,  Kleist, and Miltzow show that the problem lies in \VER\ and conjecture that it is \VER-complete. They can show that a more restricted version, area universality for triples with partial area assignments~\ourref[Graph(s)!Area Universality (Open)]{p:areauniversaltriples}, is \VER-complete. 
\also Area Universality for triples with partial area assignments~\ourref{p:areauniversaltriples}, prescribed area~\ourref{p:prearea}.
\end{problemCGO}

\begin{problemCGO}{Segment Visibility Graph}
\label{p:segvisibilitygraph}
\mindex{Visibility Graph!Segment Visibility Graph (Open)}
\index{Segment Visibility Graph (Open)|see {Visibility Graph}}
\index{Graph(s)!Visibility Graph|see {Visibility Graph}}
\given A graph $G$.
\question Is $G$ the visibility graph of a set of segments in the plane? The visibility graph of a set of segments represents each segment by a vertex, and there is an edge between two vertices if the corresponding segments can see each other, that is, if there is a line-segment connecting the two segments that does not intersect any other segment. 
\complexity In \ER. 
\comments There is also the endpoint segment visibility graph\index{Graph(s)!Endpoint Segment Visibility Graph (Open)}, which is the visibility graph of the endpoints of the segments, but the full segments are still the visibility obstacles.  
\also Point visibility graph~\ourref{p:visibilitygraph}, visibility graph of polygon with holes~\ourref{p:polygonholesvisibility},
internal/external polygon visibility graph~\ourref{p:intextvisibility}, obstacle number~\ourref{p:obstaclenumb}.
\end{problemCGO}

\begin{problemCGO}{Obstacle Number}
\label{p:obstaclenumb}
\mindex{Graph(s)!Obstacle Number (Open)}
\given A graph $G$, an integer $k \geq 0$.
\question Is $G$ the visibility graph of a set of points (in general position) in the plane with at most $k$ polygonal obstacles? Here two points can {\em see} each other if the straight-line segment connecting them does not intersect any obstacles. 
\complexity In \ER. This problem is not known to be \NP-hard, though various variants and restrictions of it are, see the summary in~\cite{GdMV18}, and~\cite{BCGGHVW24}, which shows that the problem is \NP-hard for a single, fixed obstacle.
\comments Obstacle numbers were introduced by Alpert, Koch and Liason~\cite{AKL10}. There also is a variant known as the {\em planar obstacle number} in which the visibility segments may not cross each other~\cite{GdMV18}.  
\also (Point) visibility graph~\ourref{p:visibilitygraph}, segment visibility graph~\ourref{p:segvisibilitygraph}.  
\end{problemCGO}

\begin{problemCGO}{Fiber Dimension}
\label{p:fiberdim}
\mindex{Graph(s)!Fiber Dimension (Open)}
\index{Fiber Dimension (Open)|see {Graph(s)}}
\given A graph $G$, a natural number $d\in \N$. 
\question Does $G$ have fiber dimension at most $d$? That is, is there a polytope $P \subseteq \Q^d$ and a symmetric set $M \subseteq \Z^d$ such that $G$ is isomorphic to the graph on $V = P \cap \Z^d$ with an edge $uv$ if and only if $u-v \in M$.
\complexity We are not aware of any computational complexity results.
\comments The fiber dimension of a graph was defined by Windisch~\cite{W19}. Every graph has finite fiber dimension.
\end{problemCGO}

\begin{problemCGO}{Treetope graph}
\label{p:treetopegraph}
\mindex{Graph(s)!Treetope graph (Open)}
\index{Treetope graph (Open)|see {Graph(s)}}
\given A graph $G$, an integer $k \geq 0$.
\question Is there a $k$-treetope polytope $P$ with $1$-skeleton $G$? A {\em $k$-treetope}
is a $k$-dimensional polytope with a special base facet $F$ that intersects every face of dimension at least two in at least two points. 
\complexity In dimension four Eppstein showed that the question is polynomial time solvable~\cite{E20}.
The complexity of this problem is open for dimension five and above.
\comments Eppstein states: ``\textit{Recognizing the graphs of pyramids over arbitrary 4-polytopes, and therefore also
recognizing the graphs of 5-treetopes, is as difficult as recognizing the graphs of
arbitrary 4-polytopes, which we expect to be complete for the existential theory of the
reals}~\cite{E20}.''
\also Polytope. 
\end{problemCGO}


\subsection{Game Theory}

\begin{problemGTO}{Recursive Markov Chain Acceptance Probability}
\label{p:recMCacceptprob}
\mindex{Markov Chain!Recursive Markov Chain Acceptance Probability (Open)}
\index{Recursive Markov Chain Acceptance Probability (Open)|see {Markov Chain}}
\given A recursive Markov chain with associated monotone transition operator $P:\R^n_{\geq 0} \rightarrow \R^n_{\geq 0}$ and a probability vector $c \in \Q^b$ (for definitions, see~\cite{EY09}).
\question Is there a fixed point of $P$ below $c$, that is an $x$ with $P(x) = x$ and $x \leq c$ (in all coordinates)?
\complexity Etessami and Yannakakis~\cite{EY09} show that this (and some related problems) can be expressed in \ER, and is at least as hard as the sum of square root problem. 
\also B\"{u}chi automaton accepting recursive Markov chain trajectory~\ourref{p:BuchautoMCtraj}
\end{problemGTO}

\begin{problemGTO}{B\"{u}chi Automaton Accepting Recursive Markov Chain Trajectory}
\label{p:BuchautoMCtraj}
\mindex{Automata!B\"{u}chi Automaton Accepting Recursive Markov Chain Trajectory (Open)}
\index{B\"{u}chi Automaton Accepting Recursive Markov Chain Trajectory (Open)|see {Automata}}
\given A B\"{u}chi automaton $\mathcal{A}$ and a recursive Markov chain $\mathcal{M}$ (for definitions, see~\cite{BLW13}).
\question Is there a trajectory of $\mathcal{M}$ which is accepted by $\mathcal{A}$?
\complexity Building on work by Etessami and Yannakakis; Benedikt, Lenhardt, and Worrell~\cite{BLW13} show that the problem lies in \ER.
\also Recursive Markov chain acceptance probability~\ourref{p:recMCacceptprob}
\end{problemGTO}

\subsection{Machine Learning and Probabilistic Reasoning}

\begin{problemMLO}{Probabilistic Automaton Minimisation}
\label{p:probautomin}
\mindex{Automata!Probabilistic Automaton Minimisation}
\index{Probabilistic Automaton Minimisation|see {Automata}}
\given A probabilistic automaton $A$, a natural number $n \in \N$.
\question  Does there exists a probabilistic automaton $A'$ of size $n$ so that $A$ and $A'$ are
equivalent?
\complexity The problem was shown to be \NP-hard by Kiefer and Wachter~\cite{KW14} 
via a reduction from the restricted hypercube problem~\ourref[Pointset!Hypercube Problem (Open)]{p:hypercube}, which is known to be \NP-hard~\cite{KW14}. The problem is known to be contained in \ER~\cite{KW14}.
\comments The restricted hypercube problem is similar to the known \ER-complete nested polytope problem~\ourref[Polytope(s)!Nested Polytopes]{p:nestpolytopes}, so there is hope to strengthen the result to \ER-completeness.
\also Hypercube problem~\ourref{p:hypercube}. 
\end{problemMLO}

\subsection{Markov Chains and Decision Processes}

\begin{problemMDPO}{Universal Safety for Probabilistic Finite State Automata}
\label{p:unisafeFSA}
\mindex{Automata!Universal Safety for Probabilistic Finite State Automata}
\index{Universal Safety for Probabilistic Finite State Automata|see {Automata}}
\given A probabilistic finite state automata $\A$~\cite{AGY18} and a safety polytope $H$ given by its vertex representation.
\question Is it true that for every starting distribution there exists a strategy such that $\A$ stays in the safety polytope $H$?
\complexity  Akshay, Genest, and Vyas show that the problem lies in \ER~\cite{AGY18}.
\comments The problem has many variants. One can ask for existential safety instead of universal safety, replace probabilistic finite state automata with Markov Decision Processes, or encode the safety polytope in different ways. None of these variants are known to be \ER-complete either~\cite{AGY18}.
\end{problemMDPO}

\section{Acknowledgments}

We would like to thank the following people for their input:
Mikkel Abrahamsen,
Peter B\"{u}rgisser,
Balder ten Cate,
Jeff Erickson, 
Radoslav Fulek, 
Xavier Goac, 
Christoph Hertrich,
Jeremy Kirn,
Jan Kyn\v{c}l,
Alessio Mansutti, 
Lucas Meijer, 
Arnaud de Mesmay,  
Guillermo Alberto Perez, 
Konrad Swanepol, 
Alexander Wolff.

T. M. is generously supported by the Netherlands Organisation for Scientific Research (NWO) under project no. VI.Vidi.213.150.

\newpage
\bibliographystyle{alphaurl}
\bibliography{library/cleaned-newER}

\clearpage
\printindex

\end{document}